\documentclass[12pt]{iopart} 

\usepackage{iopams}  
\usepackage{textcomp} 
\usepackage{bm}       
\usepackage{booktabs} 
\usepackage{dcolumn}  
\usepackage{graphicx} 
\usepackage[printonlyused]{acronym} 

\usepackage[normalem]{ulem} 
\usepackage{verbatim} 

\usepackage{capt-of}
\usepackage{amsarrows}

\usepackage{relsize}

\usepackage{xcolor}
\usepackage[colorlinks,breaklinks]{hyperref}

\usepackage{multirow}
\usepackage{tabularx}

\usepackage{enumitem}
\usepackage{slashed}

\usepackage{mathrsfs,amssymb} 

\newcommand{\beq}{\begin{equation}}
\newcommand{\eeq}{\end{equation}}
\newcommand{\bea}{\begin{eqnarray}}
\newcommand{\eea}{\end{eqnarray}}

\newcommand{\eqref}[1]{(\ref{#1})}

\newcommand{\nn}{\nonumber\\}
\newcommand{\bra}{\langle}
\newcommand{\ket}{\rangle}
\newcommand{\vhat}[1]{\hat{\bm{#1}}}
\renewcommand{\vec}[1]{\bm{#1}}

\newcommand{\lis}{\ensuremath{{}^{7}\mathrm{Li}}} 
\newcommand{\lie}{\ensuremath{{}^{8}\mathrm{Li}}} 
\renewcommand{\S}[2]{{}^{#1}S_{#2}}
\renewcommand{\P}[2]{{}^{#1}P_{#2}}

\newcommand{\aone}{a_{(\S{3}{1})}}
\newcommand{\atwo}{a_{(\S{5}{2})}}

\newcommand{\eftnopi}{EFT($\slashed{\pi}$)}

\def\lsim{\mathrel{\rlap{\lower4pt\hbox{\hskip1pt$\sim$}}
    \raise1pt\hbox{$<$}}}         
\def\gsim{\mathrel {\rlap{\lower4pt\hbox{\hskip1pt$\sim$}}
    \raise1pt\hbox{$>$}}}         

\newcommand{\tensor}[1]{\stackrel{\leftrightarrow}{#1}}
\newcommand{\loarrow}[1]{\stackrel{\leftarrow}{#1}}
\newcommand{\roarrow}[1]{\stackrel{\rightarrow}{#1}}
\renewcommand{\bs}[1]{\boldsymbol{#1}}
\newcommand{\ssref}[1]{\textsuperscript{\ref{#1}}}
\newenvironment{align}{\begin{eqnarray}}{\end{eqnarray}} 

\newcommand{\mrm}[1]{\mathrm{#1}}

\newcommand{\kC}[0]{k_\mrm{C}}

\def\d{\mathrm{d}}

\def\pp{\mathbf{p}}
\def\qq{\mathbf{q}}
\def\kk{\mathbf{k}}

\def\rr{\mathbf{r}}

\begin{document}

\hfill{NSF-KITP-17-039}

\title[EFT for halo nuclei]{Effective field theory description of halo nuclei}

\author{H.-W. Hammer$^{1,2}$,C. Ji$^{3,4}$, and  
    D.~R. Phillips$^{5}$}

\address{$^{1}$ Institut f\"ur Kernphysik, Technische Universit\"at Darmstadt,
  64289\ Darmstadt, Germany}
\address{$^{2}$ ExtreMe Matter Institute EMMI, GSI Helmholtzzentrum f\"ur
Schwerionenforschung, 64291\ Darmstadt, Germany}
\address{$^{3}$ ECT*, Villa Tambosi, 38123 Villazzano (Trento), Italy}
\address{$^{4}$ INFN-TIFPA, Trento Institute for Fundamental Physics and Applications, Trento, Italy}
\address{$^{5}$ Institute of Nuclear and Particle Physics and Department 
    of Physics and Astronomy, Ohio University, Athens, OH 45701, USA}

\ead{hammer@theorie.ikp.physik.tu-darmstadt.de, ji@ectstar.eu, phillid1@ohio.edu}

\begin{abstract}
Nuclear halos emerge as new degrees of
freedom near the neutron and proton driplines. They consist of a core and
one or a few nucleons which spend most of their time in the
classically-forbidden region outside the range of the interaction.
Individual nucleons inside the core are thus unresolved in the halo configuration, and the low-energy
effective interactions are short-range forces between the core and the valence nucleons. 
Similar phenomena occur in clusters of $^4$He atoms, cold atomic
gases near a Feshbach resonance, and some exotic hadrons.
In these weakly-bound quantum systems universal scaling laws
for s-wave binding emerge that are independent of the details of
the interaction. Effective field theory (EFT) exposes these correlations
and permits the calculation of non-universal corrections to them
due to short-distance effects, as well as the extension of these ideas to
systems involving the Coulomb interaction and/or binding in higher
angular-momentum channels. Halo nuclei exhibit all these features.
{\it Halo EFT}, the EFT for halo nuclei,
has been used to compute the properties
of single-neutron, two-neutron, and single-proton halos of s-wave and p-wave
type. This review summarizes these results for halo binding energies,
radii, Coulomb dissociation, and radiative capture, as well as the
connection of these properties to scattering parameters, thereby elucidating
the universal correlations between all these observables.
We also discuss how Halo EFT's encoding of the
long-distance physics of halo nuclei can be used to check 
and extend {\it ab initio} calculations that include detailed modeling of their short-distance dynamics.
\end{abstract}

\newpage
\setcounter{tocdepth}{2} 
\tableofcontents

\section{Introduction} \label{sec:introduction}
The emergence of new degrees of freedom is an
intriguing aspect of the physics of nuclei away from the valley of stability.
In particular, certain nuclei near the neutron and proton driplines form halo states which consist of a tightly bound
core and a few halo nucleons that are weakly bound to the
core~\cite{Zhukov:1993aw,Hansen:1995pu,Jonson:2004,Jensen:2004zz,Riisager:2012it}.
Neutron halos were discovered in the 1980s at radioactive beam
facilities and are characterized by an unusually large interaction
radius~\cite{Tanihata:2016zgp}. Jonson and Hansen
showed that this
large radius is connected to a small separation energy of the halo
neutrons~\cite{Hansen:1987mc}. 
The emergence of the halo structures can be considered a consequence of the
quantum tunneling of halo neutrons out of the effective potential of the
core to the classically forbidden region.

The simplest example of a halo nucleus
is the deuteron, which can be considered a
one-neutron halo nucleus with a proton core. The root
mean square charge radius of the deuteron is about three times as large as the
size of the constituent proton. We shall discuss other examples of one-neutron halo nuclei, 
with more complex cores, below. Meanwhile, 
halo nuclei with two valence nucleons 
exhibit three-body dynamics. The case in which the corresponding one-nucleon
halo is beyond the dripline is particularly interesting, as then none of the two-body subsystems in the three-body system is bound. This makes the two-neutron halo an example of a Borromean three-body problem;
the name stems from
the heraldic symbol of the Borromeo family of Italy, in which
three rings are interlocked in such a way that if any one of the rings is removed all three separate.
The most carefully studied Borromean halo nuclei are $^6$He and $^{11}$Li, each of which has
two weakly bound valence neutrons \cite{Zhukov:1993aw}.
In the case of $^6$He, the core is a $^4$He nucleus. The two-neutron
separation energy for $^6$He is about 1 MeV, and thus small compared to the
binding and excitation energies of the $^4$He core which are about 
28 and 20 MeV, respectively. There is
a strong p-wave resonance in the $J^P = {3/2}^{-}$ channel of $n\alpha$-scattering---sometimes referred to as ``$^5$He'', even though the nucleus is not bound. 
This resonance is crucial for the binding of $^6$He, and so 
$^6$He can be regarded as dominated by configurations consisting of an $\alpha$-particle
and two neutrons, both of which are in a $p_{3/2}$ wave relative to the $\alpha$ core.

The separation of scales in halo nuclei can be exploited using
effective field theory (EFT)~\footnote[1]{Those desiring a pedagogical 
introduction to EFT are referred to Refs.~\cite{Georgi:1994qn,Kaplan:1995uv,Manohar:1996cq,Phillips:2002da,Burgess:2007pt,Hammer:2016xye}.}. EFT  provides a general framework to calculate the low-energy behavior
of a physical system in an expansion of short-distance over large-distance
scales.
The underlying principle is that short-distance physics is not resolved at
low energies and may be included implicitly in 
``low-energy constants'',
while long-distance physics must be treated explicitly.
For the dynamics of the halo nucleons, the substructure of the core can be
considered short-distance physics, although low-lying excited states of the core
sometimes have to be included explicitly.

Halo nuclei can be described by
extensions of the pionless EFT for few-nucleon systems
\cite{Beane:2000fx,Bedaque:2002mn,Epelbaum:2008ga,Vanasse:2016jtc}.
One assumes the core to be structureless and
treats the nucleus as a few-body system of the core and the valence nucleons,
introducing independent field operators for the halo nucleons and the core.
At leading order the core is thus a structureless object. 
Corrections from its structure appear at higher orders in the EFT expansion, and can be
accounted for in perturbation theory.
The philosophy of Halo EFT is similar to that of cluster models but EFT organizes
different cluster-model effects into a controlled expansion
based on the scale separation, thereby facilitating estimation of the theory's uncertainties.  A new facet compared to few-nucleon
systems is the appearance of resonant interactions in higher partial
waves---as in the neutron-$\alpha$ system \cite{Bertulani:2002sz,Bedaque:2003wa}. This leads to a much richer
structure of the EFT.
However, there are many halo nuclei where s-wave interactions are
dominant, such as $^{19}$C and $^{22}$C.

In order to motivate the Halo EFT approach, we start with these simpler
systems. The scattering of the core and halo nucleons
at sufficiently low energy is then determined by their s-wave scattering length
$a_0$. We consider distinguishable particles of equal mass $m$ and
degenerate pair scattering lengths $a_0$ for simplicity.
This scenario applies approximately to the triton considered as
a two-neutron halo nucleus. 
If $a_0$ is much larger than the range of
the interaction $R$, the system shows universal
properties \cite{Efimov:1971zz,Efimov:1979zz,Braaten:2004rn}.
The simplest example is the existence of a shallow 
two-body bound state with binding energy and mean square
separation
\beq
B_2 = \frac{1}{m a_0^2}\qquad\mbox{ and }\qquad \langle r^2\rangle =a_0^2 /2\,,
\label{eq:b2}
\eeq
if $a_0$ is large and positive. 
\footnote[2]{We use natural units with $\hbar=c=1$ throughout
this review.}
The leading corrections to these universal expressions
are of relative order $R/a_0$ and can be calculated using the
EFT discussed in this review. 
The deuteron binding energy is described by Eq.~\eqref{eq:b2}
to within 35\% accuracy; this improves to 12\% accuracy if the
leading range correction is included.

\begin{figure}[htb]
\centering
\includegraphics[width=6cm,clip=true]{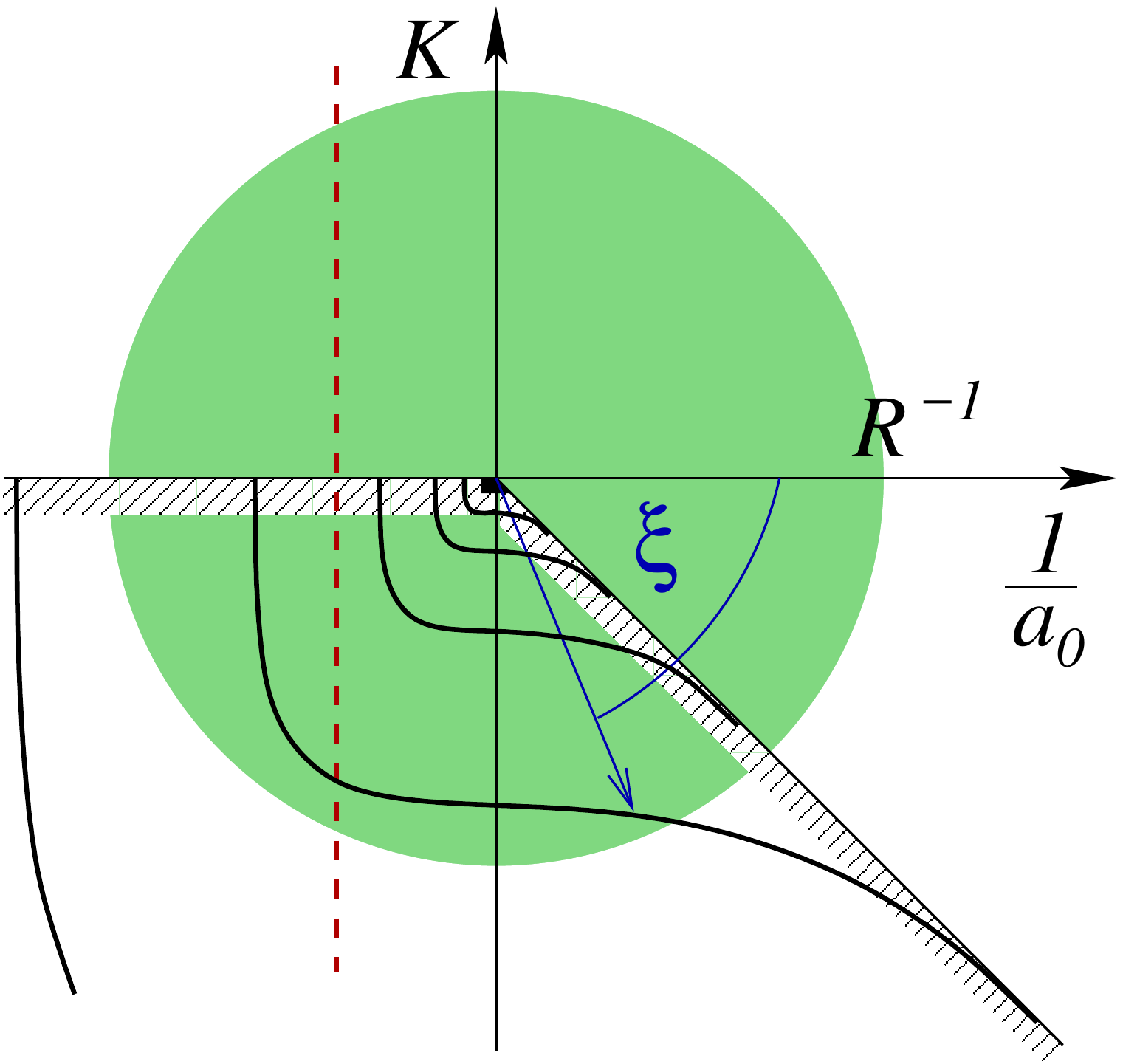}

\caption{Illustration of the Efimov spectrum: The
energy variable $K={\rm sgn}(E)\sqrt{m|E|}$ is shown as a 
function of the inverse scattering length $1/a_0$.
The shaded circular region exhibits the window of universality.
The solid lines indicate the Efimov states, while the hashed areas
give the scattering thresholds and the dashed vertical line illustrates
an exemplary system with fixed scattering length.
}
\label{fig:efiplot}
\end{figure}

If a third particle is added the two-particle s-wave scattering length no longer
determines the low-energy properties of the system. Observables such as 
the binding energy of three-body bound states and low-energy scattering phase shifts
are markedly affected by short-distance physics in the three-body system~\cite{Bedaque:1998km,Bedaque:1998kg}. 
This additional dynamics can be characterized by a single ``three-body parameter", $\kappa_*$.
All low-energy observables in the three-body system are then functions of $a_0$ and $\kappa_*$---to leading order in $R/a_0$. 
For fixed scattering length, this implies universal correlations between 
different three-body observables such as the Phillips
line \cite{Phillips:1968zze}. Moreover, the
Efimov effect \cite{Efimov:1970zz} generates a universal spectrum
of three-body bound states with binding energy
\footnote[1]{See Ref.~\protect\cite{Braaten:2002sr} for more details.}
\beq
B_3 = - \frac{1}{m a_0^2} +
\left[e^{-2 \pi n}\,f(\xi)\right]^{1/s_0} \,\frac{\kappa_*^2}{m}\,,
\label{B3-Efimov}
\eeq
where
the index $n$ labels the three-body states,
$\kappa_*$ is a three-body parameter,
$s_0 = 1.00624...$ is a transcendental number,
$f ( \xi )$ is a universal function with $f ( -\pi/2 )=1$,
and the angle $\xi$ is defined by
$\tan \xi = - (mB_3)^{1/2} \, a_0  \,$.
The leading corrections to Eq.~\eqref{B3-Efimov}
are again of relative order $R/a_0$.
The corresponding spectrum is illustrated in Fig.~\ref{fig:efiplot}
in the two-dimensional plane spanned by the momentum variable 
$K={\rm sgn}(E)\sqrt{m|E|}$ and the inverse scattering length $1/a_0$.
The shaded circular area of radius $R^{-1}$ indicates the
region where universality applies (``window of universality'').
The solid lines indicate the Efimov states while the hashed areas give 
the scattering thresholds below which the bound states can exist. 
The dashed vertical line illustrates an 
exemplary system with a fixed scattering length.
In the unitary limit $1/a_0=0$, Eq.~(\ref{B3-Efimov})
reduces to the geometric spectrum
\beq 
B_3 = e^{-2 \pi n/ s_0}\, \frac{\kappa_*^2}{m}\,,
\label{B3-Efimov-uni}
\eeq
and it is obvious that the three-body parameter $\kappa_*$ is simply
the binding momentum of the state with $n=0$.

The spectrum shown in Fig.~\ref{fig:efiplot}
is invariant under discrete scaling transformations 
by the factor $\lambda_0 = e^{\pi/ s_0}$:
\begin{eqnarray}
\kappa_* & \longrightarrow & \kappa_* \,,
\qquad
a_0  \longrightarrow  \lambda_0^{n} a_0 \,,
\qquad
B_3  \longrightarrow  \lambda_0^{-2n} B_3 \,,
\label{eq:dssym}
\end{eqnarray}
where $n$ is any integer.  This discrete scale 
invariance holds for all three-body observables.
Its manifestation in observables is
often referred to as {\it Efimov physics}.

If more particles are added, no new parameters are needed
for renormalization at leading order~\cite{Platter:2004he,Hammer:2006ct,Schmidt:2009kq,Kirscher:2015yda,Konig:2016utl}. 
As a consequence, in the universal regime all four-body
observables are also governed by the 
discrete scaling symmetry (\ref{eq:dssym}),
and can be characterized by $a_0$ and $\kappa_*$~\cite{Deltuva:2010xd,Deltuva:2011ae,Deltuva:2011ur,Deltuva:2012ms,Kirscher:2009aj}. 
A similar behavior is
expected for higher-body observables~\cite{Hanna:2006,vonStecher:2009qw,vonStecher:2011zz,Nicholson:2012zp,Bazak:2016wxm,Gattobigio:2011ey,Gattobigio:2012tk,Gattobigio:2013yda,Kievsky:2014pza,Kievsky:2014dua,vanKolck:2017}. In ultracold atoms, these
properties have now been experimentally verified for up to five particles 
\cite{Ferlaino:2011xx,Zenesini:2013zz,Naidon:2016dpf}.

The observation of this discrete scaling symmetry in the
level spectra of halo nuclei is a topic of current
research~\cite{Hammer:2016exd,Macchiavelli:2015xml},
but the contribution of higher partial waves 
and partial-wave mixing complicate the situation.
While Halo EFT naturally accommodates resonant interactions in 
higher partial waves \cite{Bertulani:2002sz,Bedaque:2003wa},
there is no Efimov effect in this case
\cite{Jona-Lasinio:2008,Nishida:2011np,Braaten:2011vf}.
Moreover, universality for resonant p-wave interactions is weaker
as two parameters, the p-wave scattering volume and effective range,
are required already at leading order in the two-body system.
In higher partial waves this pattern gets progressively worse~\cite{Bertulani:2002sz,Harada:2007ua,Braun:2016ggm}. 
Nevertheless, universality still provides powerful constraints
for the structure and dynamics of halo nuclei~\cite{Ji:2015jgm,Hammer:2016exd}.

We use the universality of resonant interactions as the starting point 
for Halo EFT. In principle, 
this framework is applicable to any system with short-range interactions and large scattering lengths.
The breakdown scale $M_{\rm core}$ of this theory is set by the lowest momentum
degree of freedom not explicitly included in the theory. 
The EFT exploits the appearance of a large scattering length
$a_0 \gg 1/M_{\rm core}$, independent of the mechanism generating it. 
In addition to nuclear halo states, examples include ultracold atoms
close to a Feshbach resonance and hadronic molecules in particle physics.
The typical momentum scale of the theory is $M_{\rm halo} \sim 1/a_0 \sim k$, which for the systems
under consideration here is usually $\sim$ tens of MeV. Meanwhile, the Halo EFT breakdown scale, $M_{\rm \rm core}$,
varies between 50 and 150 MeV, depending on the system.
The expansion is then in powers of $M_{\rm halo}/M_{\rm core}$, and for a calculation to order $n$ the omitted short-range physics should affect the EFT's answer by a fractional amount of order $(M_{\rm halo}/M_{\rm core})^{n+1}$---as long as we consider a process at a momentum $\sim M_{\rm halo}$. (A more sophisticated implementation of this prescription that employs Bayesian methodology to update the size of the error bar based on the convergence of the perturbative series can be found in Refs.~\cite{Furnstahl:2015rha,Melendez:2017phj}.)
For momenta of the order of the breakdown scale $M_{\rm core}$ or above the EFT
expansion diverges: the
omitted short-range physics is resolved and has to be treated explicitly. 

\begin{figure}[htb]
\centering
\includegraphics[width=6.5cm,clip=true]{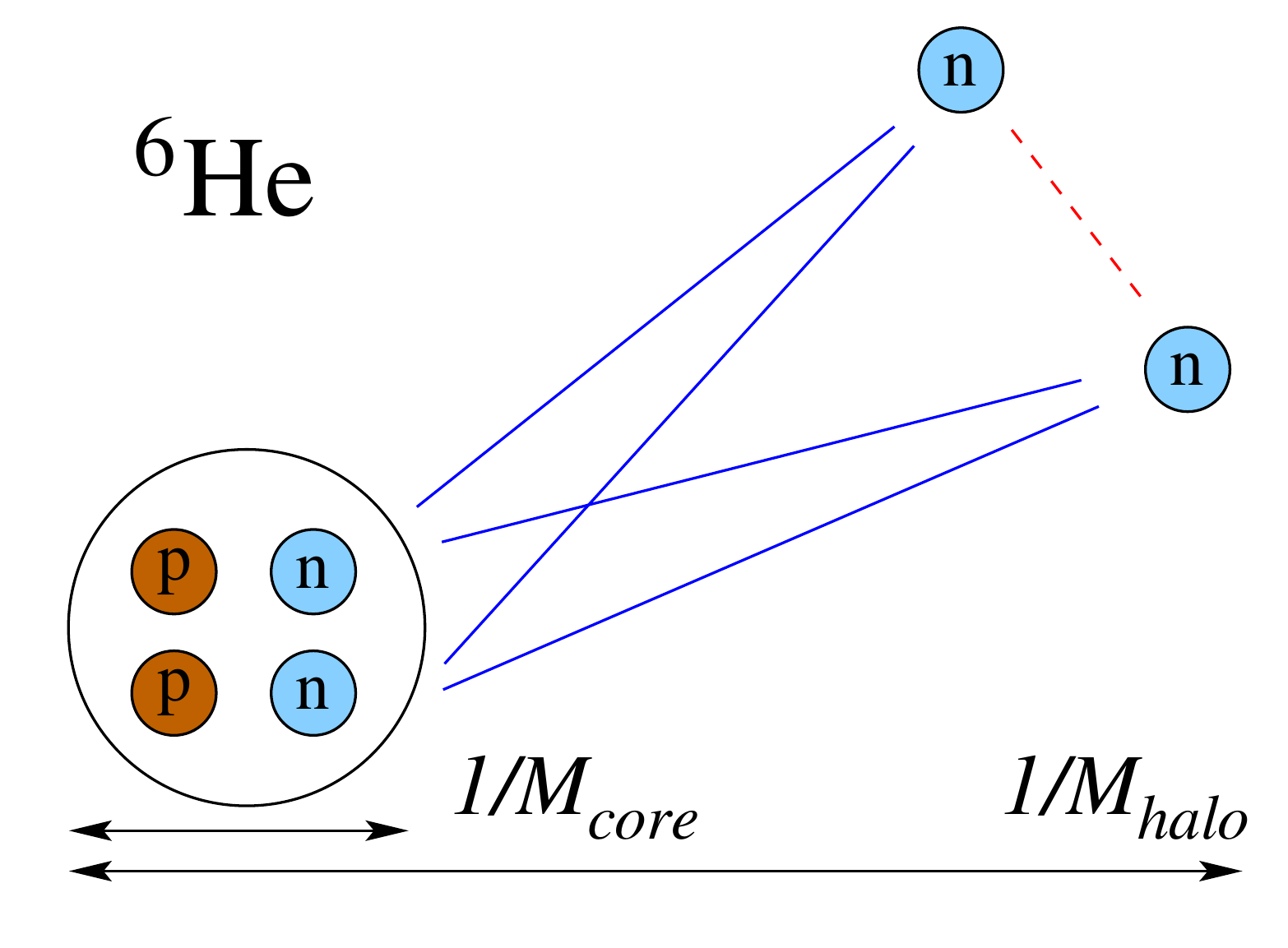}

\caption{Illustration of the antisymmetrization of neutrons in
the halo nucleus $^6$He treated as $\alpha +n+n$
in the framework of Halo EFT.
The exchange illustrated by the dashed line involves active
degrees of freedom and is antisymmetric,
while the exchanges indicated by the solid lines are not
included.
}
\label{fig:antisymm}
\end{figure}

The fact that the halo nucleons and the core are assigned
distinguishable-particle field operators in Halo EFT means
that the halo nucleons are not antisymmetrized with nucleons in the core---the
latter are not active degrees of freedom in the EFT.
(See Fig.~\ref{fig:antisymm} for an illustration in the case of $^6$He
treated as $\alpha +n+n$.). 
However, the contribution of a hypothetical configuration
where a nucleon from the core and from the halo are exchanged
to an observable is
governed by the overlap of the wave functions of the core and the halo.
Since the ranges of the core and halo wave functions are
$1/M_{\rm core}$ and $1/M_{\rm halo}$, respectively, the contribution
is determined by the standard Halo EFT expansion in $M_{\rm halo}/M_{\rm core}$.
Therefore, within the domain of applicability of Halo EFT the impact of 
 anti-symmetrization on observables is controlled and can be incorporated together with that of other short-distance effects. 
In a single-nucleon halo, these effects enter through the low-energy constants of the
nucleon-core interaction which is fitted to experimental data or
{\it ab initio} input.
In a two-nucleon halo, they also enter through a short-range
three-body force. This can be understood as follows: the full
anti-symmetrization of the wave function in a theory with active core
nucleons will result in additional nodes of the halo wave function
since some nucleons must be in excited states to obey the Pauli principle.
In a cluster model, these additional nodes are generated by including
deep unphysical bound states (ghost states) of the core and
the halo (see, e.g., Ref.~\cite{Baye:1985}). In halo EFT such deep unphysical states are not included
explicitly.
The manifestation of the corresponding physics in halo EFT can be
understood by assuming that the unphysical states have been integrated
out of the theory.
This generates a short-range three-body force between the core and the two halo nucleons (or modifies an already existing
three-body force in the theory).

Halo EFT is not meant to replace {\it ab initio} approaches to
halo nuclei, instead it complements
{\it ab initio} approaches by providing universal relations
between different halo observables. Thus it presents a unified
framework for the description of different halo nuclei and their
properties. On the one hand, these universal relations can be 
combined with inputs from {\it ab initio} theories or experiments to predict
halo properties. On the other hand, they can be used to test 
calculations and/or measurements of different observables for
their consistency. In Section~\ref{sec:universality2nhalos} we will use such relations
to elucidate tensions between different pieces of data on the ${}^{21}$C and ${}^{22}$C
isotopes, while in Section~\ref{sec:CDapplications} we will revisit the story of ${}^{19}$C, where universal scaling
relations, together with accurate Coulomb-excitation data,
diagnosed a problem with the
one-neutron separation energy assigned to that
nucleus~\cite{Nakamura:1999rp}. A similar application of universal relations to identify inconsistencies in calculations for systems of $^4$He atoms can be found
in Refs.~\cite{Braaten:2002jv,Platter:2006ev}.
Finally, Halo EFT provides a systematic framework to estimate the natural
accuracy limits of cluster models, e.g., for electromagnetic observables
it can be used to determine the order at which local gauge-invariant couplings to an
external current appear.

The review is organized as follows. In Section \ref{sec:swaves},
we review the Halo EFT formalism for s-wave halos and discuss some
key applications. The extension to halos beyond the s-wave
is presented in Section \ref{sec:pwaves}.
The inclusion of electromagnetic interactions follows in
Section \ref{sec:emprocesses}, while the case of
proton halos---where Coulomb interactions play a key role---is discussed in
Section \ref{sec:coulomb}. The body of the review ends with a discussion of Halo EFT's connection to and synergies with other approaches in Section \ref{sec:insightsconnections}. Section \ref{sec:conclusion} gives a brief summary, treats some outstanding points, and describes a few out of the many possible directions for future work.

\section{s-wave halos} \label{sec:swaves}


\subsection{Lagrangian}
\label{sec:swavelag}
In order to describe halo nuclei in a non-relativistic EFT framework, it is important to establish formulae for observables in halos that are generally suitable for all systems under consideration, and then apply those formulae to specific cases under consideration. We first focus on one- and two-neutron halos with s-wave interactions between the core and the valence neutrons. We introduce an effective Lagrangian density $\mathcal{L}$ to describe a general s-wave halo consisting of a core ($c$) with spin ${\varsigma_c}$ and mass $m_c$ and one or two valence neutrons ($n$) with spin $1/2$ and mass $m_n$. $\mathscr{L}$ is written as a sum of one-, two-, and three-body components, 
\begin{equation}
\mathscr{L} = \mathscr{L}_1 + \mathscr{L}_2 + \mathscr{L}_3.
\label{eq:swaveL}
\end{equation}
The one-body part contains the kinetic terms for the neutron spin-doublet field 
$n\equiv{n_\uparrow\choose n_\downarrow}$
and the core $(2{\varsigma_c}+1)$-component spinor field $c$ 
\begin{equation}
\label{eq:Lag-1b}
\mathscr{L}_1 = 
n^\dagger \left( i\partial_0 + \frac{\nabla^2}{2m_n} \right) n
+ c^\dagger \left( i\partial_0 + \frac{\nabla^2}{2m_c} \right) c.
\end{equation}
The two-body s-wave neutron-neutron ($nn$) and neutron-core ($nc$) short-range interactions are represented by contact terms. The two valence neutrons interact in a spin-singlet state, which has an unnaturally large scattering length.  The neutron and the core couple into states with total spin ${s}$, whose values can be ${s}_{-}=|{\varsigma_c}-1/2|$ and ${s}_{+}={\varsigma_c}+1/2$~\footnote[1]{In the case of a spinless core, the $nc$ interaction forms only one state with ${s}=1/2$.}. We write the two-body Lagrangian containing both the $nn$ and $nc$ contact interactions as
\begin{equation}
\label{eq:Lag-2b}
\fl
\mathscr{L}_2 = 
-C_{nn} 
[ n n]_{0,0}^\dagger\, [ n n]_{0,0}
-C_{nc(-)} 
[n c]_{{s}_{-}, {\beta}_{-}}^\dagger  [n c]_{{s}_{-}, {\beta}_{-}}
-C_{nc(+)} 
[n c]_{{s}_{+}, {\beta}_{+}}^\dagger [n c]_{{s}_{+}, {\beta}_{+}}.
\end{equation}
The notation $[\ ]_{\dots}$ represents the spin coupling through Clebsch-Gordan coefficients~\cite{Hammer:2011ye}. The $nn$ spin singlet is given by
\begin{equation}
[ n n]_{0,0} = \sum\limits_{\delta} \Bigg{(} \frac{1}{2} \delta\, \frac{1}{2} -\delta \Bigg{|} 0 0\Bigg{)} n_\delta n_{-\delta},
\end{equation}
with $\delta=\pm1/2$ denoting the spin projection of the neutrons.
The $nc$ pair in a spin-${s}$ state, with the projection ${\beta}$ running from $-{s}$ to ${s}$, is represented by
\begin{equation}
[ n c]_{{s}, {\beta}} = \sum\limits_{\delta} \Bigg{(} \frac{1}{2} \delta\, {\varsigma_c} {\beta-\delta} \Bigg{|} {s} {\beta}\Bigg{)} n_\delta c_{\beta-\delta}.
\end{equation}
All possible values of the repeated spin-projection indices in Eq.~\eqref{eq:Lag-2b} are summed over through a scalar product. For example,
\begin{equation}
[ n c]_{{s}, {\beta}}^\dagger [ n c]_{{s}, {\beta}} = \sum\limits_{\delta \delta' {\beta}} 
\Bigg{(} \frac{1}{2} \delta\, {\varsigma_c} {\beta-\delta} \Bigg{|} {s} {\beta}\Bigg{)}  
\Bigg{(} \frac{1}{2} \delta'\, {\varsigma_c} {\beta-\delta'} \Bigg{|} {s} {\beta}\Bigg{)}
c_{\beta-\delta}^\dagger n_\delta^\dagger n_{\delta'} c_{\beta-\delta'}.
\end{equation}

If a two-body system has a large scattering length, $a_0 \sim 1/M_{\rm halo}$, and forms either a weakly bound state or a shallow virtual state, it becomes convenient to reformulate the two-body Lagrangian by introducing a dimer field. This was first discussed in Ref.~\cite{Kaplan:1996nv}, and has been extensively applied in the description of few nucleon systems, e.g., in Refs.~\cite{Bedaque:1997qi,Bedaque:1999ve,Bedaque:1999vb,Beane:2000fi}. In this formalism, the $nn$ part of the two-body Lagrangian in Eq.~\eqref{eq:Lag-2b} is recast via the introduction of a $nn$ spin-singlet dimer field $d$:
\begin{equation}
\label{eq:Lag-nn-aux}
\mathscr{L}_{nn}^{0} =
 \, {d}^{\dagger} \left[ w_{d} \left(i\partial_0 + \frac{\nabla^2}{2 M_{d}}\right) +\Delta_{d} \right] {d} 
-\frac{g_{d}}{\sqrt{2}}
\left( {d}^{\dagger} [n n]_{0,0} + {\rm h.c.} \right),
\end{equation}
where $\Delta_{d}$ can be interpreted as the residual mass of the non-relativistic $d$ field. 
$M_{d}\equiv 2m_n$ denotes the total masses of the $nn$ system. $w_{d}$ is a parameter that is chosen to be either $0$ (for leading-order results) or $-1$.
The minus sign ensures that the effective range in the $nn$ system is positive.
If we drop the kinetic term of the dimer in Eq.~\eqref{eq:Lag-nn-aux} we may explicitly do the Gaussian integral over the dimer field in the path integral (i.e., perform a Hubbard-Stratonovich transformation)~\cite{Bedaque:1999vb}. This shows that the Lagrangian (\ref{eq:Lag-nn-aux}) is equivalent to the one with the $nn$ contact term in Eq.~\eqref{eq:Lag-2b}. 

A one-neutron halo will only exist if one of the $nc$ spin channels has a scattering length that is much larger than the range of the interaction. 
Therefore, we can introduce a $nc$ spin-${s}$ auxiliary field ($\sigma$) to the Lagrangian in the corresponding {\it unnatural} channel, where the scattering length $a_0\sim 1/M_{\rm halo}$:
\begin{equation}
\label{eq:Lag-nc-aux}
\mathscr{L}_{nc}^{{s}} =
\, \sigma^{\dagger} \left[ w_{\sigma} \left(i\partial_0 + \frac{\nabla^2}{2M_{\sigma}}\right) +\Delta_{\sigma} \right] \sigma
- g_{\sigma}
\left( \sigma_{{s},{\beta}}^{\dagger} [nc]_{{s},{\beta}}  + {\rm h.c.} \right).
\end{equation}
Similarly to Eq.~(\ref{eq:Lag-nn-aux}), $\Delta_{\sigma}$ is the residual mass of the $\sigma$ field, $M_{\sigma}\equiv m_n+m_c$ is the total mass of the $nc$ system, and $w_{\sigma}=\mp 1$ gives the appropriate sign to the $nc$ effective range in the spin-${s}$ state. The two-body Lagrangian in a $nc$ spin channel with a {\it natural} scattering length, $a_{0} \sim 1/M_{\rm core}$, retains its form in Eq.~\eqref{eq:Lag-2b}.

The three-body Lagrangian $\mathscr{L}_3$ does not contribute to a one-neutron halo nucleus but, in our EFT description of a two-neutron halo nucleus, arises from the requirement that the three-body problem be properly renormalized.
For simplicity, we express $\mathscr{L}_3$ with the explicit inclusion of a dimer field. In the s-wave $2n$ halo system, whose ground state has spin ${\varsigma_c}$, $\mathscr{L}_3$ can be represented by the coupling between $\sigma$ field with spin ${s}$ and the $n$ field with spin $1/2$, resulting in a scalar expression 
\begin{equation}
\label{eq:Lag-3b}
\mathscr{L}_3 = - h \left( [n \sigma]_{{\varsigma_c},a}^\dagger [n \sigma]_{{\varsigma_c},a} 
+ {\rm h.c.} \right),
\end{equation}
where $h$ is the three-body coupling constant. 

Based on the Lagrangian density, we build formalisms for calculating observables in halo systems. The accuracy of the calculation can be progressively improved via the systematic expansion in $M_{\rm halo}/M_{\rm core}$. In the following we use a shorthand notation, denoting calculations done at leading, next-to-leading, and $k$th (for $k\ge2$) order in this expansion as LO, NLO, and N$^k$LO.  

\subsection{Two-body amplitude}

\label{sec:2bamplitude}

Here we derive the Feynman rules and calculate the corresponding two-body amplitude.
The Feynman propagator of a free single-particle field $S_{y}$ (with $y=n,c$) is written as a function of the four-dimensional momentum $(p_0,\bm{p})$, which yields
\begin{equation}
\label{eq:prop-1b}
i S_{y}(p_0,\bm{p}) = \frac{i}{p_0 - \bm{p}^2/2m_{y} + i\epsilon}.
\end{equation}
For the two-body system in an unnaturally enhanced channel, the bare propagator of the dimer field ${d}$ or $\sigma$ is given by
\begin{equation}
iD^{(bare)}_{x}(E, \bm{P} ) = \frac{i}{w_{x}(E-\bm{P}^2/2M_{x} +i\epsilon) +\Delta_{x} },
\end{equation}
with $x={d},\sigma$ labeling the relevant dimer field. $E$ and $\bm{P}$ are, respectively, the energy and momentum of the dimer.

\begin{figure}[!t]
\centerline{\includegraphics[width=0.75\columnwidth]{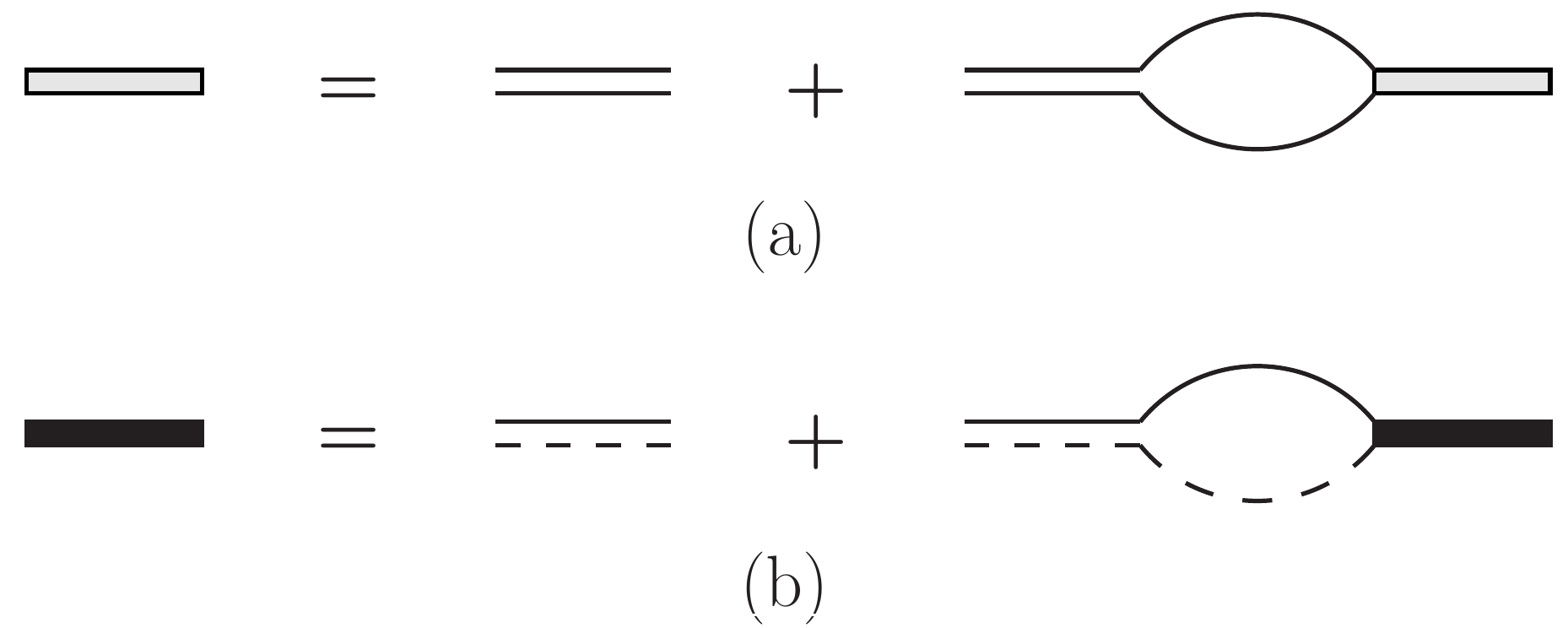}}
\caption{
The Dyson equations for the dressed $d$ (a) and $\sigma$ (b) propagators. 
The dashed line indicates the field for the core, 
and the solid line is the neutron. The solid-solid and solid-dashed double lines are, respectively, the bare 
$d$ and $\sigma$ propagators. The grey and black rectangles are, respectively, the dressed $d$ and $\sigma$ propagators.}
\label{fig:dsigma-dressed}       
\end{figure}

The dimer propagator is dressed by corrections from the s-wave pairwise interaction coupling the dimer propagator with two single particle fields. The dressed propagator $D_{x}$ is obtained by an iterative sum of the one-loop self-energy to all orders. This summation corresponds to the Dyson equation shown in Fig.~\ref{fig:dsigma-dressed}, whose solution yields
\begin{eqnarray}
    iD_{x}(E, \bm{P} )
     &=& \frac{i}{w_{x}(E-\bm{P}^2/2M_{x}+i\epsilon) +\Delta_{x} - \Sigma_{x}(E, \bm{P} )} ,
\label{eq:D2b}
\end{eqnarray}
where $\Sigma_{x}(E, \bm{P} )$ denotes the one-loop self-energy of the dimer field. 
Using dimensional regularization with power-law divergence subtraction (PDS)~\cite{Kaplan:1998tg,Kaplan:1998we,Phillips:1998uy}, $\Sigma_{x}$ is calculated as
\begin{eqnarray}
\label{eq:Sigma-x}
\Sigma_{x}(E, \bm{P}) &=& -i g_{x}^2 \left(\frac{\Lambda}{2}\right)^{4-D} \int \frac{d^D q}{(2\pi)^D}\, i S_n(q_0,\bm{q})\, i S_y(E-q_0,\bm{P}-\bm{q})
\nn
&=& \frac{\mu_{x} g_{x}^2}{2\pi} 
\left [\sqrt{2\mu_{x} \left(\frac{\bm{P}^2}{2M_{x}}- E  -i\epsilon \right)} - \Lambda\right],
\end{eqnarray}
where $y=n$ in the channel $x=d$, and $y=c$ in the channel $x=\sigma$. $\mu_{x}$ indicates the reduced mass in the two-body center-of-mass (C.M.) system. $\mu_{d}=m_n/2$ and $\mu_{\sigma} = A m_n/(A+1)$, where $A\equiv m_c/m_n$ denotes the core-neutron mass ratio.  
$\Lambda$ in Eq.~\eqref{eq:Sigma-x} represents a renormalization scale in the PDS scheme. This result is equivalent, in the large $\Lambda$ limit, to a regularization scheme using a sharp ultraviolet momentum cutoff $\Lambda$ in the integral at $D=4$, where one only needs to replace $\Lambda$ in the last line of Eq.~\eqref{eq:Sigma-x} by $2\Lambda/\pi$~\cite{Phillips:1998uy}. 

The dimer propagator in the C.M. frame can be matched to the two-body s-wave scattering amplitude in the channel $x=d,\sigma$ by
\begin{equation}
\bra \bs{k}'|t_{0,x}(E)|\bs{k}\ket =  g_{x}^2 D_{x}\left(\frac{k^2}{2\mu_{x}}, \bm{0}\right)
= \frac{2\pi}{\mu_{x}} \left[\frac{1}{a_{0,x}} -\frac{r_{0,x}}{2} k^2 +ik\right]^{-1},
\label{eq:t2b-def}
\end{equation}
where $|\bs{k}|=|\bs{k}'|=k=\sqrt{2\mu_x E}$ is the on-shell relative momentum in the C.M. frame. Eq.~\eqref{eq:t2b-def} matches the effective-range expansion (ERE) with terms of $\mathcal{O}(k^4)$ being omitted. For $k \sim M_{\rm halo}$, this approximation is accurate at orders $(M_{\rm halo}/M_{\rm core})^2$. $a_{0,x}$ indicates the s-wave scattering length, which is related, in the unnaturally enhanced channel, to the low-momentum scale by $a_{0,x} \sim 1/M_{\rm halo}$. $r_{0,x}$ denotes the s-wave effective range; it is associated with the breakdown scale, i.e., $r_{0,x}\sim 1/M_{\rm core}$.

The renormalization conditions that yield Eq.~\eqref{eq:t2b-def} require tuning $\Delta_{x}$ and $g_{x}$ in Eq.~\eqref{eq:D2b} to satisfy  
 \numparts
 \begin{eqnarray}
  a_{0,x} &=&   \left( \frac{ 2\pi \Delta_{x}} {\mu_{x} g_{x}^2} +\Lambda \right)^{-1};
  \label{eq:a-s-wave}
  \\
  r_{0,x} &=&  -w_{x}  \frac{2\pi }{\mu_{x}^2 g_{x}^2}.
  \label{eq:r-s-wave}
 \end{eqnarray}
 \endnumparts 
This also indicates that the unitary term $ik$ is invariant under regularization. In this way we obtain the renormalized dimer propagator
\begin{equation}
\fl
\label{eq:prop-2b}
D_{x}(E, \bm{P} ) = \frac{2\pi}{\mu_{x} g_{x}^2} 
\left[\frac{1}{a_{0,x}} - r_{0,x} \mu_{x} \left(E-\frac{\bm{P}^2}{2M_{x}}+ i\epsilon\right) -  \sqrt{2\mu_{x} \left(\frac{\bm{P}^2}{2M_{x}}- E  -i\epsilon \right)} \right]^{-1}.
\end{equation}

In the presence of an unnaturally large scattering length, the scattering amplitude $t_{0,x}$ can be expanded around the low-energy bound- or virtual-state pole at $k=i\gamma_{0,x}$, where $\gamma_{0,x}$ is the binding momentum of the two-body s-wave bound state ($\gamma_{0,x}>0$) or virtual state ($\gamma_{0,x}<0$). At the level of accuracy of Eq.~(\ref{eq:prop-2b}) the binding momentum is related to the scattering length $a_{0,x}$ and the effective range $r_{0,x}$ by 
\begin{equation}
\label{eq:a-gamma}
\frac{1}{a_{0,x}} = \gamma_{0,x} -\frac{1}{2}r_{0,x} \gamma_{0,x}^2.
\end{equation}
Therefore, physics of scale $k\sim|\gamma_{0,x}|$ is enhanced due to the pole structure of the scattering amplitude. The EFT is constructed based on a systematic expansion in $\gamma_{0,x} r_{0,x}$ or $r_{0,x}/a_{0,x}$. 
In the zero-range limit ($r_{0,x}= 0$ by taking $w_x=0$ in Eq.~\eqref{eq:r-s-wave}) or the unitary limit ($a_{0,x}\rightarrow \pm \infty$), we have $\gamma_{0,x}=1/a_{0,x}$: the leading order of the EFT expansion in Eq.~(\ref{eq:a-gamma}) becomes exact. 

Near the pole, the dimer propagator in the C.M. frame can be expanded about the pole  at $-\gamma_{0,x}^2/(2\mu_{x})$. Regardless of the value of $r_{0,x}$:
\begin{equation}
\label{eq:D2sigma-pole}
D_{x}(E,\bm{0}) = \frac{Z_{x}}{E+\gamma_{0,x}^2/(2\mu_{x})}+{\rm regular},
\end{equation}
with the residue of the pole obtained by
\begin{equation}
\label{eq:Z-factor}
\frac{1}{Z_{x} } = \left. \frac{\partial D_{x}^{-1}(E, \bm{0} ) }{\partial E}  \right|_{E=-\gamma_{0,x}^2/2\mu_{x}}
= \frac{\mu_{x}^2 g_{x}^2 } {2\pi \gamma_{0,x}} (1-\gamma_{0,x} r_{0,x}).
\end{equation}
$Z_{x}$ is the $x$-field wave-function renormalization in the EFT.

In a bound one-neutron halo, $Z_{\sigma}$ is connected to the asymptotic normalization coefficient (ANC) of the bound-state wave function. To see this we relate the interacting Green's function for this non-relativistic system to the two-body scattering amplitude $t(E)$ via
\begin{equation}
\label{eq:G-int}
\frac{1}{E-H}
=\frac{1}{E-H_0}
+ \frac{1}{E-H_0} t(E) \frac{1}{E-H_0}.
\end{equation}
Any bound-state pole resides in the second piece on the right-hand side of Eq.~\eqref{eq:G-int}. Just as we did for $D_\sigma$, we use the Laurent expansion for $G$ to obtain the Green's function in the vicinity of the pole at $E=-\gamma_{0,\sigma}^2/(2\mu_{\sigma})$. In the s-wave case, we insert into Eq.~\eqref{eq:G-int} the expression for $\bra \bs{k}'|t_{0,\sigma}(E)|\bs{k}\ket$ from Eq.~\eqref{eq:t2b-def} and keep only the pole part to obtain:
\begin{equation}
\langle \bs{k}' |\frac{1}{E-H} |\bs{k}\rangle
=
\frac{\psi_{0} (\vec{k}^{'})\psi_{0}^*(\vec{k})}{E+\gamma_{0,\sigma}^2/(2\mu_{\sigma})} + {\rm regular},
\label{eqn:ancandT-swave}
\end{equation}
where $\psi_{0}(\vec{k})$ is the asymptotic wave function for an s-wave neutron-core bound state, whose co-ordinate space representation is
\begin{equation}
\label{eq:psi_sigma}
\psi_{0}(\bs{r})=C_{\sigma}  Y_{00}({\hat r}) \frac{\exp(-\gamma_{0,\sigma} r)}{r} ,
\end{equation}
with $C_{\sigma}$ the ANC in the neutron-core s-wave bound state. Matching Eq.~(\ref{eqn:ancandT-swave}) to Eq.~(\ref{eq:D2sigma-pole}) shows that $C_{\sigma}$ is related to $Z_\sigma$ via
\begin{equation}
\label{eq:ANC-swave}
C_{\sigma} = \frac{\mu_\sigma g_\sigma}{\sqrt{\pi}} \sqrt{Z_\sigma} = \sqrt{\frac{2\gamma_{0,\sigma}}{1-\gamma_{0,\sigma} r_{0,\sigma}} }.
\end{equation}
Therefore, one can use the ANC $C_{\sigma}$ and the binding momentum $\gamma_{0,\sigma}$ to determine the EFT parameters $g_\sigma$ and $\Delta_\sigma$, instead of fixing them from the scattering parameters $a_{0,\sigma}$ and $r_{0,\sigma}$. At LO, $C_{\sigma, LO}=\sqrt{2\gamma_{0,\sigma}}$ is determined by the binding momentum. At NLO, the effect of a finite effective range enters and produces an 
ANC ratio different from one: 
\begin{equation}
C_{\sigma}/C_{\sigma,LO}= 1/\sqrt{1-\gamma_{0,\sigma} r_{0,\sigma} }.
\end{equation}
Note that $C_{\sigma}/C_{\sigma,LO} > 1$ if $\gamma_{0,\sigma} >0$ (i.e. the one-neutron halo is bound) and $r_{0,\sigma} > 0$. The extent to which this ratio deviates from one then indicates the short range of the neutron-core potential, compared to the extent of the bound state. Marked enhancement of the asymptotic wave function over the LO prediction correlates with a strong neutron-core potential that pushes the tail of the wave function up. 

Although $C_{\sigma}$ is not an observable that is directly measured in neutron-core scattering experiments, it can be extracted from such data by an analytic continuation of the scattering amplitude to negative energies. There $t_\sigma$ has the pole structure
\begin{equation}
\bra \bs{k}'| t_{\sigma}(E) | \bs{k}\ket =  \frac{2\pi}{\mu_\sigma }  \frac{C^2_{\sigma}/C^2_{\sigma,LO}}{\gamma_{0,\sigma}+ik} +\rm{regular}.
\end{equation}
As compared to the ERE, which is an expansion in powers of $r_{0,\sigma}/a_{0,\sigma}$ around $k=0$, this parameterization in terms of an ANC, dubbed the ``{\it z-parameterization}'' in Ref.~\cite{Phillips:1999hh}, is a more convenient choice for bound-state calculations. Using this parameterization, the pole at $k=i\gamma_{0,\sigma}$ is exactly reproduced at each order, and the residue of the scattering amplitude, $C^2_{\sigma}/C^2_{\sigma,LO}$, is expanded into a LO piece $=1$ and an NLO piece $=(C^2_{\sigma}/C^2_{\sigma,LO} -1)$. N$^2$LO and higher corrections to the ANC are then zero by definition. The $z$-parameterization of the scattering amplitude  is accurate at relative orders $(M_{\rm halo}/M_{\rm core})^2$, beyond which a new ERE parameter at $\mathcal{O}(k^4)$ enters.

Here we illustrate the utility of the $z$-parameterization in the calculation of the matter form factor of one-neutron halos. The neutron-core 
form factor is the Fourier transform of the coordinate-space probability density distribution:
\begin{equation}
\label{eq:Fnc-matter}
F_{nc}(|\bs{q}|) 
= \int d^3 r | \psi_\sigma(\bm{r})|^2 \exp(i \bm{q}\cdot \bm{r}).
\end{equation}
At LO, we use the zero-range two-body wave function by inserting $C_{\sigma,LO}=\sqrt{2 \gamma_{0,\sigma}}$ in Eq.~\eqref{eq:psi_sigma} and obtain 
\begin{equation}
\label{eq:Fnc0-matt}
F_{nc}^{(0)}(|\bs{q}|) 
= \frac{2\gamma_{0,\sigma}}{|\bs{q}|} \arctan\left( \frac{|\bs{q}|}{2\gamma_{0,\sigma}}\right).
\end{equation}

The NLO $F_{nc}$ is calculated from Eq.~\eqref{eq:Fnc-matter} using the full $C_{\sigma}$, with an additional insertion of a constant piece that ensures the matter form factor is properly normalized~\cite{Phillips:1999hh,Chen:1999tn}, i.e., $F_{nc}(0)=1$. Therefore the NLO correction to $F_{nc}$ is~\cite{Phillips:1999hh}
\begin{equation}
\label{eq:Fnc1-matt}
F_{nc}^{(1)}(|\bs{q}|) 
= - (C^2_{\sigma}/C^2_{\sigma,LO}-1)
\left[1-\frac{2\gamma_{0,\sigma}}{|\bs{q}|} \arctan\left( \frac{|\bs{q}|}{2\gamma_{0,\sigma}}\right)\right].
\end{equation}

Since the low-momentum expansion of the form factor in the one-neutron halo is related to the mean squared distance between the neutron and the core $\bra r_{nc}^2 \ket$ via
\begin{equation}
F_{nc}(|\bs{q}|) = 1-\frac{1}{6} \bra r_{nc}^2 \ket \bs{q}^2 + \mathcal{O}(\bs{q}^4),
\end{equation}
we obtain $\bra r_{nc}^2 \ket$ by calculating the first-order derivative of $F_{nc}$ with respect to $q^2$ at zero. $\bra r_{nc}^2 \ket^{1/2}$ at NLO is then
\begin{equation}
\label{eq:rnc}
\bra r_{nc}^2 \ket^{1/2} = \frac{C_{\sigma}/C_{\sigma,LO}} {\sqrt{2}\gamma_{0,\sigma}}.
\end{equation}

With the neutron-core radius in hand we can calculate the matter radius, which is defined, in the point-nucleon limit, as the average distance-squared from all nucleons in a halo nucleus to the center of mass~\cite{Tanihata:2013jwa}:
\begin{equation}
\label{eq:rm-1nhalo}
\bra r_{m}^2 \ket_{1n\rm-halo}  = \frac{A}{(A+1)} \bra r_{m}^2 \ket_{\rm core} + \frac{A}{(A+1)^2} \bra r_{nc}^2 \ket,
\end{equation}
where the first term is the correction from the matter radius of the core.

A more rigorous method of keeping the form factor normalized to $F_{nc}(0)=1$ involves imposing gauge invariance of the Lagrangian in the presence of an external gauge field. We defer detailed discussion of this approach to Sec.~\ref{sec:emprocesses} when we consider the coupling of electromagnetic fields to the halo system. 

\subsection{Applications: one-neutron s-wave halos}

Here we demonstrate several examples of one-neutron s-wave halos, whose properties measured in experiments or predicted by Halo EFT are listed in Table~\ref{tab:1n-halo}.

\begin{table}
   \centering
   \begin{tabular}{c c c c c} \hline
           & $^2$H
           & $^{11}$Be   
           & $^{15}$C 
           & $^{19}$C   \\   \hline
           Experiment  &&&&\\  
      $J^P$ 
           & $1^{+}$
           & $1/2^{+}$ 
           & $1/2^{+}$
           & $1/2^{+}$ \\ 
      $S_{1n}$ [MeV] 
           & 2.224573(2)
           & 0.50164(25) 
           & 1.2181(8)
           & 0.58(9) \\ 
      $E^*_{c}$ [MeV] 
           & 293
           & 3.36803(3)
           & 6.0938(2) 
           & 1.62(2) \\      
      $\bra r^2_{nc}\ket^{1/2}$ [fm]
           &  3.936(12)\ssref{d-1}
           &  6.05(23)\ssref{Be-1}
           &  4.15(50)\ssref{C-1}
           &  6.6(5)\ssref{C-1}  \\ 
           &  3.95014(156)\ssref{d-2} 
           &  5.7(4)\ssref{Be-2}
           &  7.2$\pm$4.0\ssref{C-2}
           &  6.8(7)\ssref{C-3} \\
           &  
           &  5.77(16)\ssref{Be-3}
           &  4.5(5)\ssref{C-3}
           &    5.8(3)\ssref{C-4} \\ \hline
      EFT &&&&\\  
      $M_{\rm halo}/M_{\rm core}$
           & 0.33
           & 0.39
           & 0.45
           & 0.6 \\  
      $r_{0,\sigma}/a_{0,\sigma}$
           & 0.32
           & 0.38
           & 0.43
           & 0.33\\   
      $C_{\sigma}/C_{\sigma, LO}$
           &  1.295
           &  1.44\ssref{Be-z}
           &  1.63\ssref{C15-z}
           &  1.3 \\
      $r_{0,\sigma}$ [fm]
           &  1.7436(19)\ssref{d-z}
           &  3.5\ssref{Be-z}
           &  2.67\ssref{C15-z}
           &  2.6\ssref{C19-z} \\ 
      $\bra r^2_{nc}\ket^{1/2}_{\rm theo}$ [fm]
           &  3.954
           &  6.85
           &  4.93
           &  5.72 \\  \hline
   \end{tabular}
   \caption[list=off]{Properties of one-neutron halos. $S_{1n}$ is the one-neutron separation energy from AME2012~\cite{Audi2012,Wang2012}. The first core excitation energies $E^*_{c}$ for $A>1$ halos are taken from the TUNL database~\cite{Ajzenberg:1987,Tilley:2004zz,AjzenbergSelove:1991zz}. $M_{\rm halo}$ and $M_{\rm core}$ are estimated using $S_{1n}$ and $E^*_{c}$, except for the deuteron, where we take $M_{\rm core}=140$ MeV. 
   \begin{enumerate}[label=\arabic*\,] 
    \item \label{d-1} From the structure radius deduced from electron-deuteron scattering~\cite{Herrmann:1997ia} ($\times2$).
    \item \label{d-2} From the structure radius determined by isotope shift spectroscopy~\cite{Parthey:2010aya} ($\times2$).
    \item \label{Be-1} Extracted by Ref.~\cite{Tanihata:2013jwa} using matter radii~\cite{Ozawa:2000gx} determined from $\sigma_I$.
    \item \label{Be-2} From GSI Coulomb dissociation data~\cite{Palit:2003av}
    \item \label{Be-3} From RIKEN Coulomb dissociation data~\cite{Fukuda:2004ct}
    \item \label{C-1} Extracted by Ref.~\cite{Kanungo:2016tmz} using reanalyzed matter radii from $\sigma_I$ data in~\cite{Ozawa:2001hb}.
    \item \label{C-2} Extracted by Ref.~\cite{Kanungo:2016tmz} using charge radii determined from $\sigma_{cc}$ data in~\cite{Kanungo:2016tmz}.
    \item \label{C-3} From the halo radius obtained in Ref.~\cite{Kanungo:2016tmz} by fitting both $\sigma_{I}$ and $\sigma_{cc}$ ($\times(A+1)/A$).
    \item \label{C-4} From the halo radius obtained in RIKEN Coulomb dissociation data~\cite{Nakamura:1999rp} ($\times(A+1)/A$).
    \item \label{d-z} From an analysis of $np$ scattering data~\cite{Hackenburg:2006qd}.
    \item \label{Be-z} Value of $\mathcal{C}_{\sigma}/\mathcal{C}_{\sigma}^{\rm LO}$ (or $r_{0,\sigma}$) obtained in an {\it ab initio} calculation~\cite{Calci:2016dfb}.
     \item \label{C15-z}  Using EFT~\cite{Rupak:2012cr} to fit $\mathcal{C}_{\sigma}/\mathcal{C}_{\sigma}^{\rm LO}$ (or $r_{0,\sigma}$) to the neutron capture  data~\cite{Nakamura:2009zzc}.
    \item \label{C19-z}  Using EFT~\cite{Acharya:2013nia} to fit $r_{0,\sigma}$ to the E1  data~\cite{Nakamura:1999rp,Nakamura:2003cyk}.
   \end{enumerate}
   }  
   \label{tab:1n-halo}
\end{table}

The deuteron has a spin-triplet ground state ($J^P=1^{+}$), which is dominated by an s-wave component. The deuteron binding energy as determined by the 2012 Atomic Mass Evaluation (AME2012)~\cite{Audi2012,Wang2012} is $S_{1n}=2.224573(2)$ MeV. (Here we use $S_{1n}\equiv \textrm{sgn}(\gamma_{0,\sigma}) \gamma_{0,\sigma}^2/(2\mu_{\sigma})$ to denote the neutron-core separation energy in a one-neutron halo, with $S_{1n}>0$ ($<0$) corresponding to a bound (virtual) s-wave state.)
In the language of Halo EFT, the low scale here is $M_{\rm halo}\sim \gamma_{0,\sigma}=\sqrt{m_n S_{1n}}= 45.7$ MeV. A naive estimation of the high scale is set by the excitation of the nucleon into the $\Delta(1232)$ isobar resonance, i.e., $M_{\rm core}\sim m_\Delta-m_n\approx293$ MeV. However, the nucleon-nucleon interaction has a range that is set by the exchange of pions among nucleons; it brings the high scale down to $M_{\rm core}\approx 140$ MeV: the pion mass is lower than the $m_\Delta-m_n$ scale~\cite{Pascalutsa:2002pi}. But, for physics at scales well below the pion mass, nuclear potentials can still be considered as short ranged by integrating out the pion degrees of freedom. The pionless EFT calculation of the deuteron is then based on the expansion parameter $M_{\rm halo}/M_{\rm core}\approx 1/3$.

The low energy physics of the deuteron can also be related to $np$ scattering data. In the s-wave spin-triplet channel, the scattering parameters $a_{0,\sigma}=5.4112(15)$ fm and $r_{0,\sigma}=1.7436(19)$ fm are determined from an analysis of $np$ elastic scattering data~\cite{Hackenburg:2006qd}. Their values indicate that the effective range is of natural size ($r_{0,\sigma}\sim 1/M_{\rm core}$) while the scattering length is unnaturally large, i.e., $a_{0,\sigma}\sim 1/M_{\rm halo}$. An EFT expansion based on $r_{0,\sigma}/a_{0,\sigma}$ is highly consistent with the $1/3$ estimated above. Using Eq.~\eqref{eq:ANC-swave} we obtain the ANC for the deuteron s-wave wave function~\cite{Phillips:1999hh} to be $C_{\sigma} =1.295$. The deuteron structure radius in the point-nucleon limit, equivalent to $0.5\bra r^2_{np}\ket^{1/2}$, is thus calculated from Eq.~\eqref{eq:rnc} to be $1.977$ fm, which overlaps with the value extracted from elastic electron-deuteron scattering~\cite{Herrmann:1997ia} and agrees with calculations based on realistic nucleon-nucleon potentials~\cite{Friar:1997js}.

Another example of a one-neutron halo is $^{19}$C, whose ground state was determined from the Coulomb dissociation spectrum~\cite{Nakamura:1999rp} to be $J^P=1/2^+$, with a separation energy $S_{1n}=0.53(13)$ MeV between the $^{18}$C core ($J^P=0^+$) and the last neutron. This result is consistent with $S_{1n}=0.65(15)$ MeV from one-neutron knock out reactions~\cite{Maddalena:2001bn}, and $S_{1n}=0.58(9)$ MeV in AME2012~\cite{Audi2012,Wang2012}. The first excitation energy of $^{18}$C is $E^*_{c}=1.62(2)$ MeV~\cite{Ajzenberg:1987}. These values suggest a separation of low and high scales by $M_{\rm halo}/M_{\rm core} \sim \sqrt{S_{1n}/E^*_{c}}\approx0.6$. Ref.~\cite{Acharya:2013nia} performed an EFT analysis on the $^{19}$C Coulomb dissociation data~\cite{Nakamura:1999rp,Nakamura:2003cyk}, where the theoretical constraints on $S_{1n}$ and the ANC were determined from those data.
We will describe how the ANC is determined from Coulomb dissociation or radiative capture processes using EFT analysis in Sec.~\ref{sec:emprocesses}.
Ref.~\cite{Acharya:2013nia} extracted $C_\sigma/C_{\sigma,LO}=1.31$, together with $S_{1n}=0.575(55)(20)$ MeV---the latter in agreement with the extraction in Ref.~\cite{Nakamura:1999rp} and the evaluation of Refs.~\cite{Audi2012,Wang2012}.  These correspond to ERE parameters $a_{0,\sigma}=7.75(35)(30)$ fm and $r_{0,\sigma}=2.6^{+0.6}_{-0.9}\pm0.1$ fm. The ratio $r_{0,\sigma}/a_{0,\sigma}=0.33$ suggests that the EFT converges faster than the naive dimensional estimate, $M_{\rm halo}/M_{\rm core} \approx 0.6$. For all of these ${}^{19}$C observables the first and second errors come from, respectively, the $1\sigma$ statistical uncertainty of the data fitting and the N$^3$LO systematic uncertainties of the EFT, which are estimated to be of relative size $(r_{0,\sigma}/a_{0,\sigma})^3$.

The above ${}^{19}$C results imply an s-wave binding momentum $\gamma_{0,\sigma}=32.0$ MeV  in $^{19}$C. This, together with the extracted ANC, yields the neutron-core distance from Eq.~\eqref{eq:rnc} to be $\bra r_{nc}^2\ket^{1/2}=5.72$ fm, which agrees with values deduced by the E1 sum rule of Coulomb dissociation~\cite{Nakamura:1999rp} and extracted from charge-changing cross section~\cite{Kanungo:2016tmz} measurements (see Table \ref{tab:1n-halo}). 

Other examples of one-neutron halos are $^{11}$Be and $^{15}$C. Their ground states both have spin-parity quantum numbers $J^P=1/2^{+}$, with one valence neutron attached to the $^{10}$Be and $^{14}$C cores ($J^P=0^{+}$). The one-neutron separation energies $S_{1n}$ of $^{11}$Be and $^{15}$C are determined in the atomic mass evaluation~\cite{Audi2012,Wang2012}; while the first excitation energies of the cores $E_c^*$ are obtained from the nuclear data evaluation~\cite{Tilley:2004zz,AjzenbergSelove:1991zz} (see Table \ref{tab:1n-halo}). 

Based on naive dimensional analysis the EFT expansion parameter in $^{11}$Be  is $M_{\rm halo}/M_{\rm core} \sim \sqrt{S_{1n}/E^*_{c}}=0.39$. The ANC in the $^{11}$Be ground state was recently obtained in an {\it ab initio} calculation that used the No-Core Shell Model with Continuum approach (described further in Sec.~\ref{sec:insightsconnections}) as $C_\sigma/C_{\sigma,LO}=1.44$~\cite{Calci:2016dfb}. This corresponds to $r_{0,\sigma}=3.5$ fm, which yields $r_0/a_0=0.38$ in $^{11}$Be, in agreement with the expansion parameter inferred from $M_{\rm halo}/M_{\rm core}$. ${}^{11}$Be will be discussed further in Sec.~\ref{sec:pwavehaloapplications}, after we have developed the formalism to describe the p-waves that play a prominent role there.

EFT calculations for $^{15}$C were performed in Refs.~\cite{Rupak:2012cr,Fernando:2015jyd}. By fitting the Halo EFT neutron capture cross section to experiment~\cite{Nakamura:2009zzc}, Rupak, Fernando, and Vaghani determined the ANC ratio $C_\sigma/C_{\sigma,LO}=1.63$, or equivalently, an effective range $r_{0,\sigma}=2.67$ fm. Their calculation suggested an unnatural scale for the effective range, with $r_{0,\sigma}\sim 1/M_{\rm halo}$~\cite{Rupak:2012cr}, implying that the z-parameterization, $C_\sigma^2/C_{\sigma,LO}^2-1$, becomes non-perturbative in this system. However, this conclusion may be associated with the specific choice of power counting in the $n$-$^{14}$C p-wave channel used in Ref.~\cite{Rupak:2012cr}, as we will explain in further detail in Sec.~\ref{sec:pwaves}. If we adopt the ANC extracted in Ref.~\cite{Rupak:2012cr} we obtain $r_0/a_0=0.43$, which, while it is a somewhat large effective range, is still consistent with $M_{\rm halo}/M_{\rm core}\approx 0.45$.

\subsection{Applications: unbound s-wave neutron-core systems}
Halo-like features also exist in unbound neutron-core systems, if such systems display a large negative scattering length. In the $np$ spin singlet state, the parameters $a_{0,d}^{(np)}= -23.7148(43)$ fm and $r_{0,d}^{(np)} =2.750(18)$ fm are determined from at analysis of low-energy $np$ elastic-scattering data~\cite{Hackenburg:2006qd}. The $nn$ singlet state is also thought to be unbound, with a scattering length $a_{0,d}^{(nn)}= -18.6(5)$ fm 
obtained from the neutron time-of-flight spectrum in the reaction $\pi^- d \rightarrow \gamma nn$~\cite{Chen:2008zzj}.
The similar, large values of the spin-singlet scattering length for $np$ and $nn$ pairs suggest that isospin is a good approximate symmetry for nuclear forces. Isospin symmetry and the $r_{0,d}/a_{0,d}$ expansion mean that pionless EFT can predict universal features shared by the $np$ and $nn$ singlet states. 

It was argued in Ref.~\cite{Hammer:2014rba} that the pion-deuteron capture experiment~\cite{Chen:2008zzj} only determines the magnitude of $a_{0,d}^{(nn)}$ but is insensitive to its sign, which raises the possibility that a weakly bound dineutron state exists. Using pionless EFT at NLO, Ref.~\cite{Hammer:2014rba} analyzed the constraint on the value of $a_{0,d}^{(nn)}$ from the $^3$H-$^3$He binding energy difference and the neutron-deuteron s-wave doublet scattering phase shifts. (See Ref.~\cite{Kirscher:2011zn} for an earlier, LO, treatment.) Hammer and K\"onig concluded that a bound dineutron cannot be excluded by an NLO pionless EFT analysis of these data. In fact, in the EFT changing the scattering length from a large positive value to a large negative value, referred to as the unitary crossing, requires 
only slight variation of $\Delta_d$ in Eq.~\eqref{eq:a-s-wave} around the pole in $a_{0,d}$. The continuity of $\Delta_d$ at the unitary limit indicates that isospin symmetry can still be a good approximation even if $a_{0,d}^{(nn)}$ and $a_{0,d}^{(np)}$ have opposite signs. This argument is consistent with the finding of Ref.~\cite{Hammer:2014rba}: that the $^3$H-$^3$He binding energy difference is a continuous function of $1/a_{0,d}^{(nn)}$ around $1/a_{0,d}^{(nn)}=0$. The unitary crossing is also an important idea in atomic physics, where it is referred to as the BCS-BEC crossover, and generates important many-body phenomena. The unitary crossing will play an important role in our later discussion of three-body problems in halo nuclei, see Sec.~\ref{sec:universality2nhalos}. 

$^{11}$Li, whose ground state has spin-parity $J^P=3/2^-$ was one of the first halo nuclei~\cite{Tanihata:1986kh} beyond the few-nucleon systems to be discovered. $^{11}$Li is a Borromean two-neutron halo, where the neutron-core is unbound. Here we focus on the separation of scales in $^{10}$Li and refer to later sections for properties of $^{11}$Li. The ground state of the $^9$Li core has $J^P=3/2^-$ and a first excitation energy $E^*_{c}=2.691(5)$ MeV~\cite{Tilley:2004zz}, which sets the $M_{\rm core}$ scale. The $M_{\rm halo}$ scale is associated with the $^{10}$Li ground state, which is an unbound s-wave neutron-core virtual state ($J^P=1^-$ or $2^-$) with $S_{1n}=-25(15)$ keV~\cite{Tilley:2004zz}. The EFT expansion parameter for $^{10}$Li is estimated as $M_{\rm halo}/M_{\rm core}\sim \sqrt{|S_{1n}|/E^*_{c}}\approx10\%$. A proton removal reaction experiment~\cite{Smith:2015bpa} observed two resonance states of $^{10}$Li at energies $E_{2,1+}=110(40)$ keV and $E_{2,2+}=500(100)$ keV above the neutron-core threshold. These are expected to be p-wave states. As such they enter at higher orders in the EFT compared to the s-wave virtual state, whose unnaturally large scattering length promotes it to LO. 

$^{21}$C is another unbound neutron-core system~\cite{Langevin:1985ior}. The ratio between the one-neutron separation energies of $^{21}$C~\cite{Audi2012,Wang2012} and $^{20}$C~\cite{Ajzenberg:1987} provides a valid expansion parameter in EFT.
The neighboring isotope $^{22}$C has recently been identified as a weakly-bound two-neutron halo and is the dripline nucleus of carbon isotopes. A Glauber-model analysis of the reaction cross section of $^{22}$C on a hydrogen target~\cite{Tanaka:2010zza} suggests that $n-^{20}$C is preferentially in $1/2^{+}$ configuration. This finding is supported by the measurement of the two-neutron removal reaction on the $^{22}$C target~\cite{Kobayashi:2011mm}. The two-neutron halo structure of $^{22}$C implies that $^{21}$C occupies an s-wave virtual state near the unitary limit. However, a recent study of the $n-^{21}$C decay spectrum via one-proton removal from the $^{22}$N beam~\cite{Mosby:2013bix} imples that the $n-^{20}$C scattering length is limited to $|a_{0,\sigma}|<2.8$ fm (or equivalently $S_{1n}<-2.8$ MeV): much smaller than expected in an unnaturally enhanced channel. Therefore, further studies on the properties of $^{21}$C are needed.

\subsection{The three-body amplitude}

\label{sec:3Bswave}

Here we describe two-neutron halos as a neutron-neutron-core three-body system. We use the Jacobi momentum plane-wave state $|\boldsymbol{p},\boldsymbol{q}\ket_i$ to represent the kinematics of the three-body system in the C.M. frame. The index $i$ indicates
that these momenta are defined in the two-body fragmentation channel $(i,jk)$,  in which particle $i$ is the spectator and  $(jk)$ the interacting pair. 
Based on this definition, $\boldsymbol{p}$ represents the relative momentum in the pair $(jk)$; while $\boldsymbol{q}$ denotes the relative momentum between the spectator $i$ and the $(jk)$ pair.
The plane-wave states are normalized as~\cite{Glockle:1983}:
\begin{equation}
{}_i\bra \bs{p} \bs{q} |
\bs{p}' \bs{q}'\ket_i = (2\pi)^6 \delta^{(3)}(\bs{p}-\bs{p}') \delta^{(3)}(\bs{q}-\bs{q}').
\label{eq:Jacobi-1}
\end{equation}
The Jacobi momenta are related to the momenta in the direct product of three single-particle states $|\bs{k}_1,\bs{k}_2,\bs{k}_3\ket$ in the C.M. frame (i.e., $\bs{k}_1+\bs{k}_2+\bs{k}_3=0$) by
\begin{eqnarray}
\fl
\bra \bs{k}_1,\bs{k}_2,\bs{k}_3 |  \bs{p} \bs{q}\ket_i &=& 
(2\pi)^6 \delta^{(3)}\left(\bs{p}_i -\mu_{jk}\left[\frac{\bs{k}_j}{m_j}-\frac{\bs{k}_k}{m_k}\right]\right)\,
\delta^{(3)}\left(\bs{q}_i -\mu_{i(jk)}\left[\frac{\bs{k}_i}{m_i}-\frac{\bs{k}_j+\bs{k}_k}{M_{jk}}\right]\right),
\nn
&&
\label{eq:Jacobi-2}
\end{eqnarray}
where $M_{jk}=m_j+m_k$, $\mu_{jk}=m_j m_k /M_{jk}$, and $\mu_{i(jk)}= m_i M_{jk}/(m_i+M_{jk})$. From Eqs.~\eqref{eq:Jacobi-1} and \eqref{eq:Jacobi-2}, the projection between different partitions must obey 
\numparts
\begin{eqnarray}
{}_n\bra \bs{p} \bs{q} |\bs{p}' \bs{q}'\ket_c 
&= (2\pi)^6 
\delta^{(3)}\left(\bs{p}+\bs{\pi}_1(\bs{q}',\bs{q})\right) \,
\delta^{(3)}\left(\bs{p}'-\bs{\pi}_2(\bs{q},\bs{q}')\right),
\label{eq:Jacobi-nc}
\\
{}_n\bra \bs{p} \bs{q} |\mathcal{P}|\bs{p}' \bs{q}'\ket_n 
&= (2\pi)^6 
\delta^{(3)}\left(\bs{p}-\bs{\pi}_3(\bs{q}',\bs{q})\right) \,
\delta^{(3)}\left(\bs{p}'+\bs{\pi}_3(\bs{q},\bs{q}')\right),
\label{eq:Jacobi-nn}
\end{eqnarray}
\endnumparts
where $\mathcal{P}$ denotes the permutation between the two valence neutrons. The momenta $\bs{\pi}_1$, $\bs{\pi}_2$ and $\bs{\pi}_3$ are defined as
\numparts
\begin{eqnarray}
\bm{\pi}_1(\bm{q},\bm{q}') = \bm{q} + A \bm{q}'/(A+1),
\\
\bm{\pi}_2(\bm{q},\bm{q}') = \bm{q} + \bm{q}'/2,
\\
\bm{\pi}_3(\bm{q},\bm{q}') = \bm{q} + \bm{q}'/(A+1).
\end{eqnarray}
\endnumparts

To discuss the spin and parity of a halo nucleus, we introduce the partial-wave-decomposed representation. The relative orbital angular momentum and the spin of the pair $(jk)$ are defined as $l_i$ and $s_i$. They are coupled to form the total angular momentum $j_i$ in the pair. We also define the spin of the spectator $i$ as ${\varsigma}_i$, the relative orbital angular momentum between the spectator $i$ and the pair $(jk)$ as $\lambda_i$, and the corresponding total angular momentum as $I_i$. The overall orbital angular momentum, spin and total angular momentum of the three-body system are denoted by $L_i$, $S_i$ and $J$. We then have:
\numparts
\begin{eqnarray}
\bs{L}_i =& \bs{l}_i+\bs{\lambda}_i,
\\
\bs{S}_i =& \bs{s}_i+\bs{\varsigma}_i,
\\
\label{eq:J-LS}
\bs{J} =& \bs{L}_i+\bs{S}_i = \bs{j}_i+\bs{I}_i.
\end{eqnarray}
\endnumparts

Knowing the spin and orbital-angular-momentum quantum numbers, we can construct an eigenstate of a three-body system with respect to the spin and orbital-angular-momentum operators. Note that $J$ for a given three-body eigenstate is a conserved quantum number, which is independent of the choice of partition representations given in Eq.~\eqref{eq:J-LS}. We decompose the Jacobi momenta 
with respect to these spin and orbital- and total-angular-momentum quantum numbers by~\cite{Glockle:1983}
\begin{equation}
\label{eq:quantum-number}
\fl
\left|p,q;\Omega_i\right\rangle_{i} = \sum\limits_{L_i S_i}\sqrt{\widehat{j}_i \widehat{I}_i \widehat{L}_i \widehat{S}_i}\,
\left\{\begin{array}{ccc}
  l_i & s_i &  j_i \\

  \lambda_i & \varsigma_i &  I_i \\
  L_i     & S_i     & J 
\end{array}\right\}\,
\left| p,q;(l_i,\lambda_i)L_i\,; \left(s_i,\varsigma_i\right)S_i\,;(L_i S_i)J \right\rangle_{i},
\end{equation}
where $\widehat{j}_i$ denotes $2{j}_i+1$ (the same holds for $\widehat{I}_i$, $\widehat{L}_i$ and $\widehat{S}_i$), $p\equiv|\bs{p}|$, and $q\equiv|\bs{q}|$. The collective symbol $\Omega_i$ represents all conserved spin, orbital- and total-angular-momentum quantum numbers in the  partition  $(i, jk)$.

In this section we focus on s-wave two-neutron halos, where $l_i=\lambda_i=L_i=0$ and the values of the total angular momenta are equal to their corresponding spins. Therefore, one can solely use spins to represent the decomposed plane-wave state in s-wave $2n$ halos as 
$| p,q; (s_i,\varsigma_i)S\rangle_{i}$, where the three-body total spin $S$ is the same in different partitions. In the $(c,nn)$ partition, $S=\varsigma_c$ since the two-neutron pair is spin singlet ($s_c=0$). Therefore, in the $(n,nc)$ partition, the neutron-core pair with spin $s_n=|\varsigma_c\pm1/2|$ couples with the second neutron with spin $\varsigma_n=1/2$ to form the three-body total spin $S=\varsigma_c$.

\begin{figure}[!t]
\centerline{\includegraphics[width=0.8\columnwidth]{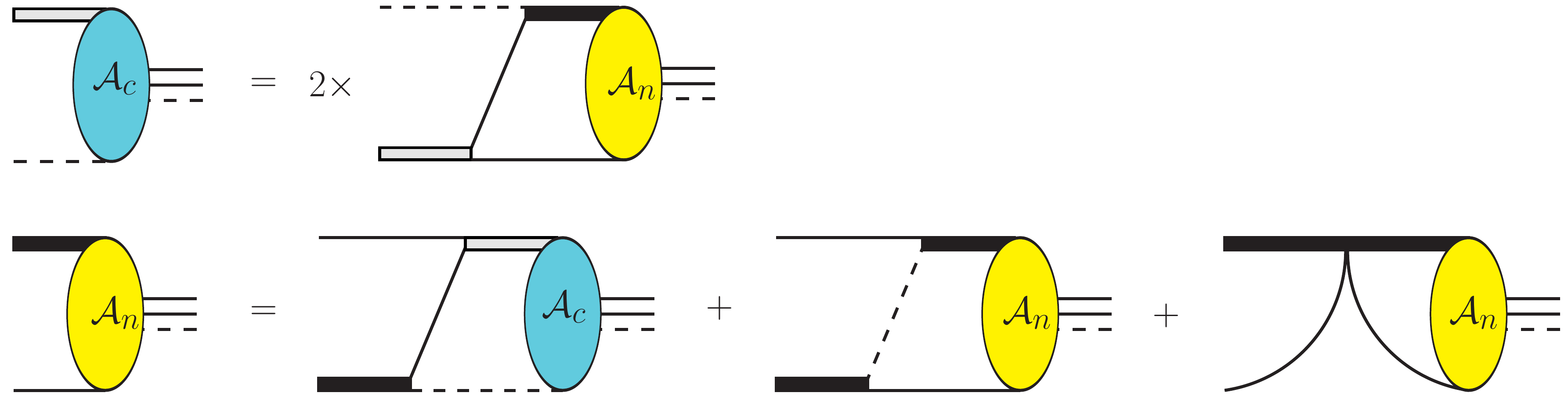}}
\caption{
The transition amplitudes $\mathcal{A}_c$ and $\mathcal{A}_n$ in coupled integral equations. The last piece represents the contribution from the $nnc$ three-body force.}
\label{fig:faddeev-eq}       
\end{figure}

Here we assume that the neutron-core states with $s_n=|\varsigma_c\pm1/2|$ are degenerate and have equal scattering lengths. Under this assumption, the three-body formalism in s-wave halos with a spin-zero core becomes general for an arbitrary s-wave two-neutron halo with spin $\varsigma_c$. 
Just as in the Faddeev formalism~\cite{Faddeev:1960su, Glockle:1983, Afnan:1977pi}, the three-body wave function is decomposed into components corresponding
to different partitions. In the EFT, $2n$-halo nuclei are described by the transition amplitudes, $\mathcal{A}_c$ and $\mathcal{A}_n$, connecting the spectator and the interacting pair to the three-body bound state. $\mathcal{A}_c$ and $\mathcal{A}_n$ are represented by functions of the Jacobi momentum $\bs{q}$ between the spectator and the pair, and are the solution of coupled-channel homogeneous integral equations~\cite{Canham:2008jd,Acharya:2013aea} which are illustrated with Feynman diagrams in Fig.~\ref{fig:faddeev-eq}. 

At leading order, one can take $r_{0,x}\rightarrow 0$ for $x=d,\sigma$ and rewrite the dimer propagator $D_{x}$ as 
\begin{equation}
\label{eq:Dx-lo}
D_{x}^{(0)}(E, \bm{P} ) = - \frac{2\pi}{\mu_{x} g_{x}^2} 
\frac{1}{-\gamma_{0,x}  +  \sqrt{2\mu_{x} \left(\bm{P}^2/2M_{x}- E  -i\epsilon \right)} }.
\end{equation}
Therefore, the integral equations for LO $\mathcal{A}_n$ and $\mathcal{A}_c$ are constructed to be
\numparts
\begin{eqnarray}
\fl
i \mathcal{A}_c(B_3,\bm{q}) &=&  \int \frac{d^4 q'}{(2\pi)^4} iS_n(q_0',\bm{q}') 
\left[ -\sqrt{2} g_{d} g_{\sigma} iS_n(-B_3-q_0'-\frac{\bm{q}^2}{2m_c},-\bm{q}-\bm{q}')\right]
\nn
\fl
&&\times iD_{\sigma}^{(0)}(-B_3-q_0',-\bm{q}')\, i \mathcal{A}_n(B_3,\bm{q}')
\label{eq:intorig-Ac}
\\
\fl
i \mathcal{A}_n(B_3,\bm{q}) &=& \int \frac{d^4 q'}{(2\pi)^4} iS_c(q_0',\bm{q}') 
\left[ -\sqrt{2} g_{d} g_{\sigma} iS_n(-B_3-q_0'-\frac{\bm{q}^2}{2m_n},-\bm{q}-\bm{q}')\right]
\nn
\fl
&&\times iD_{d}^{(0)}(-B_3-q_0',-\bm{q}')\, i \mathcal{A}_c(B_3,\bm{q}')
\nn
\fl
&&+\int \frac{d^4 q'}{(2\pi)^4} iS_n(q_0',\bm{q}') 
\left[ -g_{\sigma}^2 iS_c(-B_3-q_0'-\frac{\bm{q}^2}{2m_n},-\bm{q}-\bm{q}')
- ih \right]
\nn
\fl
&&\times iD_{\sigma}^{(0)}(-B_3-q_0',-\bm{q}')\, i \mathcal{A}_n(B_3,\bm{q}').
\label{eq:intorig-An}
\end{eqnarray}
\endnumparts

We then rescale the amplitudes to $\tilde{\mathcal{A}}_c(\bm{q}) \equiv \sqrt{2} g_{d}^{-1} \mathcal{A}_c(B_3,\bm{q})$ and $\tilde{\mathcal{A}}_n(\bm{q}) \equiv g_{\sigma}^{-1} \mathcal{A}_n(B_3,\bm{q})$ to absorb the explicit dependence on the two-body coupling constants. After integrating out the time component of the loop momentum, we obtain 
\numparts
\begin{eqnarray}
\label{eq:inteq-Ac}
\fl
\tilde{\mathcal{A}}_c(\bm{q}) &=& 2 \int \frac{d^3 q}{4\pi^2} \, G_0^c \left(\pi_2(\bm{q}',\bm{q}),q;B_3\right) \,
\tau_{\sigma}(q';B_3) \, \tilde{\mathcal{A}}_n(\bm{q}')
\\
\label{eq:inteq-An}
\fl
\tilde{\mathcal{A}}_n(\bm{q}) &=& \int \frac{d^3 q'}{4\pi^2} \, G_0^c \left(\pi_2(\bm{q},\bm{q}'),q';B_3\right) \,
\tau_{d}(q';B_3) \, \tilde{\mathcal{A}}_c(\bm{q}')
\nn
\fl
&&+\int \frac{d^3 q'}{4\pi^2} \, \left[G_0^n \left(\pi_3(\bm{q}',\bm{q}),q;B_3\right) +\frac{H(\Lambda)}{\Lambda^2} \right]\,
\tau_{\sigma}(q';B_3) \, \tilde{\mathcal{A}}_n(\bm{q}'),
\end{eqnarray}
\endnumparts
where $G_0^n$ and $G_0^c$ are the three-body Green's functions expressed in two different partitions:
\numparts
\begin{eqnarray}
G_0^n(p,q;B_3) &=& \left( m_n B_3 + \frac{A+1}{2A} p^2 + \frac{A+2}{2(A+1)} q^2 \right)^{-1},
\\
G_0^c(p,q;B_3) &=& \left( m_n B_3 +  p^2 + \frac{A+2}{4A} q^2 \right)^{-1}.
\end{eqnarray}
\endnumparts
The dimer propagators, embedded in the three-body integral equations~\eqref{eq:inteq-Ac} and \eqref{eq:inteq-An}, are rescaled via $\tau_x(q;B_3)\equiv-(m_n g_x^2/2\pi) D_x^{(0)}$ for $x=d,\sigma$, which yields
\numparts
\begin{eqnarray}
\tau_{d}(q;B_3) 
&=& \frac{2}{-\gamma_{0,d} + \sqrt{m_n B_3 + \frac{A+2}{4A} q^2 }},
\\
\tau_{\sigma}(q;B_3) 
&=&   \frac{(A+1)/A}{-\gamma_{0,\sigma} + \sqrt{ \frac{A}{A+1} \left(2 m_n B_3 + \frac{A+2}{A+1} q^2 \right)}}.
\end{eqnarray}
\endnumparts
In Eq.~\eqref{eq:inteq-An}, $H(\Lambda)\equiv -\Lambda^2 h/(m_ng_{\sigma}^2)$ is the dimensionless three-body force parameter. 

By projecting the three-body transition amplitudes to s-waves, we simplify the integral equations to the expressions given in Refs.~\cite{Canham:2008jd,Acharya:2013aea}
\numparts
\begin{eqnarray}
\tilde{\mathcal{A}}_c(q) &=& \frac{2}{\pi} \int^\Lambda_0 d q' \, q'^2 X_{00}^n(q',q;B_3) \,
\tau_{\sigma}(q';B_3) \, \tilde{\mathcal{A}}_n(q') \label{eq:faddeevswave1}
\\
\tilde{\mathcal{A}}_n(q) &=& \frac{1}{\pi}  \int^\Lambda_0 d q' \, q'^2 X_{00}^n(q,q';B_3) \,
\tau_{d}(q';B_3) \, \tilde{\mathcal{A}}_c(q')
\nn
&&+\frac{1}{\pi}  \int^\Lambda_0 d q' \, q'^2 \left[X_{00}^c(q',q;B_3) +\frac{H(\Lambda)}{\Lambda^2} \right]\,
\tau_{\sigma}(q';B_3) \, \tilde{\mathcal{A}}_n(q')
\label{eq:faddeevswave2}
\end{eqnarray}
\endnumparts
where the kernel functions are
\numparts
\begin{eqnarray}
X_{00}^n(q,q';B_3) 
&=& \frac{1}{2} \int_{-1}^{1} d (\vhat{q}.\vhat{q}') G_0^c \left(\pi_2(\bm{q},\bm{q}'),q';B_3\right)
= - \frac{1}{qq'} \textrm{Q}_0 (z_{nc}), 
\label{eq:Xnc-s}
\\
X_{00}^c(q,q';B_3) 
&=& \frac{1}{2} \int_{-1}^{1} d (\vhat{q}.\vhat{q}') G_0^n \left(\pi_3(\bm{q},\bm{q}'),q';B_3\right)
= - \frac{A}{qq'} \textrm{Q}_0 (z_{nn})
\label{eq:Xnn-s}
\end{eqnarray}
\endnumparts
with $\textrm{Q}_l$ the Legendre function of the second kind, which is related to the Legendre polynomial $P_l$ by $\textrm{Q}_l(z) \equiv \frac{1}{2} \int^1_{-1} dx\, P_l(x)/(z-x)$.
The arguments $z_{nn}$ and $z_{nc}$ of $\textrm{Q}_l$ in Eqs.~\eqref{eq:Xnc-s} and \eqref{eq:Xnn-s} are defined by
\numparts
\begin{eqnarray}
\label{eq:znn-znc}
z_{nc} &=& -\frac{1}{qq'} \left(m_n B_3 + q^2 + \frac{A+1}{2A} q'^2\right),\\
z_{nn} &=& -\frac{A}{qq'} \left(m_n B_3 + \frac{A+1}{2A} (q^2+q'^2)\right).
\end{eqnarray}
\endnumparts
For bound states
$B_3>0$, so all $z$'s are $< -1$ and no singularities of the $\textrm{Q}_l$'s are encountered. 
The superscript of $X_{00}^{y}$, $y=n,c$, indicates the particle $y$ exchanged between the dimers in two different partitions; the subscripts denote the angular momenta of the spectator particles in the incoming/outgoing partitions. This notation is different from $G_{0}^{y}$, where $y$ represents the spectator in the individual (either the incoming or outgoing) partition.  

\begin{figure}[!t]
\centerline{\includegraphics[width=0.75\columnwidth]{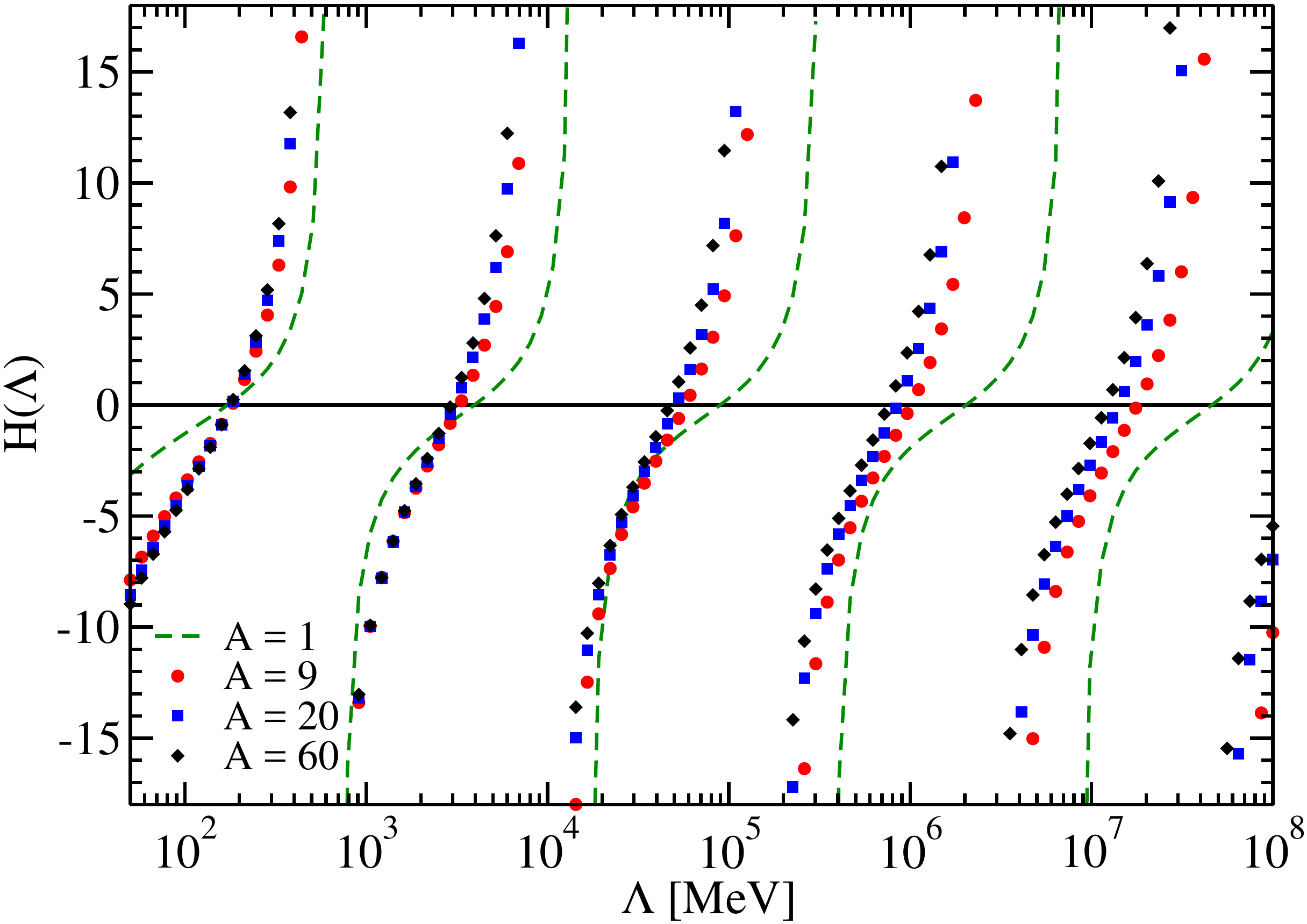}}
\caption{
(Color online) The running of $H(\Lambda)$ as a function of the cutoff $\Lambda$ for systems with $A=1$ (green dashed line), $A=9$ (red circles), $A=20$ (blue squares), and $A=60$ (black diamonds). The numerical results are obtained by setting $\gamma_{0,d},\gamma_{0,\sigma}=0$ and $B_3=1$ MeV.}
\label{fig:H03b-A}       
\end{figure}

To solve the coupled integral equations, we look for an energy $E=-B_3$ where the eigenvalue of the integral-equation kernel is one.  In Eqs.~\eqref{eq:faddeevswave1} and \eqref{eq:faddeevswave2}, regularization with a sharp ultraviolet cutoff is introduced in the momentum integration. To keep calculated low-energy observables regularization invariant the parameter $H(\Lambda)$ is tuned so that one three-body observable, such as the binding energy $B_3$, is held fixed as $\Lambda$ is varied. In two-neutron halos with three pairs resonantly interacting in s-waves, the resulting asymptotic running of $H$ is characterized by a limit cycle~\cite{Wilson:1970ag,Albeverio:1981zi,Bedaque:1998km,Bedaque:1998kg,Barford:2004fz,Mohr:2005pv}. In particular, the discrete scale invariance of this problem in the ultra-violet results in $H(\Lambda)$ being a log-periodic function of $\Lambda$. Our numerical results for various $A$ are well described by: 
\begin{equation}
\label{eq:H0-3bA}
H(\Lambda) = c_A \frac{ \sin\left(s_0\log(\Lambda/\Lambda_*) + \arctan(s_0) +d_A \right) }
{ \sin\left(s_0\log(\Lambda/\Lambda_*) - \arctan(s_0) \right) } + e_A.
\end{equation}
Here $c_A$, $d_A$, and $e_A$ are constants that depend on the core/neutron mass ratio $A$ with $d_A=e_A=0$ when $A=1$ and $\Lambda_*$ is the renormalization parameter determined by one observable in a given two-neutron halo. The period of the limit cycle is $\exp(\pi/s_0)$, where $s_0$ is the solution of a transcendental equation~\cite{Braaten:2004rn,Nielsen20013}:
\begin{equation}
\label{eq:s0-3bA}
\cosh^2\left(\frac{\pi s_0}{2}\right)
-\cosh \left(\frac{\pi s_0}{2}\right)
 \frac{2\sinh(\theta_{1}s_0)}{s_0\sin(2 \theta_{1})}
-\frac{8 \sinh^2(\theta_{2}s_0)}{s_0^2\sin^2(2 \theta_{2})} = 0
\end{equation}
where $\theta_{1}=\arcsin(1/(1+A))$, $\theta_{2}=\arcsin\sqrt{A/(2+2A)}$. Eq.~\eqref{eq:H0-3bA} was first derived in systems of three equal-mass particles~\cite{Bedaque:1998km,Bedaque:1998kg}, where $s_0=1.00624$ is obtained from Eq.~\eqref{eq:s0-3bA}. This corresponds to a discrete scaling factor $\exp(\pi/s_0)=22.694$ and reveals the presence of the Efimov effect in that case~\cite{Efimov:1971zz}. 
The log-periodicity of $H(\Lambda)$ persists when the core and neutron masses are not equal. 
The running of $H(\Lambda)$, shown in Fig.~\ref{fig:H03b-A}, clearly indicates that the limit-cycle behavior is present for values of $A > 1$. The numerical results shown there are obtained by setting $\gamma_{0,d}=\gamma_{0,\sigma}=0$ (i.e., working in the unitary limit) and $B_3=1$ MeV, but the log periodicity persists for finite but small two-body binding momenta, i.e., as long as $\gamma_{0,d},\gamma_{0,\sigma}\ll \Lambda$.

\begin{figure}[!t]
\centerline{\includegraphics[width=0.9\columnwidth]{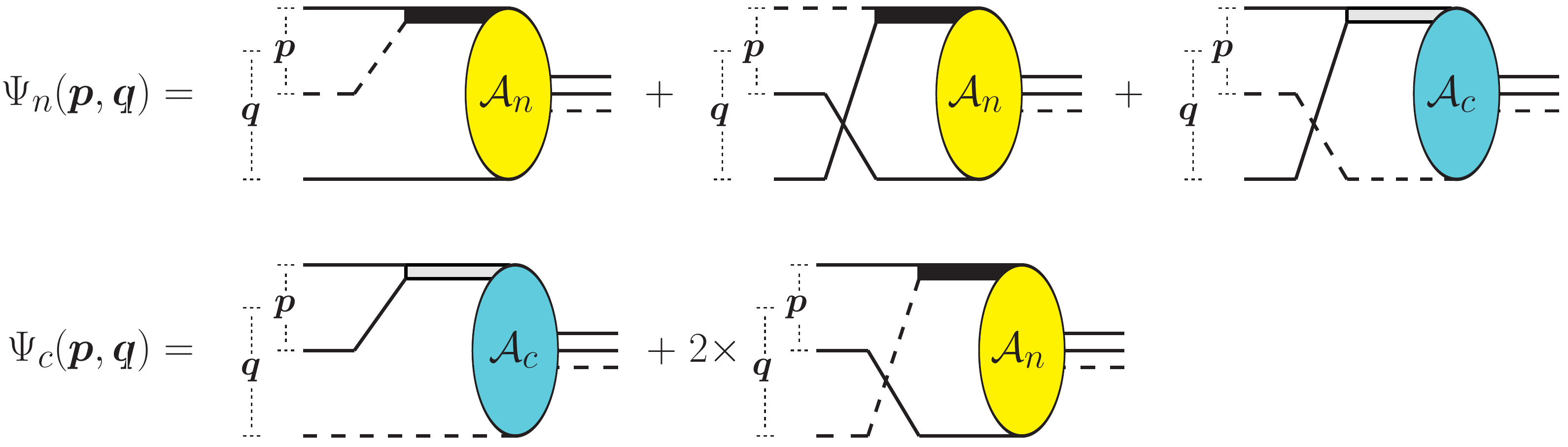}}
\caption{
The bound-state $2n$ halo wave function in Jacobi momentum space representations.}
\label{fig:Psi-3b}       
\end{figure}

The three-body wave function $\Psi$ can 
then be obtained by connecting  the three-body transition amplitudes with external one-body propagators and dimer propagators, see the Feynman diagrams in Fig.~\ref{fig:Psi-3b}. The wave function can be represented in two different Jacobi partitions labeled by the spectator $n$ or $c$. In  s-wave two-neutron halos, we obtain~\cite{Canham:2008jd,Acharya:2013aea}
\begin{eqnarray}
\label{eq:Psin-2n}
\fl
\Psi_n(p,q)
&=&
G_0^n(p,q;B_3) \left[ \tau_\sigma(q;B_3)\tilde{A}_n(q)
+\frac{1}{2} \int_{-1}^1 d\left(\vhat{p}.\vhat{q}\right)\tau_\sigma(\pi_3(\vec{p},-\vec{q});B_3)\tilde{A}_n(\pi_3(\vec{p},-\vec{q}))\right.
\nn
\fl
&&+\frac{1}{2} \left.\int_{-1}^1 d\left(\vhat{p}.\vhat{q}\right) \tau_{d}(\pi_1(\vec{p},-\vec{q});B_3)\tilde{A}_c(\pi_1(\vec{p},-\vec{q}))\right],
\end{eqnarray} 
and with the core as a spectator:
\begin{eqnarray}
\label{eq:Psic-2n}
\fl
\Psi_c(p,q)=G_0^c(p,q;B_3) \left[ \tau_{d}(q;B_3)\tilde{A}_c(q)
+\int_{-1}^1 d\left(\vhat{p}.\vhat{q}\right) \tau_\sigma(\pi_2(\vec{p},-\vec{q});B_3)\tilde{A}_n(\pi_2(\vec{p},-\vec{q}))\right].
\nn
\end{eqnarray} 

With the wave function, we calculate the one-body matter-density form factors in s-wave $2n$ halos by
\begin{eqnarray}
\label{eq:Fy-matter}
F_y(|\bs{k}|)=\int d^3 p \int d^3 q\,
\Psi_y(p,q)~\Psi_y(p,|\vec{q}-\vec{k}|),
\end{eqnarray} 
with $y=n,c$ depending on the Jacobi partitions. For normalized wave functions $F_y(0)=1$. The mean-square distance between the valence neutron and the center of mass of the neutron-core pair, $\langle r_{n-nc}^2 \rangle$, can be extracted from the form factor $F_n$ via
\begin{equation}
F_n(|\bs{k}|)=1-\frac{1}{6}\bs{k}^2 \langle r_{n-nc}^2 \rangle+ \ldots,
\end{equation} 
and the mean-square distance between the core and the center of mass of the two-neutron pair, $\langle r_{c-nn}^2 \rangle$, is determined by
\begin{equation}
F_c(|\bs{k}|)=1-\frac{1}{6}\bs{k}^2\langle r_{c-nn}^2 \rangle+ \ldots.
\end{equation}
The geometry of the neutron-neutron-core three-body system then leads to the following formula for the matter radius in a $2n$ halo:
\begin{equation}
\label{eq:rm-3b}
\bra r_m^2 \ket_{2n\rm-halo} 
= \frac{2(A+1)^2}{(A+2)^3} \bra r_{n-nc}^2\ket
+ \frac{4A}{(A+2)^3} \bra r_{c-nn}^2\ket + \frac{A}{A+2} \bra r_m^2 \ket_{\rm core},
\end{equation}
where the last term is the correction from the finite matter radius of the core.

\subsection{Applications: Efimov states and matter radii}

\label{sec:universality2nhalos}

In the zero-range limit, long-distance observables in three-body systems are correlated by few-body universality. One example is the Efimov effect, which is characterized by discrete scale invariance in the three-body system. In Eq.~\eqref{eq:H0-3bA}, the running of the three-body coupling, which is a log-periodic function of the ultraviolet cutoff, is characterized by a limit cycle with a period $\exp(\pi/s_0)$. As a consequence of the limit cycle, the three-body s-wave bound states in the unitary limit display a geometric progression. The ratio of three-body binding energies in two consecutive states is given by $\exp(2\pi/s_0)$.

Discrete scale invariance has been observed in experiments on ultracold atomic gases, where the atom-atom scattering length is tuned using a magnetic field in the vicinity of a Feshbach resonance. Near the unitary limit, the scattering lengths associated with threshold features in atom-dimer collisions and three-atom recombination are also correlated through the scaling factor $\exp(\pi/s_0)$ (see Refs.~\cite{Braaten:2004rn,Chin:2010} for reviews). 

\begin{figure}[ht]
\centerline{\includegraphics*[width=0.65\linewidth,angle=0,clip=true]{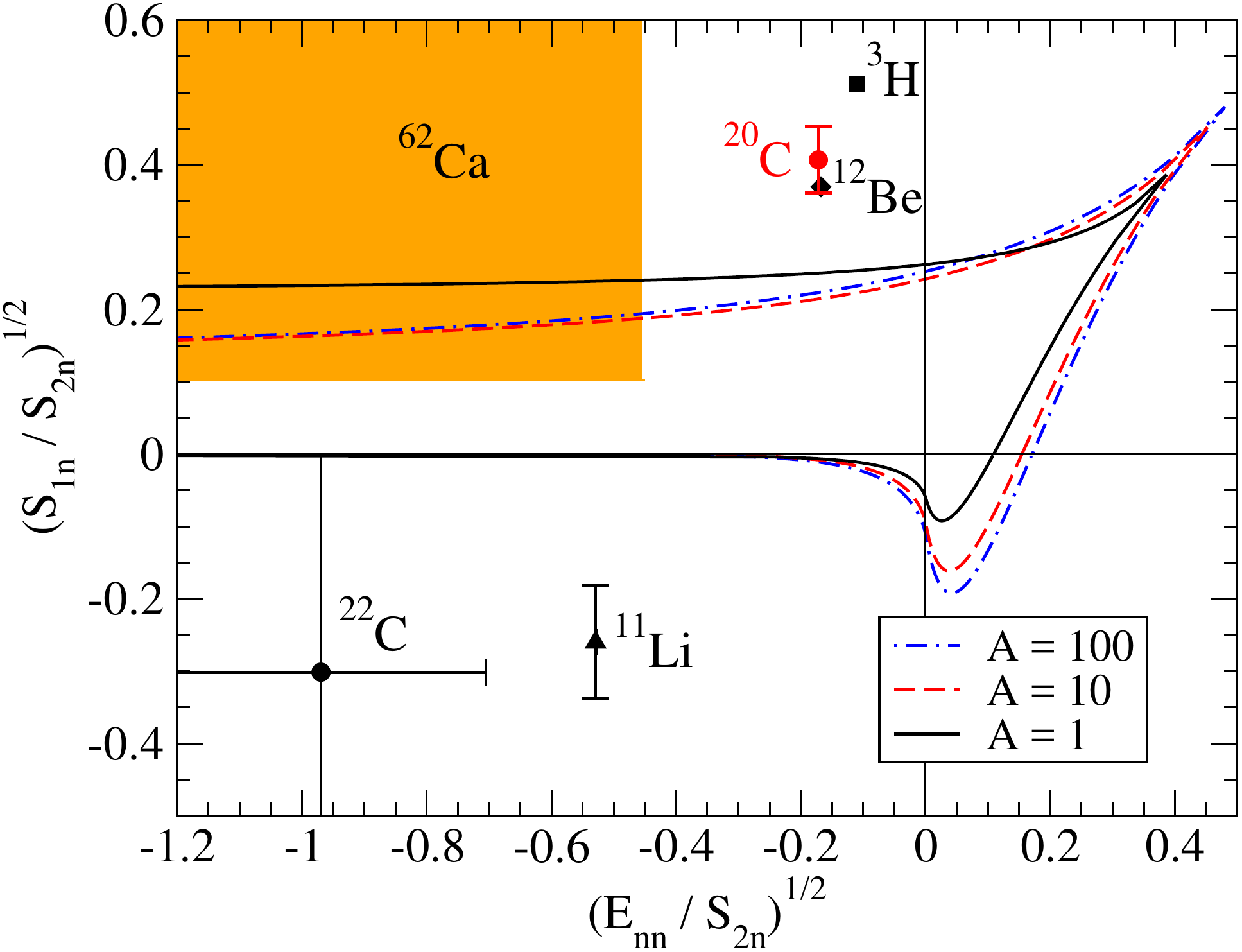}}
\caption{The contour plot in $\textrm{sgn}(E_{nn})\sqrt{|E_{nn}|/S_{2n}}$ versus $\textrm{sgn}(S_{1n})\sqrt{|S_{1n}|/S_{2n}}$ for the ground-state $2n$ halos with core-neutron mass ratios $A=1,10,100$. The hypothetical bound di-neutron regime with $E_{nn}>0$ is also included in the theoretical calculation to complete the contour.}
\label{pic:halo-efimov}
\end{figure}

In an s-wave $2n$ halo nucleus, $a_{0,d}$ and $a_{0,\sigma}$ are large but finite, and the three-body binding energy is characterized by the two-neutron separation energy, i.e., $S_{2n} = B_3$. In such systems,
the number of possible Efimov-like halo states is determined by the two ratios $E_{nn}/S_{2n}$ and $S_{1n}/S_{2n}$, where $E_{nn}=-\gamma_{0,d}^2/(2\mu_d)$ is the neutron-neutron virtual energy.
Frederico, Tomio and collaborators suggested to use a universal function of the ratios $ E_{nn}/S_{2n}$ and $S_{1n}/S_{2n}$ to explore possible Efimov states in halo nuclei, and carried out this study in a zero-range three-body model~\cite{Amorim:1997mq,Yamashita:2007ej}. Following this approach, Canham and Hammer applied EFT to explore the Efimov scenario in s-wave $2n$ halos~\cite{Canham:2008jd}.
They tuned the three-body coupling $H(\Lambda)$ so that there was an excited state of the two-neutron halo at threshold, i.e. $B_3^*=\max\{0,E_{nn},S_{1n}\}$. The $S_{2n}$ of the two-neutron halo is then predicted as all LO two-body and three-body EFT couplings are fixed. At this value of $S_{2n}$ LO EFT predicts the existence of an Efimov excited state at threshold in the halo system. This, in turn, defines (for fixed $A$) a contour in the $(S_{1n}/S_{2n})$ versus $(E_{nn}/S_{2n})$ plane:
\begin{equation}
g^{(LO)}\left(\frac{E_{nn}}{S_{2n}},\frac{S_{1n}}{S_{2n}};A\right)=1.
\end{equation}
Inside the contour the three-body bound state is deep enough, or the two-body $nc$ system is near enough to unitarity, for an Efimov state to appear in the two-neutron halo. 
This 
region is depicted in Fig.~\ref{pic:halo-efimov}, and is in good agreement with an analogous study in a zero-range model~\cite{Frederico:2012xh}.
The curves for different values of core-neutron mass ratio $A$ quickly accumulate to the same region when $A$ increases. 
The specific cases of $^3$H, $^{11}$Li, $^{12}$Be, $^{20,22}$C and $^{62}$Ca are indicated by mapping their up-to-date experimental data from AME2012~\cite{Audi2012,Wang2012} onto this two-dimensional plane. 

In Fig.~\ref{pic:halo-efimov}, we highlight the case of $^{20}$C since the EFT analysis there is updated from Ref.~\cite{Canham:2008jd} with new experimental data. 
Because of the large uncertainty in $S_{1n}$ at the time of publication, Ref.~\cite{Canham:2008jd} suggested that $^{20}$C may display an excited Efimov state. However, experimental data on the Coulomb dissociation of $^{19}$C ($S_{1n}=0.53(13)$ MeV~\cite{Nakamura:1999rp}) included in AME2012 ($S_{1n}=0.58(9)$ MeV~\cite{Audi2012,Wang2012}) now preclude this possibility.

The Efimov-effect correlation between bound and excited-state energies is just one example of the way in which universality imposes relations between different three-body observables. The point matter radius $\bra r_m^2\ket_{\rm pt}$ of a ground-state two-neutron halo, defined by subtracting the core-size contribution from the radius of the halo in Eq.~\eqref{eq:rm-3b}, is also determined by a universal function of $S_{1n}$ and $S_{2n}$ at LO:
\begin{eqnarray}
\fl
\qquad \qquad
\bra r_m^2\ket_{\rm pt}
\equiv \bra r_m^2\ket_{2n\rm-halo} 
-\frac{A}{A+2} \bra r_m^2 \ket_{\rm core}
=\frac{1}{m_n S_{2n}} f^{(LO)}\left(\frac{E_{nn}}{S_{2n}},\frac{S_{1n}}{S_{2n}};A\right).
\end{eqnarray}
Such correlations have been investigated using EFT for different s-wave two-neutron halos~\cite{Canham:2008jd,Acharya:2013aea,Hagen:2013jqa}.

$^{22}$C is a Borromean two-neutron halo. The matter radius of $^{22}$C was determined in the reaction cross section measurement on a proton target to be $\langle r_m^2\rangle^{1/2}_{2n\rm-halo} =5.4(9)$~fm~\cite{Tanaka:2010zza}. A recent interaction cross-section measurement on a carbon target obtained a more precise result of $\langle r_m^2\rangle^{1/2}_{2n\rm-halo}=3.44(8)$~fm~\cite{Togano:2016wyx}, suggesting a smaller halo configuration in $^{22}$C. 

The $2n$ separation energy of $^{22}$C is not yet directly constrained by experiment. 
In order to obtain an indirect constraint, Acharya {\it et al.}~\cite{Acharya:2013aea} performed an EFT calculation of the correlations among $\langle r_m^2\rangle^{1/2}_{2n\rm-halo}$, $S_{1n}$, and $S_{2n}$ of $^{22}$C. Using the data $\langle r_m^2\rangle^{1/2}_{2n\rm-halo} =5.4(9)$~fm~\cite{Tanaka:2010zza} known at the time of publication, they predicted an upper bound on $S_{2n}$ of  $0.1$ MeV. This is consistent with the AME2012. It is also $20\%$ lower than the
calculation in a zero-range three-body model~\cite{Yamashita:2011cb}, and $50\%$ lower than other model-dependent calculations~\cite{Fortune:2012zzb}. (Note that the same matter-radius data from Ref.~\cite{Tanaka:2010zza} was also used in Refs.~\cite{Yamashita:2011cb,Fortune:2012zzb} as input.)

\begin{figure}[ht]
\centerline{\includegraphics*[width=0.65\linewidth,angle=0,clip=true]{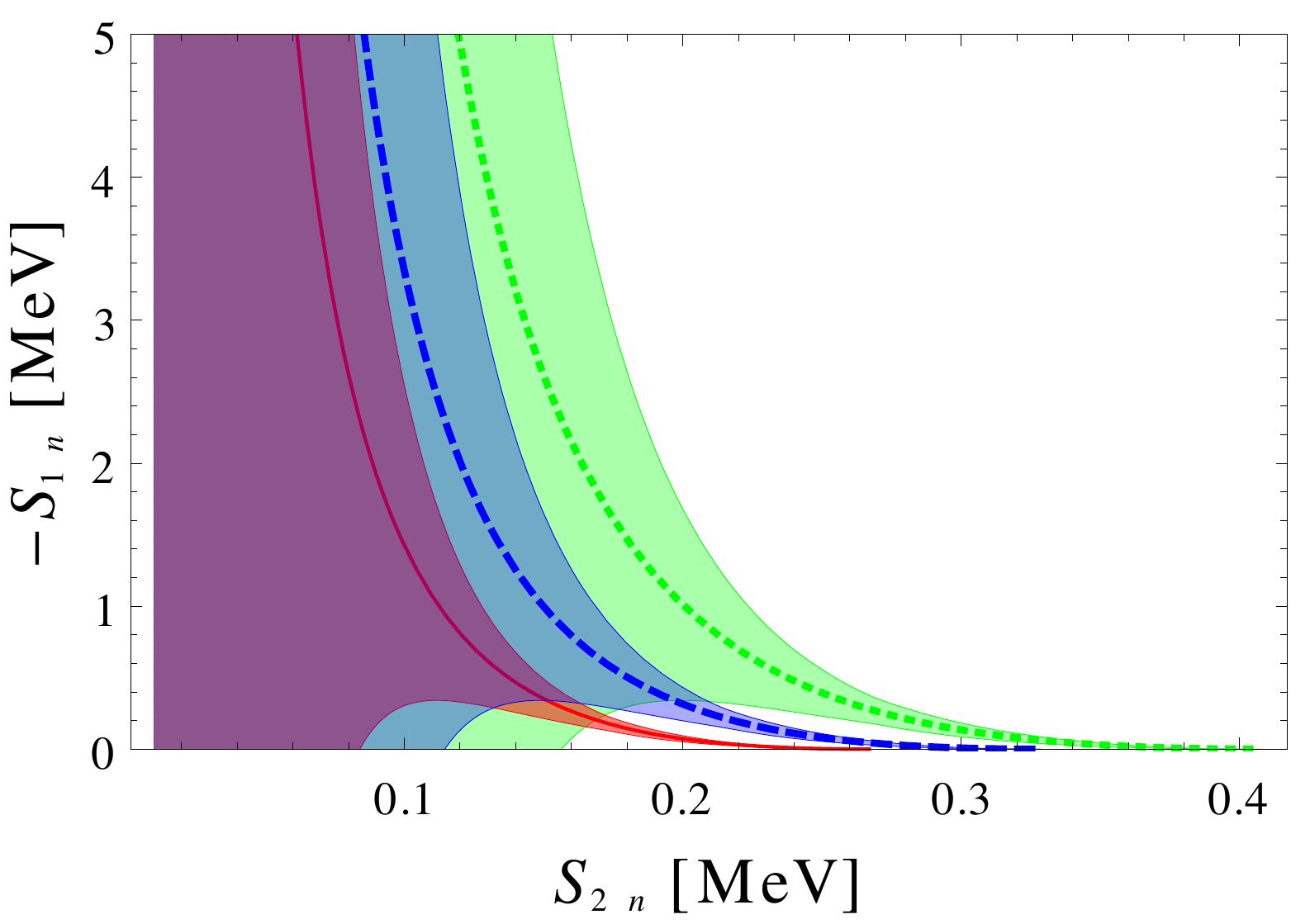}}
\caption{The correlation curve of $S_{2n}$ versus $S_{1n}$ in $^{22}$C with fixed values of the $2n$-halo point matter radius at
$\bra r_m^2\ket_{\rm pt} =$ 3.81~fm$^2$ (blue, dashed), 4.63~fm$^2$ (red, solid), and 3.10~fm$^2$ (green, dotted).
The shaded bands indicate the theoretical errors based on estimates of higher-order EFT corrections beyond LO.}
\label{pic:22C-contour}
\end{figure}

Using the recent data $\langle r_m^2\rangle^{1/2}_{2n\rm-halo}=3.44(8)$ fm and the matter radius of the $^{20}$C core, $\langle r_m^2\rangle^{1/2}_{\rm core}=2.97^{+0.03}_{-0.05}$ fm~\cite{Togano:2016wyx}, we infer a 
point matter radius squared \mbox{$\bra r_m^2\ket_{\rm pt} =3.81^{+0.82}_{-0.71}$ fm$^2$}.
With this, updated, input to the LO Halo EFT calculation of Ref.~\cite{Acharya:2013aea} we obtain the correlation between $S_{2n}$ and $-S_{1n}$ shown in Fig.~\ref{pic:22C-contour}. The three shaded regions are obtained by fixing the three-body datum 
$\bra r_m^2\ket_{\rm pt}$
to its experimental $1\sigma$ lower bound, mean value, and upper bound respectively. The bands then depict the estimated uncertainty of our LO calculation in each case. Here we have assumed that the ${}^{20}$C rms radius sets the scale $1/M_{\rm core}$, and then the relative error of this LO calculation is determined by the maximum of $1/(a_{0,d}^{(nn)} M_{\rm core})$, $\sqrt{2 m_n |S_{1n}|}/M_{\rm core}$, and $\sqrt{2 m_n S_{2n}}/M_{\rm core}$.
This contour plot determines an upper limit of $S_{2n}\le0.4$ MeV in $^{22}$C, given the constraint that $^{22}$C is a Borromean system. This value suggests a more deeply bound $^{22}$C compared to the conclusion in Ref.~\cite{Acharya:2013aea}. If the information on neutron-core energy $|S_{1n}| \ge 2.8$ MeV, determined by Mosby {\it et al.} from the proton-removal reaction~\cite{Mosby:2013bix} is accurate, $S_{2n}$ Fig.~\ref{pic:22C-contour} further constrains $S_{2n}$ to be $\leq 0.18$ MeV.

The correlations unveiled in the EFT analysis allow consistency checks among different experimental results for $^{21}$C and $^{22}$C: it was already clear from the correlation published in Ref.~\cite{Acharya:2013aea} that a ${}^{22}$C radius as large as that claimed by Tanaka {\it et al.} could only be consistent with Mosby {\it et al.}'s constraint on $|S_{1n}|$ if ${}^{22}$C was incredibly weakly bound ($S_{2n} < 0.02$ MeV). The picture is clearer with the new radius measurement of $\langle r_m^2\rangle^{1/2}_{2n\rm-halo}=3.44(8)$ fm, but  to determine whether ${}^{22}$C is bound by the $< 0.2$ MeV that is required for consistency of that number with the $|S_{1n}|$ measurement from Mosby {\it et al.}~\cite{Mosby:2013bix}, or perhaps lies closer to the upper limit around $S_{2n} \approx 0.4$ MeV in Fig.~\ref{pic:22C-contour},
requires more precise experiments on these drip-line carbon isotopes.

$^{62}$Ca, highlighted in Fig.~\ref{pic:halo-efimov}, is a special case where data are not known experimentally but are instead predicted in theory.
EFT correlations were applied in Ref.~\cite{Hagen:2013jqa} to 
investigate $^{62}$Ca as a candidate for the heaviest $2n$ halo nucleus. In their work, the $n$-$^{60}$Ca scattering parameters were extracted from the $n$--$^{60}$Ca s-wave scattering phase shift, obtained in a coupled-cluster calculation based on chiral two- and three-nucleon interactions. This {\it ab initio} calculation indicated a large scattering length $a_{0,\sigma}=54(1)$ fm and an effective range $r_{0,\sigma}=9.0(2)$ fm for the $n$-$^{60}$Ca system, where the error is estimated based on the spread of the coupled-cluster calculations at two small harmonic oscillator frequencies~\cite{Hagen:2013jqa}. This would make ${}^{61}$Ca a very shallow ($S_{1n} \approx 7$ keV) one-neutron halo. 
Using $a_{0,\sigma}$ and $r_{0,\sigma}$ as input parameters~\footnote[2]{A negative $a_{0,\sigma}$ obtained from calculations at a larger harmonic oscillator frequency was not considered in the EFT analysis in Ref.~\cite{Hagen:2013jqa}, because that result displayed strong infrared regulator effects in these threshold parameters.}, Ref.~\cite{Hagen:2013jqa} performed an EFT analysis on $^{62}$Ca as a $n$-$n$-$^{60}$Ca two-neutron halo and searched for possible signatures of Efimov states. The EFT expansion parameter in this system is estimated from the {\it ab initio} results to be $M_{\rm halo}/M_{\rm core}\sim r_{0,\sigma}/a_{0,\sigma} =1/6$. Ref.~\cite{Hagen:2013jqa} analyzed the LO Halo EFT correlation between the $n$-$^{61}$Ca scattering length and the two-neutron separation energy $S_{2n}$ of the ground state of $^{62}$Ca. The conclusion was that if $S_{2n} \ge 230$ keV ${}^{62}$Ca will have an excited bound state of Efimovian character: for this value of $S_{2n}$, an excited state of $^{62}$Ca appears at the $n$-$^{61}$Ca threshold. Correlations among $S_{2n}$ and the electric and matter radii of $^{62}$Ca were also computed in Ref.~\cite{Hagen:2013jqa}.

\subsection{Range corrections in three-body systems}
Beyond the leading-order prediction, universal physics in two-neutron halos is affected by the finite effective range which enters EFT calculations at next-to-leading order. To calculate range effects, the three-body transition amplitudes $\mathcal{A}_y$ (for $y=n,c$) are solved in integral equations~\eqref{eq:intorig-Ac} and \eqref{eq:intorig-An}, with the LO dimer propagator $D_x^{(0)}$ (for $x=d,\sigma$) in Eq.~\eqref{eq:Dx-lo} replaced by the range-modified propagator $D_x$~\eqref{eq:prop-2b}. However, the $D_x$ of Eq.~\eqref{eq:prop-2b} contains a spurious pole at momenta $\approx 2/r_{0,x}\sim M_{\rm core}$---in addition to the physical bound- or virtual-state pole at momentum $\sim M_{\rm halo}$. The spurious pole is outside the domain of validity of Halo EFT, but it is encountered in the integrals in the three-body equation, which probe the two-body amplitude at arbitrarily large negative energies. 
It can be avoided by choosing a small integration cutoff in Eqs.~\eqref{eq:intorig-Ac} and \eqref{eq:intorig-An}. However, regulation effects at such a small cutoff then complicate systematic uncertainty estimates in  EFT predictions. An alternative approach is to expand $D_x$ in powers of $r_{0,x}$. The expression at NLO is then
\begin{eqnarray}
\label{eq:Dx-expanded}
D_{x,NLO}(E, \bm{P} ) =& - \frac{2\pi}{\mu_{x} g_{x}^2} 
\frac{1}{-\gamma_{0,x}  +  \sqrt{2\mu_{x} \left(\bm{P}^2/2M_{x}- E  -i\epsilon \right)} }
\nn
&\times\left[1 +\frac{r_{0,x}}{2}\left(\gamma_{0,x}  +  \sqrt{2\mu_{x} \left(\bm{P}^2/2M_{x}- E  -i\epsilon \right)}\right)\right].
\end{eqnarray}
The amplitudes $\mathcal{A}_y$ at NLO are then solved in the range-modified integral equations by the insertion of Eq.~\eqref{eq:Dx-expanded} into Eqs.~\eqref{eq:intorig-Ac} and \eqref{eq:intorig-An}. This partial resummation technique was used in studies of range effects for the triton~\cite{Bedaque:2002yg} and was adopted in Ref.~\cite{Canham:2009xg} to investigate range corrections in two-neutron halos.~\footnote[1]{Strictly speaking, instead of Eq.~\eqref{eq:Dx-expanded}, which is a result for an infinitely large cutoff, $D_{x}$ in Ref.~\cite{Canham:2009xg} was obtained by regularizing the neutron-neutron and neutron-core interactions with a gaussian regulator of finite width.}
The partial resummation formalism iterates both the LO and NLO parts of the dimer propagators in the integral equations, and thereby arbitrarily includes higher-order range corrections above NLO. These higher-order corrections are small if the regularization cutoff $\Lambda$ is kept below or close to $M_{\rm core}$ (see e.g., Refs.~\cite{Bedaque:2002yg,Platter:2006ev,Ji:2012nj} for detailed discussions on this issue).

The rescaled amplitudes at NLO accuracy, $\tilde{\mathcal{A}}_{y,NLO}$, are then the solution of Eqs.~\eqref{eq:inteq-Ac} and \eqref{eq:inteq-An}, but with the $\tau_x$'s replaced by their NLO expressions:
\numparts
\begin{eqnarray}
\fl
\tau_{d,NLO}(q;B_3) 
=& \frac{2 \left(1+r_{0,d}\gamma_{0,d}\right) }{-\gamma_{0,d} + \sqrt{m_n B_3 + \frac{A+2}{4A} q^2 }}
+r_{0,d},
\\
\fl
\tau_{\sigma,NLO}(q;B_3) 
=&   \frac{\left(1+r_{0,\sigma}\gamma_{0,\sigma}\right) (A+1)/A}{-\gamma_{0,\sigma} + \sqrt{ \frac{A}{A+1} \left(2 m_n B_3 + \frac{A+2}{A+1} q^2 \right)}}
+\frac{A+1}{2A}r_{0,\sigma}.
\end{eqnarray}
\endnumparts
Then, in order to calculate the two-neutron halo wave functions $\Psi_y$ at NLO, one again replaces $\tau_x$ and $\tilde{A}_y$ in Eqs.~\eqref{eq:Psin-2n} and \eqref{eq:Psic-2n} with the corresponding NLO quantities, i.e.,
$\tau_d \tilde{A}_c \rightarrow \tau_{d,NLO} \tilde{A}_{c,NLO}$ and $\tau_\sigma \tilde{A}_n \rightarrow \tau_{\sigma,NLO} \tilde{A}_{n,NLO}$. 

An alternative, fully perturbative, EFT calculation of range corrections was recently carried out for two-neutron halos at NLO by Vanasse~\cite{Vanasse:2016hgn}. This implemented the method for calculating such perturbative insertions that he introduced and applied up to N$^3$LO in the three-nucleon system~\cite{Vanasse:2013sda,Margaryan:2015rzg,Vanasse:2015fph}. Results for charge and matter form factors and radii in two neutron halos were obtained with good accuracy. We refer to Refs.~\cite{Vanasse:2016hgn,Vanasse:2015fph} for detailed discussions of this rigorous perturbative treatment of range effects in three-body systems.

Using the perturbative approach, Vanasse calculated the point matter radii, $\bra r_m^2 \ket_{\rm pt}= \bra r_m^2 \ket_{2n\rm-halo} - \frac{A}{A+2}\bra r_m^2 \ket_{\rm core}$ at NLO accuracy~\cite{Vanasse:2016hgn}. 
In Table~\ref{tab:rm-2n}, we quote Vanasse's LO and NLO point matter radii and compare with experimental results for $^{11}$Li, $^{14}$Be, and $^{22}$C~\cite{Ozawa:2000gx,Ozawa:2001hb,Tanaka:2010zza,Togano:2016wyx}. Note that to obtain his NLO numbers Vanasse estimated $r_{0,\sigma}$ by assuming it is set by the pion mass, i.e. $r_{0,\sigma} \approx 1.4$ fm. Errors were then determined as $(r_{0,\sigma} \sqrt{2 m_n S_{2n}})^j$, where $j=1$ ($j=2$) for the LO (NLO) results shown in the first (second) column. 
 The matter radii in two-neutron halos were previously calculated
to LO in Halo EFT by Canham and Hammer~\cite{Canham:2008jd}. Using partial resummation, they also predicted the average neutron-core and neutron-neutron distances in the neutron-neutron-core configuration at NLO accuracy~\cite{Canham:2009xg}. The results in Refs.~\cite{Canham:2008jd,Canham:2009xg,Vanasse:2016hgn} are consistent with each other.

\begin{table}
   \centering
   \begin{tabular}{rlccc} \hline
      &
      & $^{11}$Li   
      & $^{14}$Be 
      & $^{22}$C 
      \\  \hline
      \multirow{4}{*}{$\bra r_m^2 \ket_{\rm pt}$ [fm$^2$]}
      & EFT$_{\rm LO}$  
      &$5.76\pm2.13$ &$1.23\pm0.96$ &$8.99^{+\infty}_{-5.01}$  
      \\ 
      & EFT$_{\rm NLO}$ 
      &$6.16\pm0.84$ &$1.40\pm0.85$ &$9.28^{+\infty}_{-5.17}$
      \\ \cline{2-5}
      & \multirow{2}{*}{expt}
      & $5.34\pm0.15$~\cite{Ozawa:2001hb} & $4.24\pm2.42$~\cite{Ozawa:2001hb} & $21.1\pm9.7$~\cite{Ozawa:2000gx,Tanaka:2010zza} \\
      & & & $2.90\pm2.25$~\cite{Ozawa:2001hb} & $3.81^{+0.82}_{-0.71}$~\cite{Togano:2016wyx} 
      \\ \hline
   \end{tabular}
   \caption{The point matter radius $\bra r_m^2 \ket_{\rm pt}= \bra r_m^2 \ket_{2n\rm-halo} - \frac{A}{A+2}\bra r_m^2 \ket_{\rm core}$ from EFT predictions at LO and NLO, and from experiments.} 
\label{tab:rm-2n} 
\end{table}

\section{Halos beyond the s-wave} \label{sec:pwaves}

\subsection{Lagrangian}
\label{eq:pwavelag}
First we consider a one-neutron halo involving p-wave interactions between the neutron and a core of spin $\varsigma_c$ and mass $m_c$.  (The case of proton halos is dealt with in Sec.~\ref{sec:coulomb}.) Once again writing
\begin{equation}
\mathscr{L} = \mathscr{L}_1 + \mathscr{L}_2 + \mathscr{L}_3~,
\label{eq:pwaveL}
\end{equation}
the one-body part is exactly the same as in Eq.~(\ref{eq:Lag-1b}). 

To construct the two-body part we first note that in the case of p-wave interactions between the core and the neutron we must consider operators involving the combination of fields 
\begin{equation}
n_\delta (i \tensor{\partial}_i) c^{(\varsigma_c)}_a,
\label{eq:pwaveinterpolatingfield}
\end{equation}
where we have labeled the $c$ fields according to its spin, so---as in Sec.~\ref{sec:swavelag}---its subscript runs over $a=-\varsigma_c, \ldots, \varsigma_c$.
We employ the Galilean invariant derivative,
$\tensor{\partial}_i \equiv [(\roarrow{m} \loarrow{\nabla} - \loarrow{m}\roarrow{\nabla})/(\loarrow{m}+\roarrow{m})]_i$ ($i=-1,0,1$) with $\loarrow{m}$ (or $\roarrow{m}$) the mass of the field operated on by $\loarrow{\nabla}$ (or $\roarrow{\nabla}$). By doing so we ensure that interactions constructed out of this interpolating field are invariant under Galilean boosts. 
We also include a factor of $i$ in this construction so as to guarantee that, if the fields have standard time-reversal properties, it is straightforward to write down the Hermitian conjugate of interactions involving this field. 
The interpolating field (\ref{eq:pwaveinterpolatingfield}) is constructed out of one derivative as well as the $c$ and $n$ fields, and so is the product of three irreducible representations of the rotation group. In consequence it has pieces corresponding to irreducible representations with values of $j$ from $j=|\varsigma_c-3/2|$ to $j=\varsigma_c+3/2$. In the non-relativistic systems we consider here the total spin $s$ is also a conserved quantum number, so we decompose the product field (\ref{eq:pwaveinterpolatingfield}) into irreducible representations of the rotation group corresponding to particular values of $j$ and $s$ using Clebsch-Gordon coefficients~\cite{Hammer:2011ye}:
\begin{equation}
\fl
[n (i \tensor{\partial}) c]^{(sj)}_{m}=\sum_{\delta=-1/2}^{1/2} \sum_{i=-1}^1 \sum_{a=-\varsigma_c}^{\varsigma_c} \Bigg{(}{1 \over 2} \, \delta \; \varsigma_c  \, a \Bigg{|} s \, m-i\Bigg{)} \Bigg{(}1 \, i \; s \, m - i \Bigg{|}j m \Bigg{)} n^{(1/2)}_\delta (i \tensor{\partial}_i) c^{(\varsigma_c)}_a.
\end{equation}

In order to account for channels in which the p-wave interaction is enhanced  at low energies we introduce p-wave dimer fields in those channels. For example, the one-neutron halo ${}^{11}$Be has a spin-zero core and an enhanced p-wave interaction for the $j=1/2$ channel, so the leading-order p-wave $\mathscr{L}_2$ is~\cite{Hammer:2011ye}:
\begin{eqnarray}
\fl
\mathscr{L}_2^{(1/2)}=\pi^{(1/2) \, \dagger}_\delta \left[w_\pi \left(i \partial_0 + 
\frac{\nabla^2}{2M_\sigma}\right) + \Delta_\pi^{(1/2)}\right] \pi^{(1/2)}_\delta \nonumber\\
- \frac{g_{\pi}^{(1/2)}}{2} \left( \pi^{(1/2) \, \dagger}_\delta  \left[n (i \tensor{\partial}) c\right]^{(1/2 \; 1/2)}_\delta
+ \left[n (i\tensor{\partial}) c \right]^{(1/2 \; 1/2) \, \dagger}_\delta \pi_\delta^{(1/2)}\right).
\label{eq:Be11LOL}
\end{eqnarray}
Note that repeated indices are implicitly summed, and that since the total mass is unaffected by the angular-momentum channel in which the dimer is formed $M_\sigma=m_n + m_c$ appears again here. 

Channels where the p-wave interaction is of natural size will generally be suppressed by several orders in the EFT expansion. For example, the $n$-$^{10}$Be p-wave interaction in the $j=3/2$ channel is described by
\begin{equation}
\mathscr{L}_2^{(3/2)}=-\frac{C^{(3/2)}}{4} {[n (i \tensor{\partial})
c]^{(1/2 \; 3/2) \, \dagger}_\beta} [n (i \tensor{\partial}) 
c]^{(1/2 \; 3/2)}_\beta
\end{equation}
where the possible values of the repeated index $\beta=-3/2, \ldots, 3/2$ are summed over. 

In Sec.~\ref{sec:3Bpwave} we consider the case of a three-body halo nucleus that is bound by p-wave interactions: we take ${}^6$He as it is the prototypical example. In that case the dominant low-energy resonance is in the ${}^2$P$_{3/2}$ channel, in direct contrast to the situation in ${}^{11}$Be. For this system we therefore have a four-component p-wave dimer, $\pi^{(3/2)}_\beta$. The three-body Lagrangian, $\mathscr{L}_3$, then describes a p-wave contact interaction between this dimer and the other neutron in the system:
\begin{equation}
\mathscr{L}_3 
= -h \left[ \pi^{(3/2)} [\tensor{\partial} n]^{(1/2 \; 3/2)} \right]^{0 \,\dagger} 
\left[\pi^{(3/2)} [\tensor{\partial} n]^{(1/2 \; 3/2)}\right]^0,
\label{eq:L3p}
\end{equation}
where the outer square brackets with superscript 0 indicate that the two spin-$3/2$ fields are coupled to form a rotational scalar. 

\subsection{Two-body amplitude}

We now wish to compute the propagator of a dimer field of total spin $j$, $D^{(j)}_\pi(p)_{m' m}$. We proceed as we did for the s-wave dimer in Sec.~\ref{sec:swaves}: 
assuming the Lagrangian (\ref{eq:Be11LOL}) the dimer 
obeys a Dyson equation analogous to that depicted in Fig.~\ref{fig:dsigma-dressed}.
Rather than computing the self-energy of the $\pi$ field directly it is easier
to compute the self-energy for the product field $n_\delta (i \tensor{\partial}_i) c_a$ and then project out the piece that corresponds to the $j$ of interest. Starting with the one-loop self-energy
$\Sigma_\pi(E,{\bs P})_{i'i,\delta' \delta,a'a}$, we have
\begin{equation}
\Sigma_\pi(E,{\bs P})_{i'i,\delta' \delta,a'a}=\delta_{i'i}  \delta_{\delta' \delta} \delta_{a'a} \Sigma(E,{\bs P})\,,
\end{equation} where 
the scalar function:
\begin{equation}
\fl
\Sigma(E,{\bs P})=
-\frac{\mu_\sigma (g_{\pi}^{(j)})^2}{6 \pi} 2\mu_\sigma \left(E - \frac{{\bs P}^2}{2 M_\sigma}\right) 
\left[i \sqrt{2\mu_\sigma \left(E - \frac{{\bs P}^2}{2 M_\sigma}+i\epsilon\right)}
+ \frac{3}{2} \Lambda \right].
\end{equation}
Here $\mu_\sigma$ is, as above, the reduced mass in the $nc$ channel. We have also used PDS (with scale denoted by $\Lambda$) and performed the momentum
traces in three dimensions~\footnote[1]{Here we have chosen to treat the factor of $1/(D-1)$ that
 arises in the replacement $p_i p_j \rightarrow \frac{\delta_{ij}}{D-1}  p^2$ as part of the function that is analytically continued to $D=3$ space-time dimensions when PDS is implemented. This differs from what was done
 in Ref.~\cite{Hammer:2011ye}.}. 

Since $\Sigma$ is independent of $\delta$, $a$ and $i$ 
 completeness of the 
Clebsch-Gordon coefficients guarantees that it is also diagonal in $m$.  That is:
\begin{eqnarray}
D^{(j)}_\pi(E,{\bs P})_{m' m}=\delta_{m'm} D_\pi(E,{\bs P}) \nonumber\\
D_\pi(E,{\bs P})=
\frac {1}{\Delta_\pi^{(j)} + w_\pi[E - {\bs P}^2/(2 M_\sigma)] - \Sigma(p)}\,.
\end{eqnarray}
Even though it does not appear in PDS, the fact that $\Sigma$ has a cubic divergence means that both parameters, 
 $\Delta_\pi^{(j)}$ and $g_{\pi}^{(j)}$, are needed for renormalization~\cite{Bertulani:2002sz}.

Coupling the p-wave dimer to $nc$ initial and final states generates a core-neutron amplitude in the ${}^{2s+1}$P$_{j}$ channel that is (in the center-of-mass frame):
\begin{equation}
\langle {\bs k}'|t^{(s)}_1(E)|{\bs k} \rangle={g_{\pi}^{(j)}}^2 {\bs k}' \cdot {\bs k} D_\pi(E,{\bs 0})
=\frac{6 \pi}{\mu_\sigma} \frac{k^2 \cos \theta}{1/a^{(j)}_1 - \frac{1}{2} 
r^{(j)}_1 k^2 + i k^3}\,,
\label{eq:t1}
\end{equation}
with $\cos \theta=\hat{\bs k}' \cdot \hat{\bs k}$ and $k=\sqrt{2 \mu_\sigma E}=|{\bs k}'|=|{\bs k}|$   for on-shell scattering. So we have reproduced the p-wave effective-range expansion up to $\mathcal{O}(k^2)$, with $a^{(j)}_1$ the scattering volume, and $r^{(j)}_1$ the p-wave 
``effective range" in the channel with total angular momentum $j$. Note that if there are two spin channels corresponding to the same $j$ then the couplings of  fields $[n (i \tensor{\partial}) c]^{(sj)}$ for those different values of $s$ to a spin-$j$ dimer will, in general, be different. This, in turn, leads to different ERE parameters in those channels. This is the situation in, e.g., ${}^8$Li, but there is only one spin channel for a spin-zero core, as is the case in ${}^{11}$Be. 
The renormalization conditions that produce Eq.~(\ref{eq:t1}) are:
\begin{eqnarray}
\frac{1}{a^{(j)}_1}&=&\frac{6 \pi \Delta_\pi^{(j)}}{\mu_\sigma (g_{\pi}^{(j)})^2};\\
r^{(j)}_1&=&-w_\pi \frac{6 \pi}{\mu_\sigma^2 (g_{\pi}^{(j)})^2 } - 3 \Lambda.
\end{eqnarray}
Note that $w_\pi > 0 \Rightarrow r_1^{(j)} < 0$. 

\subsection{Properties of the p-wave two-body bound state}

As in the s-wave case (see Eq.~(\ref{eq:a-gamma})) we can use Eq.~(\ref{eq:t1}) to determine the positions of poles in our amplitude, and so infer the energy of any bound states. 
We have:
\begin{equation}
D_\pi(E,{\bs P})=- \frac{6 \pi}{\mu_\sigma^2 (g_{\pi}^{(j)})^2}\frac{1}{r^{(j)}_1 + 3 \gamma_1} 
\frac{1}{E - {\bs P}^2/(2 M_\sigma) + B_1}~\mbox{$+$ regular}\,.
\label{eq:Dpidressed}
\end{equation}
Here $\gamma_1=\sqrt{2 \mu_\sigma B_1}$ is the solution of
\begin{equation}
\frac{1}{a_1^{(j)}} + \frac{1}{2} r_1^{(j)} \gamma_1^2 + \gamma_1^3=0\,.
\label{eq:gamma1}
\end{equation}
The wave-function renormalization for the dressed $\pi$ propagator
can be read off as the residue of the pole:
\begin{equation}
Z_\pi=-\frac{6 \pi}{\mu_\sigma^2 {g_{\pi}^{(j)}}^2}\frac{1}{r^{(j)}_1 + 3 \gamma_1}\,.
\label{eq:Zpi}
\end{equation}

Now, proceeding as we did for the s-wave amplitude (see discussion around Eq.~(\ref{eqn:ancandT-swave})) we can connect this wave-function renormalization to the asymptotic normalization coefficients (ANCs) of the p-wave bound state. This time the Laurent expansion of the interacting Green's function $G(E)$ around $E=-B_1$ takes the form:
\begin{equation}
\langle {\bs k}' |\frac{1}{E-H} |{\bs k}\rangle \stackrel{E\rightarrow-B}{\longrightarrow}{C_1}^2 \times \sum_m  \frac{\phi_{m} (\vec{k}')\phi_m^*(\vec{k})}{E+B_1} + \ldots\ , \label{eqn:ancandT}
\end{equation}
where $C_1$ is the ANC and $\phi_m(\vec{k})$ is the asymptotic
wave function for a p-wave state whose eigenvalue of $J_z$ is $m$. Ignoring spin for the time being, $\phi$'s co-ordinate space representation is:
\begin{equation}
\phi_{m}({\bs r})=\left(1 + \frac{1}{\gamma_1 r}\right) Y_{1m}({\hat r}) \frac{e^{-\gamma_1 r}}{r},
\end{equation}
while\begin{equation}
{C_1^{(j)}}^2={g_{\pi}^{(j)}}^2 \frac{\gamma_1^2 \mu_\sigma^2}{3 \pi} Z_\pi=
-\frac{2\gamma_1^{2}}{r_1^{(j)}+3\gamma_1} \ .
\end{equation}
As a result, the EFT parameters, $\Delta_\pi^{(j)}$ and $g_{\pi}^{(j)}$---or equivalently the scattering parameters $a_1^{(j)}$ and
$r_1^{(j)}$---can be fixed using the p-wave state's two-body binding energy, $B_1$ and the ANC, $C_1$. 
Similar connections between ANCs and effective-range parameters have previously been obtained without the use of EFT, 
e.g. in deriving $S$-factor parameterizations of radiative-capture cross sections \cite{Yarmukhamedov:2002tab,Baye:2000ig}.

In principle {\it ab initio} calculations can be used to determine both  $B_1$ and $C_1$. However, observables in halo systems are extremely sensitive to the two-body binding energy, so
it is always wise to fix $B_1$ from data. Ref.~\cite{Zhang:2013kja} did obtain ANCs from an {\it ab initio} method: Variational Monte Carlo was used there, but any underlying theory of the neutron-core bound
state could be employed. ANCs can also be measured in transfer reactions, see, e.g., Ref.~\cite{Trache:2003ir}.

\subsection{Power counting, pole structure, and bounds from causality}

\label{sec-pwavepc}

In order to analyze the physical meaning of these poles we must specify the sizes and signs of the ERE parameters $a_1$ and $r_1$. For causal scattering there are 
bounds on the parameter $r_1$ which imply that $r_1$ is always negative~\cite{Wigner:1955zz,Madsen:2002zz,Hammer:2009zh,Hammer:2010fw}. Therefore, throughout what follows, we assume $r_1 < 0$. Both signs are possible for $a_1$. 
$a_1$ and $r_1$ have dimensions, respectively, of $1/M^3$ and $M$.
 If natural scattering occurs in the p-waves then the only $M$ that enters here is the high scale $M_{\rm core}$. But, in this circumstance, all three poles of the t-matrix $t_1$ occur at momenta $\sim M_{\rm core}$, and so are inaccessible to the EFT. There is thus no utility to the introduction of dimer fields in this case: all p-wave interactions are perturbative (see Sec.~\ref{sec:CD1nswave} below for a specific example of the order at which they enter a physical process). 

For there to be a low-energy bound state or resonance, one or both of these ERE parameters must contain a power of the low scale, $M_{\rm halo}$. The first paper on Halo EFT assumed 
\begin{equation}
a_1 \sim \frac{1}{M_{\rm halo}^3}, \quad r_1 \sim M_{\rm halo},
\label{eq:Bertulanietal}
\end{equation}
 i.e., that two fine tunings, one in each of these effective-range parameters, were present~\cite{Bertulani:2002sz}. In this situation all three poles in the t-matrix occur at momenta of order $M_{\rm halo}$. These poles are the solutions of Eq.~(\ref{eq:gamma1}). If $a_1 <  0$ then two of them have the form $\pm k_R + i k_I$. They represent a broad, low-energy resonance that occurs at an energy $\sim \frac{M^2_{\rm halo}}{2 \mu_\sigma}$ and has a width of order that energy. The third  pole is a (shallow) bound state. (For full details on the pole structure
 under the power counting (\ref{eq:Bertulanietal}) we refer the reader to Ref.~\cite{Bertulani:2002sz}.) 
 
Reference~\cite{Bedaque:2003wa} pointed out that a low-energy bound state or resonance could still occur even with only one fine tuning. They suggested a scaling:
\begin{equation}
a_1 \sim \frac{1}{M_{\rm halo}^2 M_{\rm core}}, \quad r_1 \sim M_{\rm core}.
\label{eq:Bedaqueetal}
\end{equation}
In fact, the causality arguments referred to above guarantee that, at least for the single-channel case, $r_1$ is negative and at least of size $M_{\rm core}$, since the range of the $nc$ interaction in the underlying theory will typically be $1/M_{\rm core}$. This power counting is consistent with that bound. It is then useful to rewrite $t_1$ as:
\begin{equation}
t_1({\bs k}',{\bs k};E)
=\frac{12 \pi}{\mu_\sigma |r_1|} \frac{k^2 \cos \theta}{k^2  - k_R^2 + i \frac{2 k^3}{|r_1|}} \,,
\label{eq:t1rewritten}
\end{equation}
where we have assumed $r_1 < 0$ and introduced $k_R^2=\frac{2}{a_1 r_1} \sim M_{\rm halo}^2$ as the momentum characteristic of the low-energy pole.
 If $a_1 > 0$ then this is a bound-state pole, and, up to higher-order corrections in $M_{\rm core}/M_{\rm halo}$, $k_R=i \gamma_1$. If $a_1 < 0$ it corresponds to a resonance, centered at $k=k_R$, and with width $\Gamma \approx \frac{2 k_R^3}{\mu_\sigma |r_1|}$. Note that this width---whose presence ensures the amplitude is unitary---is parametrically small: it is of order $\frac{M_{\rm halo}}{M_{\rm core}}$ higher than the resonance energy $E_R\sim \frac{M_{\rm halo}^2}{2 \mu_\sigma}$~\cite{Sakurai}. Irrespective of the sign of $a_1$ the expression (\ref{eq:t1rewritten}) also exhibits a bound-state pole at a momentum $k \approx i |r_1|/2$. We note that this is a real, not a virtual bound state, but that since $r_1 \sim M_{\rm core}$, it is a deep bound state that is outside the domain of validity of Halo EFT. Full formulae for the pole positions in terms of $a_1$ and $r_1$ can be found in Refs.~\cite{Bertulani:2002sz} and \cite{Ji:2014wta}. 

Reference~\cite{Bedaque:2003wa} then argued (following Ref.~\cite{Pascalutsa:2002pi} for the $\Delta(1232)$ resonance in chiral EFT) that the power counting for the amplitude (\ref{eq:t1rewritten}) depends fundamentally on whether $k$ is close to $k_R$ or not. If $|k-k_R| \gg M_{\rm halo}^2/M_{\rm core}$, as will generically be the case, then the LO amplitude becomes
 \begin{equation}
t_1({\bs k}',{\bs k};E)
=\frac{12 \pi}{\mu_\sigma |r_1|} \frac{ k^2 \cos \theta}{k^2  - k_R^2}.
\label{eq:t1LO}
\end{equation}
Note that the spurious deep pole that was present in Eq.~(\ref{eq:t1rewritten}) has disappeared. As long as $|k-k_R| \gg M_{\rm halo}^2/M_{\rm core}$, the imaginary piece of the amplitude, that generates a width for the resonance at $k=k_R$, enters the p-wave amplitude only at NLO: it can be treated in perturbation theory. However, if $|k-k_R|$ is smaller, i.e., we consider a point that is kinematically close to the LO pole, then a perturbative treatment fails, because there $i \frac{2 k^3}{|r_1|}$ is not small compared to $k^2 - k_R^2$. In other words, within a region around the LO poles in the $k$-plane whose radius is $\sim M_{\rm halo}^{3/2}/M_{\rm core}^{1/2}$, this unitarity piece of the amplitude must be resummed. 

If $a_1 > 0$ this region of the $k$-plane lies within the bound-state region. Thus, when scattering is calculated the p-wave interactions can be taken as perturbative (see Eq.~(\ref{eq:delta1pert}) below). This is the situation for the $n$-${}^{10}$Be system. Conversely, if $a_1 < 0$ the pole is in the scattering region, and can be neglected for bound-state calculations. That situation prevails in computations of ${}^6$He that assume the power counting (\ref{eq:Bedaqueetal}). However, in either case, unitarity corrections are essential in certain situations. If $a_1 > 0$ they must be resummed when bound-state properties are computed. Similarly, in, for example, ${}^5$He-neutron scattering, which enters in calculations of resonances in the ${}^6$He system  (cf. Ref.~\cite{Romero-Redondo:2014fya}), it is necessary to resum the unitarity correction, and so generate the finite width of ${}^5$He, within a window around $k=k_R$. Such a resummation moves the resonance pole away from the real axis, thereby rendering the scattering amplitude finite for real $k$. Crucially, this resummation is not necessary for large $k$, and so, if implemented carefully, it need not generate any spurious (deep) poles. The power counting (\ref{eq:Bedaqueetal}) therefore reproduces the pole structure in all of the examples we will consider:  $^5$He ($a_1 < 0$), ${}^8$Li and ${}^{11}$Be (both $a_1 > 0$). Unless otherwise stated, we will assume it throughout the remainder of our article. 

Finally, we give the formulae for scattering phase shifts, in both the resummed and non-resummed cases. Eq~(\ref{eq:t1rewritten}) implies
\begin{equation}
k^3 \cot \delta_1=\frac{1}{2} r_1 (k^2 - k_R^2)\,,
\label{eq:kcotdelta1}
\end{equation}
if no expansions are made. 
Since $|r_1| \gg k, k_R$ we generically have $\cot \delta_1$ large, which implies 
that $\delta_1$ is approximately zero. Indeed
\begin{equation}
\delta_1=\frac{2}{r_1} \frac{k^3}{k^2 - k_R^2} + O\left(
\frac{M_{\rm halo}}{M_{\rm core}}\right)\,,
\label{eq:delta1pert}
\end{equation}
again, p.v. $|k-k_R| \gg M_{\rm halo}^2/M_{\rm core}$. Small phase shifts imply 
weak scattering, which is why the imaginary part of $t_1^{-1}$ 
can be treated perturbatively.

\subsection{Universality and the unitary limit in p-waves}
\label{subsec:wignerP}
An intriguing theoretical aspect of the s-wave ERE is the ability to take the limit $r_0 \rightarrow 0$. This produces an amplitude that is ``universal'', in that it depends only on the two-body scattering length and not on short-distance details of the two-particle potential.
In the p-wave case, however, the causality bound
\cite{Madsen:2002zz,Hammer:2009zh,Hammer:2010fw}
guarantees that---for short-range, energy-independent interactions---$r_1$ is negative and of order the breakdown scale. This
is in accordance with a Wilsonian renormalization group analysis of the p-wave
case~\cite{Harada:2007ua}.

In particular, the causality
bound on $r_1$ in the limit $|a_1|\rightarrow\infty$ reads $r_1 \leq - 2/R$,
where $R$ is the range of the interaction~\cite{Hammer:2009zh,Hammer:2010fw}. 
As a consequence, for short-range, energy-independent interactions,
the ``unitary limit'' $|a_1|\rightarrow\infty$ and $r_1\rightarrow 0$
 is not allowed by
causality.
To let $|a_1|\rightarrow\infty$,
we can take the scale $M_{\rm halo} \rightarrow 0$, which corresponds to sending $k_R \rightarrow 0$. The on-shell amplitude will then assume the form:
\begin{equation}
t_1({\bs k}',{\bs k};E)
=\frac{6 \pi}{\mu_\sigma}\frac{ \cos\theta}{\frac{|r_1|}{2}  + i k} \,,
\label{eq:t1kRgoestozero}
\end{equation}
for $k$ close to zero. The arguments of the previous subsection make it clear that---at least in our preferred power counting---there is no physical justification for taking a purely unitary amplitude as a starting point for analyses of p-wave systems.

In the power counting of Ref.~\cite{Bertulani:2002sz} it seems that an amplitude with $a_1 \rightarrow 0$, $r_1 \rightarrow 0$ might be physically realizable~\cite{Braaten:2011vf}. If we first take the limit $k_R \rightarrow 0$ then, in this power counting, the form (\ref{eq:t1kRgoestozero}) applies throughout the domain of validity of the EFT. The problem with this is that the pole at $k=\frac{i|r_1|}{2}$ then corresponds to a real bound state, and its residue is computed from Eq.~(\ref{eq:t1kRgoestozero}) to be $Z_\pi^{\rm deep}=-\frac{24 \pi}{\mu_\sigma r_1^2} < 0$. The fact that this pole has a negative residue means it does not correspond to a normalizable state~\cite{Nishida:2011np}  (see also Ref.~\cite{Jona-Lasinio:2008}).  For poles that are outside the domain of validity of the theory (i.e. those at momenta $\sim M_{\rm core}$) this does not have any physical consequences. However,
if $r_1 \sim M_{\rm halo}$, then, once $k_R \rightarrow 0$, the effective-range expansion (\ref{eq:t1}) generates a low-energy pole with a negative residue. It follows that the limit $k_R \rightarrow 0$ can only yield a physical scattering amplitude if we have $r_1 \sim M_{\rm core}$, and the amplitude (\ref{eq:t1kRgoestozero}) emerges only in a small region near $k=0$. Requiring that all (physical) poles in the quantum-mechanical theory be normalizable can thus be understood as an alternative (bound-state) version of Wigner's causality argument for p-waves. Eq.~(\ref{eq:Zpi}) also implies that this requirement, when imposed on the shallow bound-state pole, fixes $r_1 < 0$ and $|r_1| > 3 \gamma_1$.

\subsection{Applications: ${}^{11}$Be and ${}^8$Li}

\label{sec:pwavehaloapplications}

The ${}^{11}$Be nucleus has both a shallow s-wave state ($J^P=\frac{1}{2}^+$) and a shallow p-wave bound state ($J^P=\frac{1}{2}^-$). 
The binding energy of the later is the one-neutron separation energy from this state, i.e., $S_{1n}(\frac{1}{2}^-)=B_1$.
The extended nature of these states ensures that ${}^{11}$Be has
a strong enhancement of the E1 transition strength at low excitation energies~\cite{Fukuda:2004ct}.
In this section we examine the EFT treatment of the p-wave bound state, first developed in Ref.~\cite{Hammer:2011ye}. We then discuss the similarities and differences to the case of ${}^8$Li, where the p-wave bound state is the ground state~\cite{Zhang:2013kja,Fernando:2011ts,Rupak:2011nk}. 

Using the value $S_{1n}(\frac{1}{2}^-)=180$ keV from Ref.~\cite{Ajzenberg-Selove:1990}, we 
infer $\gamma_1=17.5$ MeV.
The natural  scale for the core in this system is $M_{\rm core} \approx \hbar c/(2~{\rm fm})=100$ MeV, see Table \ref{tab:1n-halo}. 
The scattering volume $a_1$ of n-${}^{10}$Be scattering in the $\frac{1}{2}^-$ channel extracted in Ref.~\cite{Typel:2004zm}:
\begin{equation}
a_1^{(1/2)}=(457 \pm 67)~{\rm fm}^3,
\label{eq:pwavescattlength}
\end{equation}
is then of the expected size: $a_1^{(1/2)} \sim M_{\rm halo}^{-2} M_{\rm core}^{-1}$.

The LO amplitude (\ref{eq:t1LO}) relates the value of $r_1^{(1/2)}$ to the values of $a_1^{(1/2)}$ and $\gamma_1$:
\begin{equation}
r_1^{(1/2)}=-\frac{2}{\gamma_1^2 a_1^{(1/2)}}
\label{eq:gamma1lo}
\end{equation}
and produces
$r_1^{(1/2)}=(-0.54\pm 0.08)$ fm$^{-1}$. This number should, however, be taken as 
indicative only, although it already shows that $r_1^{(1/2)} \sim M_{\rm core}$. 
At NLO, Eq.~(\ref{eq:gamma1lo}) is corrected to:
\begin{equation}
r_1^{(1/2)}=-\frac{2}{\gamma_1^2 a_1^{(1/2)}} - 2 \gamma_1,
\end{equation}
which, again using Eq.~(\ref{eq:pwavescattlength}), alters $r_1^{(1/2)}$ to $(-0.72\pm 0.08)$ fm$^{-1}$. Such a $\approx$ 30\% 
correction is in line with the anticipated expansion parameter of Halo EFT 
in the ${}^{11}$Be system. 

In the case of ${}^8$Li the core, $\lis$, is a $\frac{3}{2}^-$ nucleus, and so there are two possible spin channels that contribute to the formation of ${}^8$Li: $s=1$ and $s=2$. 
${}^8$Li is a $2^+$ state, which is bound by $2.03$ MeV with respect to the $n$-$\lis$ threshold. This implies the bound-state pole momentum $\gamma_1=58$ MeV. 
$M_{\rm core}$ is associated with the core breakup $\lis\rightarrow
t+{}^{4}\mathrm{He}$ which requires $2.5$ MeV of energy; this translates to $M_{\rm core} \approx 90$ MeV.  

In fact, $\lie$ also has a p-wave excited state ($J^P=1^+$),
$\lie^*$, that is bound by only $1.05$ MeV with respect to the $n$-$\lis$ threshold, and so is an even better halo than the ground state: its bound-state pole momentum is 
 $\tilde{\gamma}_1=42$ MeV. In this system
$\gamma_1$ and $\tilde{\gamma}_1$ are then considered small with respect to $M_{\rm core}$, which yields a nominal expansion parameter
$M_{\rm halo}/M_{\rm core} \approx 0.5$. Note, however, that the result found in Ref.~\cite{Zhang:2013kja}, and reported below, is that the p-wave effective ranges in $j=1$ $\lis$-$n$ scattering are $r_1^{(2)} \approx -1.4$ fm$^{-1}$; this
suggests 
a higher $M_{\rm core}$ and hence a more convergent EFT expansion. 

\begin{figure}[htb]
\begin{center}
\includegraphics[width=0.7\textwidth]{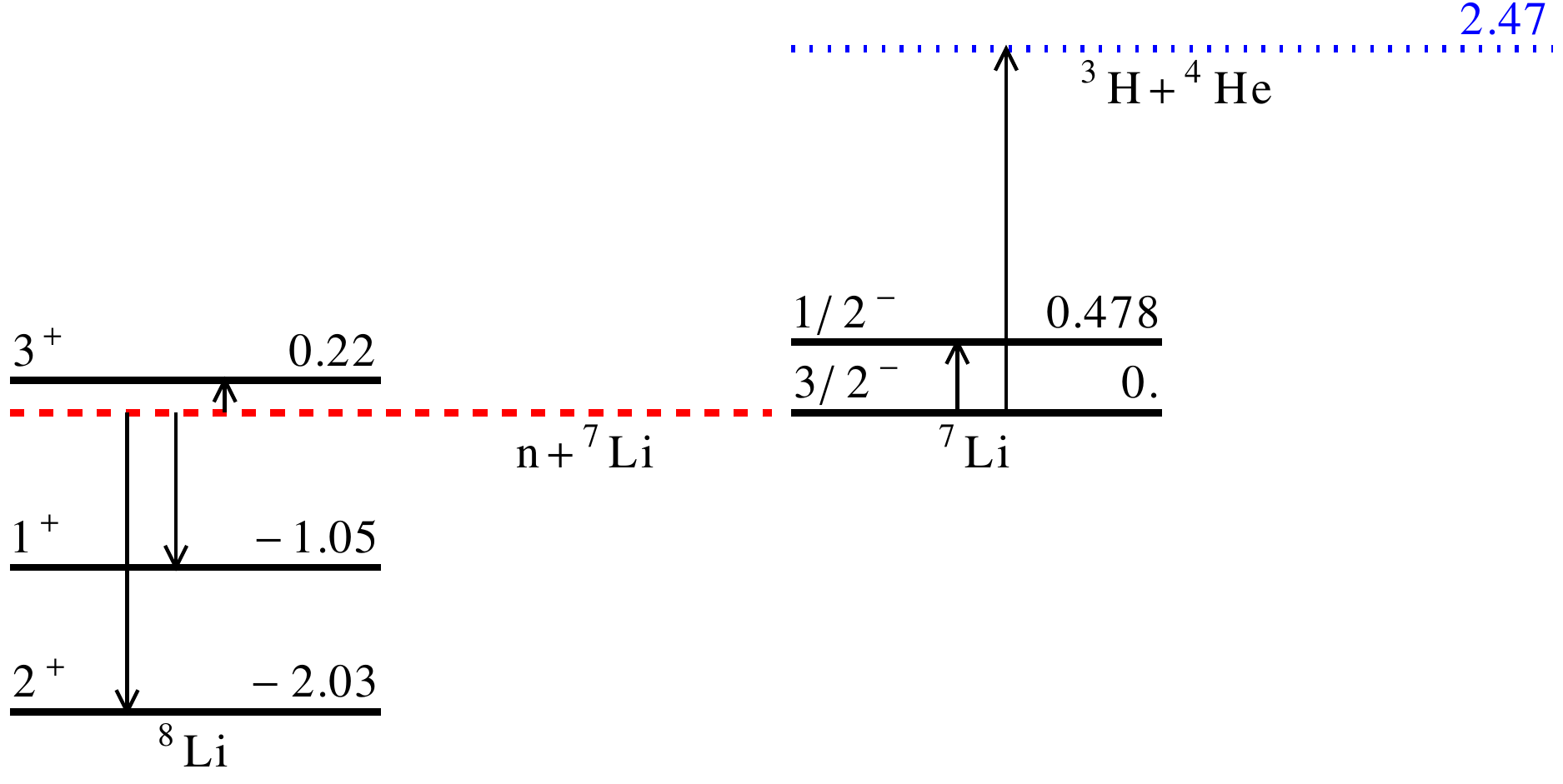}
\end{center}
\caption{Our level scheme for the coupled $n + \lis$-$\lie$ system. Data from Ref.~\cite{TUNL}, with all energies in MeV and measured relative to the $n + \lis$ threshold. The $n-\lis$ threshold is indicated by the red dashed line. The ${}^3$H-${}^4$He threshold in the $\lis$ system is indicated by the blue dotted line. Figure constructed using the LevelScheme scientific figure preparation package~\cite{Caprio:2005dm}.}
\label{fig:lilevelscheme}
\end{figure}

The EFT is designed to work for bound-state energies of 1--2 MeV so
it should also describe $n$-$\lis$ scattering for neutron energies 
in this range. However, at approximately 
$0.22$ MeV above the 
 n-$\lis$
threshold there is a $3^{+}$ resonance.
This
resonance dominates the total cross section in a narrow window around the resonance energy, but does not affect threshold capture into either of the $\lie$ bound states. The 
  $3^{+}$ resonance can be added as an additional dynamical degree of freedom
  in the EFT~\cite{Fernando:2011ts}. The set of states relevant for the low-energy dynamics of the coupled $n + \lis$-$\lie$ system is depicted in Fig.~\ref{fig:lilevelscheme}.
  
Another complication in the $\lie$ system is that the first excited state of $\lis$ ($J^P=1/2^-$) has an excitation energy of only $E_*=0.478$ MeV. 
 Refs.~\cite{Rupak:2011nk,Fernando:2011ts} argued that this state, which we denote here by $\lis^*$, could be integrated out of the EFT. In contrast, Ref.~\cite{Zhang:2013kja} included $\lis^*$ as an explicit degree of freedom in the EFT for ${}^8$Li, since its 
excitation energy is small compared with
the neutron separation energy of $\lie$. 
Ref.~\cite{Zhang:2013kja} argued that the importance of this core excitation is borne out by the ANCs for 
$\lie$ in the asymptotic channels
$n+\lis$ and into $n+\lis^{*}$: they are of similar size.  

However, spin and angular-momentum considerations mean that $\lis^*$ can only contribute in the $s=1$ channel. The effective-range expansion (\ref{eq:t1}) therefore governs $n$-$\lis$ scattering in the ${}^5$P$_2$ channel. 
But, in the ${}^3$P$_2$ channel that conventional ERE only prevails when $k \ll \sqrt{2 \mu_\sigma E_*}$. Ref.~\cite{Zhang:2013kja} derived a modified ERE that accounts for the opening of the ${}^7$Li$^*$-$n$ channel (see also Ref.~\cite{Rakityansky:2011xa}). This modifies the formulae for the ANCs of $\lie$, introducing terms that depend on the difference of $\gamma_1$ and $\sqrt{2 \mu_\sigma E_*}$. It also means that we need three ANCs to completely understand the structure of the ${}^8$Li ground state: the $s=1$ and $s=2$ ANCs for the $n$-$\lis$ asymptotic state, here denoted $C_{(\P{3}{2})}$ and $C_{(\P{5}{2})}$, together with an $s=1$ $\lis^*$-$n$ ANC, which we write as $C_{(\P{3}{2}^*)}$.~\footnote{In the case of $\lie$ we replace the notation $C_1^{(j)}$ for the p-wave ANC by spectroscopic-notation subscripts to make explicit the spin of the different $n$-$\lis$ and $n$-$\lis*$ channels.} $\lie^*$ requires four ANCs for a full description; we indicate these excited-state ANCs by $\tilde{C}_{X}$ where $X$ is a particular asymptotic state. 
 
{\it Ab initio} results for all seven of these ANCs were reported in 
Refs.~\cite{Nollett:2011qf,Zhang:2013kja}, and are reproduced in Table~\ref{tab:lieANCs}.  They were
computed from a Hamiltonian consisting of the Argonne $v_{18}$ two-nucleon
terms \cite{Wiringa:1994wb} and Urbana IX three-nucleon terms
\cite{Pudliner:1995wk} and using the Variational Monte Carlo (VMC) method.  VMC wave functions are not as precise as Green's function Monte
Carlo (GFMC) wave functions, but are quite accurate for many
purposes.  The ANCs in Table~\ref{tab:lieANCs}
typically have an error of $<5\%$ due to Monte Carlo sampling.  There
is a possibly larger but unknown error from the accuracy of the wave
functions and underlying Hamiltonian.  However, limited testing with other {\it ab initio} wave functions suggests 
that this error is no 
larger than the experimental errors \cite{Nollett:2011qf,Nollett:2012fw}.
On the experimental side, $n$-$\lis$ ANCs were  measured in Ref.~\cite{Trache:2003ir}, and these numbers are also given in Table~\ref{tab:lieANCs}. 
By using the theory ANCs (first line of Table~\ref{tab:lieANCs}) the effective range at NLO in the $j=2$ channel where the $\lie$ bound state lives
was determined to be 
$r_1^{(2)}=-1.43(2)~\mathrm{fm}^{-1}$. 
This implies that the radius of convergence of the EFT that describes ${}^8$Li is larger than the naive estimate of 90 MeV. 
However, it is important to note that $|r_1^{(2)} + 3 \gamma_1|$ is markedly smaller than $|r_1^{(2)}|$, and so NLO corrections in the ANCs can give large effects, even though they are nominally of order $\gamma_1/|r_1^{(2)}| \sim 1/5$.  
\begin{table}
   \centering
   \footnotesize
   \begin{tabular}{|c|c|c|c|c|c|c|c|} \hline
           & $C_{(\P{3}{2})}$
           & $C_{(\P{5}{2})}$   
           & $C_{(\P{3}{2}^{*})}$ 
           & $\tilde{C}_{(\P{3}{1})}$ 
           & $\tilde{C}_{(\P{5}{1})}$
           & $\tilde{C}_{(\P{1}{1}^{*})}$
           & $\tilde{C}_{(\P{3}{1}^{*})}$\\  \hline
      VMC 
           & $-0.283(12)$
           & $-0.591(12)$ 
           & $-0.384(6)$
           & $0.220(6)$ 
           & $0.197(5)$
           & $-0.195(3)$ 
           & $-0.214(3)$ \\  \hline
      EXP
           & $ -0.284(23)$
           & $-0.593(23)$ 
           & 
           & $0.187(16)$ 
           & $0.217(13)$
           &  
           &  \\  \hline  
   \end{tabular}
   \caption{${}^8$Li ANCs ($\mathrm{fm}^{-\frac{1}{2}}$) for different channels. For the {\it ab initio} (VMC) ANCs, the $\lis+n$ ANCs can be found in Ref.~\cite{Nollett:2011qf}, while $\lis^{\ast}+n$ ANCs are from Ref.~\cite{Zhang:2013kja} The measured (EXP) $\lis+n$ ANCs are from Ref.~\cite{Trache:2003ir}. The first three ANCs pertain to the ground state of $\lie$, and the last four to $\lie^*$.} \label{tab:lieANCs}
\end{table}

\subsection{Application: low-lying ${}^2$P$_{3/2}$ resonance in ${}^5$He}

\label{sec:2P3/2}

The $n \alpha$ interaction has a low-energy resonance in the ${}^2$P$_{3/2}$ partial wave, as well as an enhanced phase shift in the ${}^2$P$_{1/2}$ where the resonance is somewhat broader. The first EFT treatment of $n \alpha$ scattering was carried out by Bertulani {\it et al.}~\cite{Bertulani:2002sz}, who took both the ${}^2$P$_{3/2}$ scattering volume $a_1^{(3/2)}$, and the corresponding effective ``range", $r_1^{(3/2)}$, as unnaturally enhanced. In contrast, Bedaque {\it et al.}~\cite{Bedaque:2003wa} showed that the ${}^2$P$_{3/2}$ phase shift could be described using the power counting of Ref.~\cite{Pascalutsa:2002pi}, where the resonance's width is only re-summed in its immediate vicinity. 
And indeed, 
the $n \alpha$ ${}^2$P$_{3/2}$ scattering parameters $a_1^{(3/2)}=-62.951(3)$
fm$^3$ and $r_1^{(3/2)}=-0.8819(11)$ fm$^{-1}$~\cite{Arndt:1973ssf} are consistent with the low- and
high-momentum scales $M_{\rm halo} \approx 30$ MeV and $M_{\rm core} \approx
\sqrt{m_n E^*_{\alpha}}=140$ MeV in ${}^6$He. (A more recent analysis of n$\alpha$ data gives $a_1^{(3/2)}=-65.7$ fm$^3$, $r_1^{(3/2)}=-0.84$ fm$^{-1}$~\cite{GHale}, but this barely affects the scales inferred.)
In the ${}^2$P$_{1/2}$ channel we have $a_1^{(1/2)}=-13.821(68)$ fm$^3$ and $r_1^{(1/2)}=-0.419(16)$ fm$^{-1}$.
Ref.~\cite{Bedaque:2003wa} took these parameters to be natural, i.e. $a_1^{(1/2)} \sim 1/M_{\rm core}^3$, $r_1^{(1/2)} \sim M_{\rm core}$.

\begin{figure}[htb]
\begin{center}
\includegraphics[width=5in]{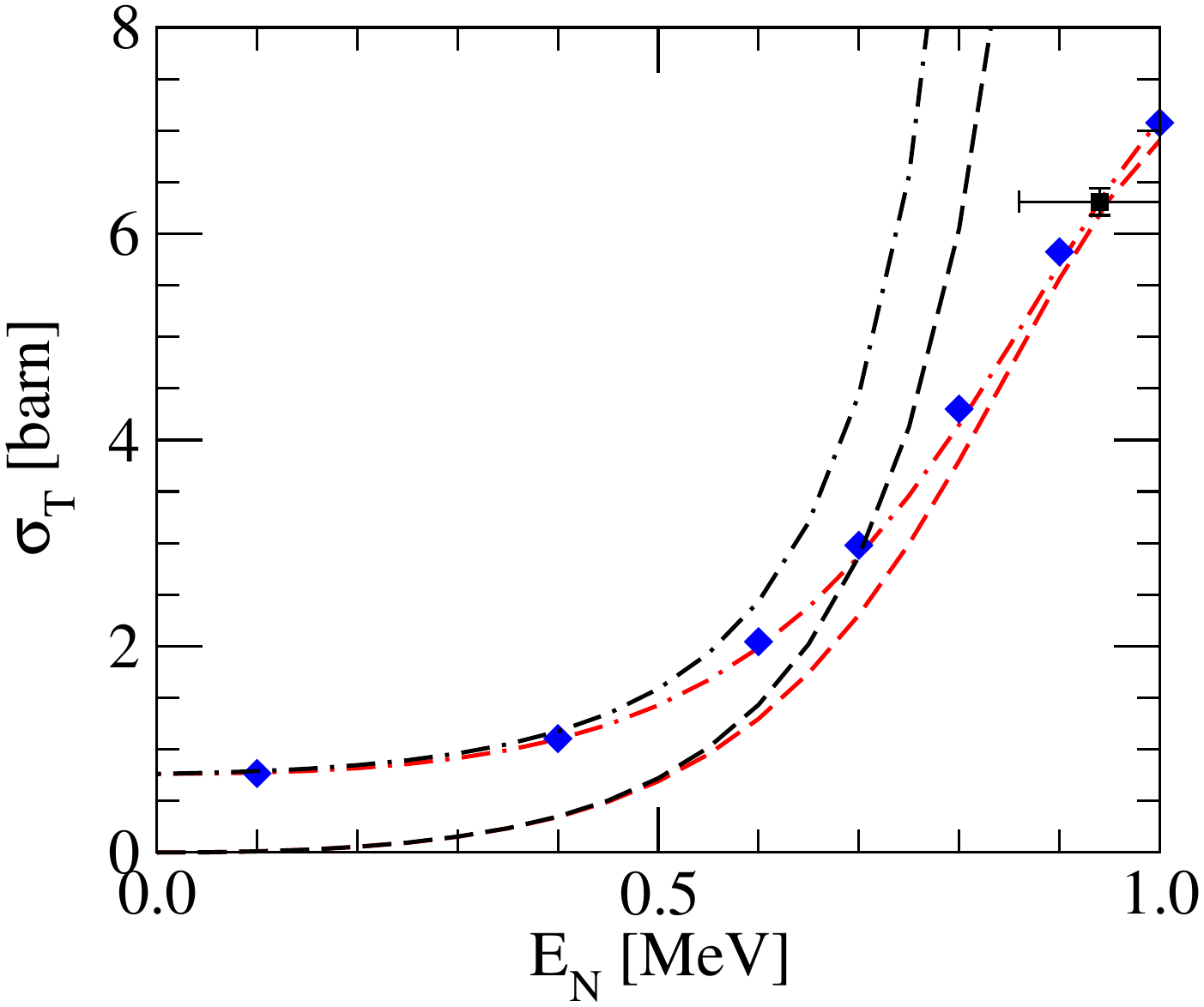}
\end{center}
\vspace*{-0.5cm} 
\caption{The total cross section for $n\alpha$ scattering
 (in barns) as a function of the neutron kinetic energy in the
 lab. frame (in MeV). The diamonds are evaluated data from
 Ref.~\cite{BNL}, and the black squares are experimental 
 data from Refs.~\cite{Haesner:1983zz,Battat:1959}.
 The dashed and dash-dotted black lines show the result in the power counting where $r_1 \sim M_{\rm core}$ at LO and NLO respectively. 
The unitarity part of the p-wave amplitude must be resummed in the vicinity of the resonance, which leads to the  
 red dashed and dash-dotted lines~\cite{Bertulani:2002sz} at LO and NLO. Figure adapted from Ref.~\cite{Bedaque:2003wa}}
\label{fig:sigtot}
\end{figure}

The $M_{\rm halo}/M_{\rm core}$ expansion then works quite well for
some low-energy quantities.
Fig.~\ref{fig:sigtot} shows the results for the total cross section
as a function of the neutron kinetic energy in the
lab. frame.
The power counting discussed above produces the dashed black line at LO and the dash-dotted line at NLO.
The LO calculation incorporates only the ${}^2$P$_{3/2}$ channel, and omits the unitarity correction there. 
It does not do a good job of describing the scattering data, in part because it has the wrong threshold behavior. 
This is remedied at NLO, where the scattering length in the 
$^2$S$_{1/2}$ channel enters, and the unitarity correction in the ${}^2$P$_{3/2}$ channel is a perturbative effect. 
The data are then reproduced up to neutron
energies of about $E_n=0.5$ MeV.
At this energy one approaches the ${}^2$P$_{3/2}$ resonance, and so the $n \alpha$ loop effect $\sim i k^3$ can no longer
 be treated in perturbation theory. In this region we need to resum these loops, thereby generating the resonance width.
We show the first two orders of the resummed result
as the red dashed (LO) and dash-dotted (NLO) lines. At NLO this reproduces the 
NLO calculation of  Ref.~\cite{Bertulani:2002sz}.
The resonance shape is well described by this calculation, as is the differential cross section. 

\subsection{The three-body equations in the case of ${}^6$He}

\label{sec:3Bpwave}

\subsubsection{Scales and quantum numbers}

The strong interaction between the neutron and ${}^4$He in the ${}^2$P$_{3/2}$ channel leads us to consider a three-body system consisting of an alpha particle and two neutrons. This is a Borromean system, in that 
no pair of particles of these three forms a bound state, and yet there is a three-body bound system, namely ${}^6$He, which is bound by 0.975 MeV. ${}^6$He was treated using the Gamow Shell Model in Halo EFT in Ref.~\cite{Rotureau:2012yu}, but that work used the power counting (\ref{eq:Bertulanietal}), i.e. it assumed a fine-tuned $r_1^{(3/2)}$. Ref.~\cite{Ji:2014wta} derived the three-body equations for this system for the case (\ref{eq:Bedaqueetal}), i.e. $r_1^{(3/2)} \sim M_{\rm core}$, and it is those results we present here. Note that Ref.~\cite{Ji:2014wta} took $n \alpha$ interactions in all other partial waves, including the ${}^2$S$_{1/2}$, to be zero at leading order, as was also assumed in Sec.~\ref{sec:2P3/2}. This is different from the power counting of Ref.~\cite{Bedaque:2003wa}, where s-wave neutron-$\alpha$ scattering is a LO effect. See also Ref.~\cite{Ryberg:2017tpv} for a recent treatment of ${}^6$He in Halo EFT, which considers other, different formulations of the problem at LO. 

The ground-state of $^6$He has total angular momentum and parity $J^P=0^+$, and so we project our three-body equations onto that state.
(For the three-body equations in other total-$J$ channels see Refs.~\cite{Braaten:2011vf,Jona-Lasinio:2008}.)
As in Sec.~\ref{sec:3Bswave} we use Jacobi-momenta $\boldsymbol{q}_i$, and $\boldsymbol{p}_i$ to represent the internal
kinematics of the three-body system in the center-of-mass frame. The relative orbital angular momentum, spin, and total angular momentum of the pair $(jk)$, as well as the relative orbital angular momentum and spin between the spectator $i$ and the pair $(jk)$ are also all denoted as in Sec.~\ref{sec:3Bswave}. 
With the $\alpha$-core as the spectator, we obtain $l_\alpha=s_\alpha=j_\alpha=0$, since the $nn$ interaction is dominated by the $^1$S$_0$ virtual state.
Furthermore, at LO $\lambda_\alpha=\varsigma_\alpha=0$ and it is then straightforward to determine that $S_\alpha=L_\alpha=0$ in the $(\alpha, nn)$ partition. 
Alternatively, if we choose a neutron as the spectator, the $n\alpha$ interaction is dominated by the $^2$P$_{3/2}$ resonance, which means $l_n=1$, $s_n=1/2$ and $j_n=3/2$. In the positive parity $^6$He ground state, the spectator neutron must also interact with the $n\alpha$ pair in a p-wave. This results in $\lambda_n=1$, $\varsigma_n=1/2$. In the 
$(n, n\alpha)$ partition, the spin-spin and orbit-orbit couplings then produce two possibilities that contribute to the ${}^6$He $0^+$ ground state: the overall orbital angular momentum and overall spin can either be both zero ($L_n=S_n=0$) or both one ($L_n=S_n=1$). This defines the states in the basis (\ref{eq:quantum-number}) that are of relevance for our calculation of ${}^6$He. 

\subsubsection{Equations for the p-wave three-body bound state}

To obtain the equations that describe the ${}^6$He bound state we follow the development of Sec.~\ref{sec:3Bswave}. Diagrammatically the equations for the ${}^6$He problem are exactly as in Fig.~\ref{fig:faddeev-eq}. The key difference is technical: this time we want to project one of our equations---that for $\tilde{\mathcal A}_n({\bs q})$---onto a relative p-wave between the neutron and the $n \alpha$ pair. 
By projecting the Faddeev components $\tilde{\mathcal A}_\alpha$ and $\tilde{\mathcal A}_n$ onto the partial-wave-decomposed states 
in their respective partitions we obtain  coupled-channel integral equations for the $^6$He ground state that are formally identical to   Eqs.~(\ref{eq:faddeevswave1}) and (\ref{eq:faddeevswave2}):
\begin{eqnarray}
\label{eq:Fc-faddeev}
\fl
\tilde{\mathcal A}_{\alpha}(q) &=& \frac{2}{\pi} \int^\Lambda_0 q'^2 dq'\, X_{10}^n(q,'q;B_3)\,  \tau_\pi(q';B_3)\, \tilde{\mathcal A}_{n}(q')~;
\\
\label{eq:Fn-faddeev}
\fl
\tilde{\mathcal A}_{n}(q) &=& \frac{1}{\pi} \int^\Lambda_0 q'^2 dq'\, X_{10}^n(q,q';B_3)\, \tau_d(q';B_3)\, \tilde{\mathcal A}_{\alpha}(q')
\nonumber\\
\fl
&&+ \frac{1}{\pi} \int^\Lambda_0 q'^2 dq'\, \left[X_{11}^\alpha(q,q';B_3) + \frac{q q'}{\Lambda^2}{H}(\Lambda)\right]\, \tau_\pi(q';B_3)\, \tilde{\mathcal A}_n(q')~.
\end{eqnarray}
Here the ultraviolet cutoff, $\Lambda$, is introduced for regularization. Note that a three-body force appears here, just as it did in the s-wave case, although now it corresponds to Eq.~(\ref{eq:L3p}), and so is a p-wave interaction
between the neutron and the ${}^5$He dimer. An $nn\alpha$ interaction of this form is  consistent with the Pauli exclusion principle.

The s-wave and p-wave equations differ only in the expressions  for the dimer propagators $\tau$ and for the functions $X^{\alpha}$ and $X^n$ that encode the exchange of particles between different clusters. 
Since we are no longer dealing with a purely s-wave problem the exchange functions now include non-trivial recoupling coefficients $\,_{i}\langle p, q; \Omega_i |p', q'; \Omega_j\rangle_{j}$~\cite{Glockle:1983} as well as a factor(s) of the pair relative momentum if one or both vertices involves a p-wave state. But the result is still essentially an angular average of the pertinent free-particle Green's function. The result of that calculation (see Ref.~\cite{Ji:2014wta} for details) is:
\begin{eqnarray}
\fl
X_{10}^n(q,q';B_3) 
=& -\sqrt{2}\, \left[\frac{A}{A+1}\, \frac{1}{q'}\, \textrm{Q}_0 (z_{n\alpha}) + \frac{1}{q}\, \textrm{Q}_1 (z_{n\alpha})\right],\nonumber
\\
\fl
X_{11}^\alpha(q,q';B_3) 
=& -A\, \left[\frac{A^2+2A+3}{(A+1)^2}\,\textrm{Q}_0(z_{n n})
+\frac{2}{A+1} \frac{q^2+q'^2}{qq'}\,\textrm{Q}_1(z_{n n})
+ \textrm{Q}_2(z_{n n})\right],
\label{eq:faddeev-X}
\end{eqnarray}
where  $\textrm{Q}_l$ are the Legendre functions of the second kind, defined in Sec.~\ref{sec:3Bswave}. $z_{nn}$ is also the same here as in that section, see Eq.~(\ref{eq:znn-znc}), and $z_{n \alpha}=z_{nc}$ for the special case of an $\alpha$ core. 

Lastly, the dimer propagator $\tau_d$ is the intermediate-state propagator for the $nn$ state ($\alpha$ spectator) and is defined as in 
Sec.~\ref{sec:3Bswave}. Meanwhile, $\tau_\pi$ is the intermediate-state propagator of the p-wave dimer ($n$ spectator). It takes the form:
\begin{eqnarray}
\tau_\pi(q;B_3) &=& -\frac{2}{r_1}\left(\frac{A+1}{A}\right)\frac{1}{\frac{2A}{A+1}\left(m_n B_3 + \frac{A+2}{2(A+1)}q^2\right) +k_R^2}~.
\end{eqnarray}

\begin{figure}[ht]
\centerline{\includegraphics[width=14cm,angle=0,clip=true]{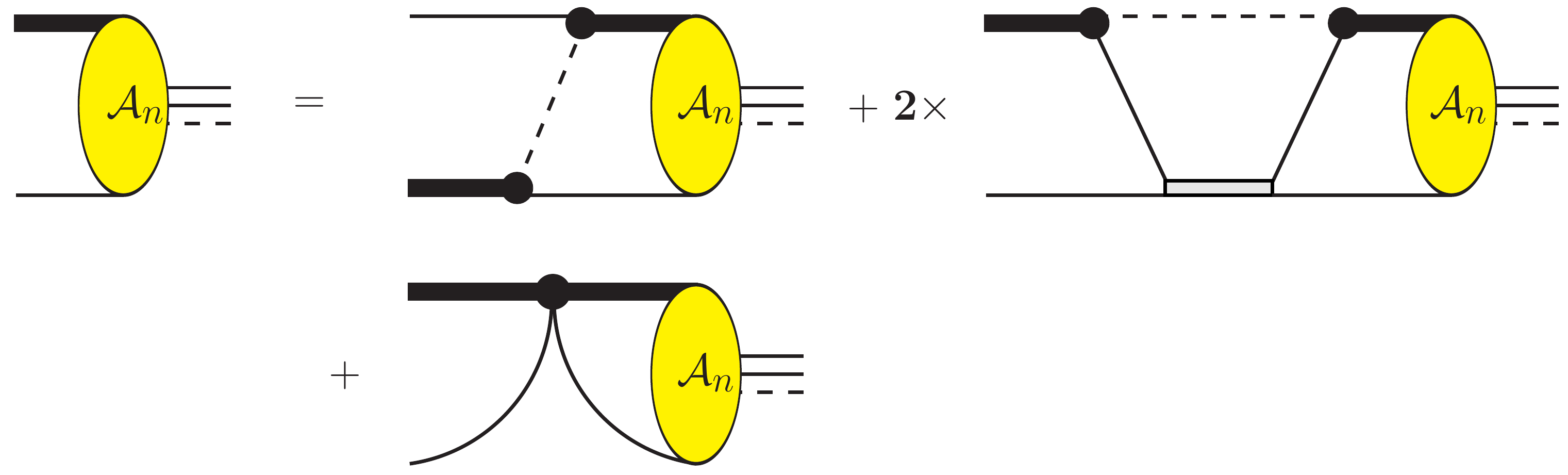}}
\caption{(Color online) The single-channel Faddeev equation for the $^6$He bound state with the addition of a $nn\alpha$ counterterm in the spectator-$n$ partition.}
\label{pic:single-3bf}
\end{figure}

Inserting Eq.~\eqref{eq:Fc-faddeev} into~\eqref{eq:Fn-faddeev} produces a single-channel integral equation that includes only the Faddeev component $\tilde{\mathcal A}_{n}$: 
\begin{eqnarray}
\label{eq:single-Fn}
\fl
&\tilde{\mathcal A}_{n}(q) = \frac{1}{\pi} \int^\Lambda_0 q'^2 dq'\, \left[X_{11}^\alpha(q,q';B_3) + \frac{q q'}{\Lambda^2}{H}(\Lambda)\right]\, \tau_\pi(q';B_3)\, \tilde{\mathcal A}_n(q')
\nonumber\\
\fl
&\; \; +\frac{2}{\pi}  \int^\Lambda_0 q'^2 dq'\, \left[\frac{1}{\pi} \int^\Lambda_0 q''^2 dq''\, X_{1 0}(q,q'';B_3)\, \tau_d(q'';B_3)X_{10}(q',q'';B_3)\right]
\tau_\pi(q';B_3)\, \tilde{\mathcal A}_n(q').\nonumber\\
\fl&
\end{eqnarray}
The resulting integral equation is llustrated in Fig.~\ref{pic:single-3bf}.

Eq.~(\ref{eq:single-Fn}) allows us to point out a possible inconsistency of this treatment. In our power counting both $X_{11}^\alpha$ and $\int X_{1 0} \tau_d X_{1 0}$ are of order $Q^0$ while $\tau_\pi$ scales as $M_{\rm core}^{-1} Q^{-2}$, with $Q$ the typical momentum in the loop integral. It follows that each iterate of the integral equation  is suppressed by one power of $Q/M_{\rm core}$ compared to the previous one. If  the three-body equation is properly renormalized, i.e., only momenta of order $M_{\rm halo}$ contribute to the loop integrations, this then leads to the conclusion that there are no ${}^6$He bound states. 

Clearly this conclusion is not correct, since ${}^6$He exists.  The power counting of Ref.~\cite{Bertulani:2002sz}, which requires two fine-tunings in the $n\alpha$ sector, does not produce this dilemma in the three-body sector. But, in the power counting of Ref.~\cite{Bertulani:2002sz} the leading-order ${}^2$P$_{3/2}$ dimer propagator includes the unitarity piece of the $n \alpha$ amplitude. The corresponding calculation for ${}^6$He was carried out in Ref.~\cite{Rotureau:2012yu}. 

\subsubsection{Renormalization of the $^6$He ground state}

\begin{figure}[ht]
\centerline{\includegraphics[width=12cm,angle=0,clip=true]{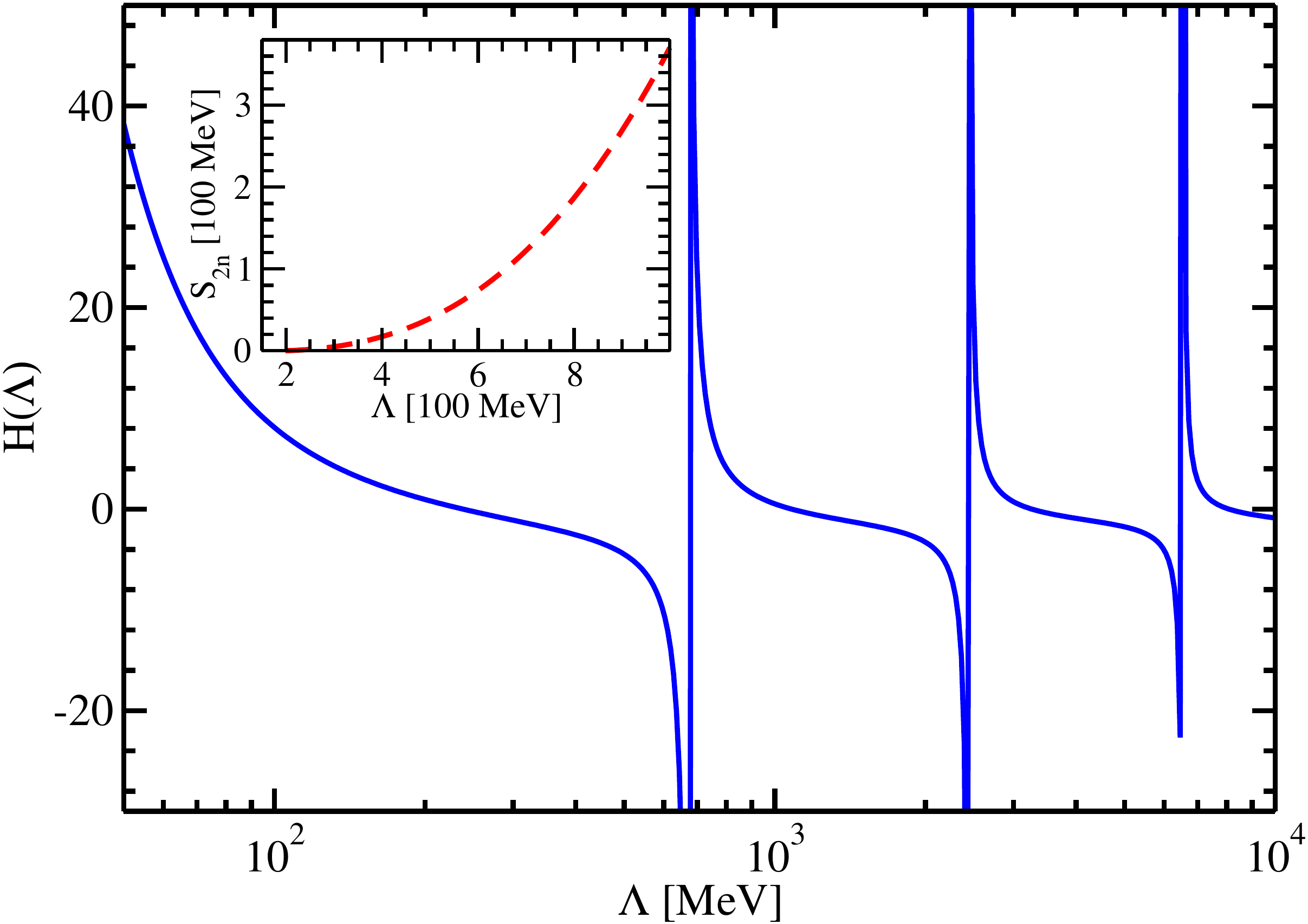}}
\caption{(Color online) The $nn\alpha$-counterterm parameter ${H}$ as a function of the cutoff $\Lambda$. ${H}$ is tuned to reproduce $S_{2n}=0.975$ MeV at each value of $\Lambda$.  The inner panel shows the result for the $^6$He two-neutron separation energy $S_{2n}$ as a function of the cutoff $\Lambda$, if the equations are solved with only two-body interactions. Figure adapted from Ref.~\cite{Ji:2014wta}.}
\label{pic:H0-he6}
\end{figure}

First we consider the case ${H}(\Lambda)=0$. With the hard cutoff $\Lambda$ imposed on the integrals of Eqs.~(\ref{eq:Fc-faddeev}, \ref{eq:Fn-faddeev}), Ref.~\cite{Ji:2014wta} obtained $B_3$ as a function of $\Lambda$. This cutoff dependence is illustrated in Fig.~\ref{pic:H0-he6}: $B_3$ behaves approximately as $\Lambda^{3}$ at values of $\Lambda$ that are large compared to $k_R$, $\gamma_{0,\sigma}$, and $\sqrt{2 m_n B_3}$. 

It is therefore essential for the $nn \alpha$ interaction $\sim {H}(\Lambda)$ to be present if the Halo EFT description is to yield meaningful predictions for ${}^6$He. 
The three-body force parameter ${H}(\Lambda)$ is  tuned to reproduce the $^6$He ground-state two-neutron separation energy $B_3 = 0.975$ MeV for all values of $\Lambda$. Note that this means that Halo EFT {\it cannot} predict the ${}^6$He binding energy. The final value $B_3=0.975$ MeV is influenced by physics at scale $M_{\rm core}$ to a large enough extent that it must be imposed on the theory by hand. Correlations of this binding energy---and other input parameters in the EFT---with other ${}^6$He observables can then be explored. 

 In Fig.~\ref{pic:H0-he6} we plot the ${H}(\Lambda)$ that produces the experimental ${}^6$He binding energy. The oscillatory behavior in $\log \Lambda$ is reminiscent of the three-body force's behavior in the leading-order s-wave problem (see Ref.~\cite{Bedaque:1998kg} and Sec.~\ref{sec:3Bswave}). However, here the period of ${H}(\Lambda)$ in $\log \Lambda$ decreases as $\Lambda$ increases. This difference in the behavior of ${H}$ may well arise from the $n\alpha$ p-wave interaction in the $^6$He system: the symmetry of discrete scale invariance, present in three-body systems with resonant s-wave interactions, is broken by this p-wave interaction.

\begin{figure}[ht]
\centerline{\includegraphics[width=12cm,angle=0,clip=true]{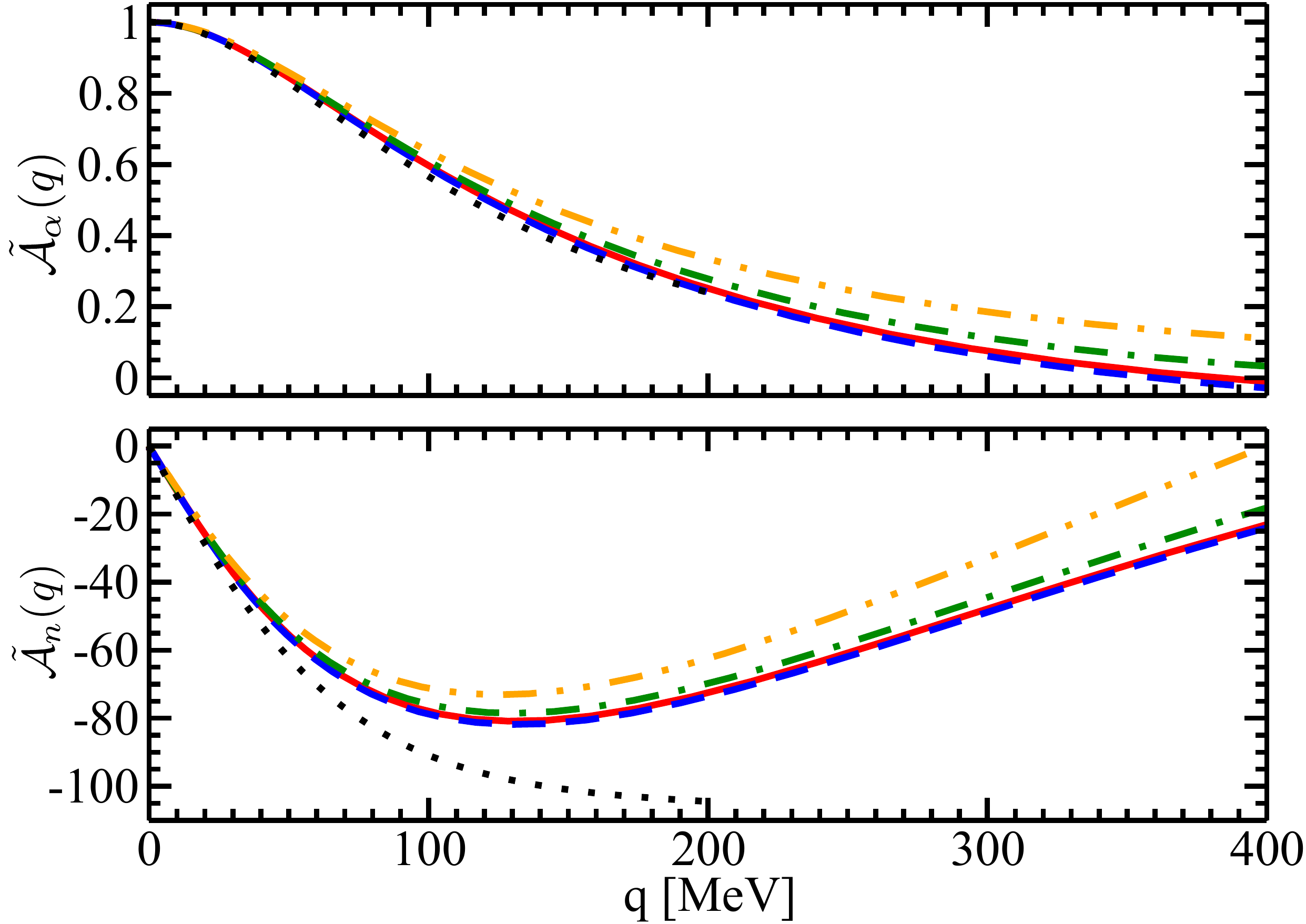}}
\caption{(Color online) The Faddeev components $\tilde{\cal A}_\alpha$ and $\tilde{\cal A}_n$ as functions of $q$, calculated with cutoff parameters $\Lambda$ at 200 MeV (black dotted line), 400 MeV (orange dot-dot-dashed line), 800 MeV (green dot-dashed line), 1.6 GeV (blue dashed line) and 3.2 GeV (red solid line). The Faddeev components are normalized to $\tilde{\cal A}_\alpha(0)=1$. Figure from Ref.~\cite{Ji:2014wta}}
\label{pic:FcFn}
\end{figure}

Once the $^6$He binding energy is renormalized the Faddeev components $\tilde{\cal A}_n(q)$, $\tilde{\cal A}_\alpha(q)$ can be calculated from Eq.~\eqref{eq:single-Fn} and Eq.~\eqref{eq:Fc-faddeev}. Fig.~\ref{pic:FcFn} shows the Faddeev components $\tilde{\cal A}_\alpha$ and $\tilde{\cal A}_n$ as functions of the momentum $q$ for different values of $\Lambda$. The cutoff dependence of the low-$q$ part of both $\tilde{\mathcal A}_\alpha(q)$ and $\tilde{\mathcal A}_n(q)$ is weak for $\Lambda > 200$ MeV.

The integral equation (\ref{eq:single-Fn}) is then renormalized. It generates a shallow bound state, with characteristic momenta $\sim M_{\rm halo}$. One three-body parameter (e.g., $S_{2n}$) is needed for renormalization of the LO equations that describe $^6$He in Halo EFT.  This conclusion remains unchanged even if a different form of the three-body force is chosen~\cite{Ryberg:2017tpv}. 

\subsection{Absence of the Efimov effect}

Resonant pairwise p-wave interactions under the power counting 
of Ref.~\cite{Bertulani:2002sz},
were also considered by Braaten {\it et al.}~\cite{Braaten:2011vf}. These authors attempted to find a scale-free situation in the two-body problem, and examine the corresponding behavior in the three-body problem. In order to do so they took a p-wave ``unitary limit" $|a_1|\rightarrow\infty$ and $r_1\rightarrow 0$. However, as discussed above, this p-wave unitary limit is not physical: it yields a two-body spectrum in which one low-energy state has negative norm. Thus the discrete scale invariance, and corresponding Efimov effect, discovered by Braaten {\it et al.} in the corresponding three-body problem cannot be realized in nature. This has been shown in Refs.~\cite{Jona-Lasinio:2008,Nishida:2011np}, implying that the claim of Ref.~\cite{Macek:2006}, that the Efimov effect can occur with p-wave interactions, is not correct. Nishida and Tan have emphasized that this is not surprising: in three dimensions the p-wave case necessarily involves the introduction of a particular scale, namely $r_1$, into the two-body interaction, thus the discrete scale invariance that is key to the Efimov effect's appearance cannot be present~\cite{Nishida:2011np,Nishida:2011ew}.
This is in line with the arguments based on the causality bound
for p-waves discussed in subsection \ref{subsec:wignerP}.

\subsection{Higher partial waves}

The power counting for resonant partial waves beyond the p-wave was
first discussed
in Refs.~\cite{Bertulani:2002sz,Bedaque:2003wa}.
Bertulani et al.~\cite{Bertulani:2002sz} performed an analysis of the
power divergences of the one-loop self energy of a dimer field
with resonant interactions and 
concluded that for $l\geq 1$ the first $l+1$ effective range parameters
are required to absorb all divergences. This
agrees with a Wilsonian renormalization group
analysis~\cite{Harada:2007ua} which found two relevant parameters for the
p-wave case and three relevant parameters for the d-wave case.

In Ref.~\cite{Bertulani:2002sz}, proper renormalization
of the $l$th partial wave was
achieved by requiring that the first $l+1$ effective range
parameters must scale with $M_{halo}$, while all others
scale with $M_{core}$. This power counting scenario requires $l+1$
fine tunings of parameters in the underlying theory
to achieve the desired scaling, which makes resonant interactions in
higher partial waves quite unlikely.
Subsequently, an alternative power counting that
requires a minimal number of fine tunings---one---in order to produce a low-energy resonance/bound state in any partial wave
was constructed~\cite{Bedaque:2003wa}. Power divergences
beyond $\Lambda^3$ are suppressed by construction in this counting,
which also leads to a simplified pole
structure of the dimer propagator. A detailed discussion of these
two scenarios in the p-wave case was given in subsection~\ref{sec-pwavepc}.

In Ref.~\cite{Braun:2016ggm}, an alternative power counting for d-wave
bound states
was proposed and applied to the description of the $J^P=5/2^+$ excited
state of $^{15}$C and its E2 transitions to the  $J^P=1/2^+$ ground
state. This power counting requires two fine tunings. If it is
extended to arbitrary $l \geq 1$, $l$ fine tunings are required
in a given partial wave.
An earlier application of Halo EFT to d-waves was carried out in
Ref.~\cite{Brown:2013zla}. The authors studied the reaction
$d + t \to n + \alpha$ and found it to be dominated by
a ${}^5$He, $J^P=3/2^+$ intermediate state which couples to
an $\alpha n$ pair with $l=2$ in the final state. However, they used
dimensional regularization with minimal subtraction which sets all
power-law divergences automatically to zero and thus may have missed
some contributions.

\section{Electromagnetic reactions on halo nuclei} \label{sec:emprocesses}
\subsection{Lagrangian: electromagnetic sector}

\label{sec:EML}

Photons are  included in the Lagrangian (\ref{eq:swaveL}) and (\ref{eq:pwaveL}) via minimal substitution:
\begin{equation}
\partial_\mu \rightarrow D_\mu=\partial_\mu + i e \hat{Q} A_\mu.
\label{eq:minimal}
\end{equation}
The charge operator $\hat{Q}$ takes different values, depending on whether 
it is acting on a $c$ field or an $n$ field.
$\hat{Q}\, n=0$ for the neutron, and we denote the eigenvalue of the 
operator $\hat{Q}$ for the $c$ field as $Q_c$.  $e^2 = 4 \pi \alpha_{em}$ defines the unit of electric charge in terms of the fine-structure constant, 
$\alpha_{em}=1/137.036$. 

In this review our focus is on electric properties of halo nuclei. For s- and p-wave halos the dominant pieces of the electric response and electric form factor can be computed
using only the Lagrangians previously introduced and the minimal substitution (\ref{eq:minimal}). Consequently these quantities are predictions---at least at LO, and in some cases for 
several orders beyond that---of the EFT. Nevertheless, eventually there will come an order in the computation where operators involving the 
electric field ${\bs E}$ and the fields 
$c$, $n$, $\sigma$, and $\pi$ which are gauge invariant by themselves 
contribute to observables.

Possible one- and two-derivative operators with one power of the photon 
field which involve two dimer fields are~\cite{Hammer:2011ye}:
\begin{eqnarray}
\fl
\mathscr{L}_{EM}^{\sigma \pi}=-L_{C0}^{(\sigma)} \sigma_\delta^\dagger (\nabla^2 A_0 - 
\partial_0 ({\bs \nabla} \cdot {\bs A}))\sigma_\delta
- L_{E1}^{(j)} \sum_{\alpha \delta} \sigma_\delta \pi^{(j) \, \dagger}_\alpha \left(\left.
\frac{1}{2} \delta j \alpha \right|1 k\right) (i \nabla_k A_0 - i \partial_0 A_k)
\nonumber\\
\fl
\qquad - L_{C0}^{(\pi)} \sum_\alpha \pi^{(j) \, \dagger}_\alpha (\nabla^2 A_0 - \partial_0 ({\bs \nabla} \cdot 
{\bs A})) \pi^{(j)}_\alpha + {\rm h.c.}. 
\label{eq:nonminimalEMdimers}
\end{eqnarray}
These operators enter in, respectively, s-wave-to-s-wave, s-wave-to-p-wave, and p-wave-to-p-wave transitions. 
We have chosen 
to write down the versions for a core spin of zero, so $\sigma$ is always a spin-$1/2$ field, and we omit its
superscript here and in what follows. 
The p-wave operators only occur in channels with a low-energy resonance or shallow bound state
so $j$ could be either $1/2$ or $3/2$, depending on the system under consideration. E1 transitions can still mediate an s-wave-to-p-wave transition even if there is no such unnatural enhancement of 
the p-wave scattering though, and in that case there is a non-minimal operator that does not involve the dimer $\pi$:
\begin{equation}
\mathscr{L}_{EM}^{\bar{j}}=- L_{E1}^{(\bar{j})} \sum_{\delta \alpha k} \sigma_{\delta} [n (i 
\stackrel{\leftrightarrow}{\partial}) c]^{(1/2 \, \bar{j}) \, \dagger}_\alpha \left(\left.
\frac{1}{2} \delta \bar{j} \alpha \right|1 k\right) (i \nabla_k A_0 - 
i\partial_0 A_k).
\label{eq:nonminimalEMnatpwave}
\end{equation}
The inclusion of the factor of $i$  in the operators in Eqs.~(\ref{eq:nonminimalEMdimers}) and (\ref{eq:nonminimalEMnatpwave}) that involve the electric field ensures that the 
Hermitian conjugation produces a time-reversal-invariant Lagrangian for real values of the couplings.
This factor of $i$ was omitted in Ref.~\cite{Hammer:2011ye}, but it makes
no difference to the final results presented there. 
Note that if we were going to consider magnetic properties we would follow a similar strategy, but in that case
would construct 
operators involving $i \nabla_i A_j - i \nabla_j A_i$ and the neutron, core, 
and bound-state fields. 

To determine the order at which these additional electric-field operators enter particular processes 
we rewrite the Lagrangian (\ref{eq:nonminimalEMdimers}) and (\ref{eq:nonminimalEMnatpwave}) in terms of 
rescaled fields~\cite{Beane:2000fi,Hammer:2011ye}. We then employ naive dimensional analysis with these fields 
for the operators that appear in $\mathscr{L}_{EM}$. This procedure generates the following results for the 
operators written above. 
\begin{eqnarray}
L_{C0}^{(\sigma)} &\sim& \frac{g_\sigma^2 \mu_\sigma^2}{M_{\rm core}^{3}} l_{C0}^{(\sigma)} \,,\label{eq:LC0sigma}\\
L_{E1}^{(j)} &\sim& \frac{g_\sigma g_\pi \mu_\sigma^2}{M_{\rm core}} l_{E1}^{(j)} \,,\\
L_{C0}^{(\pi)} &\sim& \frac{g_\pi^2 \mu_\sigma^2}{M_{\rm core}} l_{C0}^{(\pi)}\,,\\
L_{E1}^{(\bar{j})} & \sim & \frac{g_\sigma \mu_\sigma}{M_{\rm core}^4} l_{E1}^{(\bar{j})}\,,
\end{eqnarray}
where the dimensionless parameters $l_{...}^{...}$ are all of order one.

\subsection{Coulomb Dissociation of a one-neutron s-wave halo}
  
 \label{sec:CD1nswave}
 
We first derive the dipole transition strength, $B(E1)$, for the excitation of a one-neutron halo to the core + neutron continuum state. This quantity is probed in Coulomb dissociation experiments, where the one-neutron halo is accelerated to high energies and then impinges, typically peripherally, on a target with a high charge, $Q_t$. We quote  results from the reaction theory of Coulomb dissociation, which allows us to connect the distribution of B(E1) strength 
with energy to observables measured in Coulomb-dissociation experiments. Coulomb dissociation is also the time-reversed version of the radiative capture process $n + c \rightarrow \sigma + \gamma$, where the halo nucleus is formed by low-energy capture together with the emission of a photon. In the case of ${}^{15}$C this reaction is part of the carbon-nitrogen-oxygen (CNO) cycle by which helium is burnt in asymptotic-giant-branch (AGB) stars and in main-sequence stars of higher mass.  

\begin{figure}
\centerline{\includegraphics[width=0.7\textwidth]{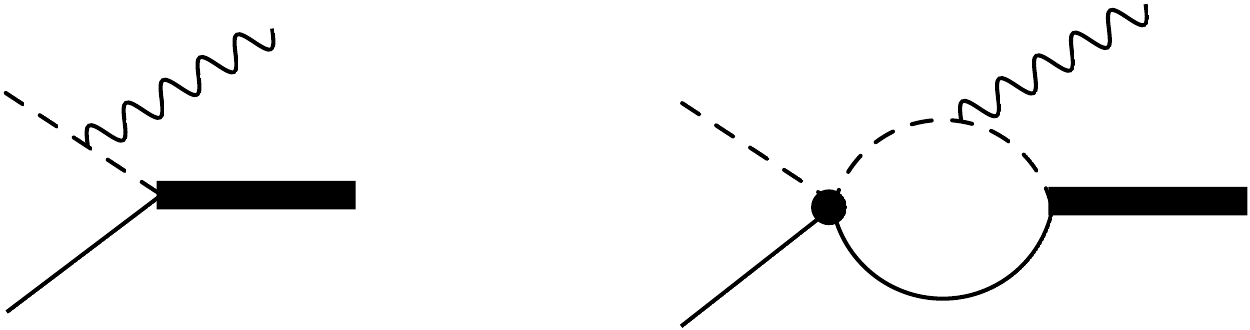}}
\caption{Diagrams contributing to the transition amplitude for the photodissociation of the halo into the core and the neutron. The first diagram represents the leading-order contribution, which is modified at next-to-leading order through multiplication by the wave-function renormalization. The second diagram occurs at N$^3$LO in the case that there is no low-energy bound state or resonance in the p-wave that is reached through the E1 transition. Note that here, as in subsequent figures containing Feynman diagrams, time runs from right to left.}\label{fig:diagrams}
\end{figure}

The following presentation here is an abbreviated version of material in Ref.~\cite{Acharya:2013nia}. 
Let us consider the transition amplitude for an E1 photon impinging on an s-wave one-neutron halo, which we will, following Sec.~\ref{sec:swaves}, encode as a field $\sigma$. The selection rules for this transition imply that the photon can break up the one-neutron halo into a $c n$ continuum state, but that the lowest partial wave excited there will be a p-wave. Initially we will treat the simplest case, 
where there are no low-energy resonances that enhance the p-wave rescattering. This is the situation, for example, in ${}^{19}$C. The breakup process for photon momenta $\sim M_{\rm halo}$ is then represented in Halo EFT via the Feynman diagrams in Fig.~\ref{fig:diagrams}. (In these, as in all Feynman diagrams in the remainder of this review, time runs from right to left.) 
The first diagram there scales as $1/Q$ because the core propagator scales as $1/Q^2$, and the photon-core vertex scales as $Q$. Rescattering in the final $nc$ state is represented by the second diagram. Assuming that there is no low-energy resonance that enhances this rescattering the (p-wave) $nc$ scattering vertex scales as $Q^2$. Including a factor of $Q$ for the $nc$ loop we find that this graph is suppressed by 
$(M_{\rm halo}/M_{\rm core})^3$
compared to the first one. Finally, we could also consider
a direct transition from the $\sigma$ state to the $nc$ continuum (now shown here, but see Fig.~\ref{fig:CaptureN5LO} for the corresponding diagram in the case of radiative capture to an s-wave proton halo). This encodes couplings of the photon to the halo nucleus at short distances and is governed by the coupling $L_{E1}^{(\bar{j})}$ of Eq.~(\ref{eq:nonminimalEMnatpwave}). This graph scales as $Q^3$, being suppressed by four powers of the 
$M_{\rm halo}/M_{\rm core}$ expansion 
relative to LO. Note that none of the diagrams containing additional couplings and/or final-state interactions appear at NLO or N$^2$LO and so the properties of the s-wave bound state determine the dissociation amplitude up to corrections that are suppressed by $\left(M_{\rm halo}/M_{\rm core}\right)^3$.

With $\bs{q}$ the momentum of the photon and $\bs{p}$ the final-state momentum of the core in the CM frame of the dissociation products, the  amplitude can be written as 
\begin{equation}
\label{eq:amplitude}
{\cal{M}}=\sqrt{\frac{2\pi\gamma_{0,\sigma}}{\mu_\sigma^{2}}} \frac{C_\sigma}{C_{\sigma, LO}} Q_c e\frac{1}{\frac{\gamma_{0, \sigma}^{2}}{2\mu_\sigma}+\frac{1}{2\mu_\sigma}\left(\bs{p}-\frac{m_n}{M_\sigma}\bs{q}\right)^{2}}.
\end{equation}
Choosing the photon momentum to be aligned with the z-axis, we obtain the matrix element of the dipole operator, $\sqrt{\alpha_{em}} \vert \bs{r} \vert Y_{1}^{0}\left(\hat{r}\right)$, by picking out the term linear in ${\bs q}$ and then dividing by $iq\sqrt{4\pi/3}$, to get
\begin{equation}
\label{eq:e1amplitude}
{\cal{M}}^{(l=1)}_{E1}=2\sqrt{6 \gamma_{0,\sigma}} \frac{C_\sigma}{C_{\sigma, LO}}  \frac{m_n}{M_\sigma}Q_c \sqrt{\alpha_{em}} \frac{p}{\left(\gamma_{0,\sigma}^{2}+p^{2}\right)^{2}}~\bs{\hat{p}}\cdot\bs{\hat{q}}.
\end{equation}

Equation~\eqref{eq:e1amplitude} gives the amplitude in the absence of neutron and core spin. 
We need to couple this result to the neutron spinor, and project to final states of good total angular momentum. Chossing $\bs{\hat{q}}$ parallel to the z-axis and using properties of the Clebsch-Gordan coefficients, we then obtain
\begin{equation}
\label{eq:ms}
{\cal{M}}^{(j=3/2)}_{E1}=\sqrt{2}{\cal{M}}^{(j=1/2)}_{E1}=4\sqrt{\gamma_{0,\sigma}} \frac{C_\sigma}{C_{\sigma, LO}}  \frac{m_n} {M_\sigma} Q_c\sqrt{\alpha_{em}} \frac{p}{\left(\gamma_{0,\sigma}^{2}+p^{2}\right)^{2}},
\end{equation}
These matrix elements are related to B(E1) by
\begin{equation}
\label{eq:be1}
d{\rm B(E1)}=\left(\vert{\cal{M}}^{(J=1/2)}_{E1}\vert^{2} + \vert{\cal{M}}^{(J=3/2)}_{E1}\vert^{2}\right) \frac{d^{3}p}{(2\pi)^3},
\end{equation}
 which yields
 \begin{equation}
\label{eq:dbe1overde}
\frac{d{\rm B(E1)}}{dE}= \frac{12}{\pi^{2}}\frac{\mu_\sigma^{3}}{m_c^{2}}Q_c^{2} \alpha_{em} \gamma_{0,\sigma} \frac{C_\sigma^2}{C_{\sigma, LO}^2} \frac{p^{3}}{\left(\gamma_{0,\sigma}^{2}+p^{2}\right)^{4}},                                             
\end{equation}
with $E=\frac{p^2}{2 \mu_\sigma}$.
The LO result is found by setting $C_\sigma=C_{\sigma, LO}$. Although here we derived the result for a spin-zero core, the same result is also found
 for arbitrary core spin $\varsigma_c$, as long as there is no enhanced p-wave channel. If we define $x$ as the dimensionless energy ratio $x \equiv E/S_{1n}$ then the dimensionless quantity $\mu_\sigma S_{1n} \frac{d {\rm B(E1)}}{dE}$ can be written
 \begin{equation}
\label{eq:dBE1scale}
\mu_\sigma S_{1n}^2 \frac{d {\rm B(E1)}}{dE} = 
\frac{C_\sigma^2}{C_{\sigma, LO}^2} \frac{3 \alpha_{em} Q_{\rm eff}^{2}}{\pi^2} \frac{x^{3/2}}{(1+x)^4},
\end{equation}
where $Q_{\rm eff}=\frac{m_n Q_c}{m_n + m_c}=\frac{\mu_\sigma Q_c}{m_c}=\frac{Q_c}{A+1}$ is the effective charge.
 
Equation~(\ref{eq:dbe1overde}) was also derived by Rupak {\it et al.} in Ref.~\cite{Rupak:2012cr}, where they considered the coupling of the core to the three-vector potential, $\bs{A}$, only  and extracted $d{\rm B(E1)}/dE$ by calculating the photo-nuclear cross-section for the E1 photon. The LO Halo EFT result (with $C_\sigma=C_{\sigma,LO}$) is also equivalent to that of Ref.~\cite{Bertulani:1988svs}, where a zero-range potential model was used for the neutron-core interaction. Eq.~\eqref{eq:dbe1overde} can also be recovered from the framework presented in Ref.~\cite{Typel:2001mx} (see also Ref.~\cite{Typel:2004us}). 

Several different experimental observables can be obtained from the E1 strength distribution $\frac{d{\rm B(E1)}}{dE}$. First, the total cross section for photodissociation of the nucleus by an E1 photon is then~\cite{Bertulani:2009zk}:
\begin{equation}
\sigma^{(E1)}(\omega)=\frac{16 \pi^3}{9} \omega \frac{d {\rm B(E1)}}{d \omega}, 
\end{equation}
with $\omega=\frac{\gamma_{0,\sigma}^2}{2 \mu_\sigma} + E$. 
The cross section for the core to capture the neutron into the halo state, with the excess energy radiated away via an E1 photon, is then found by detailed balance. That relation is~\cite{Rupak:2012cr,Bertulani:2009zk,Chen:1999tn}:
\begin{equation}
\sigma(E)=\frac{16 \pi^3}{9} \frac{\omega^3}{2 \mu_\sigma E} \frac{d {\rm B(E1)}}{dE} \frac{2 s + 1}{2 \zeta_c +1},
\end{equation}
where $\zeta_c$ is the spin of the core and $s$ denotes the spin channel in which the halo is formed~\footnote[2]{$s$ is then the spin of the halo state, since we consider only captures to $s$-wave states in this section.}. 

\begin{figure}[!t]
\centerline{\includegraphics[width=0.7\columnwidth]{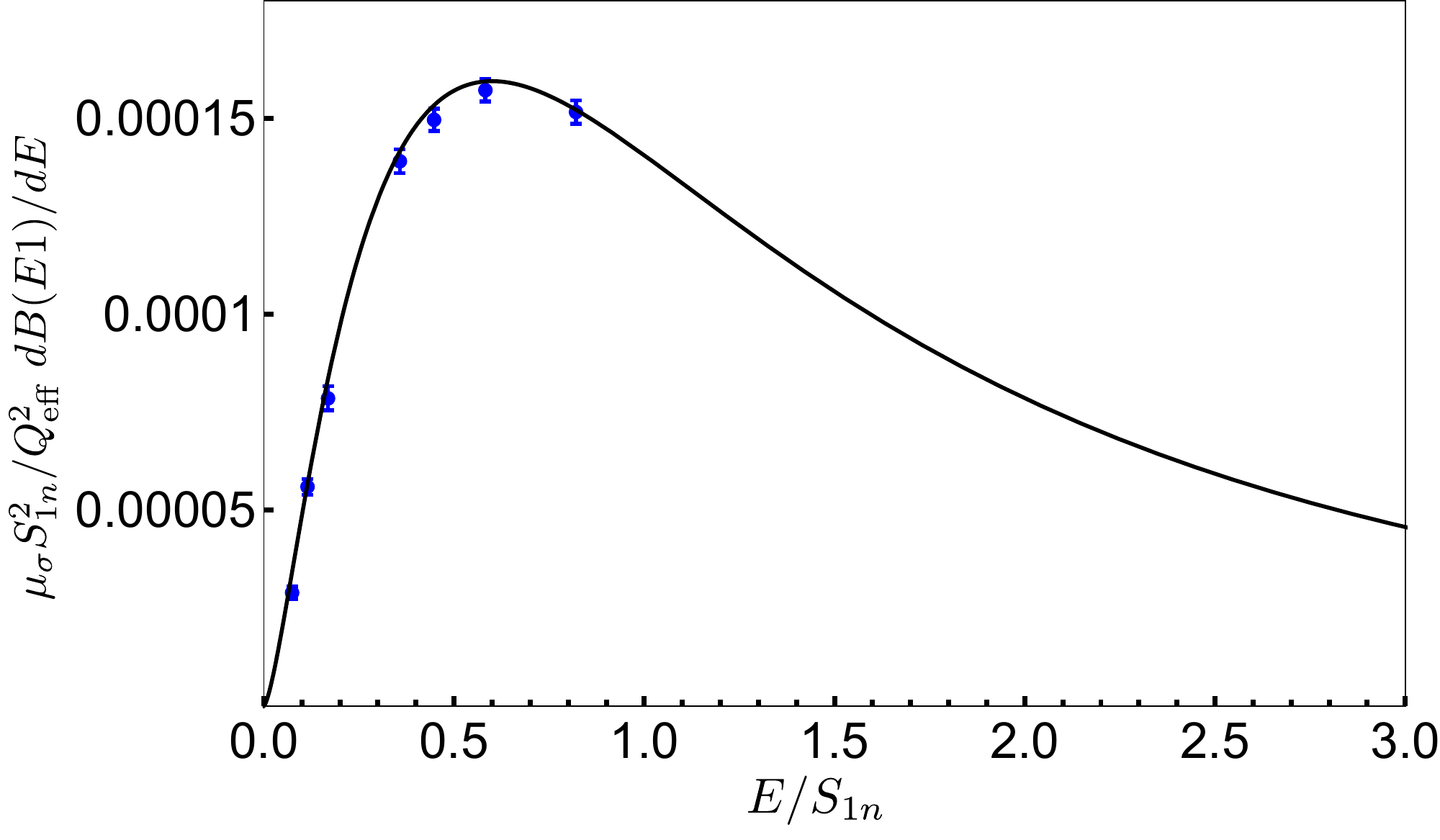}}
\caption{The universal curve (\ref{eq:dBE1scale}), compared to the E1 contribution to deuteron photo disintegration obtained from the data in Ref.~\cite{Tornow:2003ze}.}
\label{fig:universalE1}       
\end{figure}

In Fig.~\ref{fig:universalE1} we plot a rescaled version of the curve (\ref{eq:dBE1scale}), where we have divided by the factors $\left(C_\sigma/C_{\sigma, LO}\right)^2$ and $Q_{\rm eff}^2$ that are specific to a particular nucleus. As in Eq.~(\ref{eq:dBE1scale}), we have also rewritten the energies (and E1 strength) in units of the separation energy $S_{1n}$. This curve should then be universal to all one-neutron halos, up to the higher-order corrections that encode p-wave final-state interactions~\cite{Chen:1999tn,Vanasse:2014sva}. To demonstrate it works for the simplest one-neutron halo we include the E1 strength for deuterium on the plot, as extracted from the asymmetry measured in Ref.~\cite{Tornow:2003ze}~\footnote[1]{Both here and in Ref.~\cite{Tornow:2003ze}, the E1 photodissociation cross section is obtained from the measured asymmetry via multiplication by the N$^2$LO EFT cross section of Ref.~\cite{Chen:1999tn,Vanasse:2014sva}. That calculation is consistent with all extant low-energy cross section data. The errors shown for the E1 part of the cross section in Fig.~\ref{fig:universalE1} do not include any uncertainty from the EFT's N$^2$LO total cross section.}. In order to remove the deuteron-specific factor of $\left(\frac{C_\sigma}{C_{\sigma, LO}}\right)^2$ from the data we have divided it by 1.69, the value of this ratio for deuterium~\cite{Phillips:1999hh}.

Experimental observables for Coulomb dissociation are obtained by convolving the result (\ref{eq:dbe1overde}) for the differential $\mathrm{B(E1)}$ transition strength as a function of energy with the distribution of (nearly real) photons generated in the target-halo collision. Results for the latter  have been derived in Refs.~\cite{Bertulani:1987tz,Bertulani:1985fqx,Bertulani:1988svs,Bertulani:2009zk}  for the case of E1-dominated dissociation upon small-angle scattering by a charged target at high beam energy. The energy spectrum resulting from the dissociation of a halo of binding energy $S_{1n}$ is:
\begin{equation}
\label{eq:dsigmaoverde}
\frac{d\sigma}{dE}= \frac{16\pi^{3}}{9} N_{E1}(S_{1n}+E,R) \frac{d{\rm B(E1)}}{dE},
\end{equation}
where 
\begin{equation}
N_{E1}(\omega,R)= 2\frac{Q_{t}^{2}\alpha_{em}}{\pi\beta^{2}}\left[\xi K_{0}(\xi)K_{1}(\xi)-\frac{\beta^{2}}{2}\xi^{2}\left(\left(K_{1}(\xi)\right)^{2}- \left(K_{0}(\xi)\right)^{2}\right)\right],
\label{eq:photonnumb}
\end{equation} 
with $\xi=\omega R\sqrt{1-\beta^2}/\beta$,  is the virtual photon number for the E1 multipolarity, integrated over all impact parameters larger than $R$, with
$\beta$ the beam velocity, and $K_\nu$ modified Bessel functions. 

The differential cross section with respect to the center of mass angle between the dissociated fragments and the longitudinal momentum distribution of the dissociation cross section can be obtained by convolving $\frac{d {\rm B(E1)}}{dE}$ with appropriate functions representing the number of virtual photons for each case, see Ref.~\cite{Acharya:2013nia} for details. The arguments used to obtain such formulae are semi-classical, involve only first-order coupling of the nucleus to the photon field, and neglect the halo-target interaction. But all those approximations are accurate for a high-charge target, such as ${}^{208}$Pb, in the regime of small-angle/large-impact-parameter scattering~\cite{Capel:2005kj,Capel:2016}. 

\begin{figure}[!t]
\centerline{\includegraphics[width=0.9\columnwidth]{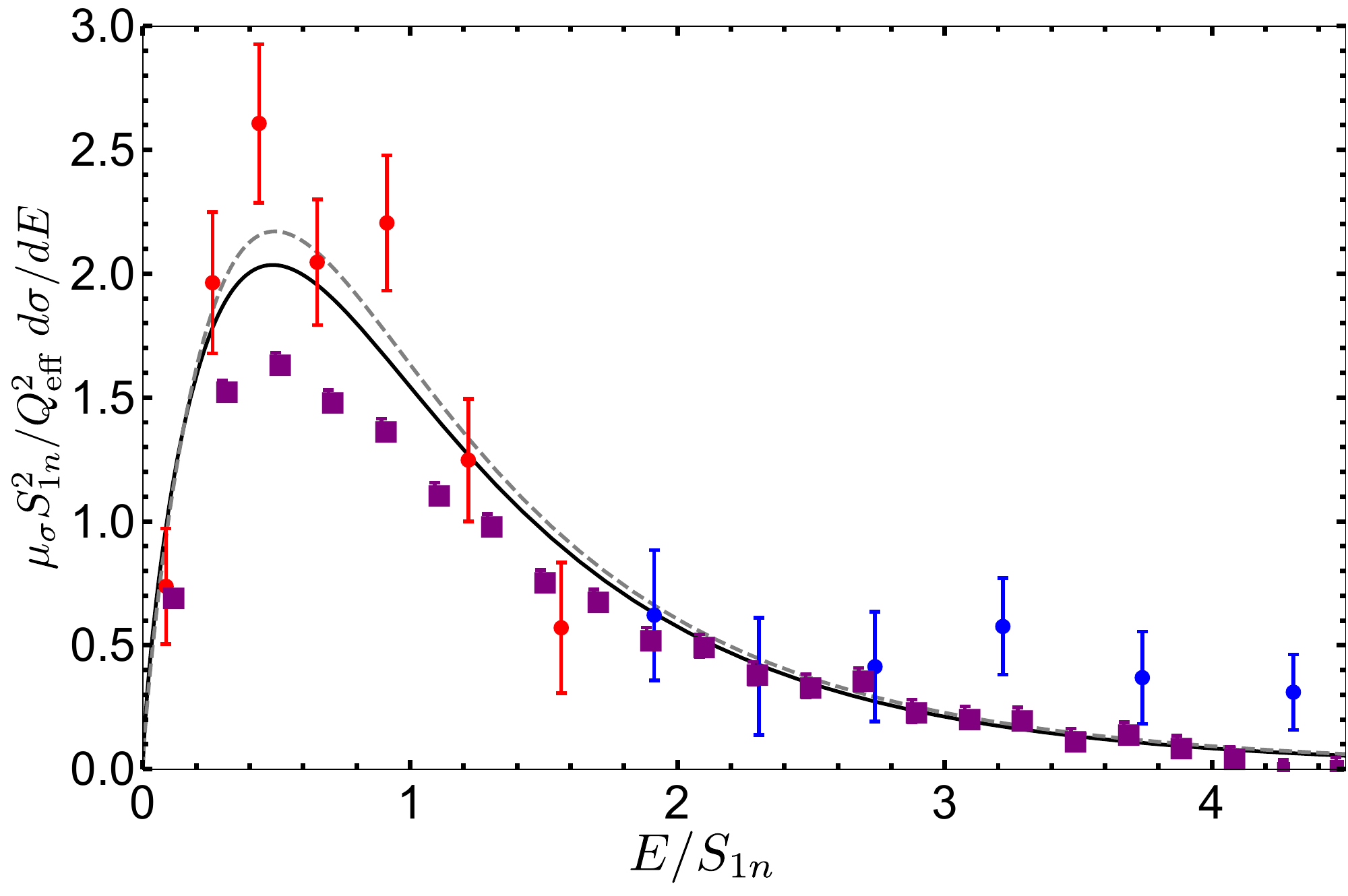}}
\caption{Rescaled Coulomb dissociation data from Refs.~\cite{Fukuda:2004ct} (for ${}^{11}$Be, purple squares) and \cite{Nakamura:2003cyk} (for ${}^{19}$C, red and blue circles), each divided by the pertinent $\left(\frac{C_\sigma}{C_{\sigma, LO}}\right)^2$, compared to the universal leading-order Halo EFT prediction for $d \sigma/dE$. Note that while the EFT prediction is universal, the experimental conditions (beam energies, detector resolution, impact parameter) for the two Coulomb-dissociation experiments were different, so the black curve shows the result for ${}^{19}$C and the grey dashed curve for ${}^{11}$Be.}
\label{fig:universalE1CD}       
\end{figure}

Refs.~\cite{Nakamura:1999rp,Nakamura:2003cyk} describe such a Coulomb-dissociation experiment for ${}^{19}$C. There ${}^{19}$C was scattered from a ${}^{208}$Pb target, and the neutron spectrum obtained as a function of both energy and angle. The data for impact parameters $R> 30$ fm, where the nuclear contribution to the breakup is negligible, is plotted (after rescaling as described above) in Fig.~\ref{fig:universalE1CD}. 

These data are in good agreement with the universal curve if we take $S_{1n}=0.58$ MeV. This shows that the older value adopted for the separation energy, $S_{1n}=0.16(11)$ MeV~\cite{Audi1993A,Audi1993B}, cannot be correct. The universal behavior of the E1 strength, in combination with these Coulomb-dissociation data, rule out such a small $S_{1n}$ for ${}^{19}$C. 

In fact, the ${}^{19}$C Coulomb-dissociation data are divided by the ratio $(C_\sigma/C_{\sigma, LO})^2$ extracted in Ref.~\cite{Acharya:2013nia} (see Sec.~\ref{sec:CDapplications}) using the red data points. The Halo EFT result is then a prediction for the blue data points. Ref.~\cite{Fukuda:2004ct} discusses an analogous experiment for ${}^{11}$Be, where p-wave interactions play a larger role---see the next subsection for discussion. In this case we take $S_{1n}=0.50$ MeV and the data are divided by the ratio  $(C_\sigma/C_{\sigma, LO})^2$ computed using the {\it ab initio} No-Core Shell Model with Continuum approach~\cite{Calci:2016dfb} that is described further in Sec.~\ref{sec:insightsconnections}. The agreement between the ${}^{19}$C Halo EFT result and the data is excellent, although it must be remembered that part of the data was used to fix the ANC in this case. The rescaled Halo EFT curve also does quite a good job in describing the shape of the ${}^{11}$Be data, although there the p-wave interactions that generate the excited bound state of ${}^{11}$Be in the $j=1/2$ channel reduce the cross section from the universal curve, as we now discuss. 

\subsection{The role of p-wave final-state interactions}

The formulae given in the previous section apply to any one-neutron halo without a low-energy p-wave resonance or bound state. ${}^{19}$C fits this description, see Table~\ref{tab:1n-halo}, and its Coulomb dissociation was calculated in Ref.~\cite{Acharya:2013nia}. In contrast, some halo systems (e.g., ${}^{15}$C and ${}^{11}$Be) have significant final-state p-wave interactions that affect their dipole-strength distribution. These p-wave FSIs modify the above formulae. However, in the power counting we have adopted in this review the modification occurs only at NLO, because the p-wave phase shift is perturbatively small unless the scattering energy is close to the bound state or resonances, see Eq.~(\ref{eq:delta1pert}). That is not the power counting that was adopted in Ref.~\cite{Rupak:2012cr} for ${}^{15}$C:  there the power counting in which both $a_1^{(1/2)}$ and $r_1^{(1/2)}$ were fine tuned was employed for the ${}^2$P$_{3/2}$ channel (we note that $r_1^{(1/2)}$ was also taken to be positive in that work). Here we focus on the case of ${}^{11}$Be, where a perturbative treatment of the FSI is well established. 

In ${}^{11}$Be the ${}^2$P$_{3/2}$ channel can be taken as non-interacting at NLO accuracy. Meanwhile, the ${}^2$P$_{1/2}$ channel of ${}^{10}$Be-neutron scattering has effective-range parameters that obey $a_1^{(1/2)} \sim M_{\rm core}^{-1} M_{\rm halo}^{-2}$ and $r_1^{(1/2)} \sim M_{\rm core}$ (see Sec.~\ref{sec:pwaves}). This means the the next-to-leading-order formula for the E1 strength has a term resulting from explicit calculation of the second diagram in Fig.~\ref{fig:diagrams}. The $j=1/2$ amplitude can be written as:
\begin{eqnarray}
{\cal M}_{E1}^{(j=1/2)}=&\frac{C_\sigma}{C_{\sigma, LO}} 
Q_{\rm eff} \sqrt{2 \gamma_{0,\sigma}\alpha_{em}} 
e^{i \delta_1^{(1/2)}(p)}  
\nn
&\times \frac{2 {p}^3 \cos(\delta_1^{(1/2)}(p)) + 
(\gamma_{0,\sigma}^3 + 3  {p}^2 \gamma_{0,\sigma})\sin(\delta_1^{(1/2)}(p))}{{p}^2(\gamma_{0,\sigma}^2+{p}^2)^2} ,
\label{eq:ME11/2}
\end{eqnarray}
where $p=\sqrt{2 \mu_\sigma E}$ is the final-state momentum of the neutron-core state in its center-of-mass frame.
Eq.~(\ref{eq:ME11/2}) is true regardless of the power counting employed for the phase shift $\delta_{1/2}$.
We convert this result to the $j=1/2$ contribution to  $d{\mathrm B(E1)}/dE$ via Eq.~(\ref{eq:be1}), and then use (\ref{eq:delta1pert}) to expand the result in powers of $M_{\rm halo}/M_{\rm core}$ and retain only the leading- and next-to-leading-order terms in that expansion. This yields:
\begin{equation}
\fl
\frac{d{\rm B(E1)}}{dE}^{(j=1/2),  {\rm NLO}}=
Q_{\rm eff}^2 \mu_\sigma \alpha_{em}
\frac{\gamma_{0,\sigma}}{\pi^2}
\frac{4 p^3}{(p^2 + \gamma_{0,\sigma}^2)^4} \left(\frac{C_\sigma^2}{C_{\sigma,LO}^2} 
+ \frac{2 \gamma_{0,\sigma}}{r_1^{(1/2)}}\frac{\gamma_{0,\sigma}^2  + 3 p^2}{p^2 + \gamma_1^2}
\right) \, .
\label{eq:dBdE1NLO}
\end{equation}
We note that the first term here is present even if there is no p-wave FSI. It tends to increase the cross section. Meanwhile, the second term opposes this enhancement, since $r_1^{(1/2)} < 0$. It depends on the parameters of the p-wave FSI: the effective ``range" $r_1^{(1/2)}$, and the pole position $\gamma_1$. This result must then be added to the contribution to the E1 strength from the $j=3/2$ channel, cf. Eq.~(\ref{eq:ms}). 

\subsection{Application to ${}^{19}$C and ${}^{11}$Be}

\label{sec:CDapplications}

Since FSI is only a NLO effect in this power counting it follows that both nuclei have a leading-order E1-strength distribution given by Eq.~(\ref{eq:dbe1overde}) with $C_\sigma=C_{\sigma,LO}$. At NLO $C_\sigma$ receives corrections that make it different from unity, and in ${}^{19}$C these are the only NLO effects. Eq.~(\ref{eq:dbe1overde}) then gives the $d{\rm B(E1)}/dE$ strength distribution once $\gamma_{0, \sigma}$ and $C_\sigma$ are known. The result is accurate up to FSI corrections, and these are suppressed by $\left(M_{\rm halo}/M_{\rm core} \right)^3$ compared to leading. In fact, these FSI corrections can be calculated using the approach of the previous subsection if the p-wave parameters in the pertinent channel are known. The first counterterm in the E1 dissociation of an s-wave halo does not occur until N$^4$LO (see term proportional to $L_{E1}^{(3/2)}$ in Eq.~(\ref{eq:nonminimalEMnatpwave})). 

Ref.~\cite{Acharya:2013nia} fit the parameters $\gamma_{0, \sigma}$ and $C_\sigma$ to the data of Ref.~\cite{Nakamura:2003cyk} taken at impact parameters $R > 30$ fm and final-state $n-{}^{18}$C energies $< 1$ MeV. The $1\sigma$ confidence intervals obtained were $(520,630)$ keV for $S_{1n}$ and $(1.7,3.2)$ fm for the effective range. The mean values $r_{0,\sigma}=2.6$ fm and $S_{1n}=575$ keV were then used in the formula 
(\ref{eq:ANC-swave})
to remove the nucleus-specific factor $(C_\sigma/C_{\sigma,LO})^2$ from the Coulomb-dissociation data in Fig.~\ref{fig:universalE1CD}. 

Meanwhile ${}^{11}$Be is governed by the NLO formula (\ref{eq:dBdE1NLO}). The p-wave inputs $r_1^{(1/2)}$ and $\gamma_1$---as well as $C_\sigma$ and $\gamma_{0,\sigma}$---are needed for a prediction here, 
Eq.~(\ref{eq:dBdE1NLO}) receives corrections at N$^2$LO. Some of these are from N$^2$LO terms we simply dropped in deriving the result. But 
 the strong p-wave FSI also means that $L_{E1}^{(1/2)}$ affects the E1 dissociation of these s-wave halos already at N$^2$LO. Since this counterterm also enters the $1/2^+$ to $1/2^-$ E1 transition amplitude (see Sec.~\ref{sec:pwaveboundstateobservables}) it can, in principle, be fixed there. However, only NLO accuracy has been achieved for ${}^{11}$Be Coulomb dissociation in Halo EFT.
 
\subsection{Electric radius for s-wave halos}

In this section we compute the electric form factor for an s-wave halo. Measurement of any part of this
form factor beyond the ${\bs q}^2$ part, i.e. the electric radius, is beyond present experimental capabilities. However,
experiments such as ELISE (Electron-Ion Scattering in a Storage Ring Experiment) at FAIR~\cite{Antonov:2011zza}  or SCRIT (Self-Confining Radioactive Ion Target) at RIKEN~\cite{Suda:2012mi}  should be able to perform collisions between electrons
and exotic nuclei, thus mapping out the form factor's ${\bs q}$-dependence is not beyond the bounds of possibility (cf. Ref.~\cite{Bertulani:2006xv}). 
Of course, for the lightest s-wave one-neutron halo, deuterium, this has already been done extensively~\cite{Phillips:2016}. 

\begin{figure}[!t]
\centerline{\includegraphics[width=0.5\columnwidth]{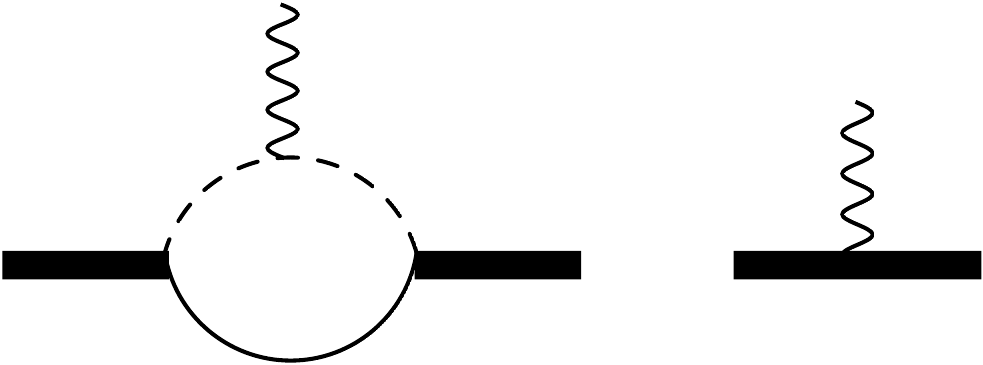}}
\caption{The LO and NLO contribution to the irreducible vertex for an $A_0$ photon 
to couple to the dimer representing an s-wave one-neutron halo. The same diagrams arise
for a p-wave one-neutron halo, although in that case they both occur at leading order.}
\label{fig:formfactor}       
\end{figure}

The s-wave form factor is computed by calculating the $\sigma$-irreducible vertex function 
for $A_0 \sigma \sigma$ interactions. This object represents a photon of four-momentum $q$ coupling to 
the $\sigma$ state. After application of wave-function renormalization it is equal to $-i e Q_c F_E^{(\sigma)}(|{\bs q}|)$, where ${\bs q}$ 
is the three-momentum of the virtual photon in the Breit frame. At LO it turns out
there is only one contribution to this vertex function, the first diagram in Fig.~\ref{fig:formfactor}. A straightforward 
calculation yields the LO---and hence universal---result for the electric form factor of an s-wave one-neutron halo: 
\begin{equation}
F_E^{(\sigma)}(|{\bs q}|)= \frac{2 \gamma_{0,\sigma}}{f|{\bs q}|} \arctan
\left(\frac{f|{\bs q}|}{2 \gamma_{0,\sigma}}\right)\,,
\label{eq:Gc}
\end{equation}
with $f=m_n/M_\sigma=\mu_\sigma/m_c=\frac{1}{A+1}$. Note that $F_E^{(\sigma)}(0)=1$, as it should. For the deuteron, 
we have $f=1/2$, and this reduces to the LO result of Ref.~\cite{Chen:1999tn}. 

The electric radius of the s-wave state can be extracted according to:
\begin{equation}
F_E^{(\sigma)}(|{\bs q}|) \equiv 1 - \frac{1}{6}\langle r_E^2 
\rangle^{(\sigma)}{\bs q}^2 + \ldots\,,
\label{def:rcqs}
\end{equation}
and an expansion of Eq.~(\ref{eq:Gc}) in powers of $|{\bs q}|$ then yields 
\begin{equation}
\langle r_E^2 \rangle^{(\sigma)}=\frac{\mu_\sigma^2}{2 m_c^2 \gamma_{0,\sigma}^2} + \langle r_{E}^2 \rangle_c + Q_c^{-1}\langle r_E^2 \rangle_n\,.
\label{rcqs}
\end{equation}
Here we have added the electric radius of the core to the right-hand side of the equation---as well as the typically much smaller radius of the neutron---since the observable
that is computed in Halo EFT is the electric radius of the halo ground state taking the core and neutron as point particles. In atomic spectroscopy experiments the measured isotpoe shift is, in fact, proportional to $\langle r_E^2 \rangle^{(\sigma)} -  \langle r_{E}^2 \rangle_c$,
but in order to facilitate comparison with the way radii are usually quoted it is sensible to write the result as in Eq.~(\ref{rcqs}). 

At NLO there are two additions to this calculation: first, the NLO wave-function renormalization, not the LO one, must be used in calculating
the contribution of Fig.~\ref{fig:formfactor} to $F_E^{(\sigma)}$. This  increases the size of $F_E^{(\sigma)}$ by a factor of $(C_\sigma/C_{\sigma, LO})^2$. The second addition is that we must consider the
operator associated with gauging the dimer kinetic term, i.e. the piece of $w_\sigma \sigma^\dagger D_0 
\sigma$ in Eq.~(\ref{eq:Lag-nc-aux}) that is proportional to $A_0$. This generates a constant shift in the form factor that guarantees $F_E^{(\sigma)}(0)=1$ (see right-hand diagram in Fig.~\ref{fig:formfactor}). 
This second effect does not, however, affect the $|{\bs q}|$ dependence of $F_E^{(\sigma)}$. The overall result is an increased point electric radius at NLO, as long as $r_0 > 0$, cf. Refs.~\cite{Beane:2000fi,Phillips:2002da,Hammer:2011ye}:
\begin{equation}
\langle r_E^2 \rangle^{(\sigma)}_{\rm pt} \equiv
\langle r_E^2 \rangle^{(\sigma)} - \langle r_E^2 \rangle_{c} - Q_c^{-1} \langle r_E^2 \rangle_{n}=
\frac{C_\sigma^2 f^2}{2 C_{\sigma, LO}^2 \gamma_{0,\sigma}^2}\,.
\label{eq:chargeradiusNLO}
\end{equation}

The astute reader will already have noticed that the right-hand side of Eq.~(\ref{eq:chargeradiusNLO}) is of a very similar form to the Halo EFT result for the matter radius $\langle r_{nc}^2 \rangle$, Eq.~(\ref{eq:rnc}). In fact, the only difference is that 
$\langle r_E^2 \rangle^{(\sigma)}_{\rm pt}$
is smaller than $\langle r_{nc}^2 \rangle$ by an overall factor of $f^2=1/(A+1)^2$. This factor arises because only the core has an electric coupling, and the electric radius is suppressed by a concomitant mass factor. However, the two radii probe the same physics: that of the tail of the s-wave halo state's wave function. 

As emphasized in Sec.~\ref{sec:2bamplitude} the Halo EFT formula for the matter radius can be obtained from a quantum-mechanical calculation with a zero-range s-wave wave function. A more rigorous way of obtaining the same result is to consider the form factor for a current that couples to the mass, rather than to the electric charge, and then imitate the field-theoretic presentation given in this section. 

The result (\ref{eq:chargeradiusNLO}) for the  
point electric radius $\langle r_E^2 \rangle^{(\sigma)}_{\rm pt}$,
is accurate up to corrections of relative order $\left(M_{\rm halo}/M_{\rm core}\right)^3$. The first correction to it comes only from the operator multiplying the constant $L_{C0}^\sigma$ in Eq.~(\ref{eq:nonminimalEMdimers}). After wave-function renormalization this gives an effect that is parameterically $\sim M_{\rm halo}/M_{\rm core}^3$, and so is markedly suppressed compared to the leading $1/M_{\rm halo}^2$ piece of the difference of squared radii. Of course, in the case of the electric radii this counting assumes that the suppression by $f^2$ also enters this counterterm. If that is not the case, e.g., the counterterm represents a change in the structure of the core induced by the presence of the neutron, then this short-distance physics will have a larger impact than indicated by this simple power-counting estimate.   

\subsection{Application:  electric radii of one-neutron halos}

We have demonstrated that both $\langle r_E^2 \rangle$ and $\langle r_{nc}^2 \rangle$ are sensitive to the same halo physics. Table~\ref{tab:1n-halo} collected a number of measurements of matter radii and compared them to predictions from Halo EFT. Good agreement was seen in almost all cases. However, matter radii can be difficult to extract from data. In contrast, the measurement of an isotope shift provides much cleaner information on $\langle r_E^2 \rangle$. Unfortunately though, with one exception, the electric radius is not as sensitive to the halo physics as the matter radius, because of the suppression of the halo contribution by $f^2=1/(A+1)^2$. 

The one exception is deuterium. From the measurement of Ref.~\cite{Parthey:2010aya} of the hydrogen-deuterium isotope shift of the 1S-2S transition we extract $\langle r_E^2 \rangle_{\rm pt}^{1/2}=1.97556(57)$ fm (using a neutron charge-radius squared of $\langle r_n^2 \rangle=-0.1161(22)$ fm$^2$~\cite{Olive:2016xmw}.). This is slightly discrepant with, but more accurate then, the deuteron point-charge radius we obtain from combining the CREMA collaboration's measurements of the 2S-2P transitions in muonic deuterium and muonic hydrogen,  $\langle r_E^2 \rangle_{\rm pt}^{1/2}=1.97335(102)$ fm~\cite{Pohl:2016glp}~\footnote{For the purposes of this review we make no distinction between charge radii and electric radii.}. Both are in remarkable agreement with the Halo EFT (or \eftnopi) prediction~\cite{Chen:1999tn,Phillips:1999hh}, $\langle r_E^2 \rangle_{\rm pt}^{1/2}=1.977$ fm. 

Applying Eq.~(\ref{eq:chargeradiusNLO}) to the ground state of ${}^{11}$Be, using as input the numbers for $S_{1n}$ and $C_\sigma/C_{\sigma, LO}$ from Table~\ref{tab:1n-halo}, we obtain
$\langle r_E^2 \rangle^{(\sigma)}_{{\rm pt},{}^{11}{\rm Be}}=0.39~{\rm fm}^2$ at NLO.
This is a post-diction of the observable associated with the isotope shift, and extracted via precise atomic spectroscopy: 
$\langle r_E^2 \rangle^{(\sigma)}_{{\rm pt},{}^{11}{\rm Be}}=0.54(17)~{\rm fm}^2$~\cite{Nortershauser:2008vp}. The NLO Halo EFT result is consistent with this measurement.
It assumes the ANC of the ${}^{11}$Be ground-state obtained in the NCSMC calculation of Calci {\it et al.}~\cite{Calci:2016dfb}. For deuterium the ANC can be obtained by analytic continuation of the neutron-proton ${}^3$S$_1$ amplitude measured in scattering experiments to the bound-state pole. 

In the case of ${}^{19}$C Ref.~\cite{Acharya:2013nia} extracted the ${}^{19}$C ANC listed in Table~\ref{tab:1n-halo} from Coulomb dissociation data (cf. Sec.~\ref{sec:CDapplications}). Taking that prediction and the above formula for the radius of an s-wave one-neutron halo yields
\begin{equation}
\langle r_E^2 \rangle^{(\sigma)}_{{\rm pt},{}^{19}{\rm C}}
=0.09^{+0.02}_{-0.03} {\rm fm}^2. 
\end{equation}
Most of the error in this (very small) predicted isotope shift comes from uncertainties in the ${}^{19}$C binding energy and ANC extracted from the Coulomb dissociation data. A bound can be placed on the 
point electric radius of ${}^{19}$C using the extractions of point-proton radii
of ${}^{19}$C and ${}^{18}$C
from the charge-changing reaction cross sections reported in Ref.~\cite{Kanungo:2016tmz}. Assuming the errors quoted there are uncorrelated that bound is 
$\langle r_E^2 \rangle^{(\sigma)}_{{\rm pt},{}^{19}{\rm C}} \leq 0.38~{\rm fm}^2$
(at the $1 \sigma$ level). 

To our knowledge. for none of the other s-wave one-neutron halos we have discussed has an electric radius been measured.

\subsection{Sum rules for the dipole strength}

The non-energy weighted sum rule (NEWSR) relates the two observables considered so far in this section: the distribution of E1 strength, and the electric radius of the one-neutron halo. Its quantum-mechanical derivation rests on writing 
\begin{equation}
\langle \sigma| {\bs r}^2|\sigma \rangle=\sum_n \langle \sigma|{\bs r}|n\rangle \langle n|{\bs r}|\sigma \rangle,
\end{equation}
where $|\sigma \rangle$ is the ground state of the $nc$ halo and $|n \rangle$ is any complete set of states. In order to relate $\langle \sigma| {\bs r}^2|\sigma \rangle$ to $d {\rm B(E1)}/dE$ the states $|n \rangle$ must be eigenstates of the Hamiltonian that governs the neutron-core interaction. For a one-neutron halo 
without p-wave enhancement,
the NEWSR is usually written:
\begin{equation}
\int dE \frac{d {\rm B(E1)}}{d E}=\frac{3}{4 \pi} \alpha_{em} 
Q_{\rm eff}^2
\langle r_{nc}^2 \rangle.
\end{equation}
Direct integration of Eq.~(\ref{eq:dbe1overde}) indeed produces a result that is in accord with Eq.~(\ref{eq:rnc}).

An energy-weighted sum rule can also be derived---for any local potential---by consideration of the matrix element of $[{\bs r},[H,{\bs r}]]$, and use of the fundamental commutator $[{\bs r}_i,{\bs p}_j]=i \delta_{ij}$. This sum rule is:
\begin{equation}
\int_0^\infty (E+S_{1n}) \frac{d {\rm B(E1)}}{dE}=
\frac{9 \alpha_{em} Q_{\rm eff}^2}{8 \pi \mu_\sigma}.
\label{eq:EWSR}
\end{equation}
It is important to note that the energy that appears in the integrand here is $E+S_{1n}$, and not just $E$~\cite{Acharya:2015}. The LO Halo EFT result for $ \frac{d {\rm B(E1)}}{dE}$, Eq.~(\ref{eq:dbe1overde}) with $C_\sigma=C_{\sigma,LO}$, obeys Eq.~(\ref{eq:EWSR}). At NLO the EWSR is {\it not} satisfied: the non-zero effective range is not generated by a local potential in Halo EFT. 

\subsection{Bound-state observables for p-wave halos}

\label{sec:pwaveboundstateobservables}

We now discuss a variety of electromagnetic observables that can be measured in p-wave one-neutron halo nuclei. We begin by examining the electric radius and the $B(E1)$ value for a transition from an s-wave to a p-wave halo state, both of which were first calculated in Halo EFT in Ref.~\cite{Hammer:2011ye}. We examine how these results apply to ${}^{11}$Be, where such a transition occurs. 

As in the s-wave case, we get the p-wave form factor by calculating contributions to the $\pi$-irreducible vertex function for 
$A_0 \pi \pi$ 
interactions. There are two diagrams at 
LO; they have the topologies depicted in Fig.~\ref{fig:formfactor}. The first diagram is 
analogous to the LO one for the s-wave state, although in the p-wave case the dimer-neutron-core vertices involve a derivative coupling that is not indicated in the figure. The second again represents a direct 
coupling of the photon  
to the $\pi$ field, but, in contrast to the s-wave case, it occurs already at LO, since the p-wave dimer
$\pi$ must be dynamical at leading order.

In the Breit frame the coupling of an $A_0$ photon 
to the $1/2^-$ state can be written:
\begin{eqnarray}
\fl
\langle \pi^{(1/2)}_\gamma ({\bs p'})| J^0 | \pi^{(1/2)}_\gamma ({\bs p})\rangle =&-i e 
Q_c \sum_{\delta i j} \left(\frac{1}{2} \delta 1 i\right|\left.
\frac{1}{2} \gamma\right) \left(\frac{1}{2} \delta 1 j\right|\left.\frac{1}{2} \gamma \right)
\nonumber\\
& \qquad
\times \Biggl[ F_E^{(\pi)}(|{\bs q}|) \delta_{ij}  +\frac{1}{2M_{\sigma}^2} F_Q^{(\pi)}(|{\bs q}|) \left(q_i q_j -
\frac{{\bs q}^2 \delta_{ij}}{3} \right)\Biggr],
\label{def:p-waveFF}
\end{eqnarray}
where ${\bs q}={\bs p'}-{\bs p}$ is the three-momentum of the virtual photon. 
Here we have expressed the vertex function in terms of the electric and quadrupole 
form factors of a vector field, although
the quadrupole form factor is, in fact, unobservable in the $1/2^-$ state.
 Choosing ${\bs q}=|{\bs q}| \hat{z}$ once again and performing
a straightforward calculation yields: 
\begin{equation}
\fl
\quad F_E^{(\pi)}(|{\bs q}|)= \frac{1}{r_1^{(1/2)}+3 \gamma_1}\left[r_1^{(1/2)} +\frac{1}{|{\bs q}| f}
\left(
2 |{\bs q}| f \gamma_1 +(|{\bs q}|^2 f^2+2 \gamma_1^2) \arctan
\left(\frac{f|{\bs q}|}{2 \gamma_1}\right)\right)\right],
\label{eq:Gc-p}
\end{equation}
where again $f=m_n/M_{\sigma}=1/(A+1)$. For a strict LO result
$r_1^{(1/2)}+3 \gamma_1$ should be replaced by $r_1^{(1/2)}$ in Eq.~(\ref{eq:Gc-p}).

$F_E^{(\pi)}(0)=1$, as required by charge conservation.
The electric radius of the p-wave state relative to the $^{10}$Be ground state
can be extracted according to Eq.~(\ref{def:rcqs}), and we obtain 
\begin{equation}
\langle r_E^2 \rangle^{(\pi)}_{\rm pt}=-\frac{5 f^2}{2 \gamma_1 r_1^{(1/2)}}
\label{eq:LOpradresult}
\end{equation}
Note that, once the suppression by $f^2$ is accounted for, 
this radius is $\sim M_{\rm halo}^{-1} M_{\rm core}^{-1}$, i.e. the p-wave halo is intrinsically more compact than an s-wave one.

This can be easily understood using simple quantum-mechanical arguments and asymptotic wave functions for the p-wave state.
From this point of view the LO radius of the p-wave state is:
\begin{equation}
\langle r_E^2 \rangle^{(\pi)}_{\rm pt}=\frac{2 \gamma_1^2}{-r_1^{(1/2)}} f^2 \int_0^\infty \, dr \,r^2 
\left(1 + \frac{1}{\gamma_1 r}\right)^2 e^{-2 \gamma_1 r}\,.
\end{equation}
which reproduces Eq.~(\ref{eq:LOpradresult}). 
Since the integral is finite, we can compute the contribution to it from 
values of $r \ll 1/\gamma_1$, and deduce that the short-distance part of the integral contributes 
to the total result according to:
\begin{equation}
\frac{\langle r_E^2 \rangle^{(\pi)}_{\rm SD}}{\langle r_E^2 \rangle^{(\pi)}_{\rm pt}} \sim \frac{M_{\rm halo}}{M_{\rm core}}.
\end{equation}
Such arguments using the asymptotic form of the
co-ordinate space wave function can be formalized using the renormalization group~\cite{Valderrama:2014vra}. 
The parametric dependence of this short-distance contribution on $1/M_{\rm core}$ 
agrees with the size of the counterterm obtained from naive dimensional analysis in Sec.~\ref{sec:EML}. 

It might seem counterintuitive that there is a short-distance contribution 
to $\langle r_E^2 \rangle^{(\pi)}$ already at NLO---especially when the 
corresponding effect does not occur in $\langle r_E^2 \rangle^{(\sigma)}$ 
until N$^3$LO (see Eq.~(\ref{eq:LC0sigma}) and Ref.~\cite{Chen:1999tn}). The physics 
of this is associated with the p-wave neutron's tendency to 
get caught between the 
attractive potential that produces the excited state of ${}^{11}$Be and the 
centrifugal barrier. Observables associated with a shallow p-wave bound state 
will, therefore, generically exhibit counterterms at lower order than their s-wave counterparts. 

\begin{figure}[!h]
\centerline{\includegraphics[width=0.6\columnwidth]{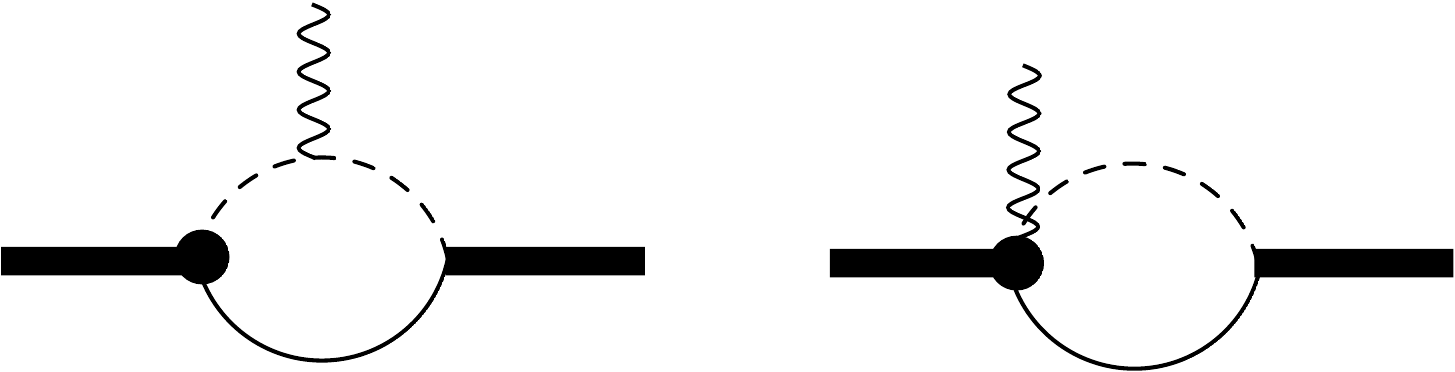}}
\caption{The two diagrams needed for the irreducible vertex that governs the s-to-p-state transition, 
$\Gamma_{j \mu}$ in Halo EFT at leading order.}
\label{fig:Gammajmu}       
\end{figure}

A similar effect enhances the importance of short-distance physics in the E1 transition 
 from the $1/2^+$ state to the $1/2^-$ state (for a spin-zero core). Recall that in Sec.~\ref{sec:EML} we already
 determined that the first counterterm would enter this observable at NLO. 
The irreducible vertex for the s-wave-to-p-wave transition is depicted in 
Fig.~\ref{fig:Gammajmu}. We compute the transition for a photon of arbitrary 
four momentum $k=(\omega,{\bs k})$, and the sum of diagrams yields 
the transition vertex function
$-i \Gamma_{s' s  \mu}$ where $s'$ ($s$) is the spin projection of the $1/2^-$ 
($1/2^+$) state and $\mu$ is the polarization index of the photon. 

The two diagrams depicted in Fig.~\ref{fig:Gammajmu} 
are both divergent, but the divergences cancel, as they must since gauge 
invariance precludes us from writing down any contact interaction that 
contributes to this observable at leading order. We also find that the sum of the two
diagrams yields a conserved current~\cite{Phillips:2010dt}:
\begin{equation}
k^\mu \Gamma_{s's \mu}=0\,.
\label{eq:currentcons}
\end{equation}
If only the long-distance E1 mechanism on the left-hand side of 
Fig.~\ref{fig:Gammajmu} is considered, as was done, for example, in 
Ref.~\cite{Typel:2008bw}, then Eq.~(\ref{eq:currentcons}) is not satisfied.

Since we are considering electric properties, and the spin of the neutron is 
not affected by the photon interaction, we can choose the photon to be traveling in the 
$\hat{z}$ direction, {\it i.e.}, $\vec{k}=|\vec{k}| \hat{z}$,
and 
it then follows,
using the definition of B(E1) strength (see, e.g.~\cite{Typel:2004zm}), that the transition strength is related
to the reduced, renormalizable
irreducible vertex ${\Gamma}_{++3}$ by:
\begin{equation}
{\rm B(E1)}=\frac{3}{4 \pi} \left(\frac{{\Gamma}_{++ 3}}{\omega}\right)^2.
\label{eq:BE1formula}
\end{equation}

Current conservation (\ref{eq:currentcons}) provides an alternative 
way to calculate $\Gamma_{++3}$, since it relates the third component of the pertinent current to the zeroth component:
\begin{equation}
\omega \Gamma_{++0}=|{\bs k}| \Gamma_{++3}\,.
\end{equation}
But, for $\Gamma_{++0}$, the diagram on the right of Fig.~\ref{fig:Gammajmu} 
needs not be considered, and so
\begin{equation}
\Gamma_{++0}({\bs k}) \propto  \int d^3r \frac{e^{-\gamma_1 r}}{r} \left(1 + \frac{1}{\gamma_1 r}\right) Y_{10}(\bs{\hat{r}}) 
e^{i {\bs k} \cdot {\bs r}} \frac{e^{-\gamma_{0,\sigma} r}}{r}\,,
\label{eq:Gamma0j}
\end{equation}
where there is a constant of proportionality here that we have omitted. 
As $|{\bs k}| \rightarrow 0$ Eq.~(\ref{eq:Gamma0j}) reduces to
the canonical form of the 
E1 matrix element. 

Evaluating the integral in either momentum or co-ordinate space yields:
\begin{equation}
{\rm B(E1)}=\frac{Q_c^2 e^2 f^2}{3 \pi} \frac{\gamma_{0,\sigma}}{-r_1^{(1/2)}} \left[
\frac{2 \gamma_1 + \gamma_{0,\sigma}}{(\gamma_{0,\sigma} + \gamma_1)^2}\right]^2
\label{eq:BE1}
\end{equation}
as the LO Halo EFT result.
No cutoff parameter is needed in order to get this finite result for B(E1): 
our value is finite without regularization, c.f. Ref.~\cite{Typel:2008bw}. We note that the result 
(\ref{eq:BE1}) is ``universal" in the sense that it applies to any E1 
s-to-p-wave transition in a one-neutron halo nucleus. Once $r_1^{(1/2)}$, $\gamma_1$, 
and $\gamma_{0,\sigma}$ are known for a given one-neutron halo the prediction 
(\ref{eq:BE1}) is accurate up to corrections of order $M_{\rm halo}/M_{\rm core}$.

At the next order in the expansion there are corrections to the s-wave and p-wave ANCs. Both tend to increase B(E1) over the 
LO prediction. 
Short-distance effects also enter B(E1) at NLO. The B(E1) 
($1/2^+ \rightarrow 1/2^-$) transition therefore cannot be predicted at 
NLO. This can be seen either from the 
presence of the operator $\sim L^{(1/2)}_{E1}$ in Eq.~(\ref{eq:nonminimalEMdimers}), or from 
a co-ordinate space argument similar to the one made above 
for the $1/2^-$ state's charge radius.

\subsection{Application: radius of and E1 transition to $1/2^-$ state in ${}^{11}$Be}

We choose to fix the value of the effective ``range" $r_1^{(1/2)}$ in the $1/2^-$ channel in ${}^{11}$Be by demanding that the experimental number for B(E1) obtained
in Ref.~\cite{Kwan:2014dha}: 
\begin{equation}
{\rm B(E1)}=(0.098 \pm 0.004) \, \, e^2 {\rm fm}^2
\end{equation}
is reproduced by the LO expression (\ref{eq:BE1}). This gives:
\begin{equation}
r_1^{(1/2) \, \rm LO}=-0.71~{\rm fm}^{-1},
\label{eq:r1LO}
\end{equation}
where we do not bother to propagate the error from the experiment, 
since NLO effects are presumably a much larger source of uncertainty. 

Numerical evaluation of the LO expression (\ref{eq:LOpradresult}) then leads to 
the prediction for the charge radius of the 
$^{11}$Be p-wave state relative to the $^{10}$Be ground state
$\langle r_E^2 \rangle^{(\pi)}_{\rm pt}=0.32$ fm$^{2}$ at LO.
The NLO correction to the p-wave ANC $C_1^{(1/2)}$ produces a 20\% increase in
$\langle r_E^2 \rangle^{(\pi)}_{\rm pt}$,
in agreement with the expectation from the power counting for ${}^{11}$Be.
However, we remind the reader that there is no prediction for the charge radius at NLO, since the
operator  
$\sim {\pi_\alpha^{(j)\,\dagger}} (\nabla^2 A_0) \pi_\alpha^{(j)}$ enters at that order. Thus the only prediction we can 
offer here is a leading-order one. Using again the experimental result for the 
$^{10}$Be charge radius \cite{Nortershauser:2008vp}, we
predict the (full, with core-radius included) electric radius of the $1/2^-$ state as:
\begin{equation}
\langle r_E^2 \rangle^{1/2}_{^{11}{\rm Be}^*}=(2.42 \pm 0.1)\, {\rm fm}
\label{eq:rE2pi}
\end{equation}
with the error solely from the anticipated size of NLO effects. To our knowledge there is, as yet, no experimental determination of the 
electric radius of the $1/2^-$ state in ${}^{11}$Be.

\subsection{Radiative capture into a p-wave halo}

We now turn our attention to radiative capture from an s-wave neutron-core state to a shallow p-wave bound state, as occurs in the transition ${}^7$Li $+ n \rightarrow {}^8$Li $+ \gamma$. In this section we present a LO calculation of this process, as reported in Ref.~\cite{Zhang:2013kja}.

\begin{figure}[ht]
\centerline{\includegraphics[scale=0.8, angle=0]{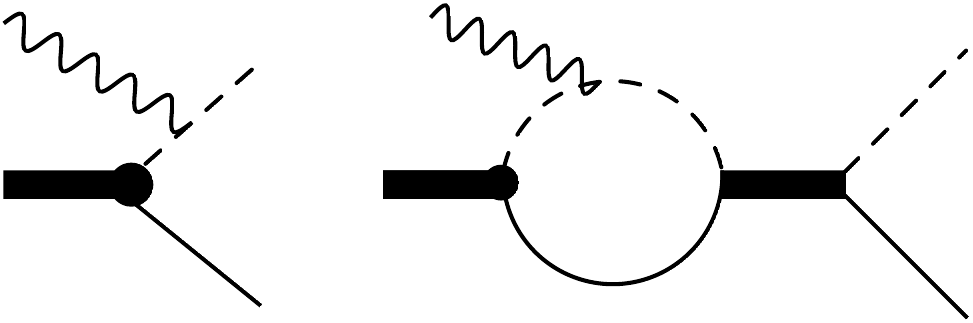}}
\caption{Tree and loop diagrams for neutron-capture to $\lie$ and $\lie^{*}$. The first diagram
is LO for both the initial spin channels:  $s=2$ and $s=1$. The dominant components in the initial state are $\S{5}{2}$ and $\S{3}{1}$, but d-wave components also contribute. In the second diagram the incoming particles scatter in the s-wave, with that scattering amplitude encoded in the EFT via an s-wave dimer propagator. Only the $s=2$ channel contributes at LO, while perturbative initial-state scattering effects in the $s=1$  channel enter at NLO.} \label{fig:ncapture}
\end{figure}

The LO tree-level diagrams for the capture reaction are shown in Fig.~\ref{fig:ncapture}. The general incoming $n$-${}^7$Li state can be decomposed to states of initial spin $s=2$ and $s=1$. Both spin channels play a role in the formation of the p-wave ${}^8$Li state, as discussed in Sec.~\ref{sec:pwavehaloapplications}. Therefore, in each spin channel both s- and d-wave initial-state components can contribute to the reaction. For the second, loop, diagram in Fig.~\ref{fig:ncapture} only the incoming s-wave contributes, and then only for $s=2$, where there is an unnaturally large scattering length $\atwo=-3.63(5) \ \mathrm{fm}$ \cite{Koester:1983}. The $s=1$ loop diagrams  do not appear until NLO because the scattering length in that channel is natural: $\aone=0.87(7) \ \mathrm{fm}$~\cite{Koester:1983}.

The loop diagram thus generates an initial-state interaction that affects only the ${}^5$S$_2 \rightarrow {}^8 {\rm Li} (2^+) + \gamma$ partial cross section. The Lagrangian (\ref{eq:nonminimalEMdimers}) implies that there is a contact term associated with this E1 $\S{5}{2} \rightarrow \lie$ transition already at NLO. However, the LO capture amplitude in this channel is a prediction of the EFT. 

The differential cross section for radiative capture of two non-relativistic particles into a non-relativistic bound state is
\begin{equation}
\frac{\d\sigma}{\d\Omega}=\frac{\mu_\sigma\omega}{8\pi^2p}\sum_i\left| \bs{\epsilon}_i\cdot\mathcal{{\bs M}}\right|^2~,
\end{equation}
where $\omega$ is the energy of the outgoing photon, $p=\sqrt{2 \mu_\sigma E}$ is the
relative momentum of the incoming particle pair and $\bs{\epsilon}_i$ are
the photon polarization vectors. The vector amplitude $\mathcal{{\bs M}}$ is
for the capture process with a vector photon $A_i$ being emitted. Any spin indices associated with the incoming particles have been implicitly averaged over, and those associated with the bound state are implicitly summed. Note
that we are working in Coulomb gauge, where the relation
\begin{equation}
{\bs \epsilon}_i\cdot\qq=0~,
\label{eq:Ward1}
\end{equation}
for a real photon with momentum $\qq$, is fulfilled.

The diagrams in Fig.~\ref{fig:ncapture} then yield a partial cross section: 
\begin{eqnarray}
\fl
\sigma =& \frac{5 \pi}{2} \alpha_{em} Q_c^2 C_{({}^5P_2)}^2 \frac{\omega}{m_c^2 S_{1n} p} 
\nn
\fl
& \times \left[|1 + X(p;a_{({}^5S_2)},\gamma_1)|^2 
- \frac{4 E}{3 (E + S_{1n})} \left(\frac{S_{1n}}{E + S_{1n}} + {\rm Re}[X(p;a_{({}^5S_2)},\gamma_1)]\right)\right].
\label{eq:crosssection}
\end{eqnarray}
In Eq.~(\ref{eq:crosssection}) $S_{1n}=\frac{\gamma_1^2}{2 \mu_\sigma}$ is the one-neutron separation energy of ${}^8$Li and  $X$ 
is a function that encodes loop contributions for capture from an s-wave state of relative momentum $p$, where there is a large scattering length, $a_0$, to a p-wave state with binding momentum $\gamma_1$. It is defined as:
\begin{equation}
X(p;a_0,\gamma_1) \equiv  \frac{-i}{a_0^{-1}+ip} \left[p-\frac{2}{3}i\frac{\gamma_1^{3}-ip^{3}}{\gamma_1^{2}+p^{2}}\right]  \ . \label{eqn:xtwodef}
\end{equation}
The factors of $X$ are absent for the ${}^3$S$_1$ initial state at leading order, since the scattering length is natural there. 

\begin{figure}
\centering
\includegraphics[width=0.65\linewidth]{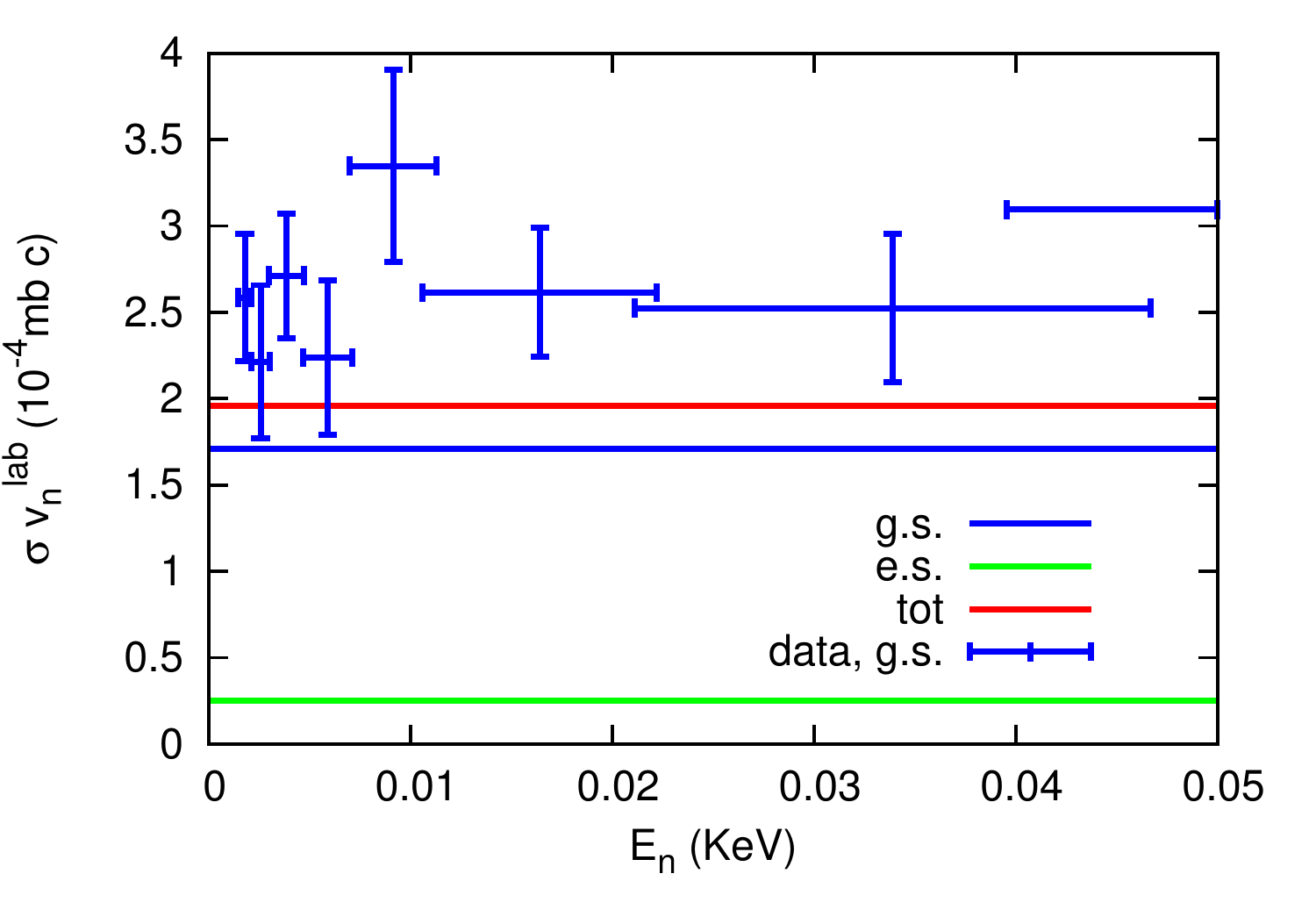}
\caption{Total cross section $\times$ neutron velocity vs.~neutron lab energy for $\lis(n,\gamma)$ at leading order in Halo EFT. ``g.s.'' and  ``e.s.'' correspond to capture to $\lie$ and $\lie^{*}$, while ``tot'' is the sum of these two; data from Ref.~\cite{Blackmon:1996zz}. Figure from Ref.~\cite{Zhang:2013kja}.} \label{fig:xsecvel}
\end{figure}

We display the LO result for capture into the ground and excited state in Fig.~\ref{fig:xsecvel}. The nominal accuracy of the LO amplitude is $\sim \gamma_1/r_1 \approx 20$\%, which translates into an uncertainty of $\approx 40$\% for the cross section. There is also a much smaller uncertainty ($< 5$\%) in the cross-section prediction due to the uncertainties in the VMC ANCs, see table~\ref{tab:lieANCs}. We find agreement between theory and experiment within the combined error bars.

There is also data on the ratio of the partial cross sections associated with 
different initial spin states as well as for the branching ratios to different final states.
LO  Halo EFT results based on VMC ANCs agree very well with these data on relative amplitudes.
Considering first the relative contributions of different initial spin states at $p=0$
gives
\begin{equation}
\frac{\sigma[({}^3{\rm S}_1)\rightarrow 2^{+}]}{\sigma[({}^5{\rm S}_2)\rightarrow 2^{+}]}=  \frac{\left(C_{(\P{3}{2})}^{\mathrm{LO}}\right)^{2}}{\left(C_{(\P{5}{2})}^{\mathrm{LO}}\right)^{2}(1-\frac{2}{3}\gamma_1 a_{(\S{5}{2})})^{2}} \ .  
\end{equation}  
which says that the partial cross section for threshold or near-threshold capture from the ${}^5$S$_2$ initial state contributes 93\% of the total. 
In Ref.~\cite{Gulko:1968} (c.f.~Ref.~\cite{Barker:1995oba}), an experimental lower bound of 86\% for this ratio has 
been reported.
Ref.~\cite{Zhang:2013kja} also predicted the corresponding ratio of $s=2$ and $s=1$ partial cross sections for capture to the excited state, but there is no experimental data on this ratio. 

Finally, we discuss the branching ratio for capture to the ground state.
Near threshold, the ratio of capture cross sections is
\begin{eqnarray}
\frac{\sigma(\rightarrow 1^+)}{\sigma(\rightarrow 2^+)}&=&\frac{3}{5} \frac{\left(\tilde{C}^{\rm LO}_{(\P{3}{1})}\right)^2 + \left(\tilde{C}^{\rm LO}_{(\P{5}{1})}\right)^2 |1 - \frac{2}{3} \tilde{\gamma}_1 \atwo |^2}{\left(C^{\rm LO}_{(\P{3}{2})}\right)^2 + \left(C^{\rm LO}_{(\P{5}{2})}\right)^2 |1 - \frac{2}{3} \gamma_1 \atwo |^2} . \label{eq:1plusto2plus}\nonumber\\
\end{eqnarray}
Numerical evaluation then predicts a branching ratio of $0.88$ for capture to the ground state, with a theory uncertainty of $\pm 0.04$~\cite{Zhang:2013kja}. 
Note that the initial-state interaction effect of the large s-wave scattering length in $s=2$ channels plays a role here. Refs.~\cite{Lynn:1991zz,Nagai:2005qu}, report
branching ratios of $0.89 \pm 0.01$ for thermal neutrons and neutrons with energies of 20--70 keV respectively.  Both of these are in excellent agreement with the Halo EFT number.

We note that this additional input information from {\it ab initio} calculations let Ref.~\cite{Zhang:2013kja} predict ratios of partial cross sections in the $\lis(n,\gamma)\lie$ and  $\lis(n,\gamma)\lie^*$ reactions as 
dynamical 
quantities related to the couplings in the EFT Lagrangian.  In
contrast, the couplings of spin channels in the final state were
assumed
for simplicity
 to be equal in Ref.~\cite{Rupak:2011nk,Fernando:2011ts} so that 
the branching
ratios there carry no link to the actual short-range physics at work in the $n$-$\lis$ system. Reference \cite{Rupak:2011nk} 
consequently failed
to satisfy the experimental lower bound on the fraction of the cross section to the ground state that comes from the ${}^5$S$_2$ initial state. 

\subsection{Electromagnetic radii of two-neutron halos}

Once the radius $\langle r_{c-nn}^2 \rangle$ has been computed as described in Section~\ref{sec:3Bswave} we can then form the point-electric radius of a two-neutron halo:
\begin{equation}
\langle r_E^2 \rangle_{\rm pt}=\left(\frac{2}{A+2}\right)^2 \langle r_{c-nn}^2 \rangle.
\end{equation}
This can be converted to a total electric radius by accounting for the finite electric radii of the core and the neutrons. Ref.~\cite{Hagen:2013xga} developed a gauge invariant formalism for the electric form factors of two-neutron halo nuclei and computed the electric radii of several halo nuclei. However, their calculation contained an error in the pre-factor of one term, as noted by Vanasse in Ref.~\cite{Vanasse:2016hgn}. In Table~\ref{tab:electricradii} we quote Vanasse's LO electric radii for ${}^{11}$Li, ${}^{12}$Be, and ${}^{22}$C, together with his earlier computation of the electric radius of ${}^3$H~\cite{Vanasse:2015fph}. The input parameters are the neutron-neutron scattering length, the neutron-core resonance energy, $S_{1n}$---which is negative for all but ${}^3$H since the other three are Borromean systems---and the two-neutron separation energy, $S_{2n}$.  The relevant $S_{1n}$'s are given in Table~\ref{tab:1n-halo} and the assumed $S_{2n}$ is shown in Table~\ref{tab:electricradii}. 

Electric radii have been measured for ${}^3$H and ${}^{11}$Li. In the latter case Ref.~\cite{Puchalski:2006zz} updated the earlier experimental result of Ref.~\cite{Sanchez:2006zz} to obtain  $\langle r_E^2 \rangle_{{}^{11}\rm Li,\, pt}=1.104(85)$ fm$^2$. 
This disagrees with the LO Halo EFT result---but the disagreement could be explained by the anticipated $\approx 40$\% NLO corrections that the point radius will receive due to range effects. Indeed, in the case of the triton the LO result is also well away from 
the experimental number of $2.55(11)$ fm$^2$. However, Vanasse's NLO calculation is in perfect agreement with this result: corrections due to the finite effective ranges in the NN system turn out to be large in this observable. Order-by-order results are:
\begin{equation}
\langle r^2_{{}^3{\rm H}} \rangle_{\rm pt}=1.30 + 1.23 + 0.096~{\rm fm}^2.
\end{equation}
However, the triton remains the only two-neutron halo for which ranges are well-enough determined for an accurate NLO calculation like this to have been completed.
An NLO computation of the ${}^{11}$Li electric radius would be an important further development---especially in terms of investigating the role that p-wave interactions in the ${}^{10}$Li system have on this observable. However, in order for a definitive NLO number for the ${}^{11}$Li electric radius to emerge the value of the effective range for s-wave $n$-${}^9$Li scattering needs to be pinned down.   

\begin{table}
   \centering
   \begin{tabular}{c c c} \hline
   	 Nucleus &       $S_{2n}$ [MeV] &         $\bra r^2_{E}\ket_{\rm pt}$ [fm$^2$]\\
	 \hline
          $^3$H   &		8.48		    &				1.30\\
          $^{11}$Li  &	      0.3693(6) 	    &				0.744\\
          $^{14}$Be &	1.27(13)	    &				0.126\\
          $^{22}$C   &	0.11(6)	    &				0.519$^{+\infty}_{-0.274}$\\
           \hline
 \end{tabular}
\caption{Two-neutron separation energies and leading-order electric radii squared for four different two-neutron halos. Adapted from Refs.~\cite{Vanasse:2015fph} and \cite{Vanasse:2016hgn}.}
   \label{tab:electricradii}
\end{table}

\subsection{Coulomb dissociation of two-neutron halos}

The E1 response of a two-neutron halo is also governed by a universal function, in close analogy to the one-neutron case discussed above. Preliminary Halo EFT calculations in this direction were reported in Refs.~\cite{Acharya:2015,Acharya:2015gpa,Hagen:2014}, following work within three-body models in Refs.~\cite{Ershov:2012fy,Kikuchi:2013ula}. In the case of ${}^{11}$Li, the calculation showed good agreement with the E1 strength extracted from Coulomb dissociation data in Ref.~\cite{Nakamura:2006zz} at transition energies within the domain of validity of Halo EFT.

\section{Halo EFT with Coulomb} \label{sec:coulomb}

\subsection{Formalism}

Halo systems with two or more charged particles also have
electromagnetic interactions. In Coulomb gauge,
the interaction can be split
into an instantaneous Coulomb interaction and the exchange of
transverse photons. In the low-energy 
regime, the dominant effect is given by the Coulomb interaction, i.e., 
a static $1/r$ potential between charged particles.
The exchange of transverse photons, as well as magnetic
interactions, are suppressed by $(\vec{p}/m)^2$ where  $\vec{p}$ is a typical
momentum and $m$ the mass of the particle.

In halo nuclei the Coulomb interaction is repulsive. Its presence 
introduces a new scale, the Coulomb momentum
$k_C$, which is given by the inverse Bohr radius of the system.  This 
scale $k_C$ is independent of the scales $M_{\rm halo}$ and
$M_{\rm core}$ and complicates the power counting. In the effective 
Lagrangian, the Coulomb interaction between the core and a halo proton
can be incorporated as a non-local term
\begin{equation}
 \mathscr{L}_{\rm Cb}
 = \frac{-e^2Q_c}{\vec{q}^2} \left.\int d^3y \,e^{-i\vec{q}\cdot(\vec{x}-\vec{y})}
 c^\dagger(x)c(x)
 p^\dagger(y)p(y)\right|_{y_0=x_0} \,,
\label{eq:L-Coulomb}
\end{equation}
where $Q_c$ is the charge of the core,
$\vec{q}$ is the three-momentum transfer,
and $c$ ($p$) denote core (proton) fields, repectively.
The full interaction between the proton and the core
then consists of the long-range
Coulomb photon exchange and the short-range strong interactions. In Halo EFT,
the latter are represented by contact interactions while the Coulomb
interaction must be treated explicitly. A central issue is the relative
importance of these contributions, which is characterized by
the Sommerfeld parameter $\eta=k_C/p$, where $k_C = Q_c \alpha_{em} \mu$
is the Coulomb momentum scale, with
$\mu$ the reduced mass of the system and $p$ a typical momentum
of order $M_{\rm halo}$. Thus, at low energies $\eta\gtrsim 1$ and 
Coulomb becomes non-perturbative. At sufficiently high (binding) energy,
however, Coulomb can be included in perturbation theory.
We will discuss several examples below.

The Coulomb interaction was first included in the framework
of pionless EFT by Kong and Ravndal for the case $\eta \gtrsim 1$
\cite{Kong:1998sx,Kong:1999sf}.
They calculated low-energy proton-proton scattering in an effective field
theory with four-proton contact interactions. Here, we focus
on halo nuclei and use a dimer formalism following the discussion
in Refs.~\cite{Ryberg:2013iga,Ryberg:2015lea}. We will also
consider the case $\eta \gtrsim 1$.

We start with s-wave interactions and later discuss the extension
to p-waves.
The Lagrangians for the strong proton-core and proton-proton
interactions are as in Eqs.~(\ref{eq:Lag-nn-aux}, \ref{eq:Lag-nc-aux}),
with $n$ replaced by $p$, respectively. Similarly, the
propagators for $c$ and $p$ are obtained from Eq.~\eqref{eq:prop-1b}.
For convenience, we will also define the noninteracting proton-core
Green's function in the center-of-mass frame,
\begin{equation}
  iG_0(E,\pp)=\frac{i}{E-\pp^2/(2 \mu_\sigma)+i\epsilon},
\label{eq:Stot}
\end{equation}
where $\mu_\sigma$ denotes the reduced mass of the proton-core system
which is represented by the $\sigma$ field. Analogously, the proton-proton
system is represented by the $d$ field.

In  the dimer propagators for the $\sigma$ and $d$ fields,
the Coulomb interaction has to be
included. Since we only consider one-proton halos, we
discuss the $\sigma$ propagator in detail. The $d$
propagator, which is relevant for the proton-proton system
and for two-proton halos, can be obtained analogously.
\begin{figure}[t]
\centerline{\includegraphics*[scale=0.6,angle=0,clip=true]{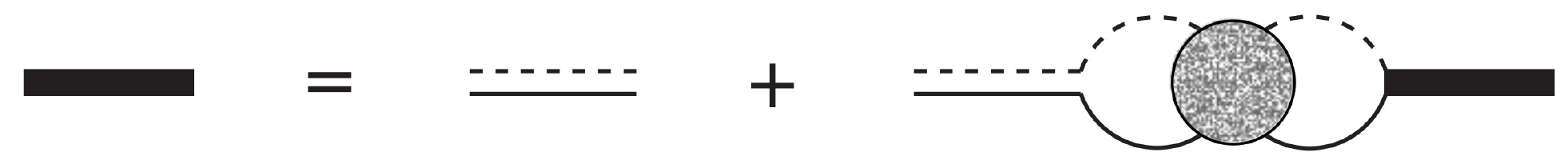}}
\caption{Integral equation for the full halo propagator (thick line).
  The double solid/dashed line
  denotes the bare halo propagator, the dashed single line denotes the
  core field, the solid single line denotes the proton field and the
  shaded blob the Coulomb four-point function $\chi$ defined in 
  Fig.~\ref{fig:fourpointchi}.}
\label{fig:FullDibaryon}
\end{figure}

The proton-core propagator at rest is given by
\begin{equation}
iD^{(bare)}_\sigma (E,{\mathbf{0}})\equiv
iD^{(bare)}_\sigma (E)=\frac{i}{\Delta_\sigma+w_\sigma\left(E+i\epsilon\right)}.
\end{equation}
The corresponding propagator for finite momentum $\pp$ can always be
obtained by replacing $E\to E-\pp^2/(2M_\sigma)$.  The power
counting for large scattering length
requires that the s-wave interaction is summed up to all orders
\cite{vanKolck:1998bw,Kaplan:1998tg,Kaplan:1998we}. The resulting full
$\sigma$ propagator is thus given by the integral equation shown in
Fig.~\ref{fig:FullDibaryon}. For a $\sigma$ field at rest, we obtain
\begin{equation}
iD_\sigma(E)=\frac{i}{\Delta_\sigma+w_\sigma\left(E+i\epsilon\right)-\Sigma(E)}.
\label{eq:FullDibProp}
\end{equation}
The irreducible self-energy, $\Sigma$, now includes strong and
Coulomb interactions and will be discussed below.

\begin{figure}[t]
\centerline{\includegraphics*[scale=1.,angle=0,clip=true]{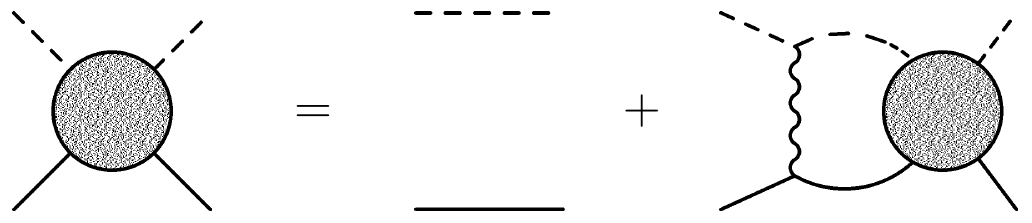}}
\caption{The four-point function $\chi$ defined iteratively. The
  wiggly line denotes a Coulomb photon exchange. 
  External propagators are amputated.
  Otherwise, the notation is as in Fig.~\ref{fig:FullDibaryon}.}
\label{fig:fourpointchi}
\end{figure}

We include the Coulomb interaction between the core and the valence
proton through the full Coulomb Green's function,
\begin{equation}
  \langle \kk | G_\mrm{C}(E)|\pp\rangle = -G_0(E,\kk)\chi(\kk,\pp;E)
   G_0(E,\pp)~,
\label{eqCGFGamma}
\end{equation}
where $\pp$ and $\kk$ are the relative 
incoming and outgoing momenta and $E$ is the energy.
The momentum-space 
Coulomb four-point function $\chi$ in the center-of-mass
frame of the proton and the core is given by the integral equation
depicted in
Fig.~\ref{fig:fourpointchi}. To distinguish coordinate-space from
momentum-space states we denote the former with round brackets,
i.e., $|\rr)$. The Coulomb Green's function can be expressed via its
spectral representation in coordinate space
\begin{equation}
(\rr|G_\mrm{C}(E)|\rr')=\int\frac{d^3p}{(2\pi)^3}
\frac{\psi_\pp(\rr)\psi^*_\pp(\rr')}{E-\pp^2/(2\mu_\sigma)+i\epsilon},
\label{eq:CGFSpectral}
\end{equation}
where the Coulomb wave function $\psi_\pp(\rr)$ is the solution of the
Schr\"odinger equation for a Coulomb potential.
We express $\psi_\pp(\rr)$ through its partial wave expansion
\begin{equation}
\psi_\pp(\rr)=\sum_{l=0}^{\infty}(2l+1)i^l\exp{(i\sigma_l)\frac{F_l(\eta,\rho)}{\rho}P_l(\hat{\pp}\cdot\hat{\rr})},
\end{equation}
where $\rho=pr$ and $\eta= k_C/p$ is the Sommerfeld parameter.
We have also introduced the pure Coulomb phase shift
$\sigma_l=\arg{\Gamma(l+1+i\eta)}$. For the Coulomb functions $F_l$
and $G_l$, we use the conventions of Ref.~\cite{Koenig:2012bv}.  The
regular Coulomb function $F_l$ can be expressed in terms of the
Whittaker M-function according to
\begin{equation}
F_l(\eta,\rho)=A_l(\eta)M_{i\eta,l+1/2}(2i\rho),
\label{eq:CoulWavefunctionF}
\end{equation}
with
\begin{equation}
A_l(\eta)=\frac{|\Gamma{(l+1+i\eta)}|\exp{\left[-\pi\eta/2-i(l+1)\pi/2\right]}}{2(2l+1)!}.
\end{equation}
We will also need the irregular Coulomb wave function, $G_l$, which is given by
 \begin{equation}
G_l(\eta,\rho)=iF_l(\eta,\rho)+B_l(\eta)W_{i\eta,l+1/2}(2i\rho)~,
\end{equation}
where $W$ is the Whittaker W-function and the coefficient $B_l$ is defined as
\begin{equation}
B_l(\eta)=\frac{\exp{(\pi\eta/2+il\pi/2)}}{\exp(i\arg{\Gamma{(l+1+i\eta))}}}.
\end{equation}
Finally, the absolute value and the argument of the $\Gamma$-function are
given by
\begin{equation}
|\Gamma{(l+1+i\eta)}|=\sqrt{\Gamma{(l+1+i\eta)}\Gamma{(l+1-i\eta)}}
\end{equation}
and
\begin{equation}
\exp(i\arg{\Gamma{(l+1+i\eta)}})=\sqrt{\frac{\Gamma{(l+1+i\eta)}}{\Gamma{(l+1-i\eta)}}}~.
\end{equation}

\begin{figure}[t]
\centerline{\includegraphics*[width=0.35\linewidth,angle=0,clip=true]{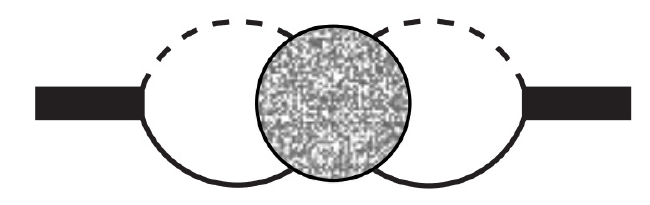}}
\caption{Feynman diagram for the irreducible self energy. The injected
  four-momentum is $(E,{\mathbf{0}})$. External legs are amputated.
  Otherwise, the notation is as in 
  Fig.~\ref{fig:FullDibaryon}.}
\label{fig:IrreducibleSelfEnergy}
\end{figure}

To obtain the full $\sigma$ propagator $D_\sigma$, which includes
strong and Coulomb interactions, we calculate the irreducible
self-energy shown in Fig.~\ref{fig:IrreducibleSelfEnergy}. 
Using Eq.~(\ref{eqCGFGamma}), it can be expressed as
\begin{equation}
-i\Sigma(E)=-ig_\sigma^2\int\frac{\d^3k_1\d^3k_2}{(2\pi)^6}~\langle \kk_2 | G_\mrm{C}(E)|\kk_1\rangle,
\end{equation}
which can be written in coordinate space using Fourier
transformations:
\bea
\Sigma(E)&=&g_\sigma^2(0|G_\mrm{C}(E)|0)\nonumber\\
&=&g_\sigma^2\left(\frac{\Lambda}{2}\right)^{4-D}
\int\frac{\d^{D-1}p}{(2\pi)^{D-1}}
\frac{\psi_\pp(0)\psi^*_\pp(0)}{E-\pp^2/(2\mu_\sigma)+i\epsilon},
\eea
where we have continued the last integral to $D-1$ spatial dimensions, and introduced the renormalization
scale $\Lambda$ to maintain the correct dimensionality of $\Sigma(E)$.
Evaluating this integral using dimensional regularization,
we obtain \cite{Kong:1999sf}
\begin{equation}
\Sigma(E)=-g_\sigma^2\frac{k_{C}\mu_\sigma}{\pi}H(\eta)-\Sigma^{div},
\end{equation}
with
\begin{equation}
H(\eta)=\psi(i\eta)+\frac{1}{2i\eta}-\log{(i\eta)},
\end{equation}
where $\psi(z)=\Gamma'(z)/\Gamma(z)$ is the logarithmic derivative
of the Gamma function and $\eta=k_C/k$. In the strong Coulomb regime with
$\eta \gtrsim 1$, the function $H(\eta)$ can be expanded as
\begin{equation}
  \label{eq:H-expand}
  H(\eta)=\frac{1}{12\eta^2}+\frac{1}{120\eta^4}+\ldots+\frac{i\pi}{e^{2\pi\eta}-1}\,,
\end{equation}
leading to a simplified treatment of the Coulomb
interaction~\cite{Higa:2008dn}. Note
that $H(\eta)$ has nothing to do with the three-body force $H(\Lambda)$
that appeared earlier. The difference between the two functions is obvious
from the context.

Using PDS~\cite{Kaplan:1998tg,Kaplan:1998we}, the divergent part of $\Sigma$ is given by
\begin{equation}
 \label{eq:app-renorm}
\Sigma^{div}=-\frac{g_\sigma^2 k_{C}\mu_\sigma}{\pi}
\Big[\frac{1}{3-D}+\log{\left(\frac{\sqrt{\pi}\Lambda}{2k_{C}}\right)}
+1-\frac{3C_{E}}{2}\Big]+\frac{g_\sigma^2\mu_\sigma\Lambda}{2\pi},
\end{equation}
where $C_E=0.5772$ is Euler's constant. Note that $\Sigma^\mrm{div}$
is independent of energy and therefore will vanish when we take the
energy derivative of the irreducible self-energy to arrive at the LSZ
residue below. Moreover, Eq.~\eqref{eq:app-renorm} absorbs two 
divergences into a single parameter: the linear dependence on $\Lambda$
from the PDS scheme and a logarithmic term in $\Lambda$ from the
one photon exchange.  An alternative scheme for the perturbative case
$\eta \ll 1$ that 
isolates the linear and logarithmic divergences has been suggested in Ref. ~\cite{Konig:2015aka}.

\subsection{Renormalization}

\label{sec:scatteringwithCoulomb}

\begin{figure}[t]
\centerline{\includegraphics*[scale=0.6,angle=0,clip=true]{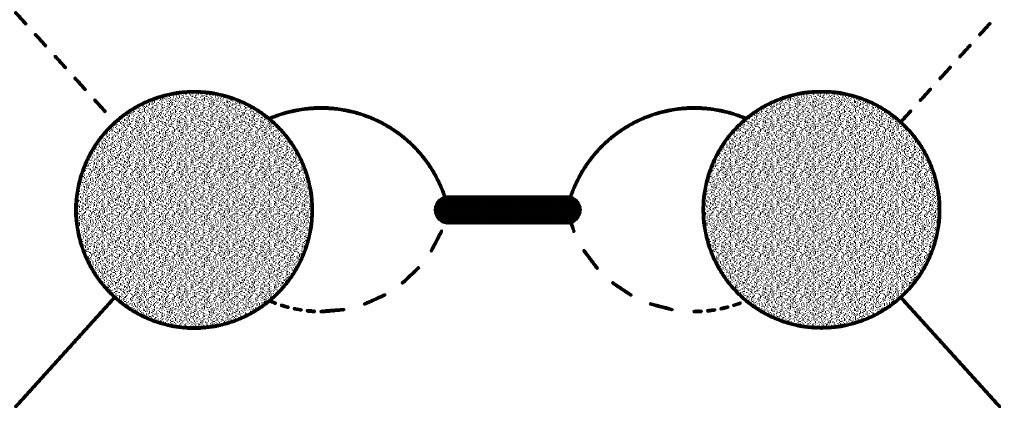}}
\caption{Scattering amplitude, $t_{CS}$, for elastic proton-core scattering. 
The notation is as in Fig.~\ref{fig:FullDibaryon}.}
\label{fig:ElasticScattering}
\end{figure}

Expressions for EFT low-energy constants such as $g_\sigma$ and $\Delta_\sigma$
defined above are frequently obtained by matching them to elastic
scattering observables. The full amplitude for elastic proton-core
scattering consists of two parts: the amplitude for pure Coulomb 
scattering $t_C$ and the Coulomb-modified strong 
scattering amplitude $t_{CS}$.
The former is essentially given by the four-point function $\chi$ depicted in
Fig.~\ref{fig:fourpointchi}. For matching to the Coulomb-modified
effective range expansion only the latter
amplitude $t_{CS}$ is required.
The amplitude $t_{CS}$ is obtained from the diagram in
Fig.~\ref{fig:ElasticScattering} and evaluates to
\begin{equation}
  it_{0,\sigma}(E)=ig_\sigma^2\exp{(2i\sigma_0)}C_\eta^2D_\sigma(E)~,
\label{eq:tMatrixA}
\end{equation}
with the Gamow-Sommerfeld factor $C_\eta^2=\Gamma{(1+i\eta)}\Gamma{(1-i\eta)}$. 
The index $CS$ has been dropped for convenience.

In the absence of the Coulomb interaction, the t-matrix is usually
expressed in terms of effective range parameters. However, it is not
possible to separate the strong interaction from the Coulomb
interaction in a model-independent way \cite{Kong:1998sx,Kong:1999sf}.
Therefore one uses the
so-called Coulomb-modified effective range expansion
\cite{Bethe:1949yr} to relate the phase shifts to redefined effective range
parameters. The t-matrix is then written
\begin{equation}
   t_{0,\sigma}(E)=-\frac{2\pi}{\mu_\sigma}\frac{C_\eta^2\exp{(2i\sigma_0)}}{kC_\eta^2(\cot{\delta_0}-i)}~,
\label{eq:tMatrixB}
\end{equation}
where the total phase shift is given by $\sigma_0+\delta_0$.  The
s-wave Coulomb-modified ERE is
\begin{eqnarray}
  kC_\eta^2(\cot{\delta_0}-i)+2\kC H(\eta)=-\frac{1}{a^C_{0,\sigma}}
  +\frac{1}{2}r^C_{0,\sigma}k^2+\ldots~,
\label{eq:CoulModERE}
\end{eqnarray}
where $a^C_{0,\sigma}$ and $r^C_{0,\sigma}$ are the Coulomb-modified scattering length and
effective range, respectively. In Eq.~\eqref{eq:CoulModERE}, the imaginary part
of $2k_\mathrm{C}H(\eta)$ exactly cancels $-ikC_\eta^2$, leaving an analytic,
real function of $k^2$.

Comparing Eqs.~(\ref{eq:tMatrixA}) and (\ref{eq:tMatrixB}), order
by order in the momentum $k$, we can express the Coulomb-modified
scattering length and
effective range in terms of the low-energy coupling constants $g_\sigma$ and
$\Delta_\sigma$
\bea
\label{eq:a-renorm}
\frac{1}{a^C_{0,\sigma}}&=&\frac{2\pi}{g_\sigma^2\mu_\sigma}\left(\Delta_\sigma+\Sigma^\mrm{div}\right)~,\\
r^C_{0,\sigma}&=&-\frac{2\pi w_\sigma}{g_\sigma^2\mu_\sigma^2}~.
\label{eq:r-renorm}
\eea
These equations are the equivalent of Eqs.~\eqref{eq:a-s-wave}, \eqref{eq:r-s-wave}
in the presence of Coulomb interactions.
Equation (\ref{eq:a-renorm}) shows how the low-energy constant
$\Delta_\sigma$ absorbs the divergent part of the irreducible self-energy
$\Sigma^\mathrm{div}$. Note that here and below LO results can be obtained
by setting $w_\sigma=0$. When this is done results in the EFT
depend only on the combination $\Delta_\sigma/g_\sigma^2$.

The residue at the bound state, $E=-B$, of the full $\sigma$
propagator defines the wave-function renormalization,
which is required for the calculation of bound-state
observables. Note that we keep the formalism general here. In
our application to proton halos below, $B$ is to be identified with
the one-proton separation energy, $S_{1p}$.
Using Eq.~\eqref{eq:Z-factor}
and the matching condition Eq.~(\ref{eq:r-renorm}), we obtain
\bea
Z_\sigma&=&
\frac{1}{w_\sigma-\Sigma'(-B)}
=\frac{6\pi\kC}{g_\sigma^2\mu_\sigma^2}\frac{-1}{\tilde{H}(\gamma,\kC)
  -3\kC r^C_{0,\sigma}},
\label{eq:LSZRes1}
\eea
In writing
Eq.~(\ref{eq:LSZRes1}), we have defined the function
\begin{equation}
\tilde{H}(\gamma,\kC)=\left.\frac{6\kC^2}{\mu_\sigma}\frac{d}{d E}H(\eta)\right|_{E=-B},
\end{equation}
with the binding momentum $\gamma=\sqrt{2\mu_\sigma B}$.
The expression Eq.~(\ref{eq:LSZRes1}) is valid at next-to-leading order
(NLO), since it includes the effective-range correction.
The corresponding expression at leading order (LO) is obtained by
setting $r^C_{0,\sigma}$ to zero in Eq.~\eqref{eq:LSZRes1}.

If the ratio between the binding momentum and Coulomb momentum
$\gamma/k_\mathrm{C}$ is small, we can use the expansion in
Eq.~\eqref{eq:H-expand} to obtain
\begin{equation}
\tilde{H}(\gamma,\kC)=1-\frac{\gamma^2}{5\kC^2}+\frac{\gamma^4}{7\kC^4}+\dots.
\end{equation}
Thus, for systems where the separation $\gamma\ll\kC$ is fulfilled we
can use $\tilde{H}(\gamma,\kC)\to1$ in all expressions~\cite{Higa:2008dn}.

The residue $Z_\sigma$ can be related to the asymptotic normalization coefficient (ANC),
 $C_\sigma$, or to experimental radiative capture data. The ANC is defined as the coefficient in the asymptotic bound
 state wavefunction
\begin{equation}
w_l(r)=C_\sigma W_{-i\eta,l+1/2}(2\gamma r),
\end{equation}
where $W$ is the Whittaker-W function. The $Z$-factor is related to the ANC according to
\begin{equation}
Z_\sigma=\frac{-\pi}{g_\sigma^2\mu_\sigma^2\left[\Gamma(1+\kC/\gamma)\right]^2}C_\sigma^2.
\label{eq:Z-ANC}
\end{equation}
At LO, the wave-function renormalization is determined solely by
$\gamma$ and $\kC$, as can be seen in Eq.~(\ref{eq:LSZRes1})
with $r^C_{0,\sigma}=0$, and as such
the ANC is predicted to LO accuracy as
\begin{equation}
C_{\sigma, LO}=\sqrt{\frac{6\kC}{\tilde{H}(\gamma,\kC)}}\Gamma(1+\kC/\gamma).
\label{eq:LOANC}
\end{equation}

At NLO, $r^C_{0,\sigma}$ contributes as well. It is
advantageous to use the $z$-parametrization and fix $Z_\sigma$
directly from the ANC. For this purpose,
it is useful to define the
NLO wavefunction renormalization in terms of the matching to the ANC,
Eq.~(\ref{eq:Z-ANC}). The ratio to the LO residue is then obtained as
\begin{equation}
\frac{Z_\sigma}{Z_{\sigma,LO}}=
\frac{C_\sigma^2}{C_{\sigma, LO}^2}~.
\label{eq:ZNLOoverZLO}
\end{equation}
It is important to note that $Z_\sigma$ and the ANC receive contributions from
all ERE parameters. As such the expression (\ref{eq:LSZRes1}) is only valid at NLO.
However, the matching (\ref{eq:Z-ANC}) itself is valid at any order in the power
counting and there is no EFT error due to the non-inclusion of higher-order contact
interactions.

For completeness, we also provide expressions for the effective range and the scattering length in the $z$-parameterization, up to corrections due to higher-order ERE parameters.
For a given one-proton separation energy $B=\gamma^2/(2\mu_\sigma)$,
the ANC determines the Coulomb-modified effective range
\begin{equation}
 \label{eq:anc-range}
  r^C_{0,\sigma}=\frac{\tilde{H}(\gamma,\kC)}{3\kC}-\frac{2 [\Gamma(1+\kC/\gamma)]^2}{C_\sigma^2}
\end{equation}
combining Eqs.~(\ref{eq:LSZRes1}) and (\ref{eq:Z-ANC}). The Coulomb-modified scattering length is then obtained from the pole position of the t-matrix (\ref{eq:tMatrixB}), that is 
\begin{equation}
a^C_{0,\sigma}=-\frac{2}{4\kC H(-i\kC/\gamma)+\gamma^2r^C_{0,\sigma}}~.
\label{eq:a0uptoP}
\end{equation}
These s-wave effective-range-expansion results are accurate up to
corrections due to the shape parameter in the ERE. It is therefore important
to note that, if the ERE parameters are to be extracted to high accuracy
from the ANC, one would need to derive the expressions
(\ref{eq:anc-range}) and (\ref{eq:a0uptoP}) to higher orders. Vice versa,
if the ANC is to be predicted accurately from elastic scattering data, then
the expression (\ref{eq:LSZRes1}) needs to be improved by including dependencies
of additional ERE parameters.

\subsection{Applications to Scattering}

Kong and Ravndal first included Coulomb interactions in the
pionless effective field theory using the PDS regularization scheme
\cite{Kong:1998sx,Kong:1999sf}.
They considered proton-proton scattering near threshold were the Coulomb
interaction is strong and needs to be included nonperturbatively
at leading order. In particular, they showed that the 
extraction of a strong proton-proton scattering length $a_{pp}$
from the standard Coulomb-modified scattering length ${a_{pp}^C}$ is
scheme-dependent and depends on the
renormalization scale $\Lambda$
\begin{equation}
  \frac{1}{a_{pp}(\Lambda)}=\frac{1}{a_{pp}^C}+\alpha_{em}m_n
\Big[\log{\left(\frac{\sqrt{\pi}\Lambda}{\alpha_{em}m_n}\right)}
+1-\frac{3C_\mathrm{E}}{2}\Big]\,,
\end{equation}
since the short-distance parts of the Coulomb and strong interactions
cannot be uniquely separated.
The effective range, however, remains unaffected by Coulomb interactions.
Complementary renormalization group treatments of the $pp$ system can be
found in Refs.~\cite{Barford:2002je,Ando:2008jb}.
An application to proton-$\alpha$ scattering is discussed in
\cite{Higa:2010zi}.
The same power counting also applies to the excited $1/2^+$ 
state of $^{17}$F. Using the formalism described
above, Ryberg and collaborators
calculated the electric radius
and the astrophysical S-factor for low-energy proton capture~\cite{Ryberg:2013iga,Ryberg:2015lea}.
They also 
provided a general discussion of the suppression of proton halos compared to 
neutron halos by the need for two fine tunings in the underlying theory.

\begin{figure}[t]
\centerline{\includegraphics[width=8.0cm]{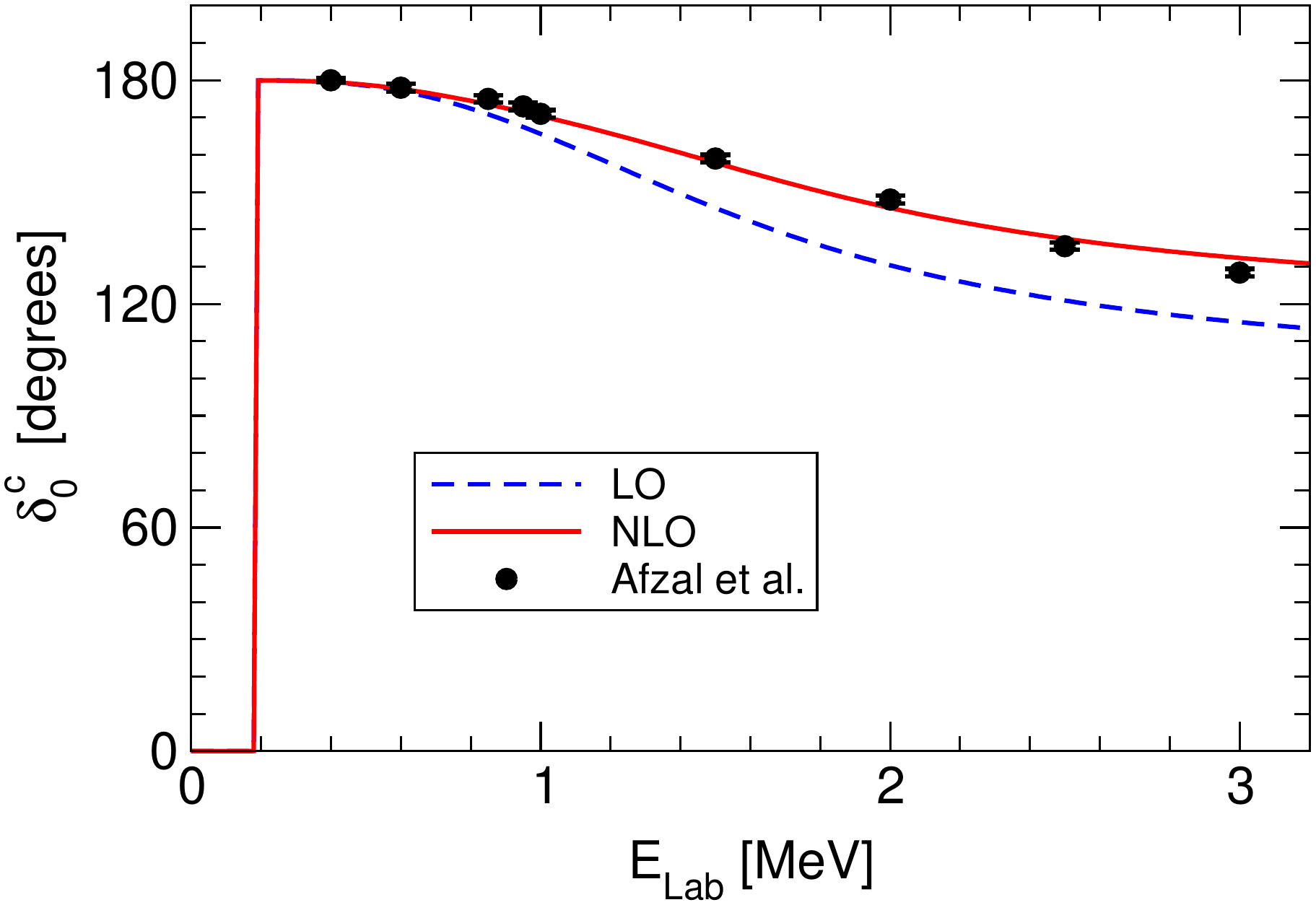}}
\vspace*{-0pt}
\caption{The $\alpha \alpha$ s-wave phase shift $\delta^c_0$ 
as a function of the laboratory energy $E_{Lab}$. 
The EFT results in LO and NLO from Ref.~\cite{Higa:2008dn}
are given by the (blue) dashed and (red) solid lines, respectively.
The experimental phase shifts are shown as (black)
solid circles with error bars \cite{AFZAL:1969zz}.
}
\label{fig:alal_pshift}
\end{figure}

This treatment of Coulomb, however, is not appropriate for all
weakly-bound systems in which Coulomb interactions are present.
An example is given by the $\alpha-\alpha$ system
at low energies which is highly fine tuned \cite{Higa:2008dn}.
Due to the delicate interplay  between 
attraction from the strong interaction and the Coulomb repulsion,
there is a narrow resonance at an energy of 
about 0.1~MeV in this system. 
The scenario of Ref.~\cite{Higa:2008dn} can be 
viewed as an expansion around the limit where,
when electromagnetic interactions 
are turned off, the $^8$Be ground state is at threshold and exhibits conformal 
invariance.  This implies treating the Coulomb momentum $k_C$ as a 
high-momentum scale and expanding observables in $k/k_C$,
using the expansion in Eq.~\eqref{eq:H-expand} such that
the inverse of the amplitude $t_{CS}$ becomes proportional to
\begin{equation}
  \label{eq:TCSalpha}
  -\frac{1}{a_{0,\sigma}^C}+r_{0,\sigma}^C k^2/2-2k_C H(\eta)+\ldots\,.
\end{equation}
The corresonding phase shifts are shown in Fig.~\ref{fig:alal_pshift}.
When Coulomb interactions are
turned off, the third term of Eq.~\eqref{eq:TCSalpha} becomes the usual
unitarity term $−ik$. Since $a_{0,\sigma}^C \sim M_{\rm core}/M_{\rm halo}^2$ and
$r_{0,\sigma}^C \sim M_{\rm core}^{-1}$, the first two terms are
subleading corrections for momenta $k\sim M_{\rm halo}$ and only the
unitarity term remains at leading order.
Therefore, at leading order the $^8$Be
system shows conformal invariance, and the corresponding 3-body system $^{12}$C
acquires an exact Efimov spectrum \cite{Efimov:1970zz}.
This is a possible realization of the unitary limit. When Coulomb is
restored, the Coulomb potential breaks scale invariance
and the three terms in  Eq.~\eqref{eq:TCSalpha} are of comparable size.
However, the fact that the $^8$Be ground state stays close to threshold
can be seen as a remnant of the now broken conformal symmetry.
This scenario provides a possible realization of the conjecture that
the Hoyle state in $^{12}$C could be considered an approximate
Efimov state of $\alpha$ particles~\cite{Efimov:1970zz}. The
properties of this state will be modified by the long-range
Coulomb interaction once it is restored~\cite{Hammer:2008ra}.

\subsection{Form Factors}

We now move on to a calculation of the electric form factor of an s-wave proton
halo to NLO. At LO there are two
loop diagrams, $\Gamma_\mrm{LO}(|\qq|)$, and at NLO a constant tree-level
diagram, $\Gamma_\mrm{NLO}$, enters. We derive and evaluate these
diagrams below. The electric form factor is then given by the sum of
diagrams
\begin{equation}
F^{(\sigma)}_{E}(|\qq|)=\frac{Z_\sigma}{e(Q_c+1)}\left[\Gamma_{LO}(|\qq|)
+\Gamma_{NLO}(|\qq|)+\dots\right]~.
\label{eq:ChargeFF3}
\end{equation}

\subsubsection{Leading order}

\begin{figure}[t]
\centerline{\includegraphics*[scale=0.6,angle=0,clip=true]{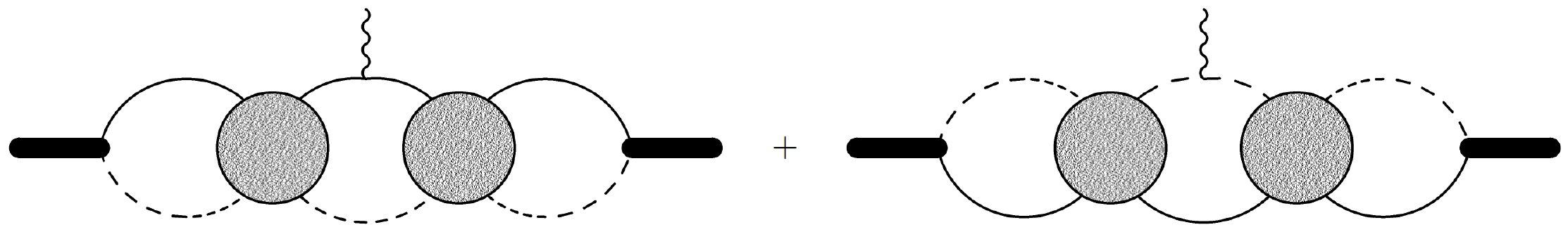}}
\caption{The diagrams contributing to the charge form factor at LO.
  The notation is as in Fig.~\ref{fig:FullDibaryon}.}
\label{fig:GammaLO}
\end{figure}

At leading order, we have to consider the diagrams shown in
Fig.~\ref{fig:GammaLO}. In contrast to neutron halos, the 
photon couples to both single-particle lines through the operator 
$\psi^\dagger A_0\psi$ where $\psi$ can be a proton field $p$ or a core field
$c$. We choose incoming and
outgoing total four-momenta as $(E,-\qq/2)$ and $(E,\qq/2)$,
respectively, where $E=-B+|\qq|^2/(8M_\sigma)$.
After carrying out the integrals over the zero
component of the loop momenta the resulting amplitude
can be simplified using the Coulomb Green's
function in Eq.~(\ref{eqCGFGamma}). This leads to
\begin{eqnarray}
i\Gamma_{LO}(|\qq|)=&ig_\sigma^2eQ_c\int\frac{\d^3k_1\d^3k_2\d^3k_3}{(2\pi)^{9}}~\langle\kk_3|G_\mrm{C}(-B)|\kk_2-f\qq/2\rangle\nonumber\\
&\times
\langle\kk_2+f\qq/2|G_\mrm{C}(-B)|\kk_1\rangle +\left[(f\to1-f),~(Q_c\to 1)\right],
\label{eq:GammaLOmom2}
\end{eqnarray}
where $f=m_n/M_\sigma=1/(A+1)$.
Fourier transforming each of the momentum-space bras
and kets, we arrive at the coordinate-space integral
\begin{eqnarray}
\fl
& i\Gamma_{LO}(|\qq|)=ig_\sigma^2eQ_c\int\frac{\d^3r}{(2\pi)^3} (0|G_\mrm{C}(-B)|\rr)~\exp{(if\qq\cdot\rr)}~(\rr|G_\mrm{C}(-B)|0)\nonumber\\
& \qquad \qquad +\left[(f\to1-f),~(Q_c\to 1)\right],
\label{eq:GammaLOcoord1}
\end{eqnarray}
which is much more convenient to use.  In Eq.~(\ref{eq:GammaLOcoord1}),
the diagram in Fig.~\ref{fig:GammaLO} is also
better visualized. It consists of two Coulomb Green's functions,
that propagate the fields from separation zero to $\rr$ and back from 
separation $\rr$ to zero, respectively, and the current operator in between
the propagators. Since the Coulomb Green's functions have one end at
zero separation, only their s-wave part will
contribute. We thus expand the Coulomb Green's function in
partial waves
\begin{equation}
(\rr'|G_\mrm{C}(E)|\rr)=\sum_{l=0}^\infty(2l+1)G^{(l)}_\mrm{C}(E;r',r)P_l(\hat{\rr}'\cdot\hat{\rr}).
\end{equation}
and obtain~\cite{Ryberg:2015lea}
\begin{align}
(0|G_\mrm{C}(-B)|\rr)=&G^{(0)}_\mrm{C}(-B;0,r)
  =-\frac{\mu_\sigma\Gamma(1+\kC/\gamma)}{2\pi}
  \frac{W_{-\kC/\gamma,1/2}(2\gamma r)}{r}.
\end{align}

The integral (\ref{eq:GammaLOcoord1}) can now be written as
\begin{eqnarray}
i\Gamma_{LO}(|\qq|)=&i\frac{g_\sigma^2eQ_c\mu_\sigma^2}{8\pi^4}
[\Gamma(1+\kC/\gamma)]^2\int\d r~j_0(f r |\qq|)[W_{-\kC/\gamma,1/2}(2\gamma r)]^2\nonumber\\
&+\left[(f\to1-f),~(Q_c\to 1)\right]\,,
\label{eq:GammaLOcoord2}
\end{eqnarray}
which can be evaluated numerically. The charge form factor is 
given by Eq.~(\ref{eq:ChargeFF3})
and the LO electric radius is obtained in terms of the loop-integral $\Gamma_{LO}(|\qq|)$ and the wavefunction renormalization $Z_{LO}$:
\begin{equation}
\langle r_{E}^2\rangle_{{\rm pt}, LO}^{(\sigma)}=-\frac{3Z_{\sigma,LO}}{e(Q_c+1)}\left.\frac{\d^2}{\d|\qq|^2}\Gamma_{LO}(|\qq|)\right|_{|\qq|=0}~.
\end{equation}

We go on to demonstrate that the electric form factor is normalized correctly
to 1 at $|\qq|=0$. Starting from the coordinate-space integral
(\ref{eq:GammaLOcoord1}) at $|\qq|=0$ and the spectral representation of
the Coulomb Green's function
\begin{equation}
(0|G_\mrm{C}(E)|\rr)=\int\frac{\d^3p}{(2\pi)^3}
\frac{\psi_\pp(0)\psi^*_\pp(\rr)}{E-\pp^2/(2\mu_\sigma)+i\varepsilon}~,
\label{eq:CGFSpectral2}
\end{equation}
we find that
\begin{eqnarray}
&\Gamma_\mrm{LO}(0)=g_\sigma^2e(Q_c+1)\int\d^3r\left|(0|G_\mathrm{C}(-B)|r)\right|^2\nonumber\\
&\qquad=g_\sigma^2e(Q_c+1)\int\frac{\d^3p}{(2\pi)^3}\frac{\psi_\pp(0)\psi^*_\pp(0)}{(-B-\pp^2/(2\mu_\sigma))^2}\nonumber\\
&\qquad =-e(Q_\mathrm{c}+1)\Sigma'(-B)~.
\label{eq:GammaLOlimit}
\end{eqnarray}
In the first step above the orthonormality of the Coulomb wavefunctions
was used.
The correct normalization of the LO electric form factor now follows from
combining Eqs.~(\ref{eq:ChargeFF3}), (\ref{eq:GammaLOlimit})
and (\ref{eq:LSZRes1}).

\subsubsection{Next-to-leading order}

At NLO the full LSZ residue $Z_\sigma$ from Eq.~(\ref{eq:Z-ANC})
must be used instead of $Z_{\sigma,LO}$ and the NLO
operator $\sigma^\dagger A_0\sigma$ enters through
the right diagram in Fig.~\ref{fig:formfactor}. Its contribution
is given by the Feynman rule for an $A_0$ photon coupling to the
$\sigma$ field
\begin{equation}
i\Gamma_{NLO}=-iw_\sigma e(Q_c+1)~.
\label{eq:GammaNLO}
\end{equation}
Thus the NLO correction (\ref{eq:GammaNLO}) is
independent of the momentum transfer $|\qq|$.

The NLO electric form factor is given by the sum of the diagrams up to
NLO, according to Eq.~(\ref{eq:ChargeFF3}).
Using Eqs.~(\ref{eq:GammaLOlimit}), (\ref{eq:LSZRes1}), and (\ref{eq:GammaNLO}), as well as
the formula (\ref{eq:ChargeFF3})
it is clear that the electric form factor at $|\qq|=0$ is still normalized to
unity. The electric radius is
then given by the order $|\qq|^2$ part of the LO loop-integral
(\ref{eq:GammaLOcoord2}) together with the full wavefunction
renormalization (\ref{eq:Z-ANC}). The resulting electric radius is
\begin{equation}
  \langle r_{E}^2\rangle_{{\rm pt}, NLO}^{(\sigma)}=-\frac{3Z_\sigma}{e(Q_c+1)}\left.\frac{\d^2}
          {\d|\qq|^2}\Gamma_{LO}(|\qq|)\right|_{|\qq|=0}~.
          \label{eq:rENLO}
\end{equation}
The NLO electric radius (\ref{eq:rENLO})
is different from the LO result by a factor 
$Z_\sigma/Z_{\sigma,LO}=C_\sigma^2/C_{\sigma,LO}^2$
(cf.~Eq.~(\ref{eq:ZNLOoverZLO})) which is
proportional to the square of the ANC.

At higher orders there are various types of corrections. Firstly,
there are local short-range operators
$\psi^\dagger\left[\nabla^2A_0-\partial_0(\nabla\cdot\mathbf{A})\right]\psi
\quad (\psi=c,p)$, 
which enter with finite-size contributions of the core and
proton fields. Similar to what was done in Sec.~\ref{sec:emprocesses},
these higher-order finite-size contributions can be added in a
straightforward way, yielding
\begin{equation}
  \langle r_E^2\rangle = \langle r_E^2\rangle_{\rm pt} + \frac{Q_c}{Q_c+1}
  \langle r_E^2\rangle_c + \frac{1}{Q_c+1} \langle r_E^2\rangle_p .
\end{equation}
Secondly, there is a local short-range operator
$\sigma^\dagger\left[\nabla^2A_0-\partial_0(\nabla\cdot\mathbf{A})\right]\sigma $,
that comes in with an undetermined short-range parameter.

\subsection{Radiative Capture}
\label{sec:radiative-capture}

In this subsection we consider radiative E1 capture of a proton into a halo
state including range corrections.
Specifically, we assume a spin zero core and capture of the proton into
an s-wave state with $J^P=1/2^+$, which requires
the incoming particle pair to have relative angular momentum $l\geq1$.
These quantum numbers are relevant for the capture process
$^{16}\mathrm{O}(\mathrm{p},\gamma)^{17}\mathrm{F}^*$, which will
be discussed as an application below. 

We present our results in terms of the astrophysical S-factor, which
is defined as
\begin{equation}
S(E)=E\exp{(2\pi\eta)}\sigma_\mathrm{tot}(E)~,
\end{equation}
with the incoming center-of-mass energy $E$ and the total cross section $\sigma_\mrm{tot}$.
This removes the otherwise dominant exponential suppression of the capture cross section at
energies below the Coulomb barrier.

\subsubsection{Leading order}

\begin{figure}[t]
\centerline{\includegraphics*[scale=0.6,angle=0,clip=true]{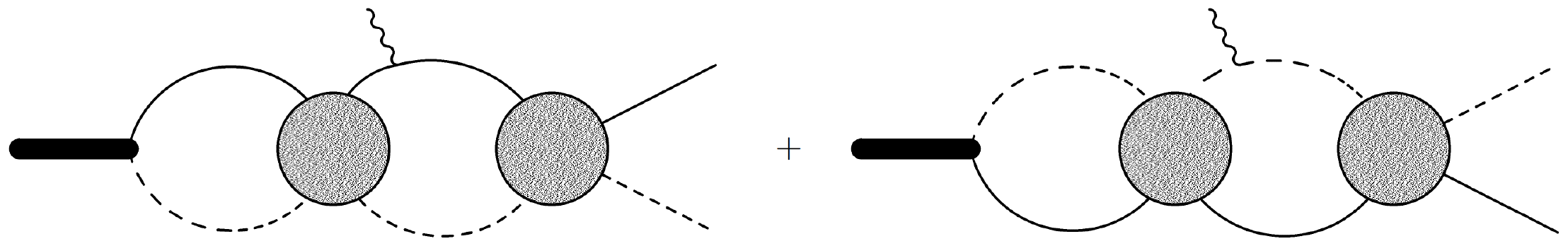}}
\caption{The diagrams for radiative capture at LO. The notation is as in
  Fig.~\ref{fig:FullDibaryon} and time flows from right to left.}
\label{fig:captureLO}
\end{figure}

At leading order, we only need to consider diagrams where the photon couples
to one of the single-particle lines, that is through the operator
$ie\,\psi^\dagger\hat{\mrm{Q}}\,\mathbf{A}\cdot\nabla\,\psi/m \quad (\psi=c,p)$.
All diagrams which have the incoming particle pair in a relative
s-wave vanish because we are considering E1 capture which changes
the angular momentum by one unit. Since the p-wave
dominates at low energies, we neglect all partial waves with $l\geq 2$.
The two radiative capture diagrams of interest are
shown in Fig.~\ref{fig:captureLO}. These diagrams are to be read from the right to the left, that is with the incoming proton-core pair on the right-hand side and the outgoing bound state to the left. We will write them using the pure
Coulomb t-matrix $T_\mrm{C}$ \cite{Chen:1972zz}. 
The four-point function $\chi$ defined in Fig.~\ref{fig:fourpointchi} is
directly proportional to the Coulomb Green's function $G_C$ and
receives contributions from the Coulomb t-matrix $T_C$ as well as from the 
free propagation of the core-proton system (cf.~Fig.~\ref{fig:fourpointchi}). 
In the center-of-mass of the halo and the radiated photon,
the momentum-space diagrams from  Fig.~\ref{fig:captureLO}
become
\begin{eqnarray}
\fl
  i\mathcal{M}=-ig_\sigma\sqrt{Z_{\sigma,LO}}\frac{eQ_cf}{\mu_\sigma}\int\frac{\d^3k_1\d^3k_2}{(2\pi)^6}~G_0(-B,\kk_2+f\qq)
 \chi(\kk_2+f\qq,\kk_1+f\qq,-B) \nonumber\\
\times G_0(-B,\kk_1+f\qq)~\kk_1 \bigg[\delta^3(\pp-\kk_1)+\left[E-\kk_1^2/(2\mu_\sigma)\right]^{-1}T_\mrm{C}(\kk_1,\pp)\bigg]\nonumber\\
-\left[(f\to1-f)~,~(Q_c\to1)\right].
\label{eq:CaptureLO01}
\end{eqnarray}
where we have separated the contributions from the  Coulomb t-matrix $T_C$
and the free propagation of the core-proton system in the initial state.
In the final state, we use the Coulomb four-point function $\chi$ as
before. The total energy flowing through the diagram is 
$E=-B+\omega+\omega^2/({2M_\sigma})=\pp^2/(2\mu_\sigma)$.

Then, using the relation \cite{Chen:1972zz}
\begin{equation}
\psi_\pp(\kk_1)=\delta^{3}(\kk_1-\pp)+\left[E-\kk_1^2/(2\mu_\sigma)\right]^{-1}T_\mrm{C}(\kk_1,\pp)
\end{equation}
for the Coulomb wave function
and replacing the four-point function $\chi$ with
Eq.~(\ref{eqCGFGamma}) the integral (\ref{eq:CaptureLO01}) can be
expressed as:
\begin{eqnarray}
i\mathcal{M}=&-ig_\sigma\sqrt{Z_{\sigma,LO}}\frac{eQ_c f}{\mu_\sigma}\int\frac{\d^3k_1\d^3k_2}{(2\pi)^6}~\langle\kk_2|G_\mrm{C}(-B)|\kk_1+f\qq\rangle~\kk_1~\psi_\pp(\kk_1)\nonumber\\
&-\left[(f\to1-f)~,~(Q_c\to1)\right]
\label{eq:CaptureLO1}
\end{eqnarray}
Performing Fourier transforms and using $\kk_1\exp{(i\kk_1\cdot\rr_2)}=-i\nabla_2\exp{(i\kk_1\cdot\rr_2)}$ we can write the integral (\ref{eq:CaptureLO1}) as 
\begin{eqnarray}
i\mathcal{M}=g_\sigma\sqrt{Z_{\sigma,LO}}\frac{eQ_c f}{\mu_\sigma}\int\d^3r~G_\mrm{C}^{(0)}(-B;0,r)~\exp{(-if\omega r \cos{\theta})}~\left(\nabla\psi_\pp(\rr)\right)\nonumber\\
-\left[(f\to1-f)~,~(Q_c\to1)\right]~.
\label{eq:CaptureLO3}
\end{eqnarray}
Finally, summing over all polarizations and  doing the angular integration,
the amplitude squared in Eq.~(\ref{eq:CaptureLO3})
evaluates to
\begin{eqnarray}
\fl
\sum_i\left|\epsilon_i\cdot\mathcal{M}\right|^2=\Bigg|\sqrt{Z_{\sigma,LO}}\sin{\theta}(\cos{\phi}+\sin{\phi})\frac{4\pi g_\sigma eQ_c f\exp{(i\sigma_1)}}{\mu_\sigma p}\nonumber\\
\fl
\qquad \times\int\d r~G_\mrm{C}^{(0)}(-B;0,r)j_0(f\omega r)\frac{\partial}{\partial r}\left[rF_1(\kC/p,pr)\right] -\left[(f\to1-f)~,~(Q_c\to1)\right]\Bigg|^2.
\end{eqnarray}
This integral can be calculated numerically, using the s-wave
projected Coulomb Green's function and the regular Coulomb wave
function.

\subsubsection{Next-to-leading order}
The higher-order ERE parameters appear with $\nabla + ie\hat{\mrm{Q}}
\mathbf{A}$ operators that, 
in principle, can give contributions to the radiative-capture amplitude. 
However, the diagrams with these higher-order ERE operators are diagrams 
with initial-wave scattering due to the strong force. Since we only have 
included the resonant s-wave part of the strong interaction these initial-wave 
scattering diagrams are identically zero, which can be understood from 
the fact that the E1 capture process changes the angular momentum by one. 
If we were to include also the non-resonant p-wave interaction explicitly
in the field 
theory, then the effective range, the shape parameter, and so on, would 
contribute through diagrams with initial p-wave scattering.
The physics of  these diagrams is implicitly 
included in local short-range operators with growing powers of the photon 
energy $\omega$. Such an operator is explicitly discussed below.

Consequently, there are no additional capture diagrams to consider
at NLO. The only contribution at NLO is due to the change in the
wavefunction renormalization. This leads to a constant factor
\begin{equation}
\frac{Z_{\sigma}}{Z_{\sigma,LO}}=\frac{C_\sigma^2}{C_{\sigma,LO}^2}=
     \frac{\tilde{H}(\gamma,\kC)}{\tilde{H}(\gamma,\kC)-3k_\mathrm{C}r_{0,\sigma}^C}
\label{eq:NLOfactor}
\end{equation}
difference compared to the LO result. It is important to note
that if $r_{0,\sigma}^C\approx\tilde{H}(\gamma,\kC)/(3\kC)$,
or equivalently if the ANC $C_\sigma$ is very
large, then the NLO correction will be large, too. We discuss this
case in more detail in Sec.~\ref{sec:finetuning}.

\subsubsection{Higher Orders}
\label{sec:N4LO}

\begin{figure}[t]
\centerline{\includegraphics*[scale=0.5,angle=0,clip=true]{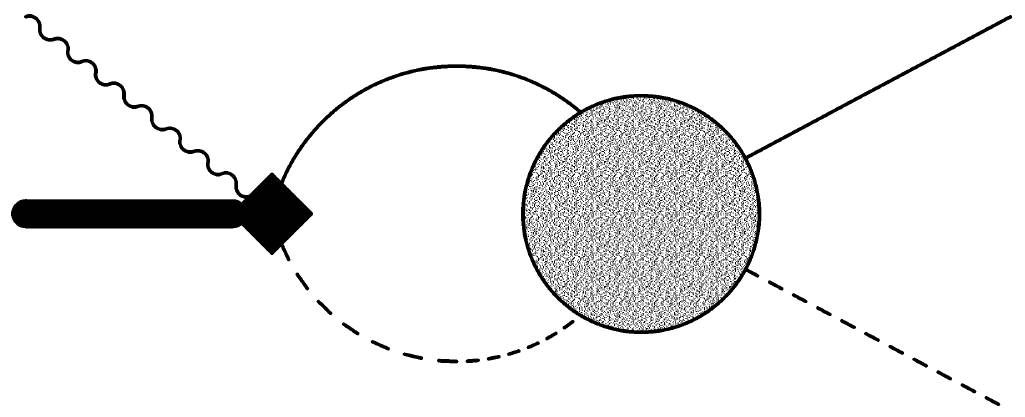}}
\caption{The capture diagram entering at N$^4$LO. The notation is as in Fig.~\ref{fig:FullDibaryon} and time flows from right to left. The diamond vertex corresponds to the short-range interaction of Eq.~(\ref{eq:nonminimalEMnatpwave}).}
\label{fig:CaptureN5LO}
\end{figure}

At N$^4$LO  the short-range E1 operator from Eq.~(\ref{eq:nonminimalEMnatpwave})
appears.
This short-range interaction is simply a contact vertex, where the
incoming proton-core pair is in a relative p-wave.
Note, that in the present case the $j=1/2$ and $j=3/2$ channels contribute.
The operator (\ref{eq:nonminimalEMnatpwave}) gives rise to the capture
diagram in Fig.~\ref{fig:CaptureN5LO}. The tree-level amplitude
is given by
\begin{equation}
\mathcal{M}=D^{(E1)}\sqrt{Z_\sigma}\omega\exp{(i\sigma_1)}\pp\sqrt{(1+\eta^2)C_\eta^2}~.
\label{eq:captureN5LO}
\end{equation}
The derivation of the amplitude (\ref{eq:captureN5LO}) involves the
evaluation of the p-wave integral
$\int\frac{\d^3k}{(2\pi)^3}\kk\psi_\pp(\kk)=-\exp{(i\sigma_1)}\pp\sqrt{(1+\eta^2)C_\eta^2}$,
where $\sigma_1$ is the pure Coulomb phase shift in the p-wave.
The symbol $D^{(E1)}$ has been introduced as a compact notation for the
total constant of proportionality and includes both spin channels.

The next relevant operator
includes an additional time derivative acting on the photon
field and enters at N$^6$LO and as such this
calculation is valid up to N$^5$LO.

It should be noted that if the strong p-wave interaction is included explicitly
into the field theory then there would also exist local non-minimal operators
of the form \cite{Hammer:2011ye} $\pi^\dagger(\partial_0A_j-\nabla_jA_0)
\pi$, where
$\pi$ is a p-wave dicluster field (the spin indices on the dicluster fields have
been suppressed). If the p-wave is enhanced this operator
would appear already at NLO. However, as has already been mentioned,
we have no reason to suspect that the p-wave interaction is enhanced.
In the following application to $^{16}\mathrm{O}(p,\gamma)^{17}\mathrm{F}^*$,
we will assume that it does not enter up to N$^5$LO.

\subsubsection{Application to
$^{16}\mathrm{O}(p,\gamma)^{17}\mathrm{F}^*$}
\label{sec:results-fluorine-17}

\begin{figure}[t]
\centerline{\includegraphics*[scale=1,angle=0,clip=true]{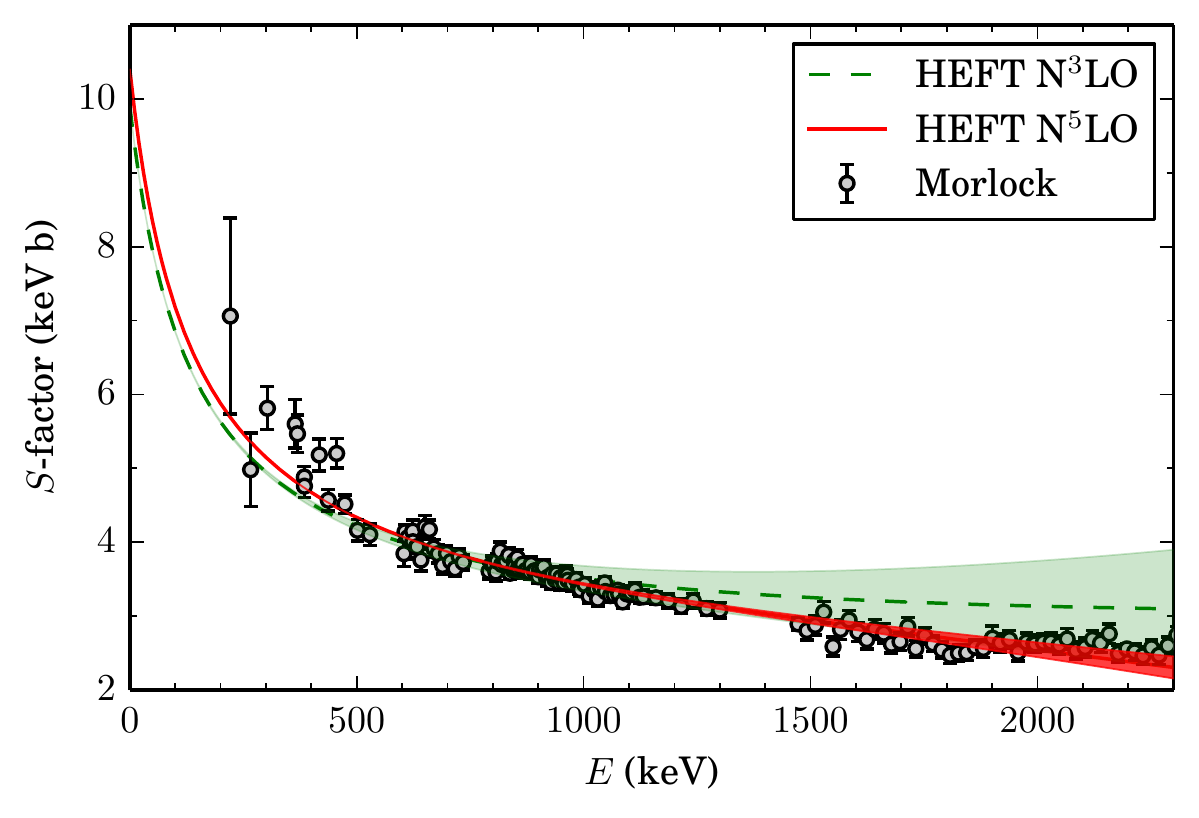}}
\caption{Energy dependent S-factor for $^{16}$O$(p,\gamma)^{17}$F$^*$
fitted to experimental data by
  Morlock {\it et al.}~\cite{Morlock:1997zz}. The error bands correspond to the
  model error from omitted terms at higher orders. See text for details.
  (Figure taken from Ref.~\cite{Ryberg:2015lea}.)}
\label{fig:SfactorNLO}
\end{figure}

In Ref.~\cite{Ryberg:2015lea}, this formalism was applied to
the excited $1/2^+$ state of $^{17}$F which is an s-wave halo state
of the valence proton and the $^{16}$O$(0^+)$ core. Their results for the
radiative proton capture reaction $^{16}\mathrm{O}(p,\gamma)^{17}\mathrm{F}^*$,
together with data by Morlock {\it et al.} \cite{Morlock:1997zz},
are shown in  Fig.~\ref{fig:SfactorNLO}. The
green dashed line is the Halo EFT result valid up to N$^3$LO. The
single free parameter of the theory is the ANC which
was fitted to the experimental data by Morlock 
{\it et al.}~\cite{Morlock:1997zz}
by minimizing an objective function that takes into account 
experimental errors and theoretical errors from the truncation of
the EFT expansion
(See Ref.~\cite{Ryberg:2015lea} for details).
Taking energy-dependent EFT errors into account allowed them to
include data up to a center-of-mass energy of $2.3~\mrm{MeV}$.
They determined the amplitude, $C_\mrm{EFT}$, of the EFT error
which scale as $(p/M_\mrm{core})^4$ by the
statistical guiding principle that the total $\chi^2$ per degree of
freedom should be unity.\footnote[1]{Note that the Morlock data has a
normalization error of $10\%$, which was subtracted in quadrature from
the total experimental error during the fitting procedure. This
normalization error was added back after the fit had been
performed.}
The constant $C_\mrm{EFT}$ is expected to be of natural
size and is determined iteratively such that the $\chi^2$ per degree
of freedom is minimized to unity. This systematical theory error at
N$^3$LO is shown as a green band in Fig.~\ref{fig:SfactorNLO}, with
$C_\mrm{EFT}=6.9$ and the breakdown scale given by
$M_\mrm{core}=76~\mrm{MeV}$. Although somewhat large, this value of  
$C_\mrm{EFT}$ is still consistent with our power counting estimate.

Similarly, the red line is given by a fit of the N$^5$LO result to the
same data set where the systematic theory error estimate scales as
$(p/M_\mrm{core})^6$. At this order we find $C_\mrm{EFT}=1.9$ with
$M_\mrm{core}=76~\mrm{MeV}$. It is clearly seen that the result
converges with increasing order of the EFT and that the S-factor value
at threshold is stable. From these fits, they extracted a threshold
S-factor
\begin{equation}
S(0)=\left\{
\begin{array}{c}
\big(9.9\pm0.1~(\mrm{stat})\pm1.0~\mrm{(norm)})~\mrm{keV~b}~,~\mrm{N^3LO}\\
\big(10.4\pm0.1~(\mrm{stat})\pm1.0~(\mrm{norm})\big)~\mrm{keV~b}~,~\mrm{N^5LO}
\end{array}\right.
\end{equation} 
with the $1\%$ error due to the EFT fit (mainly statistical error) and
the $10\%$ error from the uncertainty in the absolute
normalization of the experimental data. These results give the ANC
\begin{equation}
C_\sigma=\left\{
\begin{array}{c}
\big(77.4\pm0.2~(\mrm{stat})\pm3.8~(\mrm{norm})\big)~\mrm{fm^{-1/2}}~,~\mrm{N^3LO}\\
\big(79.3\pm0.2~(\mrm{stat})\pm3.9~(\mrm{norm})\big)~\mrm{fm^{-1/2}}~,~\mrm{N^5LO}
\end{array}\right..
\label{eq:ANCextract}
\end{equation}
The ANC for this system has also been extracted by Huang { et al.}
\cite{Huang:2008ye}, using a single-particle model fit of radiative capture
data, as $C_\sigma=77.21~\mrm{fm}^{-1/2}$ and experimentally by Gagliardi
{ et al.} \cite{Gagliardi:1998zx}, using the transfer reaction
$^{16}$O$(^{3}$He, $d)^{17}$F, as $C_\sigma=(80.6\pm4.2)~\mrm{fm}^{-1/2}$.
The ANCs are consistent with the values extracted using Halo EFT.

The electric radius of the $^{17}$F$^*$ can now be obtained by using
the extracted ANC. Using the $^{16}$O-proton ANC extracted from
the N$^5$LO radiative proton capture fit, the resulting NLO electric
radius is given by
\begin{equation}
\sqrt{\langle r_{E}^2\rangle_{NLO}^{(\sigma)}}
=(2.20\pm0.04~(\mrm{EFT})\pm0.11~(\mrm{ANC}))~\mrm{fm}~.
\label{eq:chargeradiusresult}
\end{equation}
The NLO EFT error in Eq.~(\ref{eq:chargeradiusresult}) was estimated
from the EFT expansion parameter squared,
$\left(\gamma/M_\mrm{core}\right)^2$, using a breakdown scale
$M_\mrm{core}=76~\mrm{MeV}$. The dominant error in
Eq.~(\ref{eq:chargeradiusresult}) is from the normalization error of
the Morlock data, through the extracted ANC. However, there could also
be additional EFT errors for this result due to the non-inclusion of the 
operators that are responsible for the finite-size contributions of the 
constituents \cite{rybergbiraforssen}. 
Alternatively, the result~(\ref{eq:chargeradiusresult}) can be 
interpreted as the radius relative to the $^{16}$O core.


\subsection{Fine tuning in s-wave proton halos}
\label{sec:finetuning}
\begin{figure}[t]
\centerline{\includegraphics*{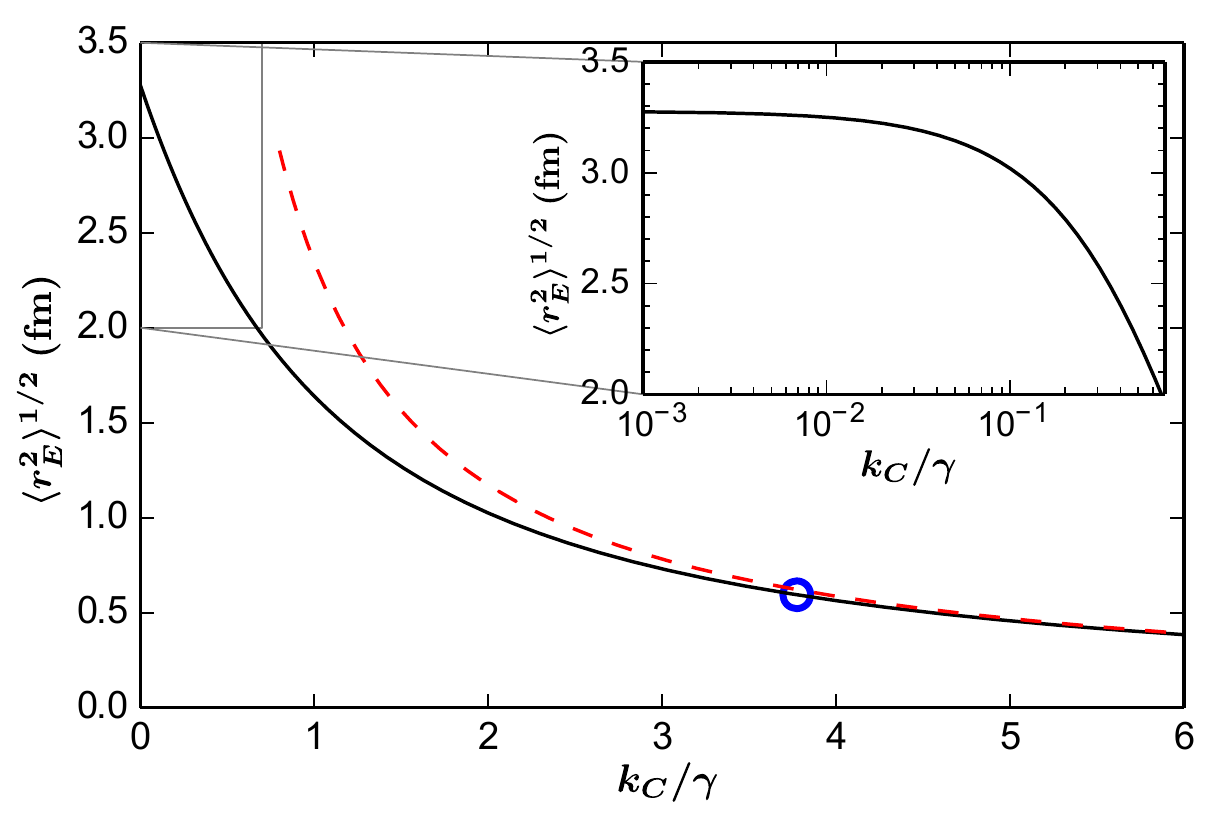}}
\caption{The dependence of the LO electric radius on $\kC/\gamma$. The solid
  black line is the LO Halo EFT result, the blue circle denotes the
  $^{17}$F$^*$ system, and the dashed red
  line is the asymptotic $1/\kC$ behavior. The inset shows the low-energy region in a semi-logarithmic scale, illustrating the hypothetical neutron halo limit~\eqref{eq:rCneutronlimit}. The curve was generated using a
  binding momentum $\gamma=13.6~\mrm{MeV}$.
(Figure taken from Ref.~\cite{Ryberg:2015lea}.)}
\label{fig:rCplotkC}
\end{figure}

Along the neutron drip-line there exist several neutron halo
states. These states are characterized by an unnaturally large
neutron-core scattering length, which brings the state very close to
threshold. However, proton halos are much more rare. In the s-wave
case this can be understood by considering the Coulomb repulsion
between the valence proton and the core. For a proton halo state to
exist, we need the attractive strong force to be almost cancelled by
the Coulomb repulsion, resulting in a threshold 
state~\cite{Zhukov:1995zz,Woods:1997cs}. This
cancellation can be seen within our formalism as an additional
fine-tuning in the effective range. Note that for an s-wave proton
halo state this means that the proton-core scattering length needs to
be unnaturally large and that the effective range must be fine-tuned
to cancel the Coulomb repulsion. The existence of proton halos is
therefore doubly suppressed by the need for two fine-tunings.

Comparing the ANCs from Eq.~(\ref{eq:ANCextract})
with the LO result (\ref{eq:LOANC}),
$C_{\sigma,LO}=21.4~\mrm{fm}^{-1/2}$, makes clear that the effective
range must be very close to the pole position $1/(3\kC)$, such that
the LSZ residue becomes large. This implies that the effective
range correction for this system is much larger than what one naively
expects. This is connected to the fact that proton halos contain
$\kC$ as a second scale  and $\kC\gg\gamma$ for this system.
Within our
framework, $\kC$ is a parameter that is independent of the
Coulomb-modified effective-range parameters. When $\kC/\gamma\gg1$, we
are in the extreme Coulomb regime. In this regime the two
particles tend to be close together, since otherwise the system is
ripped apart by the Coulomb repulsion. This limit is
realized for the $^{17}$F$^*$ system, where $\kC/\gamma=3.8$.  In
Fig.~\ref{fig:rCplotkC} the LO electric radius is shown as a function of
the Sommerfeld parameter $\kC/\gamma$, where we have used the binding
momentum for $^{17}$F$^*$, $\gamma=13.6~\mrm{MeV}$. The blue circle
is the parameter point corresponding to the physical $^{17}$F$^*$
system. It is clear that this system is almost in the extreme Coulomb
regime. Note that the resulting LO electric radius is very small for a
strong Coulomb repulsion, since it has an asymptotic $1/\kC$
behavior. At the far left, where $\kC\ll\gamma$, the system mimics
that of a neutron halo, with the only difference being that the photon
also can couple through minimal substitution to the nucleon field. The
limiting value for $\kC/\gamma\to 0$ is therefore given by
\cite{Hammer:2011ye}
\begin{equation}
\lim_{\kC/\gamma\to0}\langle r_\mrm{E}^2 \rangle^{(\sigma)}=\frac{1}{Q_c+1}\frac{Q_c f^2+(1-f)^2}{2\gamma^2}~.
\label{eq:rCneutronlimit}
\end{equation}

In the standard power counting for systems with large s-wave
scattering length the effective range enters at NLO. The 
hierarchy of scales in this case is $\gamma, \kC \ll 1/r_{0,\sigma}^C\sim M_\mrm{core}$,
where $M_\mrm{core}$ is the momentum scale set by the core. However, the
discussion above implies that we can have $\kC r_{0,\sigma}^C \sim 1$ instead
of $\kC \ll 1/r_{0,\sigma}^C$. In the
zero-range limit, the inverse Coulomb momentum sets the scale for 
the LO electric radius and the effective-range contributions will
therefore be numerically large since the  
LSZ-factor for s-wave proton halo nuclei with the effective range
included behaves as
\begin{equation}
Z_\sigma\propto\frac{1}{1-3\kC r_{0,\sigma}^C}~.
\end{equation}
It appears to be fine tuned to the pole position with
$r_{0,\sigma}^C\sim1/(3\kC)$. In the case of the $^{17}$F$^*$ system the
effective-range correction results in a factor $3.6$--$3.8$ larger
electric radius.  This hierarchy-of-scales problem can be solved by
fixing the bound-state pole position of the t-matrix at leading order
and the ANC at NLO. This procedure ensures that corrections beyond NLO
scale naturally again and is similar in spirit to the so-called
Z-parameterization as introduced in Ref.~\cite{Phillips:1999hh}.

\subsection{Radiative capture in p-wave proton halos: $^7\mathrm{Be}(p,\gamma)^8\mathrm{B}$ and solar neutrinos}

Radiative proton capture to p-wave proton-halo nuclei has also been considered
in Halo EFT. In particular, 
the radiative capture process $^7\mathrm{Be}(p,\gamma)^8\mathrm{B}$,
which determines the $^8$B neutrino flux from the sun, was
investigated~\cite{Zhang:2014zsa,Ryberg:2014exa,Zhang:2015ajn}.
The low-energy ($E_{cm} < 500$ keV) $S$-factor for
$^7\mathrm{Be}(p,\gamma)^8\mathrm{B}$ consists entirely of electric-dipole 
(E1) capture from s- and d-wave initial states to the
p-wave final states which dominate $^7\mathrm{Be}+p$ configurations
within $^8$B. Beyond $E_{cm}=500$ keV
higher-order terms in Halo EFT presumably become important, and resonances unrelated to the
$S$-factor in the Gamow peak appear.  

The Halo EFT treatment of this E1 capture
is similar to that of neutron capture reaction $^7\mathrm{Li}(n,\gamma)^8
\mathrm{Li}$ discussed in Sec.~\ref{sec:emprocesses}
and Ref.~\cite{Zhang:2013kja}.
Analogous to the  $^7\mathrm{Li}+n$ case,
the $^7\mathrm{Be}+p$ system
is described by a local quantum field theory that contains fields for the proton and the $^7\mathrm{Be}$ core
expanded in powers of their relative momentum. (The internal structure of the proton and the core are included
perturbatively at higher orders through low-energy constants matched to experiment
and/or {\it ab initio} calculations.)
However, in the $^7\mathrm{Be}+p$ case the point-Coulomb part of the
potential must be iterated to all orders when computing the scattering and bound-state wave
functions~\cite{Zhang:2014zsa,Ryberg:2014exa}. This can be done  exactly using the formalism discussed in Sec.~\ref{sec:scatteringwithCoulomb},
where we saw that the EFT result for the strong
interaction is fully determined by the order at which the Lagrangian
is truncated~\cite{Ryberg:2013iga,Zhang:2014zsa,Ryberg:2014exa,Zhang:2015ajn}.

The amplitude for E1 capture in Halo EFT is organized in an
expansion in  $M_{\rm halo}/M_{\rm core}$.
The momentum scale $M_{\rm core}$ is set
by the ${}^{7}\mathrm{Be}$ binding energy relative to the
${}^3\mathrm{He}+{}^4\mathrm{He}$
threshold, $1.59$ MeV,  so $M_{\rm core} \approx 70$ MeV, corresponding to
a co-ordinate space cutoff of approximately $3$ fm. The ${}^{8}\mathrm{B}$ ground state, which is 0.1364(10) MeV below
the ${}^{7}\mathrm{Be}-p$ scattering continuum~\cite{Audi2012,Wang2012},
is a shallow p-wave bound state in Halo EFT. In the case of $^7$Li $+ n$ we had to also include
the first excited state of ${}^7$Li as a degree of freedom in the EFT, and here too, the $J^\pi=\frac{1}{2}^-$ bound
excited state of ${}^{7}\mathrm{Be}$ is a field in the EFT Lagrangian. 
${}^{7}\mathrm{Be}^*$ is 0.4291 MeV above the ground state.
 
 Meanwhile, the
${}^{7}\mathrm{Be}$ ground state is $\frac{3}{2}^-$, and so---again as in ${}^7$Li $+n$---there are two possible
initial spin channels, denoted here by $s=1,2$. They correspond, respectively,
to $\S{3}{1}$ and $\S{5}{2}$ components in the incoming scattering state, and E1 capture links them to, respectively, the
$\P{3}{2}$ and $\P{5}{2}$ configurations in  ${}^{8}\mathrm{B}$. In this case
there are large
($\sim 10$ fm) ${}^{7}\mathrm{Be}-p$ scattering lengths in both spin channels
and these
play a key role
in the low-energy dynamics. In Eq.~(\ref{eq:crosssection}) this effect was captured through the function $X$, but
here the large scattering lengths occur in the context of an effective-range expansion that is modified by Coulomb, see Eq.~(\ref{eq:CoulModERE}) above.
For $p + {}^7$Be scattering
$k_C \approx 24$ MeV, and the
binding momentum of ${}^{8}\mathrm{B}$ is 15 MeV, so these low-momentum
scales, which we generically denoted by $M_{\rm halo}$,  are well separated from $M_{\rm core}$. We anticipate that $M_{\rm halo}/M_{\rm core} \approx 0.2$.

\subsubsection{Leading order}

The LO amplitude includes only  direct capture. And since the core is structureless at this order
only ``external direct capture" (in the terminology of models, see Sec.~\ref{sec:Be7pgmodels} below)
is possible.
The parameters that appear at LO are the two asymptotic normalization
coefficients (ANCs), $C_{s}$, for the ${}^{7}\mathrm{Be}-p$ configuration
in ${}^{8}\mathrm{B}$ in each of the spin channels, together with the
corresponding s-wave scattering lengths,
$a_{0,s}^C$~($s=1,2$)~\cite{Zhang:2013kja,Zhang:2014zsa,Ryberg:2014exa}. 
(Since there are two relevant spin channels here, we have removed the subscript $\sigma$ that was
used in Sec.~\ref{sec:swaves} and indexed each scattering length by the spin channel in which it occurs.)
Directly at threshold only the two ANCs $C_{s}, s=1,2$ contribute.

But these ANCs also enter the expression for the radius of ${}^8$B (see Eq.~(\ref{eq:NLOfactor}) in the same
combination. 
This {\it a priori} unknown combination can thus be eliminated in the $S$-factor
expression and a correlation between $S(0)$ and $\langle r_E^2 \rangle_{\rm pt}$ of ${}^8$B established~\cite{Ryberg:2014exa}.
In other words, provided that  the  one-proton  separation  energy  of ${}^{8}\mathrm{B}$
is fixed, at LO in Halo EFT  the electric  radius  is  directly
correlated with the $S$-factor at threshold at LO:
\begin{equation}
  \langle r_E^2 \rangle \approx \left(1.5 + 0.054
  \;\frac{S(0)}{\mbox{(eV b)}}\right)^2\;\mbox{fm}^2\,.
  \end{equation}
This provides another example how different
observables are related in Halo EFT.  Note however, that this
direct correlation does not exist at NLO since additional  counterterms
enter. Corrections to this correlation are thus expected to be of order 
$M_{\rm halo}/M_{\rm core}\approx 20\%$.

\subsubsection{Next-to-leading order}

This encourages us to improve the Halo EFT formula for $S(E)$ to a next-to-leading-order result, in order to obtain a precision
result. At NLO $S(E)$ is~\cite{Zhang:2015ajn}:
\begin{eqnarray}
S(E)&=&f(E) \sum_{s} C_{s}^2 
\bigg[ \big\vert \mathcal{S}_\mathrm{EC} \left(E;\delta_s(E)\right) 
  + \overline{L}_{s} \mathcal{S}_\mathrm{SD} \left(E;\delta_s(E)\right)
  \nonumber \\&& 
+ \epsilon_{s} \mathcal{S}_\mathrm{CX}\left(E;\delta_s(E)\right) \big\vert^2 
+|\mathcal{D}_\mathrm{EC}(E)|^2 \bigg] \ .
\label{eq:Sfactor7Be}
\end{eqnarray}
Here, $f(E)$ is an overall normalization composed of final-state phase space
over incoming flux ratio, dipole radiation coupling strength, and a factor
related to Coulomb-barrier penetration~\cite{Zhang:2014zsa}.
$\mathcal{S}_\mathrm{EC}$ is proportional to the  spin-$s$ E1
\cite{Walkecka1995book, Zhang:2014zsa, Zhang:2013kja} external
direct-capture matrix element between continuum ${}^{7}\mathrm{Be}-p$ s-wave
and ${}^{8}\mathrm{B}$ ground-state wave functions. $\mathcal{S}_\mathrm{CX}$ is
the contribution from capture with core excitation, i.e.~into the
${}^{7}\mathrm{Be}^*-p$ component of the ground state, whose size is parametrized by
$\epsilon_s$. (Since ${}^{7}\mathrm{Be}^*$ is spin-half this component only
occurs for $s=1$, so $\epsilon_2=0$.) Because the inelasticity in
${}^{7}\mathrm{Be}-p$ s-wave scattering is small
\cite{Navratil:2010jn, Navratil:2011sa} it is
an NLO effect. 

The short-distance contributions, $\mathcal{S}_\mathrm{SD}$, are also
NLO. They originate from NLO contact terms in the EFT Lagrangian
(\ref{eq:nonminimalEMdimers}). The size of these
is set by the parameters $\overline{L}_s$, which must be fit to data.
$\mathcal{S}_\mathrm{EC}$, $\mathcal{S}_\mathrm{SD}$, and
$\mathcal{S}_\mathrm{CX}$ are each functions of energy, $E$, but
initial-state interactions mean they also depend on the s-wave phase
shifts $\delta_s$. At NLO, $\delta_s(E)$ is parametrized by a
Coulomb-modified effective-range expansion (\ref{eq:CoulModERE}) that includes 
all terms up to that proportional to $r_{0,s}^C k^2$, where $r_{0,s}^C$ is
the effective range  for spin-channel $s$~\cite{Higa:2008dn,Koenig:2012bv}. Finally,
$\mathcal{D}_\mathrm{EC}$ is the E1 matrix element between the d-wave
scattering
state and the ${}^{7}\mathrm{Be}$ bound-state wave function.
It is not affected by
initial-state interactions up to NLO, and hence is the same for
$s=1,\,2$ channels and introduces no new parameters.

This leaves $9$ parameters in total: $C_{s}^2$, $a^C_{0,s}$ ($s=1,2$) at LO
and five more at NLO: $r^C_{0,s}$, $\overline{L}_{s}$, and
$\epsilon_1$~\cite{Zhang:2015ajn}. (We have dropped the subscript $\sigma$ for simplicity since we only consider different spin channels in the s-wave 
$p$-$c$ interaction in this application.) The NLO amplitudes
in E1 capture have been fitted to the experimental $S(E)$
data in the low-energy region in~\cite{Zhang:2015ajn}, but the nine parameters cannot be independently constrained from the
existing capture data. However, calculations of the
solar neutrino flux do not require that all parameters be known: it is
enough to determine $S$ at an energy $E=(18\pm 6)~\mathrm{keV}$.

\begin{figure}
\centerline{\includegraphics[width=10cm]{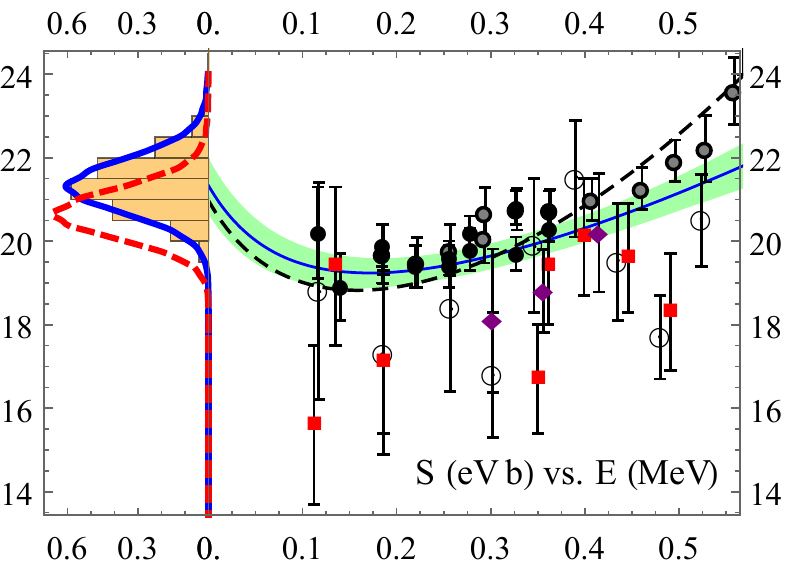}}
\caption{Right panel: NLO $S$-factor at
  different energies, including the median values (solid blue
  curve). Shading indicates the 68\% interval. The dashed line is the
  LO result.  The data used for parameter determination are also shown:
  Junghans {\it et.al.}, BE1 and BE3 \cite{Junghans:2010zz} (filled
  black circle and filled grey circle), Filippone {\it et.al.,}
 \cite{Filippone:1984us} (open circle), Baby {\it et.al.,}
\cite{Baby:2002hj, Baby:2002ju} (filled purple diamond), and Hammache
     {\it et.al.,} \cite{Hammache:1997rz, Hammache:2001tg} (filled red
     box). Left panel: 1d probability distribution functions for $S(0)$
     (blue line and
     histogram) and $S(20~\mathrm{keV})$ (red-dashed
     line). Figure taken from Ref.~\cite{Zhang:2015ajn}, {\tt doi:10.1016/j.physletb.2015.11.005} under the terms of the Creative Commons Attribution License,
     {\tt https://creativecommons.org/licenses/by/4.0/}.} \label{fig:results3}
\end{figure}

In Ref.~\cite{Zhang:2015ajn}, the 9-dimensional posterior probability distribution function (pdf)
of the NLO Halo EFT parameters was determined via a Bayesian analysis of 42 data points from all modern experiments
with more than one data point for the direct-capture $S$-factor in the
low-energy region:
Junghans {\it et.al., } (experiments ``BE1'' and ``BE3'')
\cite{Junghans:2010zz},  Filippone {\it et.al.,} \cite{Filippone:1984us},
Baby {\it et.al.,} \cite{Baby:2002hj, Baby:2002ju}, and Hammache
{\it et.al.,} (two measurements published in 1998 and 2001)
\cite{Hammache:1997rz, Hammache:2001tg}. Refs.~\cite{Zhang:2015ajn,Zhang:2015vew}
contain further details. The pdf of the Halo EFT parameters was then used to obtain 
the pdfs for the $S$-factor
at energies within the domain of validity of the theory. From these a median value (the thin solid blue
line in Fig.~\ref{fig:results3}), and 68\% interval (shaded region
in Fig.~\ref{fig:results3}) were extracted. The probability distribution functions for $S$
at $E=0$ and $20~\mathrm{keV}$ are also
shown on the left of the figure: the blue line and
histogram are for $E=0$ and the red-dashed line is for $E=20$ keV.

The $S$-factor at zero energy is constrained by this analysis
to be
\begin{equation}
  S(0)= (21.3\pm 0.7)\;\mbox{eV\,b}\,,
\end{equation}
which is an uncertainty smaller by a factor of two than
the value  $S(0)=(20.8\pm 1.6)$ \mbox{eV\,b} recommended by Adelberger
et al.~\cite{Adelberger:2010qa} (theoretical and experimental
uncertainties are added in quadrature). Zhang {\it et al.} also found particular choices of the EFT-parameter vector that correspond to 
natural coefficients, produce curves close to the median $S(E)$
curve of Fig.~\ref{fig:results3}, and have large values of the posterior
probability.

\subsubsection{Comparison to models of this reaction}

\label{sec:Be7pgmodels}

In the next section we will more broadly compare and contrast Halo EFT and other theories of  halo nuclei. But the case
 of ${}^7$Be(p,$\gamma$) provides a specific instance where we can consider what Halo EFT does (and does not) do 
 as compared to other treatments of this process. 
 
Halo EFT provides a simple, transparent, and systematic way to organize the
low-energy extrapolation of this reaction from experimentally 
accessible energies to the threshold region. Since the EFT  incorporates
all dynamics at momentum scales $< M_{\rm core}$ its radius of convergence is
larger than other efforts at systematic expansions of this
$S$-factor~\cite{WilliamsKoonin81,Baye:2000ig, Baye:2000gi,Baye:2005pxc,Jennings:1998qm,Jennings:1998ky,Jennings:1999in, Cyburt:2004jp, Mukhamedzhanov:2002du}. 

Model, descriptions of this process
are---like Halo EFT---dominated by ``external direct
capture''~\cite{christyduck,jennings98}.  These models 
involve some arbitrary choices (like Woods-Saxon shapes or matching radii)
and differ in how they
combine the tails of the final state with phase shift information and
in how they model the non-negligible contribution from short-range,
non-asymptotic regions of the wave functions.

At NLO Halo EFT includes both external direct capture and (as independent parameters)
the contribution to the E1 matrix element from the short-distance part of the E1 integral ($r <$ about 3 fm)
in both the $s=1$ and $s=2$ channel.  Compared with a potential model,
Halo EFT has about twice as many adjusted parameters.
Statistical tests imply the resulting amplitude (\ref{eq:Sfactor7Be}) can describe the experimental 
data on $S(E)$ in the
low-energy region  $E_{cm} < 500$ keV.  But, it can also represent  all the models
whose disagreement constitutes the 1.4 \mbox{eV\,b} uncertainty
quoted in Ref.~\cite{Adelberger:2010qa}---including the microscopic
calculation of Ref.~\cite{Descouvemont:2004hh}.
This is what allows NLO Halo EFT to obtain the high precision for $S(0)$ quoted above; by appropriate
choices of its nine parameters, NLO Halo EFT spans
the space of models of E1 capture in the low-energy
region~\cite{Zhang:2015ajn}. It therefore captures much of the model uncertainty in this problem, and
allows it to be quantified---and ultimately constrained---by the capture data.

\section{Comparison with other approaches} \label{sec:insightsconnections}

The EFT description of halo nuclei laid out here is complementary and supplementary to traditional shell-model and modern {\it ab initio} descriptions of nuclear structure. As such, it has much in common with phenomenological R-matrix or potential-model treatments of these systems. In this section we explore the connection of Halo EFT to these other approaches. 

In the traditional shell-model approach to these systems the nucleons are viewed as largely independent particles moving in single-particle orbitals generated by the other nucleons. Residual two-body interactions between valence nucleons are included too; these are typically fit to the spectra of nuclei in a given shell. 

This picture of nuclei is a different language to that we have used for discussing halo systems, since in it co-ordinates are typically discussed with respect to the center of the nucleus or, equivalently, the origin of the single-particle potential. Partial waves, and interactions, are thus expressed in this co-ordinate system. In contrast, our Halo EFT calculations refer to  partial-wave interactions defined in the relative co-ordinate of the core and the halo. Such an interaction may project into several such ``independent-particle basis" partial waves. In particular, in a two-neutron halo, such as ${}^6$He, the p-wave interaction between the neutron and the core can populate a variety of p-shell independent-particle levels.

The shell-model approach also describes states as a superposition of independent-particle orbitals. For example, a nuclear ground state of spin-parity $J^\pi$ could be written
\begin{equation}
|\Psi (J^\pi) \rangle=\sum_{NLS} C_{NLS} |N(LS)J \rangle\times|\Phi\rangle
\end{equation}
with $C^2_{NLS}$ ``the spectroscopic factor" of a particular single-particle state within the nuclear state $|\Psi (J^\pi) \rangle$
and $|\Phi\rangle$ the wave function of the core. Even though our description of one-nucleon halos looks like it is framed in this manner, it is not---for three reasons.  

First, the halo is, by definition, weakly bound, and the halo nucleon or nucleons have a substantial fraction of their wave function outside the nuclear mean field. In a shell-model calculation such a state necessarily involves significant ``coupling to the continuum". 
In the Halo EFT approach any bound states appear as eigenstates of the Hamiltonian of a negative energy, and the ``coupling to the continuum" is automatically
included. 

Second, Halo EFT does not emphasize, or even require, spectroscopic factors. At long distances a single partial wave dominates all the halos we have considered here. Therefore the long-distance part of the wave function can be defined unambiguously, and the Halo EFT wave function has only one component. In the shell-model representation it may well be that several single-particle states contribute, but it is clear that in a basis with the correct long-distance behavior of the halo wave function only one component is required to describe that part. Other components can---and will---enter at short distances~\cite{Capel:2010js}. But we do not consider that piece of the wave function in our treatment, and so spectroscopic factors are not needed. Instead, all the interior effects that are modeled by inclusion of many shell-model states in the wave function of the state $|\Psi(J^\pi) \rangle$ are transformed into short-distance operators, whose impact on low-energy observables is then organized according to the EFT power counting. 

These short-distance operators can also take account of the impact that deformation of the nuclear core has on some observables.
For example, comparing our calculation of ${}^{11}$Be with a shell-model treatment of the same nucleus, it 
is clear that one effect which is
subsumed into the NLO counterterm $L^{(1/2)}_{E1}$ is the transition of a 
neutron
from a $d_{5/2}$ to a $p_{3/2}$ orbital, with that neutron coupled to the
$2^+$ state of ${}^{10}$Be. This $2^+$ state is 3.4 MeV
above the ${}^{10}$Be ground state, so the dynamics associated with it takes 
place
at distances $1/M_{\rm core}$. Hence in our EFT it can only appear in 
short-distance operators
such as that multiplying $L^{(1/2)}_{E1}$ in Eq.~(\ref{eq:nonminimalEMdimers}). 
The computation of Ref.~\cite{Millener:1983zz} suggests that such a contribution
reduces the E1 matrix element by $\approx 30$\%, which is the anticipated size
of an NLO effect when the $M_{\rm halo}/M_{\rm core}$ expansion is employed in 
the ${}^{11}$Be system. We thus see that Halo EFT can use a point-like core, and avoid the traditional treatment in terms of a deformed rotator, as long
as the energy required to excite the corresponding rotational band is large compared to the energy of the halo system. Archaeology of Halo EFT's short-distance operators in terms of shell-model descriptions of the same system could be interesting, since the shell model has historically had much success describing the dynamics of nuclei at distances of order $\sim 1/M_{\rm core}$. 

A contrasting example, where low-lying excited states of the core (albeit not due to deformation) are included in the EFT is the treatment of ${}^7 {\rm Li}(n,\gamma){}^8 {\rm Li}$ in Ref.~\cite{Zhang:2013kja}. The first excited state of ${}^7$Li lies only 0.48 MeV above the ground state; the effects of this state cannot be replaced by contact operators for reactions that take place at this energy scale. Generally speaking, once energies comparable to its excitation energy are attained a state cannot be integrated out of the EFT's Fock space. Instead, these low-lying excitations of the core must be included explicitly in the calculation. Halo EFT is sufficiently flexible that this can be done if it will improve the convergence of the theory. 

The No-Core Shell Model (NCSM)  involves an exact diagonalization of a particular NN + 3N Hamiltonian, $\hat{H}$,  in the harmonic-oscillator basis~\cite{Navratil:2007we}. The choice of oscillator parameter predicates the convergence of this {\it ab initio} method to nuclei whose size is of order the oscillator length. Systems that are weakly bound compared to that length scale, and so have exponential tails that persist outside the parabolic potential, converge very slowly with basis size in the NCSM. This was recognized early in the development of the NCSM, and some early studies ameliorated this issue by matching the NCSM wave function to the asymptotic form~\cite{Navratil:2005xm,Navratil:2006tt}. The issue of incorrect asymptotics also initially precluded the treatment of scattering within the NCSM. However, over the last decade the NCSM has been successfully combined with Resonating-Group Method (RGM) ideas. The basis for diagonalization now includes the NCSM set of solutions of the $A$-body problem, together with clustered wave functions corresponding to a few low-lying two-body channels. If the clusters contain $a$ and $b$ particles respectively (with $A=a+b$) then these clustered wave functions are products of solutions of $\hat{H}$ in the $a$- and $b$-body spaces, with the wave function representing the relative degree of freedom between the two clusters initially unknown~\cite{Baroni:2013fe,Navratil:2016ycn}:
\begin{equation}
\fl
|\Psi (J^\pi) \rangle=\sum_\lambda |A \lambda J^\pi \rangle + \sum_{{\rm channels} \; ab,\nu} \frac{\gamma_\nu(r_{ab})}{r_{ab}} {\cal A}_\nu [|a \lambda_a J^\pi_a \rangle |b \lambda_b J^\pi_b \rangle]^{s} Y_{l}(\hat{r}_{ab})]^{J^\pi}.
\label{eq:NCSMC}
\end{equation}
In the first, standard NCSM, term, $\lambda$ indicates states of $A$ particles in the harmonic-oscillator basis (quantum numbers $N$, $L$, and spins)~\footnote{Note that, in contrast to the above discussion of the nuclear shell model, here the wave function is formulated in relative coordinates.}. The second term includes all relevant two-body clusterizations and has $\nu$ as a collective label that enumerates the states in the product representation of the two harmonic oscillator bases, coupled to appropriate spin, orbital angular momentum, total $J$, and isospin (not shown): $\nu=\{\lambda_a J^\pi_a; \lambda_b J^\pi_b;s,l\}$. ${\bf r}_{ab}$ is the relative co-ordinate between the centers-of-mass of the $a$ and $b$ clusters. Lastly, the anti-symmetrizer ${\cal A}_\nu$ ensures that the resulting wave function is anti-symmetric. (For a full discussion with more explicit notation, see Refs.~\cite{Baroni:2013fe,Navratil:2016ycn}.) 

{\it Ab initio} calculations that use the basis (\ref{eq:NCSMC})  are referred to as ``No-Core Shell Model with Continuum" (NCSMC). The basis is over-complete, but this is dealt with using an extension of the standard RGM procedure of constructing a norm kernel and thereby obtaining orthogonalized cluster-channel states~\cite{Navratil:2016ycn}. The division between long- and short-distance is not as straightforward as that the first, NCSM, term gives the short-distance behavior and the second, clustered, part gives the long-distance behavior. For one thing the norm kernel explicitly induces short-distance effects due to anti-symmetrization of nucleons between the two clusters. For another, the inclusion of clusters in the NCSMC basis can improve the convergence of the short-distance part of the wave function with basis size. Nevertheless, at long distances only the clustered piece of the wave function survives, and from it the scattering phase shifts between the $a$-body and $b$-body clusters can be obtained. The intimate connection between the scattering matrix and low-energy bound states that we've emphasized throughout this review then guarantees that halo states should be accessible within the NCSMC, too. And indeed, the clustered part of the NCSMC basis contains exactly the states needed to improve the convergence of the diagonalization for halos built on those clusters. In practice, an R-matrix technique is used to solve the NCSMC equations: beyond $r_0$ it is assumed that only the Coulomb potential acts between the two clusters, and the solution of $\hat{H}$ for $r < r_0$ is matched to Whittaker (or Coulomb) functions at that radius. The NCSMC (or its predecessor the NCSM + RGM) have now been successfully applied to  one-nucleon halo systems discussed in this review: ${}^5$He~\cite{Quaglioni:2008sm,Hupin:2013wsa,Navratil:2016ycn}, ${}^8$B~\cite{Navratil:2011sa}, and ${}^{11}$Be~\cite{Calci:2016dfb}. The NCSM + RGM has also recently been extended to three-body configurations, for example, ${}^4{\rm He} + n + n$~\cite{Quaglioni:2013kma,Romero-Redondo:2014fya}, and augmented with NCSM six-body eigenstates to produce both bound and scattering solutions for ${}^6$He~\cite{Romero-Redondo:2016qmc}. 

RGM-augmented NCSM wave functions of the form (\ref{eq:NCSMC}) are  not the only method that has been employed to circumvent the harmonic-oscillator basis' difficulties in describing halo nuclei. The so-called ``Gamow shell model" diagonalizes $\hat{H}$ on a basis with the correct asymptotics for shallow bound states, and has been employed to obtain the energies and widths of the low-energy p-wave resonances in the ${}^4$He + $n$ system~\cite{Papadimitriou:2013ix}. The Gamow-shell-model basis can also be combined with an RGM-type expansion \'a la Eq.~(\ref{eq:NCSMC}) to compute scattering phase shifts~\cite{Fossez:2015qxa,Jaganathen:2014mxa}. 

 Finally, we note that although we have dwelt on these issues in the context of the NCSM they are not exclusive to this {\it ab initio} technique. To take one other example, Green's Function Monte Carlo calculations of halo states have been slow to emerge in part because the method is predicated on a finite volume, and the calculation of halo states is therefore computationally (even more) challenging. GFMC calculations of the charge radius of ${}^6$He had to be tuned very carefully in order to ensure that this long-distance observable in the halo system was completely converged~\cite{Pieper:2007ax}. Halo EFT could perhaps be used to compute the finite-volume effects that complicate convergence in such cases~\cite{Konig:2017krd}. 

The ability of {\it ab initio} methods like the NCSMC to solve for structure and low-energy reactions in systems 
 like ${}^{11}$Be and ${}^6$He, that have been analyzed in Halo EFT, opens new possibilities for the EFT to take {\it ab initio} results as input, and so divorce itself from the need to fix the coefficients of contact operators from experiment. This provides consistency checks on the implementation of long-distance physics in the {\it ab initio} calculation, and insights into the EFT convergence pattern. It also means the EFT can be combined with the {\it ab initio} input to predict observables that remain challenging for direct computation in the {\it ab initio} approach.

The R-matrix employed to solve the NCSMC equations is an {\it ab initio} R-matrix, constructed from knowledge of the interior wave function. Since Halo EFT is agnostic about this portion of the nuclear state it perhaps has more in common with the phenomenological R-matrix, where, for scattering, the exterior of the wave function ($r > r_0$) is expressed in terms of phase shifts and asymptotic solutions (typically Whittaker or Coulomb functions) and the only thing known about the interior is the R-matrix that encodes the log derivatives and channel couplings at the boundary $r_{ab}=r_0$. Any resonances needed to describe scattering in the kinematic domain of interest are incorporated explicitly into the R-matrix fit---just as in Halo EFT resonances can be added as explicit degrees of freedom. In contrast to Halo EFT, the non-resonant part of the scattering is described through a combination of scattering due to a hard core of radius $r_0$
and ``background resonances" at higher energies. The effective-range expansion emerges from the R-matrix description for low scattering energies~\cite{Teichmann:1951}, but the parameters of the phenomenological R-matrix fit must be tuned to obtain specific values of the effective-range parameters. 

For capture reactions the distinction between interior and exterior contributions again makes the connection with EFT strong: for each resonance there is an ``external" ($r > r_0$) part of the capture that is computed explicitly using the asymptotic wave function of that resonance. The interior part of every resonance's capture matrix element is parameterized, and must be fit to capture data. This means that R-matrix predicts the rapidly-varying-with-energy pieces of the capture cross section (effects due to resonances, Coulomb and angular-momentum barriers, external direct capture) and fits the short-distance, more slowly varying, pieces to experiment. This is very similar to the EFT formulation of $n + p \rightarrow d + \gamma$, ${}^7{\rm Li}(n,\gamma)^8{\rm Li}$, $^7{\rm Be}(p,\gamma)^8{\rm B}$, etc. Indeed, it has been shown that---at least in the cases of $np$ scattering and $d + ^3{\rm H} \rightarrow n + {}^4{\rm He}$---Halo EFT is obtained in the limit of an R-matrix calculation with $r_0 \rightarrow 0$~\cite{Brown:2013zla,Hale:2013ama}. The extension of this result to radiative captures awaits proof, but seems plausible. All of this suggests that---at least at low energies---R-matrix analyses may be amenable to error estimates and uncertainty quantification of the type advocated for and carried out in EFT frameworks in Refs.~\cite{Furnstahl:2015rha,Melendez:2017phj,Zhang:2015ajn,Furnstahl:2014xsa,Wesolowski:2015fqa}.
Another place that Halo EFT could augment R-matrix is in the consistent treatment of three-body reactions. The phenomenological R-matrix has traditionally not been applied in such channels, and only recently have attempts been made at a phenomenological R-matrix formulation of, e.g., ${}^4 {\rm He} + n + n$~\cite{Brune:2015rqa}.

Lastly, Halo EFT also has much in common with potential models based on clusters. After all, many of the EFT Lagrangians employed here can be recast as Schr\"odinger-equation calculations with particular two-body (and three-body) potentials. The physics of Halo EFT is very much the same physics as was expressed in these potential models. However, because EFT is a minimal formulation of the halo physics at a given accuracy it can diagnose the correlations that drive the phenomenology of reactions and halo structure, and so help to explain why different potential models produce different results. For example, Rupak and Higa showed that differences in potential-model predictions for ${}^7$Li$(n,\gamma)$ were entirely due to those models having different values of the p-wave effective range $r_1$ in the channel(s) where ${}^8$Li binds: disparate predictions stemmed from variations in the bound-state ANC, and were not due to any sensitivity to the form of the potential~\cite{Rupak:2011nk}. Another key difference between EFT and potential models is the use of power counting to organize the contributions to the potentials, as well as the recognition that two-body potentials will not, in general, exhaust the physics of a system. Higher-body operators---three-body forces in multi-particle systems and two-cluster currents in electromagnetic processes---will always enter, and no amount of tuning of the two-body potential can ensure that all of them go to zero.  Once again, the systematic organization of the EFT is key, as it ensures that this set of higher-body operators does not become a plethora, but instead can be organized to make the calculation manageable. In few-nucleon systems the use of such organization---and the associated uncertainty quantification it provides---has brought renewed vigour to the study of few-nucleon systems in the past two decades. Similar possibilities await halo nuclei.

\section{Conclusion, sins of omission, and future paths} \label{sec:conclusion}
\subsection{Summary}
Effective field theory provides a robust method for the study of low-energy observables in halo nuclei. Halo EFT is based on a systematic expansion exploiting the hierarchy between the momentum scales associated with the halo ($M_{\rm halo}$) and the core ($M_{\rm core}$). In contrast to standard pionless EFT explicit fields are
introduced for tightly bound clusters of nucleons.
The uncertainty of Halo EFT calculations is estimated and improved by using the expansion in $M_{\rm halo}/M_{\rm core}$, which provides a systematic organization
for calculations of halo observables. Such calculations capture the physics that is essential to the long-distance properties of halo nuclei and clustered systems. Short-distance effects are not resolved explicitly but their
effect on long-distance observables is captured by local operators.
In particular, the EFT predicts that halo observables are connected by few-body universality: the idea that low-energy properties of weakly-bound few-body systems share properties that are independent of the details of the
interactions at short distances. 

In this review, we have discussed the Halo EFT approach for calculating structure and reaction observables in halo nuclei. In one- and two-neutron s-wave halos we built the Lagrangian and explained the formalism for calculating binding energies, matter radii, and Efimov states. 
The formalism was then extended to resonances and bound states in one- and two-neutron halos involving p-wave neutron-core interactions. By introducing external electromagnetic currents, we explained the EFT calculations of Coulomb dissociation, radiative capture, and the electric radii of halo nuclei. The Halo EFT formalism with electromagnetic interactions was then extended to proton halos,
whose structure depends on a delicate interplay of the strong and Coulomb
interactions between the proton and core. Lastly, we connected Halo EFT to other theoretical approaches for investigating weakly-bound halo nuclei and discussed possible synergies. 

\subsection{Sins of omission}
There are some frontier topics that have been treated within the Halo EFT framework but are not covered in detail by this review. Here we briefly discuss two examples: halo systems with strangeness and four-body systems.

\subsubsection{Halo systems with strangeness}
Halo structures also exist in a unique type of nuclear system produced by strangeness-exchange reactions, namely hypernuclei. In such systems, one or two of the constituent nucleons are replaced by hyperons (see, e.g., Ref.~\cite{Gal:2016boi} for a review). The lightest known hypernucleus is the hypertriton ${}^3_\Lambda$H, which consists of a proton, a neutron, and a $\Lambda$ hyperon. ${}^3_\Lambda$H is a non-Borromean system: the  $np$ pair can be bound, while the $n\Lambda$ and $p\Lambda$ pairs are unbound. The separation energy of the $\Lambda$ in $^3_\Lambda$H is $S_\Lambda=0.13(5)$ MeV~\cite{Juric:1973zq,Davis:1991zpu}, one order of magnitude smaller than the binding of the deuteron. As a consequence,
the very low energy properties of  ${}^3_\Lambda$H can be described in an
EFT with $\Lambda$ and deuteron degrees of freedom \cite{Ando:2013kba}. Above the
deuteron breakup threshold a three-body treatment is required.
Due to the limited data from $N\Lambda$ scattering experiments, and the fact they are 
available only at relatively high energy, the $N\Lambda$ interaction is not as well constrained as the NN interaction. A recent calculation based on SU(3) chiral  EFT potentials at next-to-leading order determined the $N\Lambda$ scattering lengths to be $-2.9$ fm in the singlet channel and $-1.6$ fm in the triplet channel~\cite{Haidenbauer:2013oca}. These values are consistent with phenomenological models~\cite{Rijken:1998yy,Haidenbauer:2005zh} and markedly larger than the range of the $N\Lambda$ interaction, which is driven by two-pion exchange. Therefore, we can, once again, use the $M_{\rm halo}/M_{\rm core}$ scale hierarchy and study weakly bound systems in hypernuclei using the EFT discussed in this review: the $N\Lambda$ potentials are represented by contact interactions. In this spirit, in Ref.~\cite{Hammer:2001ng} Hammer performed a leading-order EFT calculation in ${}^3_\Lambda$H, where an Efimov-like limit cycle behavior was found in the $np\Lambda$ three-body system. In addition, the low-energy phase shift parameters for the doublet $\Lambda$-$^2$H scattering was calculated through their universal correlations with the ${}^3_\Lambda$H binding energy. Hammer found $a_{\Lambda-{}^2{\rm H}}=16.8^{+4.4}_{-2.4}$ fm and $r_{\Lambda-{}^2{\rm H}}=2.3(3)$ fm, with the errors associated with the uncertainty in the ${}^3_\Lambda$H binding energy.  
Intrigued by experimental hints of a possibly bound $^3_\Lambda n$ state~\cite{Rappold:2013jta}, Ando {\it et al.} recently applied the same EFT to investigate the $nn\Lambda$ three-body system as a Borromean halo~\cite{Ando:2015fsa}. 

Light double-$\Lambda$ hypernuclei open other opportunities for studying halo/clustering phenomena. The investigation of such exotic systems is crucial for understanding the hyperon-hyperon interaction. The observation of $_{\Lambda\Lambda}^{\;\;\,6}$He in hybrid emulsion experiments determined the two $\Lambda$ separation energy, $B_{2\Lambda}\approx 7$ MeV, and the $\Lambda$-$\Lambda$ interaction energy of approximately $1$ MeV~\cite{Takahashi:2001nm,Ahn:2013poa}. These energy scales are considerably shallower than the excitation energy of the $^4$He nucleus. This suggests a halo structure $\Lambda$-$\Lambda$-$^4$He in $_{\Lambda\Lambda}^{\;\;\,6}$He. Using Halo EFT, Ando {\it et al.} investigated $_{\Lambda\Lambda}^{\;\;\,6}$He and predicted a universal correlation between $B_{2\Lambda}$ and the $\Lambda$-$\Lambda$ scattering length at leading-order accuracy~\cite{Ando:2014mqa}. Similar EFT arguments were applied to $_{\Lambda\Lambda}^{\;\;\,4}$H to describe its halo features in a $\Lambda$-$\Lambda$-$^2$H configuration~\cite{Ando:2013kba}. 

\subsubsection{Four-body systems}
Halo EFT studies have also been carried out for systems with more than three particles. These also manifest few-body universality in the large scattering length limit.
Combining an EFT formulation at leading order with Yakubovsky equations in the presence of a three-body coupling~\cite{Gloeckle:1993vr}, Platter {\it et al.} found that the discrete scaling symmetry (\ref{eq:dssym}) persists in systems of four identical bosons~\cite{Platter:2004he,Hammer:2006ct} and no four-body
force is required for renormalization at leading order in the
$M_{\rm halo}/M_{\rm core}$ expansion. They predicted that there are two four-body states (tetramers) universally tied to each three-body state (trimer), given that the momentum scales of these tetramers, $(m B_4)^{1/2}$, are much smaller than $M_{\rm core}$. This four-body EFT formalism was then extended to describe the $\alpha$-particle as a four-nucleon system, and the well-known empirical correlation between the binding energies of the triton and the $\alpha$-particle (the Tjon line) was explained by the absence of a four-body force at leading order~\cite{Platter:2004zs}.

The universal behavior of Efimov tetramers was also found by calculations in short-range potential models~\cite{vonStecher:2009,Blume:2000,Deltuva:2012xf}, and has been observed in experiments using ultracold atomic gases of $^{7}$Li~\cite{Pollack:2009}, $^{39}$K~\cite{Zaccanti:2009}, and $^{133}$Cs~\cite{Ferlaino:2009zz}. 
The long-predicted excited Efimov trimer of $^4$He atoms has been recently observed in a Coulomb explosion imaging measurement~\cite{Kunitski:2015qth}. This finding motivates further investigations searching for $^4$He Efimov tetramers. These studies on Efimov tetramers in cold atoms raise the question of whether there are analogous structures in halo nuclei.

However, it is difficult to find a three-neutron halo that is stable, due to Pauli-blocking effects in a configuration with three valence neutrons. In such a system at least one of the valence neutrons must interact with the core in a p-wave, which tends to increase the ground-state energy of the three-neutron halo compared to the corresponding two-neutron isotope. While such a system could have
universal properties it would not display an Efimov character and discrete
scale invariance.
However, neutron halos are just one example of a clustered few-body system that exhibits large-scattering-length physics. The first excited state of the $\alpha$ particle is a $0^+$ state, and is thought to be a breathing mode of the ground state~\cite{Werntz:1966,Hiyama:2004nf,Bacca:2014qva}. It
is a possible candidate for a four-nucleon Efimov resonance
and lies only 0.4 MeV above the $^3$H$+p$ threshold. However, the
structure of this state is complicated by the presence of the
long-range Coulomb interaction between protons since the Coulomb
momentum is comparable to the momentum scale of this excited state.
The excited $0^+$ state in $^{10}$Be, whose binding energy is only $1.2$ MeV below the $^6$He-$\alpha$ threshold, may also be thought of as a four-body Efimov state of the $\alpha \alpha nn$ system. Its structure is complicated not only by nonperturbative $\alpha$-$\alpha$ Coulomb interactions, but also by the presence of the p-wave resonances that dominate in the $n$-$\alpha$ subsystem at low energies. The possibility to realize---even if only approximately---aspects of Efimov physics in these and other halo four-body systems remains an important subject for future study. 

\subsection{Future paths}
The EFT formalism can be extended to new directions to further enhance understanding of halo nuclei. 
For example, a number of
radiative capture processes whose physics parallels
$^7\mathrm{Be}(p,\gamma)^8\mathrm{B}$ are important in astrophysics.
Halo EFT should be applicable to many of them. 

In nuclear reactions with exotic beams, the interplay between nuclear forces and electromagnetic interactions determines the relevant reaction mechanisms at different incident beam energies. Coulomb dissociation requires either a very low incident energy or a very forward-angle fragmentation, so that the halo projectile does not penetrate the Coulomb barrier of the high-charge nuclear target.
With a light nuclear target, the reaction mechanism is dominated by nuclear interactions, provided the momentum transferred is at least of order the Coulomb momentum between the target and projectile. Such reactions are realized, in elastic scattering (e.g., $(p,p)$~\cite{Alkhazov:1997zz,Lagoyannis:2000te,Neumaier:2002eay}), nucleon knockout reactions (e.g., $(p,pn)$~\cite{Kondo:2010zza,Aksyutina:2013rla}), and nucleon transfer reactions (e.g., $(p,d)$~\cite{Sanetullaev:2016oiu}, $(p,t)$~\cite{Tanihata:2008vw}, and $(d,p)$~\cite{Kanungo:2010zza,Schmitt:2012bt}).

Theoretical analyses of ``direct'' reaction mechanisms have been performed with the development of many independent approaches, including the distorted-wave Born approximation (DWBA), the dynamical eikonal approximation (DEA), the continuum discretized coupled channels (CDCC) method, and the Faddeev-Yakubovsky scattering calculation (see {\it e.g.} Refs.~\cite{Casal:2016fnw,Baye:2005ib,Goldstein:2006sv,Moro:2005rw,Upadhyay:2011ta,Deltuva:2015iad,Deltuva:2016his} for applications in halo nuclei).
If the momentum transfer in the reaction is near the halo scale $M_{\rm halo}$, it should be feasible to extend the existing Halo EFT calculations to describe the corresponding reaction mechanism. Using the $M_{\rm halo}/M_{\rm core}$ expansion, one can systematically construct the hierarchy of different interaction channels that contribute to the reaction. This controlled expansion may reduce or clarify the optical-potential-model dependence in traditional reaction theories. The Halo EFT is based on a Faddeev-Yakubovsky scattering calculation and uses consistent nucleon-core interactions within the halo and among projectile and target fragments. Therefore it can, in principle, provide a unified description of structure and reaction observables.

\section*{Acknowledgments}

We thank our collaborators for fruitful investigations and informative conversations on the topics discussed here.
We are especially grateful to Bijaya Acharya for providing the updated calculations that appear in Fig.~\ref{pic:22C-contour} and to Jared Vanasse for carefully reading the manuscript and supplying the cross sections used in Fig.~\ref{fig:universalE1}. We also thank Pierre Capel and Takashi Nakamura for supplying data and information relevant to Fig.~\ref{fig:universalE1CD}. We appreciated useful comments on this manuscript made by Carl Brune and Sofia Quaglioni. We thank the ECT* and the organizers of the workshops ``Three-body systems in reactions with rare isotopes " and ``Open quantum systems: from atomic nuclei to ultracold atoms and quantum optics", as well as the organizers of the KITP program ``Universality in few-body systems": all three constituted an excellent environment in which to work on this article. DRP's work at the KITP was supported in part by the National Science Foundation under Grant No. NSF PHY11-25915. This research was also supported
by the U.S. Department of Energy under 
grant DE-FG02-93ER40756,  by the German Federal Ministry of Education
and Research under contract 05P15RDFN1, 
and by the Deutsche Forschungsgemeinschaft (SFB 1245).

\section*{References}



\end{document}